\documentclass{ws-ijmpa}
\usepackage[super,compress]{cite}
\usepackage{graphicx}
\usepackage{braket}
\usepackage{bm}
\usepackage{comment}
\usepackage{here}
\usepackage{multirow}
\usepackage{color}

\def\Journal#1#2#3#4{{#1} {\bf #2}, #3 (#4)}

\def\CPC{Chin. Phys. C}

\def\EPJC{Eur. Phys. J. C}

\def\EPL{EPL}

\def\IJMPA{Int. J. Mod. Phys. A}

\def\JCAP{J. Cosmol. Astropart. Phys.}
\def\JHEP{J. High Energy Phys.}

\def\JPG{J. Phys. G} 
\def\JPGNP{J. Phys. G: Nucl. Part. Phys.} 
 
\def\MPLA{Mod. Phys. Lett. A}

\def\NJP{New. J. Phys.}
\def\NPB{Nucl. Phys. B}

\def\PLB{{Phys. Lett.} B}

\def\PRL{Phys. Rev. Lett.}
\def\PRD{Phys. Rev. D}

\def\PTP{Prog. Theor. Phys.}
\def\PTEP{Prog. Theor. Exp. Phys.}
\def\RMP{Rev. Mod. Phys.}

\def\RMF{Revista Mexicana de Fisica.}

\begin{document}
\markboth{Y. Chen, Y. Hyodo, and T. Kitabayashi}{Modified trimaximal mixing for solar and reactor neutrino mixing angles}

%
\catchline{}{}{}{}{}
%


\title{Modified trimaximal mixing for solar and reactor neutrino mixing angles}

\author{Yulin Chen}

\address{Graduate School of Science, Tokai University,\\
4-1-1 Kitakaname, Hiratsuka, Kanagawa 259-1292, Japan}

\author{Yuta Hyodo}

\address{Graduate School of Science and Technology, Tokai University,\\
4-1-1 Kitakaname, Hiratsuka, Kanagawa 259-1292, Japan}

\address{Micro/Nano Technology Center, Tokai University,
4-1-1 Kitakaname, Hiratsuka, Kanagawa 259-1292, Japan}

\author{Teruyuki Kitabayashi\footnote{Corresponding author}}

\address{Department of Physics, Tokai University,\\
4-1-1 Kitakaname, Hiratsuka, Kanagawa 259-1292, Japan\\
teruyuki@tokai-u.jp}

\maketitle

\begin{history}
\received{Day Month Year}
\revised{Day Month Year}
\end{history}

\begin{abstract}
The trimaximal mixing scheme is one of the widely studied neutrino mixing schemes. However, the predicted solar neutrino mixing angle $\theta_{12}$ and the reactor neutrino mixing angle $\theta_{13}$ cannot simultaneously realize their best-fit values. To address this issue, a minimal modification to the trimaximal mixing scheme is proposed. The modified trimaximal mixing scheme can simultaneously predict the best-fit values of $\theta_{12}$ and $\theta_{13}$ using an additional real parameter introduced via the modification.
\end{abstract}

\ccode{PACS numbers:14.60.Pq}


\section{Introduction\label{section:introduction}}

Numerous neutrino mixing matrices and textures related to flavor mass matrix have been proposed previously, including tri-bi maximal mixing (TBM) \cite{Harrison2002PLB,Xing2002PLB,Harrison2002PLB2,Kitabayashi2007PRD}, texture zeros \cite{Berger2001PRD,Frampton2002PLB,Xing2002PLB530,Xing2002PLB539,Kageyama2002PLB,Xing2004PRD,Grimus2004EPJC,Low2004PRD,Low2005PRD,Grimus2005JPG,Dev2007PRD,Xing2009PLB,Fritzsch2011JHEP,Kumar2011PRD,Dev2011PLB,Araki2012JHEP,Ludle2012NPB,Lashin2012PRD,Deepthi2012EPJC,Meloni2013NPB,Meloni2014PRD,Dev2014PRD,Felipe2014NPB,Ludl2014JHEP,Cebola2015PRD,Gautam2015PRD,Dev2015EPJC,Kitabayashi2016PRD1,Zhou2016CPC,Singh2016PTEP,Bora2017PRD,Barreiros2018PRD,Kitabayashi2018PRD,Barreiros2019JHEP,Capozzi2020PRD,Singh2020EPL,Barreiros2020,Kitabayashi2020PRD,Kitabayashi2017IJMPA,Kitabayashi2017IJMPA2,Kitabayashi2019IJMPA}, $\mu-\tau$ symmetric texture \cite{Fukuyama1997,Lam2001PLB,Ma2001PRL,Balaji2001PLB,Koide2002PRD,Kitabayashi2003PRD,Koide2004PRD,Aizawa2004PRD,Ghosal2004MPLA,Mohapatra2005PRD,Koide2005PLB,Kitabayashi2005PLB,Haba206PRD,Xing2006PLB,Ahn2006PRD,Joshipura2008EPJC,Gomez-Izquierdo2010PRD,He2001PRD,He2012PRD,Gomez-Izquierdo2017EPJC,Fukuyama2017PTEP,Kitabayashi2016IJMPA,Kitabayashi2016PRD,Bao2021arXiv,Garces2018JHEP,JuanCarlos2019JHEP,JuanCarlos2008PRD,Ge2010JCAP,He2015PLB,JuanCarlos2017IJMPA,Spinrath2012NPB,JuanCarlos2023RMF}, magic texture \cite{Harrison2004PLB,Lam2006PLB,Gautam2016PRD,Hyodo2020IJMPA,Yang2022PTEP,Channey2019JPGNP,Verma2020JPGNP,Hyodo2021PTEP, Verma2022arXiv}, and textures and mixings under discrete symmetries, e.g., $A_n$ and $S_n$\cite{Altarelli2010PMP}. 

The trimaximal mixing scheme is a widely studied neutrino mixing scheme  \cite{Lam2006PRD,He2007PLB,He2011PRD,Albright2009EPJC,Albright2010EPJC,Bjorken2006PRD,Kumar2010PRD,Kumar2013PRD,Yang2022NPB,Grimus2008JHEP,Albright2009EPJC,Grimus2010PLB,Kumar2010PRD,Morisi2011PRD,Luhn2013NPB,Rodejohann2017PRD,Gautam2018PRD,Zhao2022PRD}. There are versions of the trimaximal texture. In this paper, we focus on the so-called TM1 and TM2 mixings. The predicted values of neutrino mixing angles using TM1 and TM2 are consistent with the observations. However, the predicted solar neutrino mixing angle $\theta_{12}$ and reactor neutrino mixing angle $\theta_{13}$ cannot simultaneously realize their best-fit values. In the case of TM2, the predicted value of $\theta_{12}$ approaches the upper limit of the $3\sigma$ allowed region. Hence, the TM2 mixing scheme may soon be excluded from neutrino oscillation experiments.

Although simultaneous reproducibility of $\theta_{12}$ and $\theta_{13}$ cannot be realized using the TM1 and TM2 mixing schemes, these schemes still garner research attention. The most interesting feature of TM1 and TM2 is related to the $Z_2$ symmetry. The flavor neutrino mass matrices in the TM1 and TM2 mixing schemes are invariant under the exact $Z_2$ symmetry \cite{Yang2022PTEP}. Thus, modifying the TM1 and TM2 mixing schemes to realize the simultaneous reproducibility of $\theta_{12}$ and $\theta_{13}$ using the exact $Z_2$ symmetry is an important research topic neutrino physics. Although, we have attempted to construct new modified TM1 and TM2 mixing schemes using the exact $Z_2$ symmetry, we have not yet succeeded. However, constructing a modified version of TM1 and TM2 mixing schemes that do not adhere to the $Z_2$ symmetry is also important.

In this study, modified TM1 and TM2 mixing schemes are constructed using the following strategy:
\begin{itemize}
\item Improving the simultaneous reproducibility of $\theta_{12}$ and $\theta_{13}$ is the highest priority. 
\item $Z_2$ symmetry breaking is allowed.
\item We modify the original TM1 mixing matrix $U_1$ and TM2 mixing matrix $U_2$ by introducing a matrix $A_i$, resulting in the modified mixing matrix $\tilde{U}_i = A_iU_i$ $(i=1,2)$. Usually, the modified mixing matrix $\tilde{U}_i$ is nonunitary \cite{Blennow2017JHEP, Escrihuela2015PRD, Escrihuela2017NJP, Forero2021PRD, Gariazzo2023JCAP}. However, because the original TM1 and TM2 mixing schemes are constructed based on unitary arguments \cite{Kumar2010PRD}, the unitarity of the modified TM1 and TM2 mixing matrices needs to be maintained.
\item Minimize the number of parameters in the matrix $A_i$ to realize the minimal modification of TM1 and TM2.
\end{itemize}

We successfully derived the modified TM1 and TM2 mixing matrices. These modified matrices simultaneously predict the best-fit values of $\theta_{12}$ and  $\theta_{13}$ using the additional real parameter. 

The remaining sections of this study are structured as follows. Section \ref{sec:TM} provides an overview of model-independent useful formulas and observed data related to neutrinos. Additionally, this section highlights the key features of the TM1 and TM2 mixing schemes. Section \ref{modification} outlines the methodology employed for modifying the TM1 and TM2 mixings. The successful modification of TM1 and TM2 is presented in Sections \ref{sec:MTM1} and \ref{sec:MTM2}, respectively. Lastly, Section \ref{sec:summary} presents a summary of the obtained results and their implications.

\section{Trimaximal mixing\label{sec:TM}}
\subsection{Model-independent formulas and observations}
The neutrino mixings are parametrized based on the Pontecorvo-Maki-Nakagawa-Sakata (PMNS) mixing matrix
\cite{Pontecorvo1957,Pontecorvo1958,Maki1962PTP,PDG}

\begin{eqnarray}
U=\left(
\begin{matrix}
U_{e1} & U_{e2} & U_{e3}  \\
U_{\mu 1} & U_{\mu 2} & U_{\mu 3}  \\
U_{\tau 1} & U_{\tau 2} & U_{\tau 3}  \\
\end{matrix}
\right), 
\label{Eq:U}
\end{eqnarray}
where 
\begin{eqnarray}
U_{e1} &=& c_{12}c_{13}, \quad U_{e 2} = s_{12}c_{13}, \quad U_{e 3} = s_{13} e^{-i\delta},  \label{Eq:UPMNS_elements} \\
U_{\mu 1} &=&- s_{12}c_{23} - c_{12}s_{23}s_{13} e^{i\delta}, \nonumber \\
U_{\mu 2} &=&  c_{12}c_{23} - s_{12}s_{23}s_{13}e^{i\delta}, \quad U_{\mu 3} = s_{23}c_{13}, \nonumber \\
U_{\tau 1} &=& s_{12}s_{23} - c_{12}c_{23}s_{13}e^{i\delta},\nonumber \\ 
U_{\tau 2} &=& - c_{12}s_{23} - s_{12}c_{23}s_{13}e^{i\delta}, \quad U_{\tau 3} = c_{23}c_{13}.\nonumber 
\end{eqnarray}
In the abbreviations $s_{ij}=\sin\theta_{ij}$ and $c_{ij}=\cos\theta_{ij}$, $\theta_{ij} = \{\theta_{12}, \theta_{23}, \theta_{13}\}$ denote neutrino mixing angles. $\delta$ denotes the Dirac CP phase. 

The sines and cosines of the three mixing angles can be obtained as follows:
\begin{eqnarray}
s_{12}^2 = \frac{|U_{e2}|^2}{1-|U_{e3}|^2},  \
s_{23}^2 = \frac{|U_{\mu 3}|^2}{1-|U_{e3}|^2},   \
s_{13}^2 = |U_{e3}|^2.  \label{Eq:s12s_s23s_s13s} 
\end{eqnarray}
The Jarlskog invariant $J_{\rm CP}$ \cite{Jarlskog1985PRL} is derived based on the Dirac CP phase $\delta$ and the elements of the PMNS matrix. In standard parametrization, the  Jarlskog invariant  can be expressed as follows:
\begin{eqnarray}
\label{Eq:JCP}
J_{\rm CP} = {\rm Im} (U_{e1} U_{e2}^* U_{\mu 1}^* U_{\mu 2}) =  s_{12}s_{23}s_{13}c_{12}c_{23}c_{13}^2 \sin\delta.  
\end{eqnarray}

A global analysis of the current data obtained from neutrino oscillation experiments yields the following best-fit values of the squared mass differences $\Delta m_{ij}^2 = m^2_i - m^2_j$ related to the neutrino mass eigenstates $\{ m_1, m_2, m_3 \}$ and the mixing angles for the normal mass ordering (NO), $m_1 < m_2 < m_3$, as  \cite{Esteban2020JHEP}
\begin{eqnarray} 
\frac{\Delta m_{21}^2}{10^{-5} {\rm ~ eV^2}} &=& 7.41^{+0.21}_{-0.20} \ (6.82 \rightarrow 8.03), \nonumber \\ 
\frac{\Delta m_{31}^2}{10^{-3} {\rm ~ eV^2}} &=& 2.507^{+0.026}_{-0.027} \ (2.427 \rightarrow 2.590), \nonumber \\ 
s^2_{12} &=& 0.303^{+0.012}_{-0.012} \ (0.270 \rightarrow 0.341), \nonumber \\
s^2_{23} &=& 0.451^{+0.019}_{-0.016} \ (0.408 \rightarrow 0.603), \nonumber\\
s^2_{13} &=& 0.02225^{+0.00056}_{-0.00059} \ (0.02052 \rightarrow 0.02398),\nonumber \\
\delta/^\circ &=& 232^{+36}_{-26} \ (144 \rightarrow 350),
\label{Eq:NuFit_NO}
\end{eqnarray}
where the $\pm$ represents the $1 \sigma$ region and the parentheses denote the $3 \sigma$ region. For inverted mass ordering (IO), $m_3 < m_1 \simeq m_2$, 
\begin{eqnarray} 
\frac{\Delta m_{21}^2}{10^{-5} {\rm ~ eV^2}} &=& 7.41^{+0.21}_{-0.20} \ (6.82 \rightarrow 8.03), \nonumber \\ 
\frac{\Delta m_{32}^2}{10^{-3} {\rm ~ eV^2}} &=& -2.486^{+0.025}_{-0.028} \ (-2.570 \rightarrow -2.406), \nonumber \\ 
s^2_{12} &=& 0.303^{+0.012}_{-0.011} \ (0.270 \rightarrow 0.341), \nonumber \\
s^2_{23} &=& 0.569^{+0.016}_{-0.021} \ (0.412 \rightarrow 0.613), \nonumber \\
s^2_{13} &=& 0.02223^{+0.00058}_{-0.00058} \ (0.02048 \rightarrow 0.02416), \nonumber \\
\delta/^\circ &=& 276^{+22}_{-29} \ (194 \rightarrow 344).
\label{Eq:NuFit_IO}
\end{eqnarray}

Herein, the observed values of $\theta_{12}$ are the same in the NO and IO cases. Furthermore, the observed values of $\theta_{13}$ are also nearly identical in the NO and IO cases. Therefore, we compare the model predictions of $\theta_{12}$ and $\theta_{13}$ based on these observed values in the NO case. 

\subsection{Trimaximal mixing}
In this subsection, we provide a brief review of trimaximal mixing. The trimaximal mixing scheme is one of the modifications of the TBM. Since the predicted reactor mixing angle is not consistent with the observation in the TBM scheme, some modification of the TBM is required. It is a simple way to modify the TBM that the first or second column of the mixing matrix in the TBM is unchanged and the other two columns are modified. Such modification of the mixing matrix of the TBM can be realized by multiplying it from the right-hand side by a complex (2,3) or (1,3) rotation matrix. These types of modifications lead the so-called the first or second trimaximal mixing (TM1 or TM2). 

Moreover, there is third trimaximal mixing (TM3). In the texture of TM3, the third column of the mixing matrix in the TBM is unchanged and the other two columns are modified (multiplying TBM mixing matrix from the right-hand side by a complex (1,2) rotation matrix). It is known that predicted reactor mixing angle to be zero for TM3. Therefore, the TM3 mixing is exclusion from neutrino oscillation experiments.

The mixing matrix of TM1 is given as follows:\footnote{Although the matrix $U_1$ does not have a ``tri" column or row, we refer to  $U_1$ as TM1 according to literature such as  Ref. \cite{Zhao2022PRD}. }
\begin{eqnarray}
U_1=\left(
\begin{matrix}
\sqrt{\frac{2}{3}} & \frac{\cos\theta}{\sqrt{3}} & \frac{\sin\theta}{\sqrt{3}}  \\
-\frac{1}{\sqrt{6}}  & \frac{\cos\theta}{\sqrt{3}} - \frac{e^{i\phi} \sin\theta}{\sqrt{2}}  & \frac{\sin\theta}{\sqrt{3}} + \frac{e^{i\phi} \cos\theta}{\sqrt{2}} \\
-\frac{1}{\sqrt{6}} & \frac{\cos\theta}{\sqrt{3}} + \frac{e^{i\phi} \sin\theta}{\sqrt{2}}  & \frac{\sin\theta}{\sqrt{3}} - \frac{e^{i\phi} \cos\theta}{\sqrt{2}}  \\
\end{matrix}
\right),
\label{Eq:UTM1}
\end{eqnarray}
yields the following mixing angles and the Dirac CP phase
\begin{eqnarray}
s^2_{12} &=& 1-\frac{2}{3-\sin^2\theta},  \label{Eq:s12sUTM1}\\
s^2_{23} &=& \frac{1}{2}\left(1+\frac{\sqrt{6}\sin 2\theta \cos\phi}{3-\sin^2\theta}  \right),  \label{Eq:s23sUTM1}\\
s^2_{13} &=& \frac{1}{3}\sin^2\theta,  \label{Eq:s13sUTM1}\\
\tan \delta &=&  \frac{5+ \cos2\theta}{1+5\cos2\theta}\tan\phi.  \label{Eq:deltaUTM1}
\end{eqnarray}
In the term of $\theta_{13}$, we obtain
\begin{eqnarray}
s^2_{12} &=& \frac{1-3s^2_{13}}{3(1-s^2_{13})},  \label{Eq:s12ss13sUTM1} \\
s^2_{23} &=& \frac{1}{2}\left(1+\frac{6\sqrt{2(1-3s^2_{13})}s_{13} \cos\phi}{3(1-s^2_{13}) }  \right), \label{Eq:s23ss13sUTM1} \\
\tan \delta  &=& \frac{1-s^2_{13}}{1-5s^2_{13}}\tan\phi.  \label{Eq:tandeltaUTM1}
\end{eqnarray}

For TM2 mixing, the mixing matrix
\begin{eqnarray}
U_2=\left(
\begin{matrix}
\sqrt{\frac{2}{3}}\cos\theta &  \frac{1}{\sqrt{3}} & \sqrt{\frac{2}{3}}\sin\theta  \\
-\frac{\cos\theta}{\sqrt{6}} + \frac{e^{-i\phi}\sin\theta}{\sqrt{2}}  & \frac{1}{\sqrt{3}}  & -\frac{\sin\theta}{\sqrt{6}} - \frac{e^{-i\phi}\cos\theta}{\sqrt{2}} \\
-\frac{\cos\theta}{\sqrt{6}} - \frac{e^{-i\phi}\sin\theta}{\sqrt{2}}  & \frac{1}{\sqrt{3}}  & -\frac{\sin\theta}{\sqrt{6}}  + \frac{e^{-i\phi}\cos\theta}{\sqrt{2}} \\
\end{matrix}
\right),\nonumber \\
\label{Eq:UTM2}
\end{eqnarray}
yields the following mixing angles and the Dirac CP phase
\begin{eqnarray}
s^2_{12} &=&  \frac{1}{3-2\sin^2\theta},  \label{Eq:s12sUTM2}\\
s^2_{23} &=& \frac{1}{2}\left( 1+\frac{\sqrt{3}\sin2\theta \cos\phi}{3-2\sin^2\theta} \right),  \label{Eq:s23sUTM2}\\
s^2_{13} &=& \frac{2}{3}\sin^2\theta,  \label{Eq:s13sUTM2}\\
\tan \delta &=& \frac{2+ \cos2\theta}{1+2\cos2\theta}\tan\phi,  \label{Eq:deltaUTM2}
\end{eqnarray}
as well as
\begin{eqnarray}
s^2_{12} &=& \frac{1}{3(1-s^2_{13})},  \label{Eq:s12ss13sUTM2}  \\
s^2_{23} &=& \frac{1}{2}\left(1+\frac{3\sqrt{2-3s^2_{13}}s_{13} \cos\phi}{3(1-s^2_{13} )}  \right),  \label{Eq:s23ss13sUTM2} \\
\tan \delta  &=& \frac{1-s^2_{13}}{1-2s^2_{13}}\tan\phi.  \label{Eq:tandeltaUTM2}
\end{eqnarray}
%

\begin{figure}[t]
\begin{center}
\includegraphics[scale=1.0]{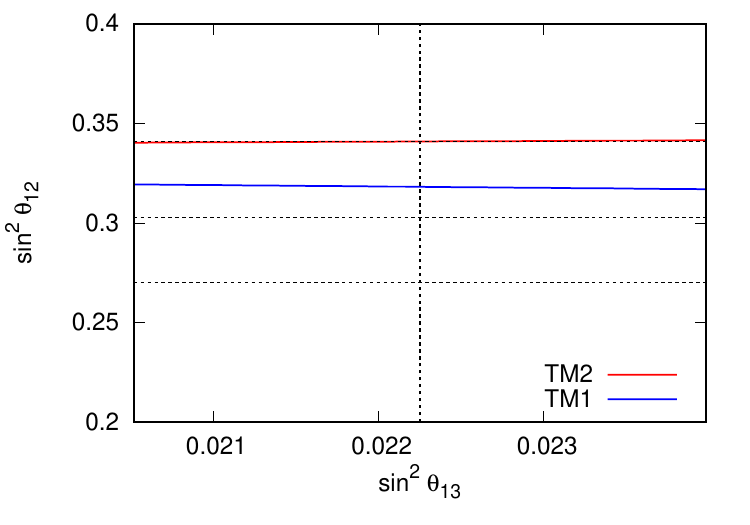}
\caption{ The predicted values of $\theta_{12}$ and $\theta_{13}$ using TM1 and TM2 mixing schemes are depicted using blue and red colors, respectively. The upper, middle, and lower horizontal dotted lines represent the $3\sigma$ upper limit, the best-fit value, and the $3\sigma$ lower limit of $\theta_{12}$ in NO case, respectively. The vertical dotted line represents the best-fit value of $\theta_{13}$ in the NO case. }
\label{Fig:TM_13_12} 
\end{center}
\end{figure}

\begin{figure}[t]
\begin{center}
\includegraphics[scale=1.0]{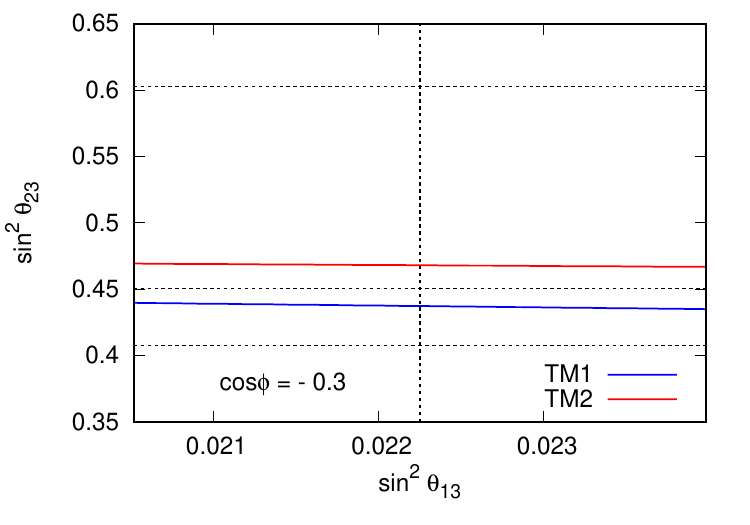}
\includegraphics[scale=1.0]{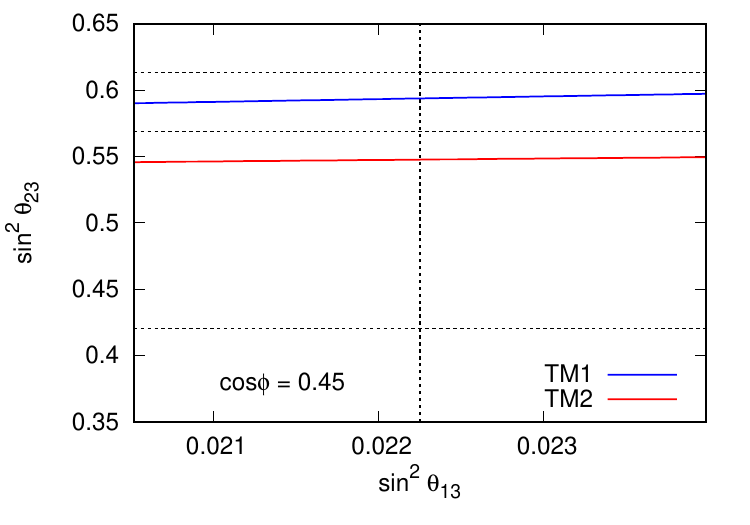}
\caption{Predicted values of $\theta_{13}$ and $\theta_{23}$ using the TM1 (blue) and TM2 (red) mixing schemes with respect to $\cos\phi=-0.3$ (upper panel) and $\cos\phi=0.45$ (lower panel) are depicted. The upper, middle, and lower horizontal dotted lines in each panel represent the $3\sigma$ upper limit, best-fit value, and $3\sigma$ lower limit of $\theta_{23}$ in the NO case for the upper panel and in the IO case for the lower panel. The vertical dotted line in both panels corresponds to the same value as depicted in Fig. \ref{Fig:TM_13_12}.
}
\label{Fig:TM_13_23} 
\end{center}
\end{figure}

Figure \ref{Fig:TM_13_12} illustrates the predicted values of $\theta_{12}$ and $\theta_{13}$ using the TM1 (blue) and TM2 (red) mixing schemes. The upper, middle, and lower horizontal dotted lines represent the $3\sigma$ upper limit, best-fit value, and $3\sigma$ lower limit of $\theta_{12}$ in the NO case, respectively. The vertical dotted line denotes the best-fit value of $\theta_{13}$ in the NO case. Neither TM1 nor TM2 can simultaneously reproduce the best-fit values of $\theta_{12}$ and $\theta_{13}$. In the case of TM2, the predicted value of $\theta_{12}$ approaches the upper limit of the $3\sigma$ allowed region. 

Figure \ref{Fig:TM_13_23} depicts the predicted values of $\theta_{13}$ and $\theta_{23}$ using the TM1 and TM2 mixing schemes with respect to $\cos\phi=-0.3$ (upper panel) and $\cos\phi=0.45$ (lower panel). The upper, middle, and lower horizontal dotted lines in each panel represent the $3\sigma$ upper limit, best-fit value, and $3\sigma$ lower limit of $\theta_{23}$ in the NO case for the upper panel and in the IO case for the lower panel. The vertical dotted line in both panels corresponds to the same value as depicted in Fig. \ref{Fig:TM_13_12}. Figure \ref{Fig:TM_13_23} suggests that TM1 and TM2 can simultaneously reproduce the best-fit values of $\theta_{23}$ and $\theta_{13}$ by choosing an appropriate value of $\cos\phi$.

Moreover, the predicted value of the Dirac CP phase $\delta$ using TM1 and TM2 is based on $\phi$. From Eq. (\ref{Eq:tandeltaUTM1}) and Eq. (\ref{Eq:tandeltaUTM2}), we obtain the following expression.
\begin{eqnarray}
\tan\delta \sim 
\begin{cases}
-6.9\tan\phi & (\rm TM1) \\
1.4\tan\phi  & (\rm TM2) \\ 
\end{cases}.
\end{eqnarray}
%

\subsection{$Z_2$ symmetry}
The flavor neutrino mass matrix with TM2 mixing $U_1$,
\begin{eqnarray}
M_1= U_1^\ast D U_1^\dag,
\end{eqnarray}
where $D = {\rm diag.}(m_1,m_2,m_3)$, satisfies the $Z_2$ symmetry
\begin{eqnarray}
S_1 M_1 S_1^{\rm T} = M_1,
\end{eqnarray}
where
\begin{eqnarray}
S_1 = \frac{1}{3}\left(
\begin{matrix}
-1 & 2 & 2  \\
2 & 2  &-1\\
2 & -1  & 2  \\
\end{matrix}
\right),
\end{eqnarray}
and $S_1^2=S_1S_1^{\rm T}=S_1^{\rm T}S_1=1$. 

Similarly, the flavor neutrino mass matrix in the TM2 mixing scheme
\begin{eqnarray}
M_2= U_2^\ast D U_2^\dag,
\end{eqnarray}
satisfies the $Z_2$ symmetry
\begin{eqnarray}
S_2 M_2 S_2^{\rm T} = M_2,
\end{eqnarray}
where
\begin{eqnarray}
S_2 = \frac{1}{3}\left(
\begin{matrix}
1 & -2 & -2  \\
-2 & 1  &-2\\
-2 & -2  & 1  \\
\end{matrix}
\right),
\end{eqnarray}
and $S_2^2=S_2S_2^{\rm T}=S_2^{\rm T}S_2=1$.

\section{Modification methods\label{modification}}
We modify the mixing matrix $U=\{U_1,U_2\}$ by introducing a matrix $A$ as follows:
\begin{eqnarray}
\tilde{U} = AU, 
\label{Eq:UMTM}
\end{eqnarray}
where 
\begin{eqnarray}
A=\left(
\begin{matrix}
\alpha_{11} & \epsilon_{12} & \epsilon_{13}  \\
\epsilon_{21} & \alpha_{22}   & \epsilon_{23}  \\
\epsilon_{31}& \epsilon_{32}  & \alpha_{33}  \\
\end{matrix}
\right),
\label{Eq:A}
\end{eqnarray}
with the conditions $|\alpha_{ii}| \le 1$ and $| \alpha_{ii} | \gg |\epsilon_{ij}| $. In the case of $\alpha_{ij} = 1$ and $\epsilon_{ij}=0$, we obtain the identity $A={\rm diag.}(1,1,1)$ and $\tilde{U} = U$. 

In order to satisfy the unitarity condition for the modified mixing matrix $\tilde{U}$, it is necessary for the matrix $A$ to be unitary: 
\begin{eqnarray}
AA^\dag = A^\dag A=1.
\end{eqnarray}

To achieve modification with the minimum number of parameters, we assume that the matrix $A$ is real. In addition, we consider scenarios where some of the elements $\epsilon_{ij}$ are set to zero. Having a greater number of zero elements in the matrix is advantageous as it reduces the number of parameters required. However, it should be noted that the case with all six zero elements is not allowed, as it would result in all $\epsilon_{ij}$ being zero, limiting the ability to modify the matrix.

Is it possible to have five zeros in the matrix $A$? For example, if $\epsilon_{13} =\epsilon_{21} = \epsilon_{23} = \epsilon_{31}=\epsilon_{32}=0$ and only $\epsilon_{12}$ is non-zero, then the matrix $A$ is
\begin{eqnarray}
A_{12} &=& \left(
\begin{matrix}
\alpha_{11} & \epsilon_{12}& 0  \\
0& \alpha_{22}   & 0 \\
0& 0  & \alpha_{33}  \\
\end{matrix}
\right),
\label{Eq:Azero12}
\end{eqnarray}
where the subscript of $A$ indicates the position of the non-zero $\epsilon_{ij}$. For $A_{12}$ to be unitary, $\epsilon_{12}=0$ is required. This is inconsistent with the five-zero condition. Thus, $A_{12}$ should be excluded. Similarly, all five-zero matrices are excluded because they cannot simultaneously satisfy the unitarity and the five-zero conditions.

Among the possible configurations, there are six matrices that satisfy both the unitarity condition and the four-zero condition:
\begin{eqnarray}
A_{12,21}^{(1)} &=& \left(
\begin{matrix}
\alpha & \epsilon & 0  \\
-\epsilon & \alpha  &0\\
0 & 0  & 1  \\
\end{matrix}
\right)
= \left(
\begin{matrix}
\alpha & \sqrt{1-\alpha^2} & 0  \\
-\sqrt{1-\alpha^2} & \alpha  &0\\
0 & 0  & 1  \\
\end{matrix}
\right),
\nonumber \\
A_{12,21}^{(2)} &=&  \left(
\begin{matrix}
\alpha & -\epsilon & 0  \\
\epsilon & \alpha  &0\\
0 & 0  & 1  \\
\end{matrix}
\right)
= \left(
\begin{matrix}
\alpha & -\sqrt{1-\alpha^2} & 0  \\
\sqrt{1-\alpha^2} & \alpha  &0\\
0 & 0  & 1  \\
\end{matrix}
\right),
\nonumber \\
\label{Eq:A12-21unitary_1_2}
\end{eqnarray}
\begin{eqnarray}
A_{23,32}^{(1)}&=& \left(
\begin{matrix}
1 & 0& 0  \\
0& \alpha   &  \epsilon \\
0&  -\epsilon  & \alpha  \\
\end{matrix}
\right)
= \left(
\begin{matrix}
1 & 0& 0  \\
0& \alpha   &  \sqrt{1-\alpha^2} \\
0&  -\sqrt{1-\alpha^2}  & \alpha  \\
\end{matrix}
\right),
\nonumber \\
A_{23,32}^{(2)} &=& \left(
\begin{matrix}
1 & 0& 0  \\
0& \alpha  &  -\epsilon \\
0&  \epsilon & \alpha  \\
\end{matrix}
\right)
= \left(
\begin{matrix}
1 & 0& 0  \\
0& \alpha   &  -\sqrt{1-\alpha^2} \\
0&  \sqrt{1-\alpha^2}  & \alpha  \\
\end{matrix}
\right),
\nonumber \\
\label{Eq:A23-32unitary_1_2}
\end{eqnarray}
and
\begin{eqnarray}
A_{13,31}^{(1)} &=& \left(
\begin{matrix}
\alpha & 0&\epsilon  \\
 0 & 1  &0\\
-\epsilon& 0  & \alpha  \\
\end{matrix}
\right)
= \left(
\begin{matrix}
\alpha & 0&\sqrt{1-\alpha^2}  \\
 0 & 1  &0\\
-\sqrt{1-\alpha^2}& 0  & \alpha  \\
\end{matrix}
\right),
\nonumber \\
A_{13,31}^{(2)} &=& \left(
\begin{matrix}
\alpha & 0& -\epsilon \\
 0 & 1  &0\\
\epsilon& 0  & \alpha  \\
\end{matrix}
\right)
= \left(
\begin{matrix}
\alpha & 0&-\sqrt{1-\alpha^2}  \\
 0 & 1  &0\\
\sqrt{1-\alpha^2}& 0  & \alpha  \\
\end{matrix}
\right),
\nonumber \\
\label{Eq:A13-31unitary_1_2}
\end{eqnarray}
where $\alpha$ denotes a real parameter ($0 \ll \alpha \le 1$) and 
\begin{eqnarray}
\epsilon = \sqrt{1-\alpha^2}.
\label{Eq:epsilon}
\end{eqnarray}
In the case of $\alpha = 1$, we obtain $\tilde{U} = U$. 

Recall that the rotation matrix for 1-2 plane could be written as follows:
\begin{eqnarray}
\left(
\begin{matrix}
 \cos \beta & -\sin\beta& 0  \\
\sin\beta & \cos \beta   & 0 \\
0& 0  & 1  \\
\end{matrix}
\right).
\label{Eq:12rotation}
\end{eqnarray}
By comparing Eq. (\ref{Eq:12rotation}) and Eq. (\ref{Eq:A12-21unitary_1_2}), we can interpret $A_{12,21}^{(1)}$ and $A_{12,21}^{(2)}$ as matrices representing rotations in the 1-2 plane. Similarly, $A_{23,32}^{(1)}$ and $A_{23,32}^{(2)}$ ($A_{13,31}^{(1)}$ and $A_{13,31}^{(2)}$) can be considered as rotation matrices for the 2-3 (1-3) plane.

In conclusion, if we impose the condition of unitarity on the modified mixing matrix $\tilde{U}$, we can achieve a minimal modification of the original mixing matrix $U$ by setting $\tilde{U} = AU$, where $A$ is a matrix representing a certain rotation and has only one real parameter.\footnote{We use the parameter $\alpha$ to estimate the magnitude of the mixing matrix modification. However, the small parameter $\epsilon$ in Eq.(\ref{Eq:epsilon}) or $\beta$ in Eq.(\ref{Eq:12rotation}) is also a  good index to represent the magnitude of the modification.}

\section{Modified TM1 mixing \label{sec:MTM1}}
\subsection{$A_{12,21}^{(1)}$ and $A_{12,21}^{(2)}$}

\begin{figure}[t]
\begin{center}
\includegraphics[scale=1.0]{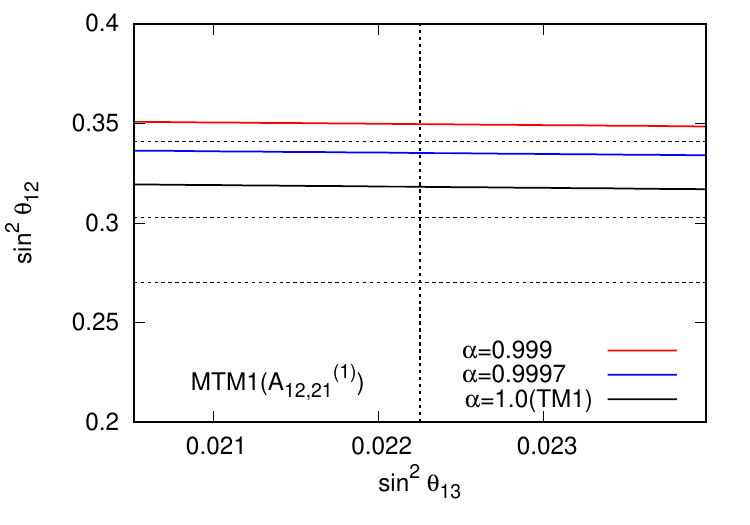}
\includegraphics[scale=1.0]{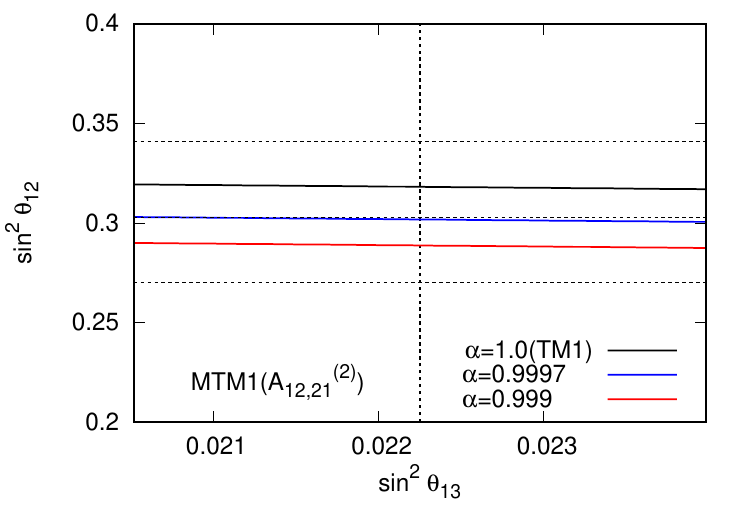}
\caption{Prediction of $\theta_{13}$ and $\theta_{12}$ in MTM1($A_{12,21}^{(1)}$) and MTM1($A_{12,21}^{(2)}$). The horizontal and vertical dotted lines in each panel are the same as in Fig. \ref{Fig:TM_13_12}}
\label{Fig:MTM1_A1221_12_13} 
\end{center}
\end{figure}


\begin{figure}[t]
\begin{center}
\includegraphics[scale=1.0]{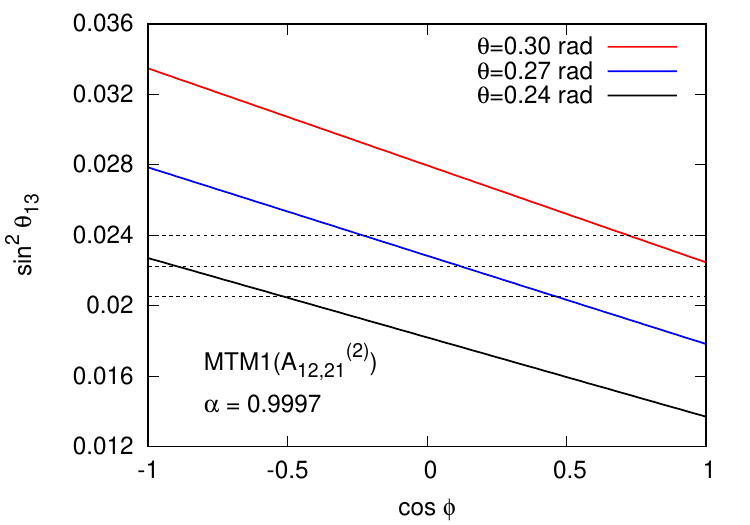}
\includegraphics[scale=1.0]{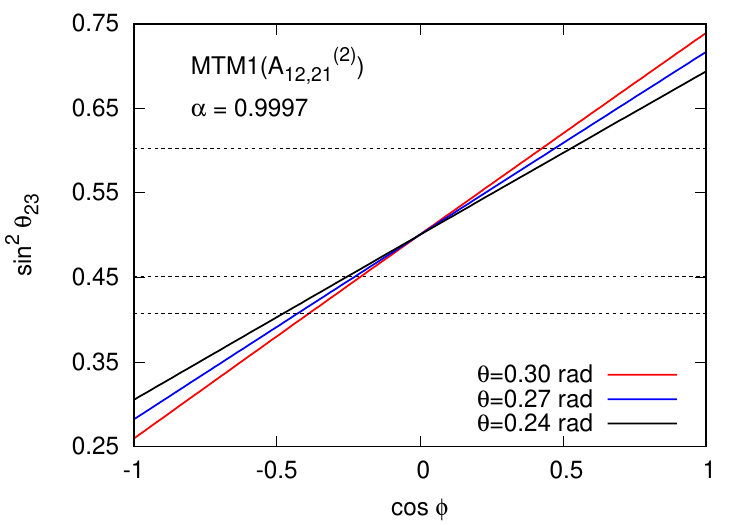}
\caption{The predicted values of $\theta_{13}$ (upper panel) and $\theta_{23}$ (lower panel) as a function of $\phi$ in the MTM1 ($A_{12,21}^{(2)}$). The upper, middle, and lower horizontal dotted lines represent the $3\sigma$ upper limit, the best-fit value, and the $3\sigma$ lower limit of $\theta_{13}$ ($\theta_{23}$) in the NO (IO) case in the upper (lower) panel. }
\label{Fig:MTM1_A1221_cosphi_13_23} 
\end{center}
\end{figure}

\begin{figure}[t]
\begin{tabular}{cc}
\begin{minipage}[t]{0.3\hsize}
\centering
\includegraphics[keepaspectratio, scale=0.5]{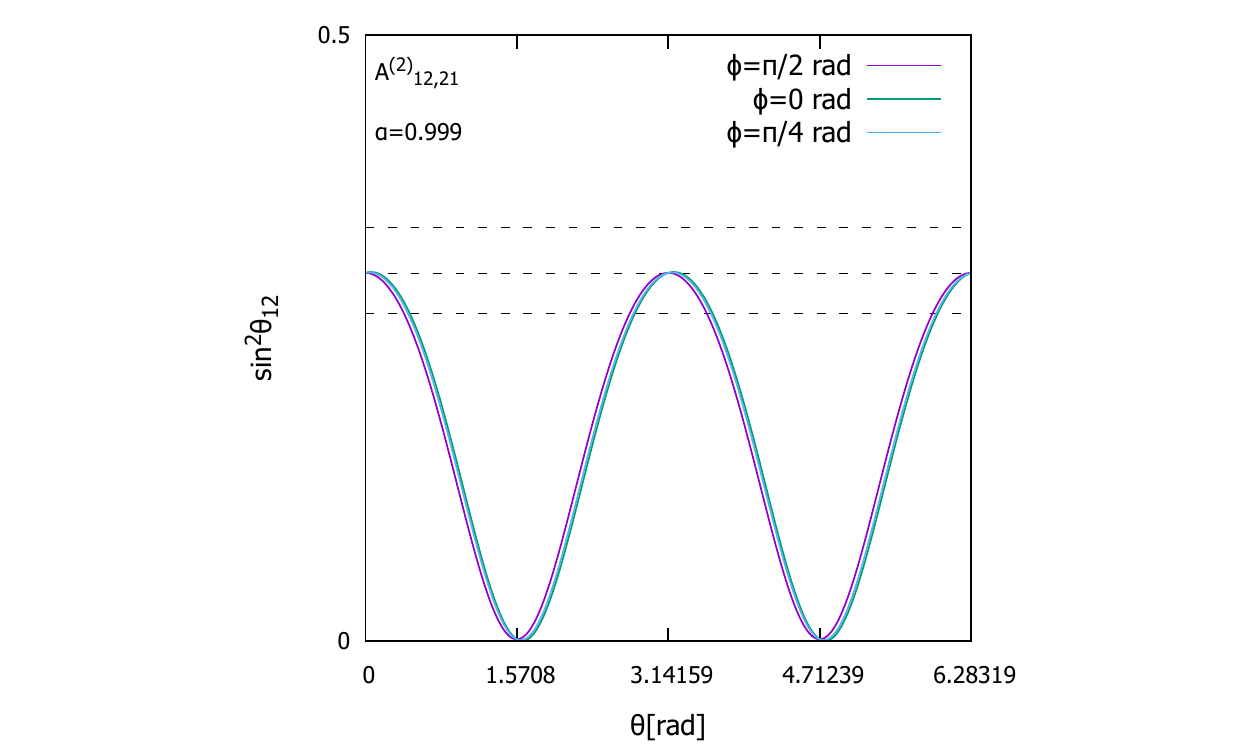}\\
\includegraphics[keepaspectratio, scale=0.5]{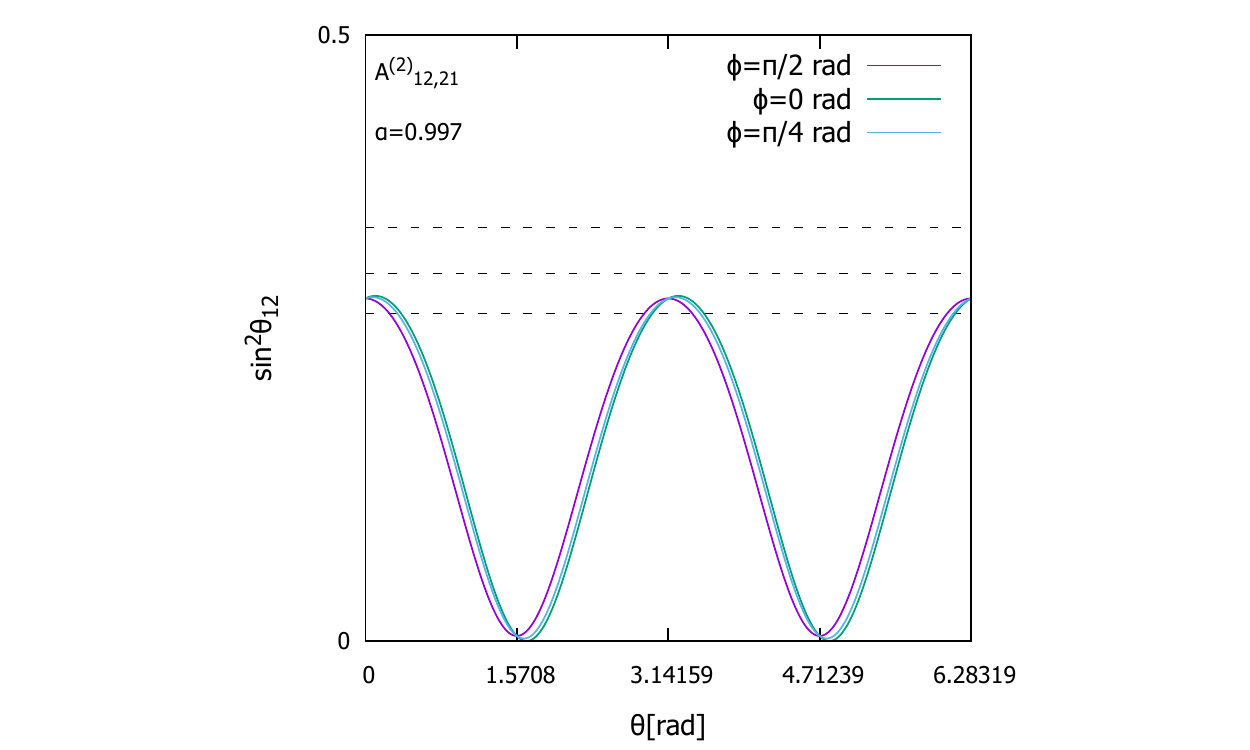}\\
\includegraphics[keepaspectratio, scale=0.5]{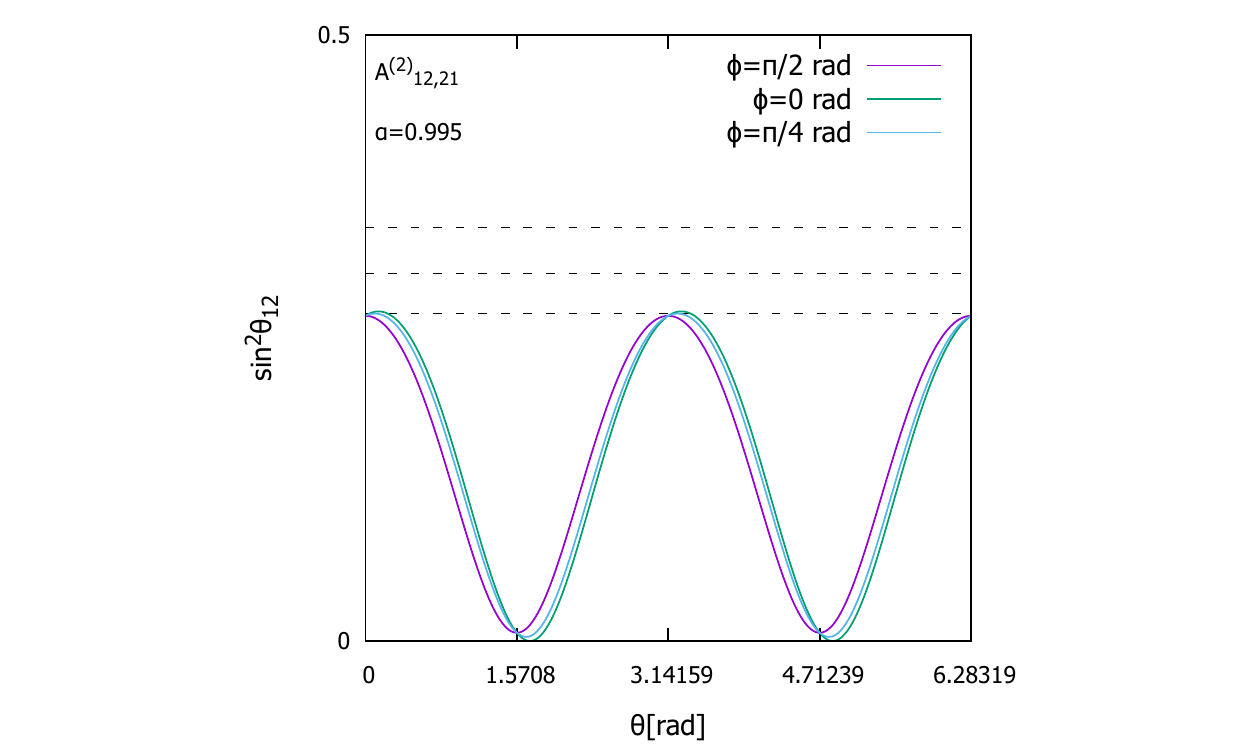}
\end{minipage}&
\begin{minipage}[t]{0.3\hsize}
\centering
\includegraphics[keepaspectratio, scale=0.5]{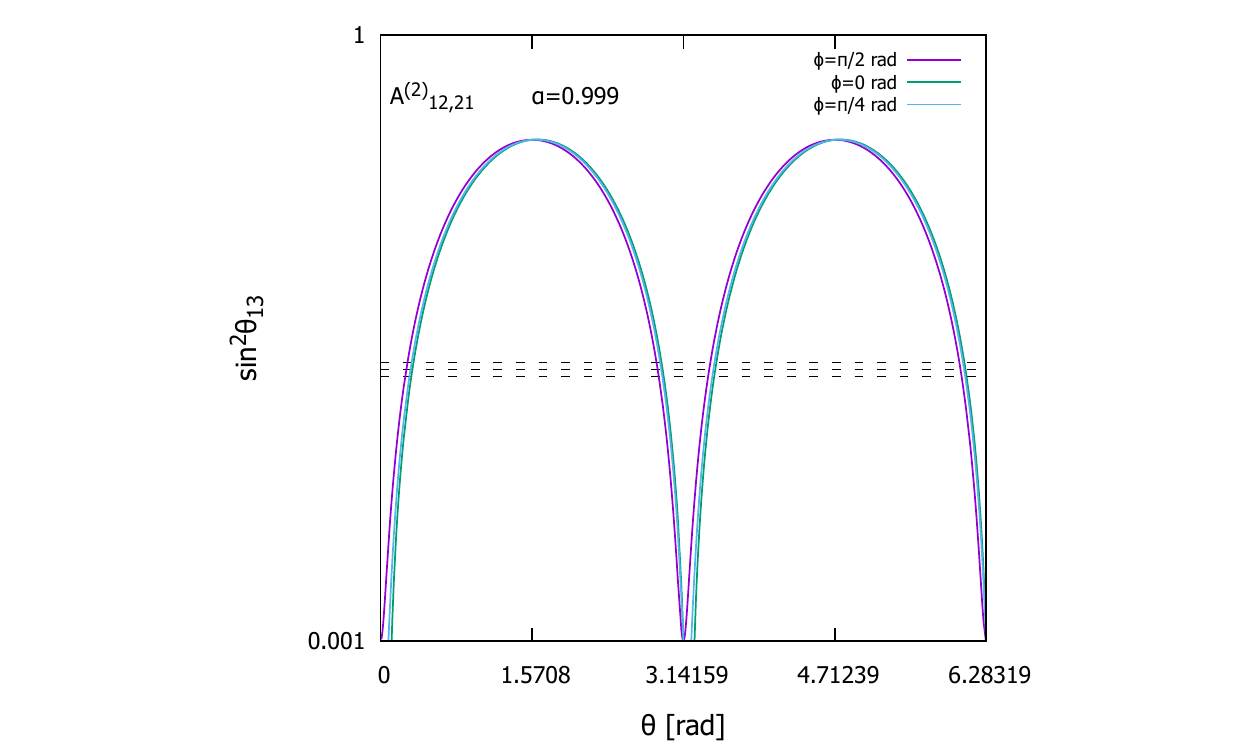}\\
\includegraphics[keepaspectratio, scale=0.5]{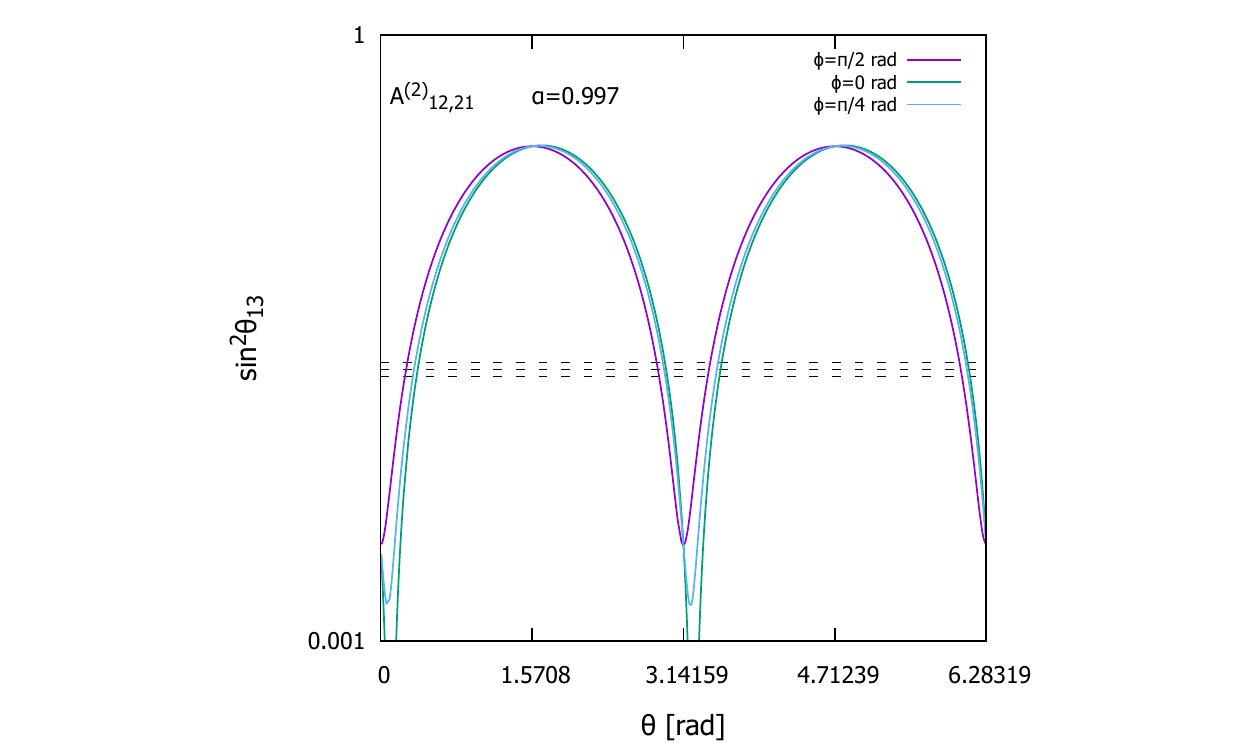}\\
\includegraphics[keepaspectratio, scale=0.5]{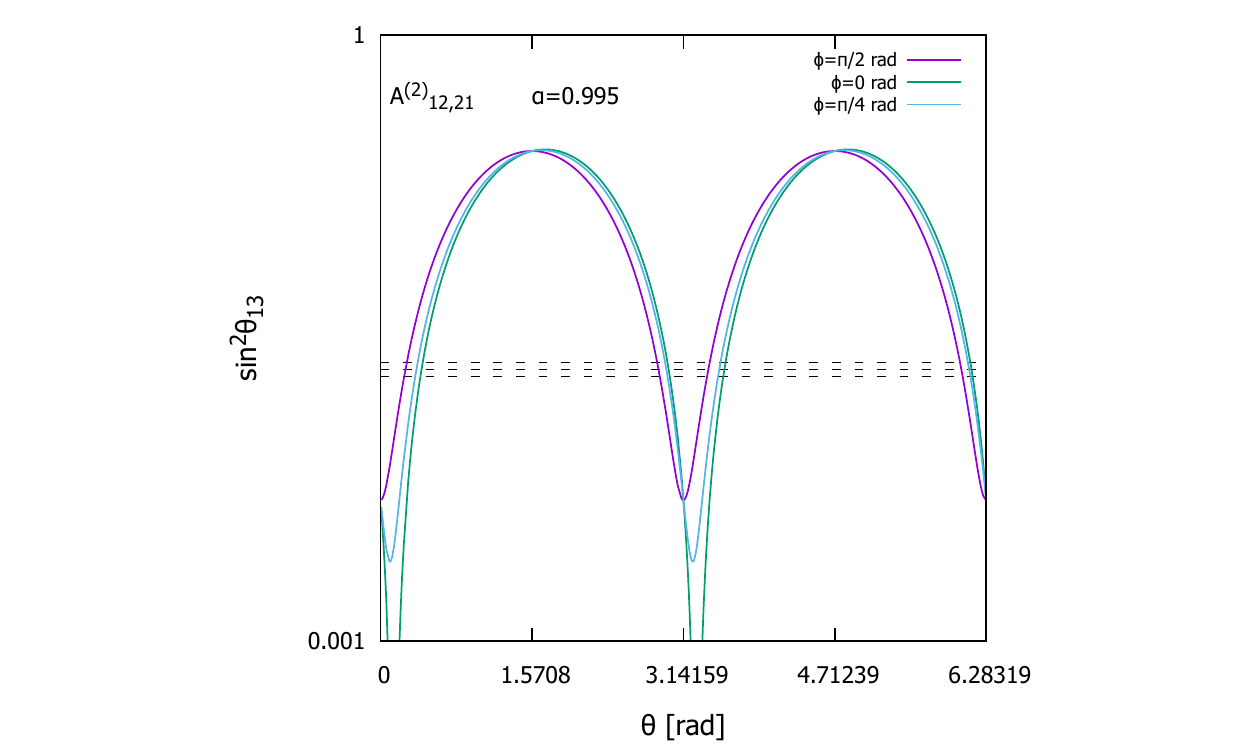}
 \end{minipage} \\
\end{tabular}
 \caption{$s_{12}^2$ (left-side panels) or $s_{13}^2$ (right-side panels) vs $\theta$ for $\phi$ [rad] $=0, \frac{\pi}{4}, \frac{\pi}{2}$ in the case of MTM1($A_{12,21}^{(2)}$).}
 \label{Fig:MTM1_A1221_2_a1p1_12_13} 
  \end{figure}

\begin{figure}[t]
\begin{tabular}{cc}
\begin{minipage}[t]{0.48\hsize}
\centering
\includegraphics[keepaspectratio, scale=0.5]{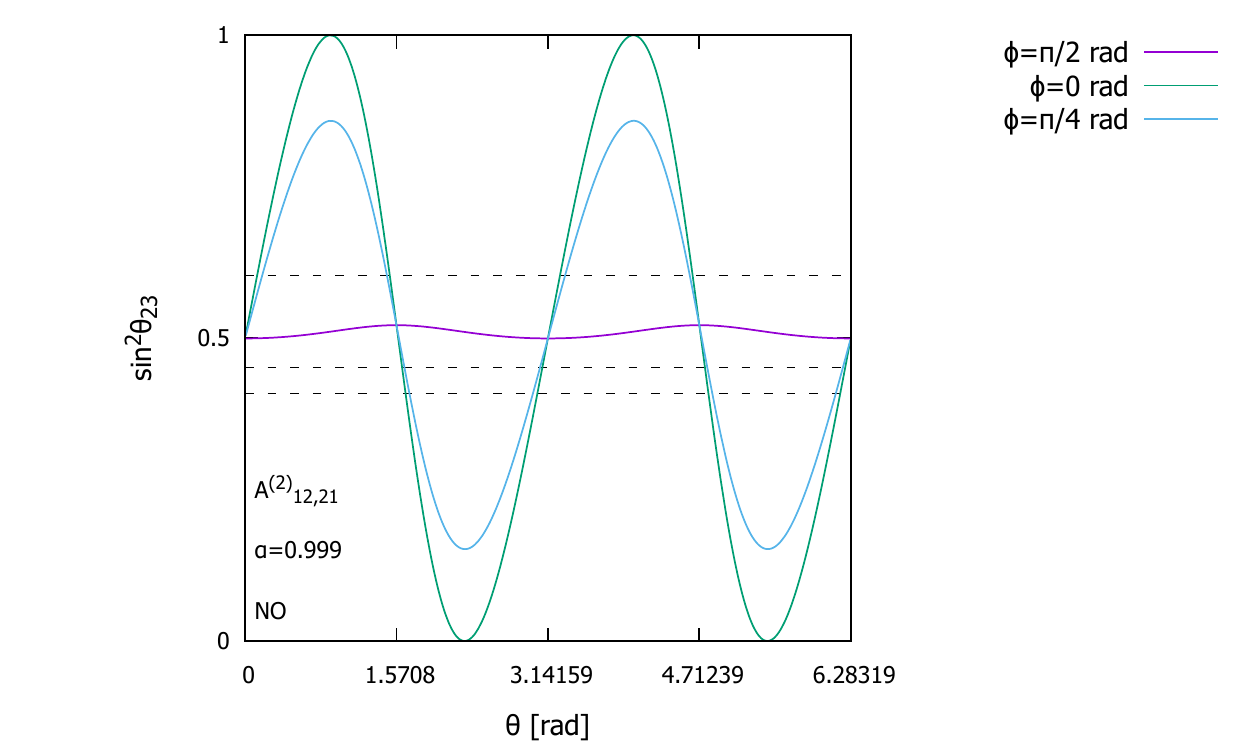}\\
\includegraphics[keepaspectratio, scale=0.5]{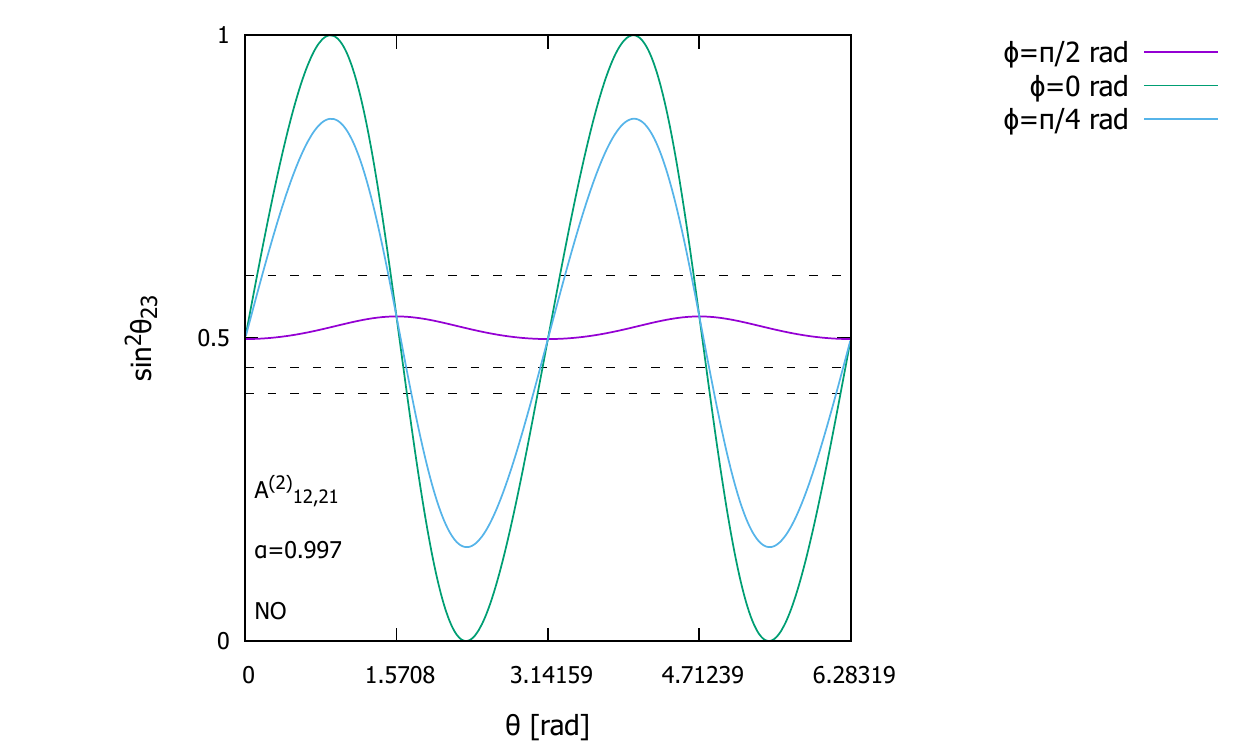}\\
\includegraphics[keepaspectratio, scale=0.5]{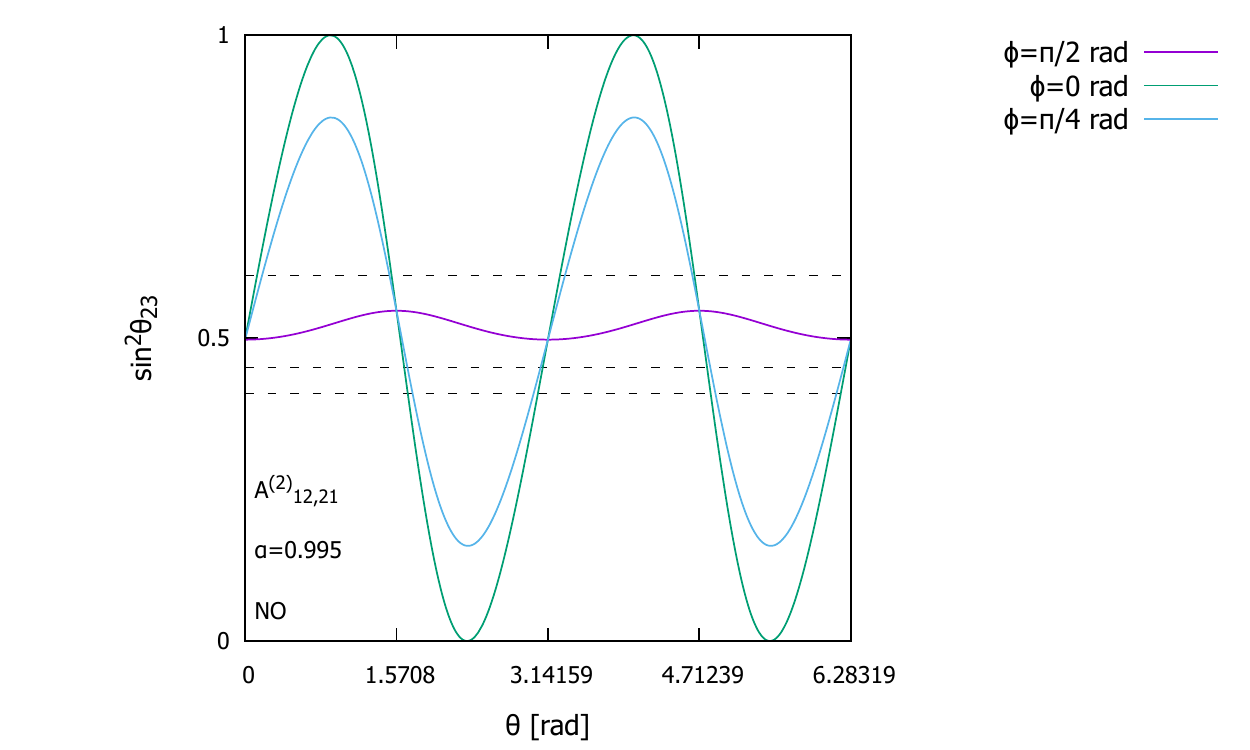}
\end{minipage}&
\begin{minipage}[t]{0.48\hsize}
\centering
\includegraphics[keepaspectratio, scale=0.5]{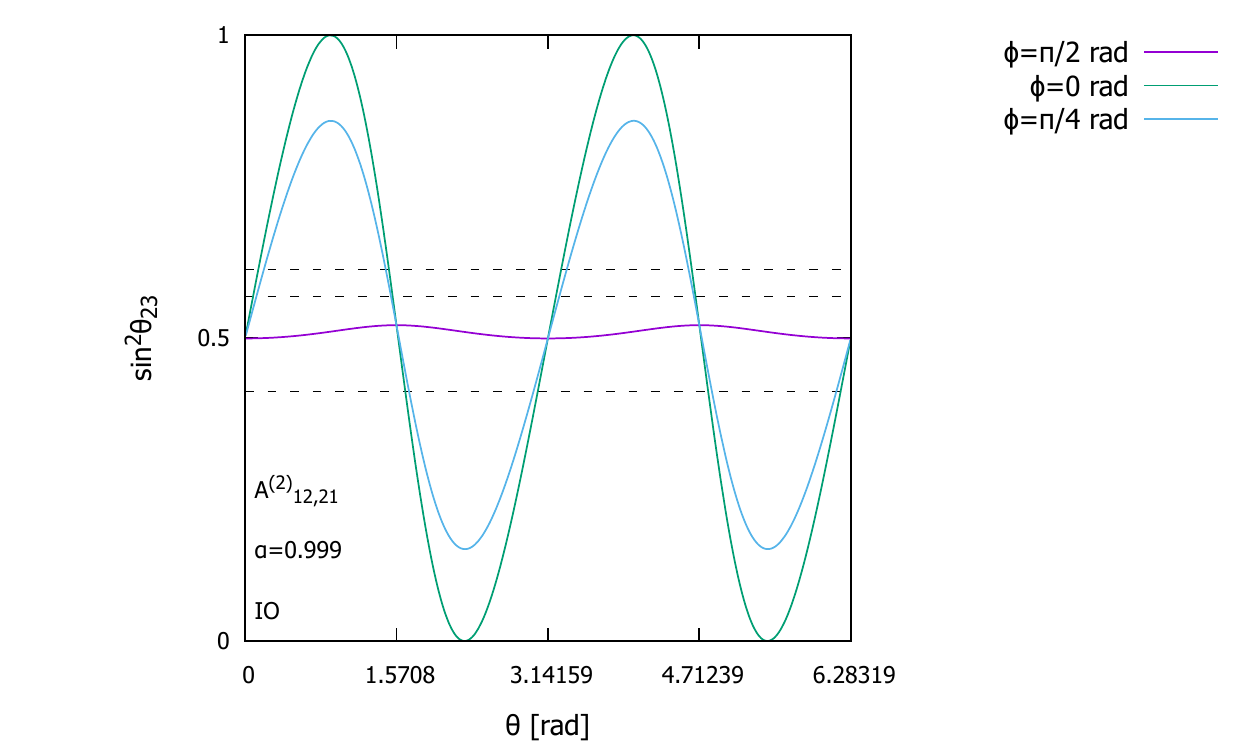}\\
\includegraphics[keepaspectratio, scale=0.5]{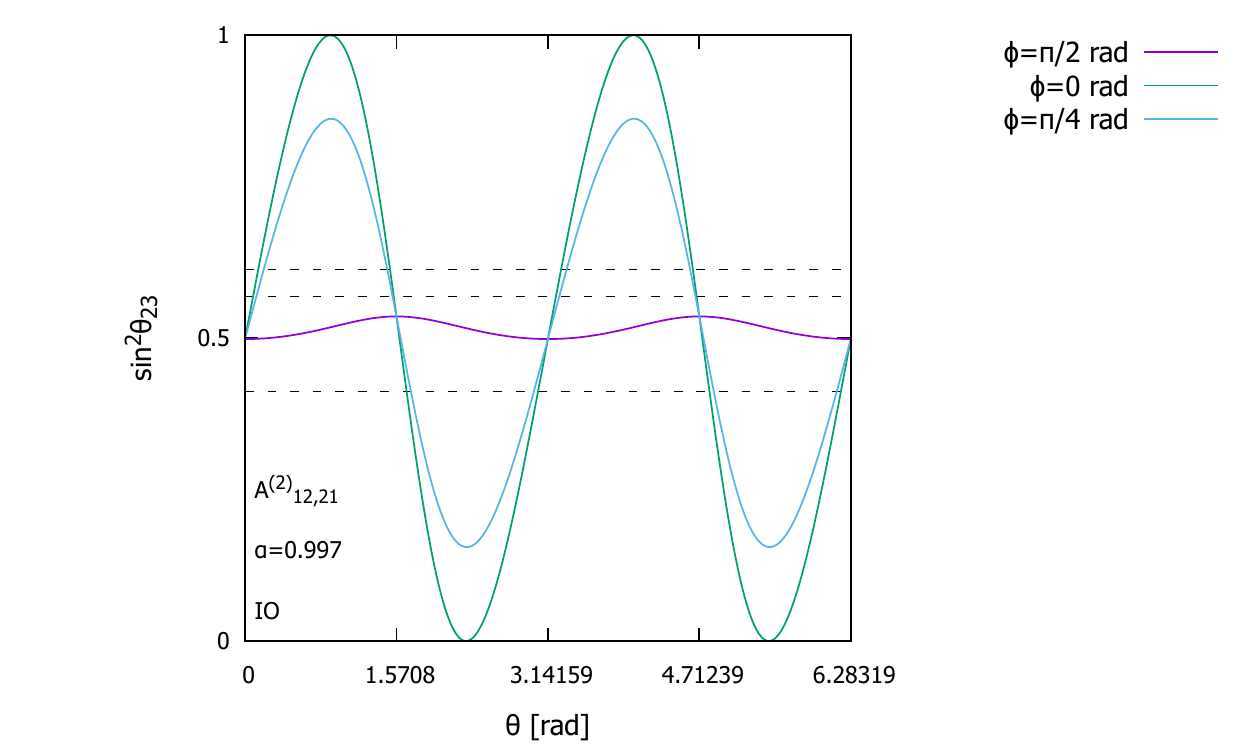}\\
\includegraphics[keepaspectratio, scale=0.5]{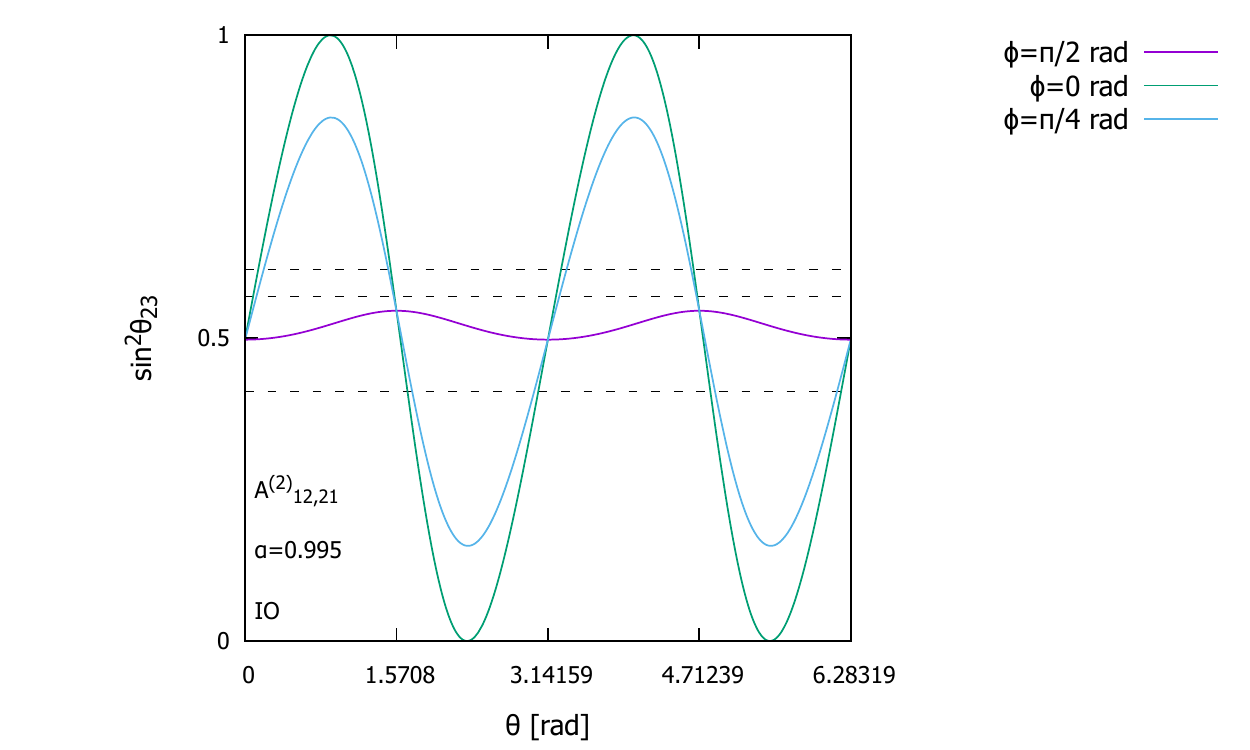}
 \end{minipage} \\
\end{tabular}
 \caption{$s_{23}^2$ vs $\theta$ for $\phi$ [rad] $=0, \frac{\pi}{4}, \frac{\pi}{2}$ in the case of MTM1($A_{12,21}^{(2)}$).}
 \label{Fig:MTM1_A1221_2_ap_23} 
  \end{figure}

\begin{figure}[t]
\begin{tabular}{cc}
\begin{minipage}[t]{0.48\hsize}
\centering
\includegraphics[keepaspectratio, scale=0.5]{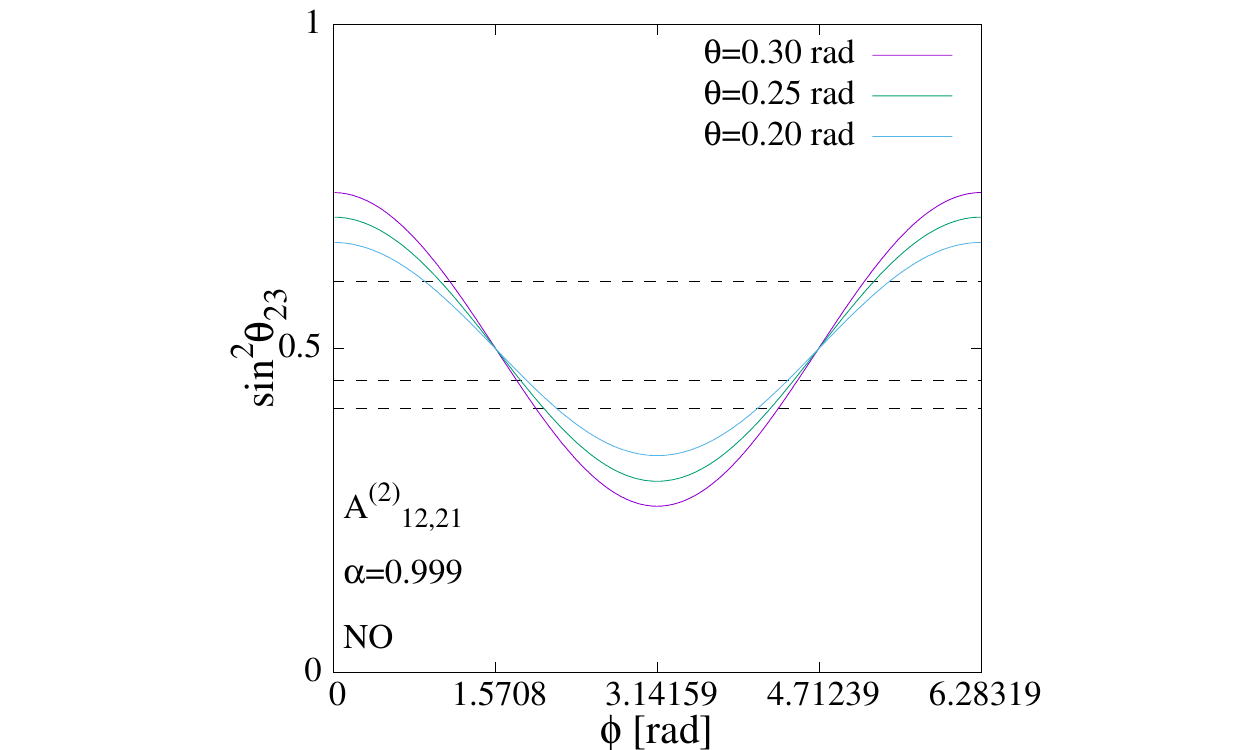}\\
\includegraphics[keepaspectratio, scale=0.5]{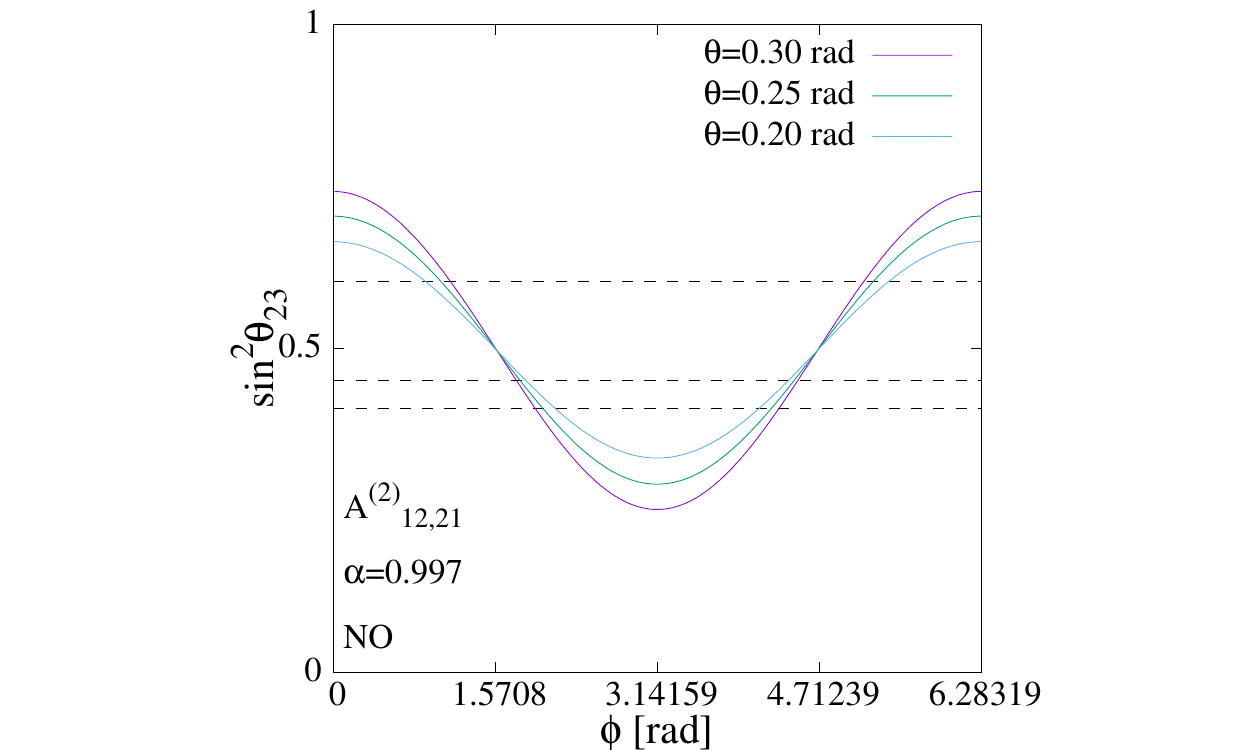}\\
\includegraphics[keepaspectratio, scale=0.5]{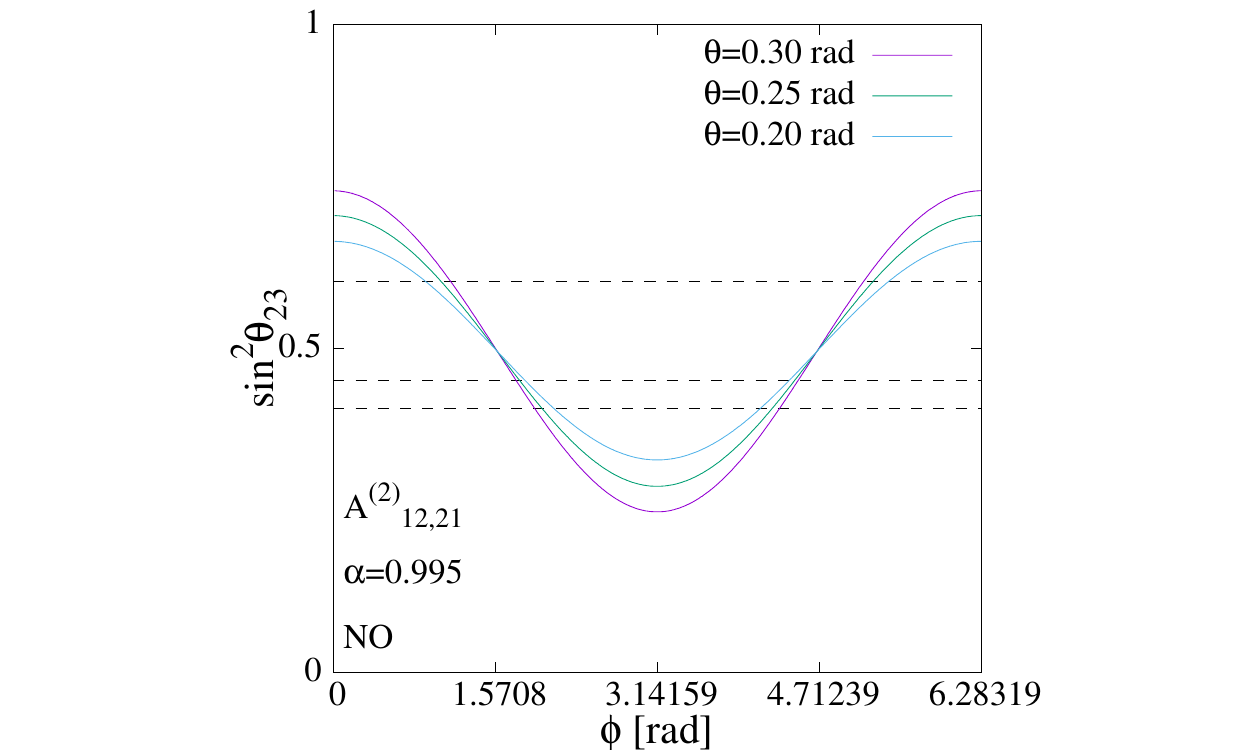}
\end{minipage}&
\begin{minipage}[t]{0.48\hsize}
\centering
\includegraphics[keepaspectratio, scale=0.5]{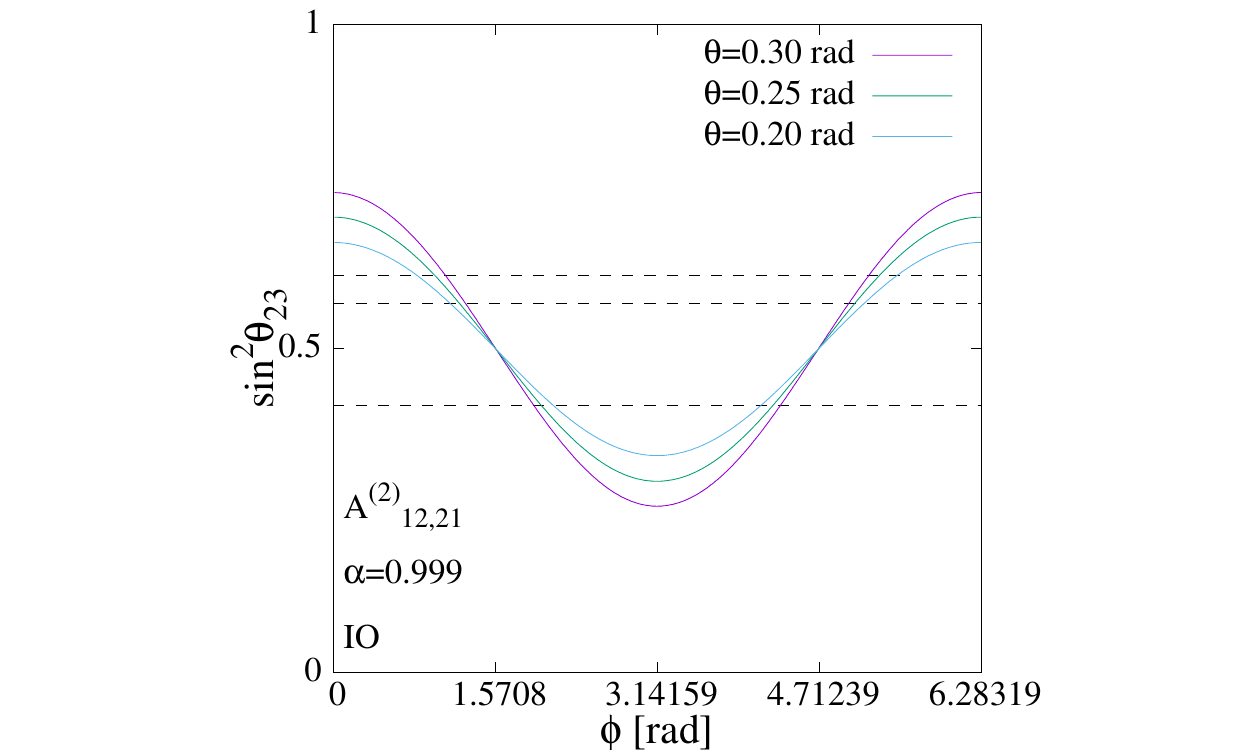}\\
\includegraphics[keepaspectratio, scale=0.5]{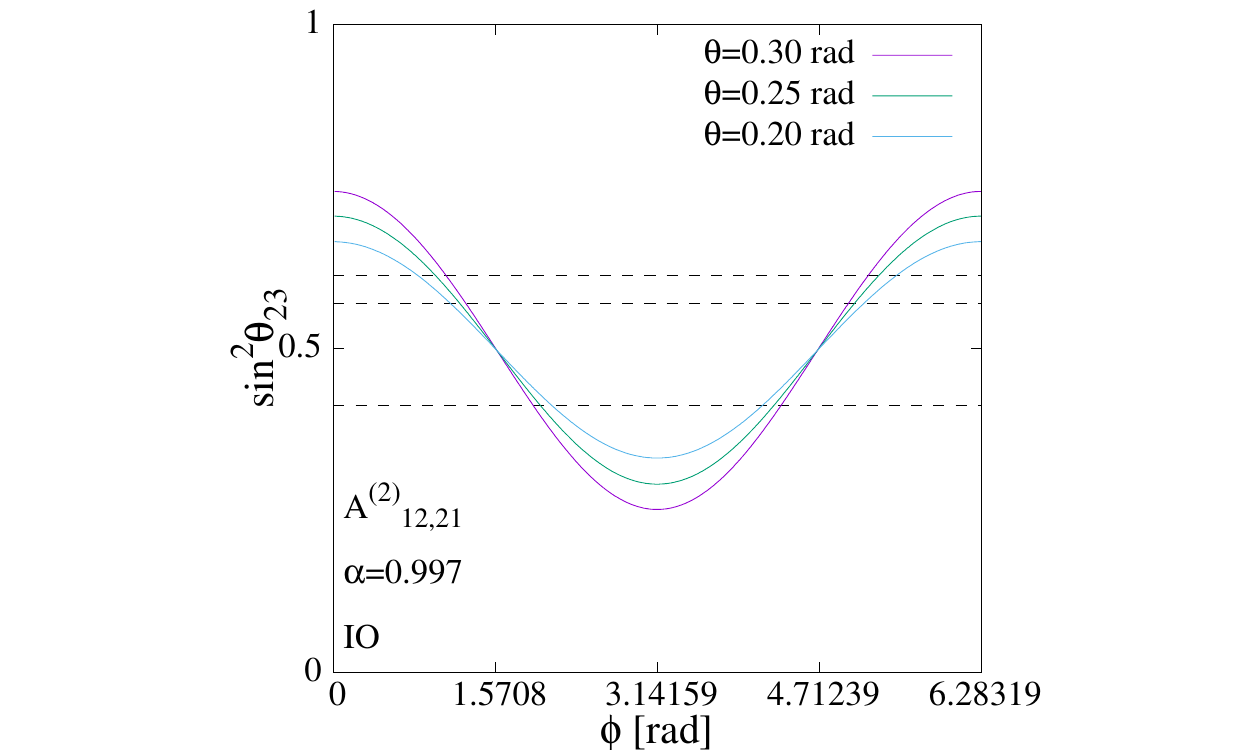}\\
\includegraphics[keepaspectratio, scale=0.5]{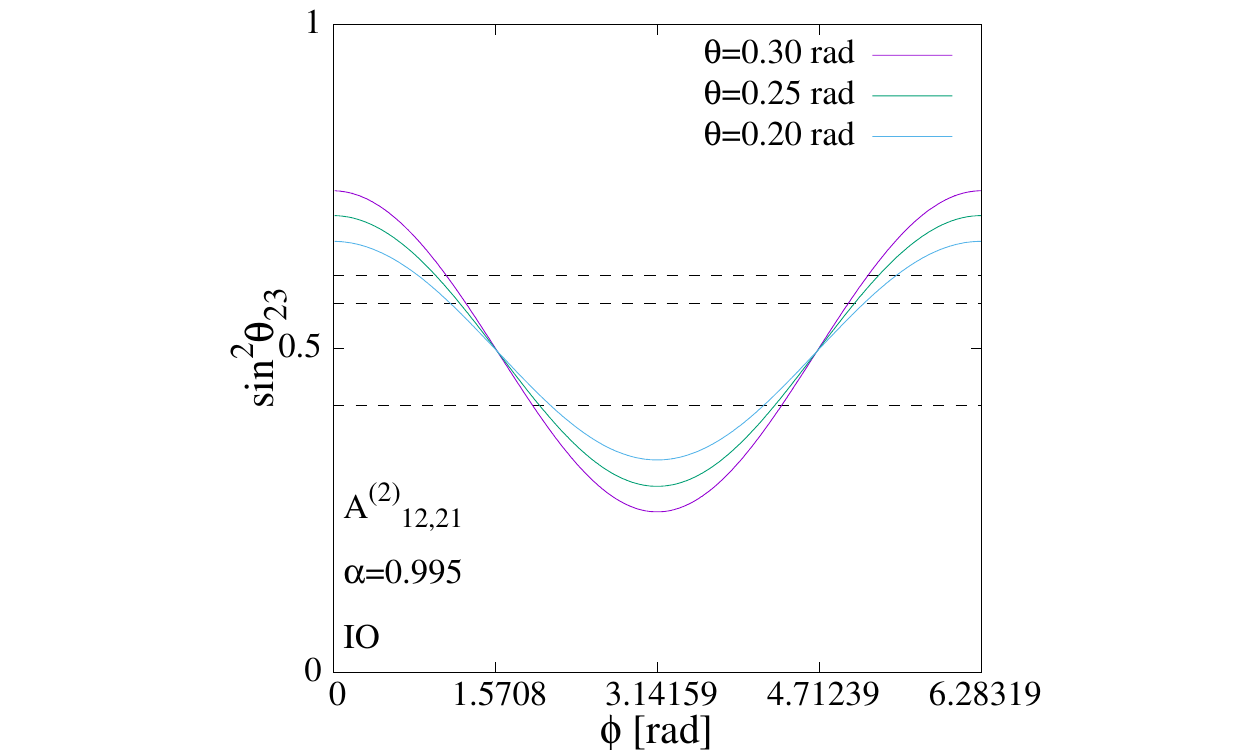}
 \end{minipage} \\
\end{tabular}
 \caption{$s_{23}^2$ vs $\phi$ for $\theta$ [rad] $=0.20, 0.25, 0.30$ in the case of MTM1($A_{12,21}^{(2)}$).}
 \label{Fig:MTM1_A1221_2_at_23} 
  \end{figure}

\begin{figure}[t]
\begin{tabular}{cc}
\begin{minipage}[t]{0.48\hsize}
\centering
\includegraphics[keepaspectratio, scale=0.5]{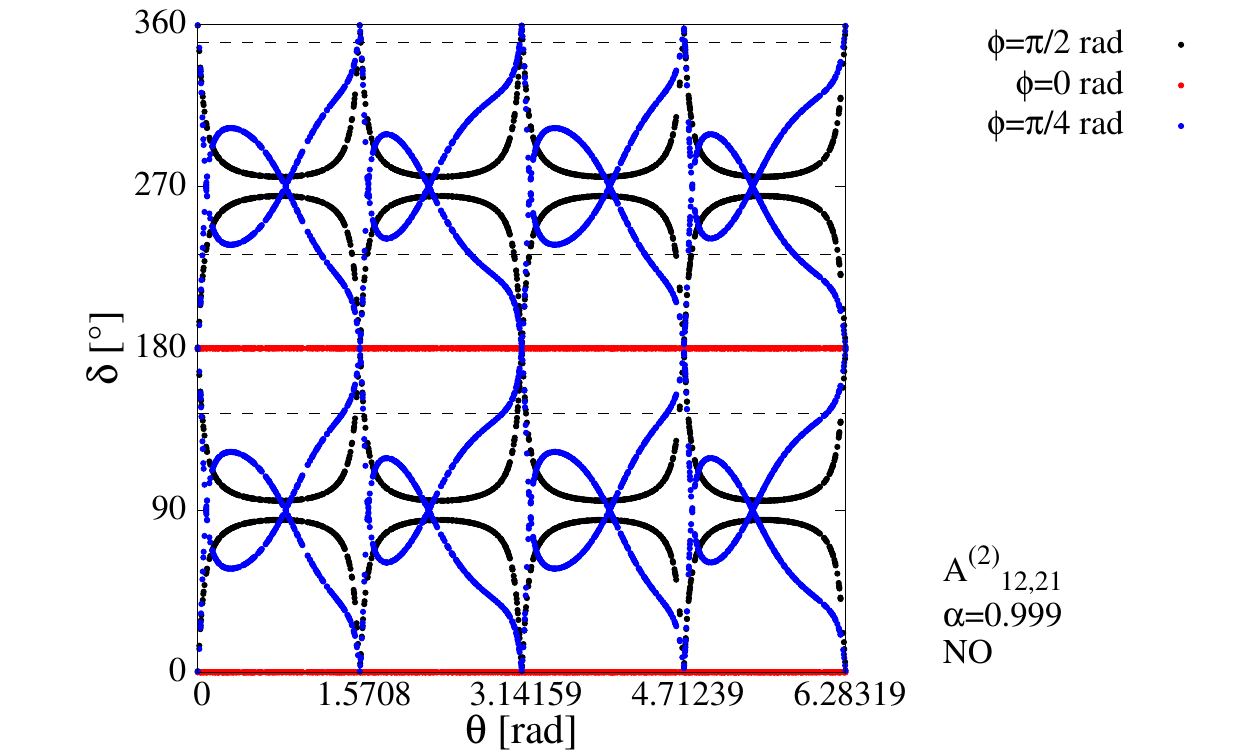}\\
\includegraphics[keepaspectratio, scale=0.5]{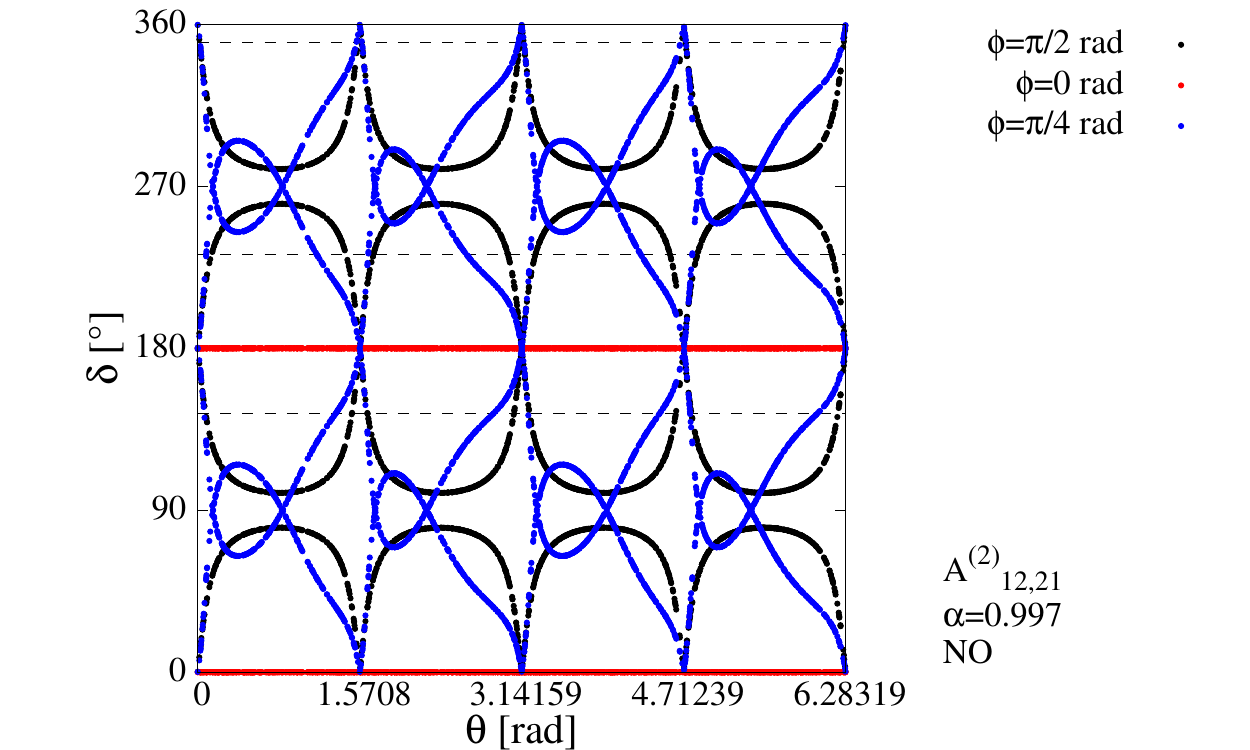}\\
\includegraphics[keepaspectratio, scale=0.5]{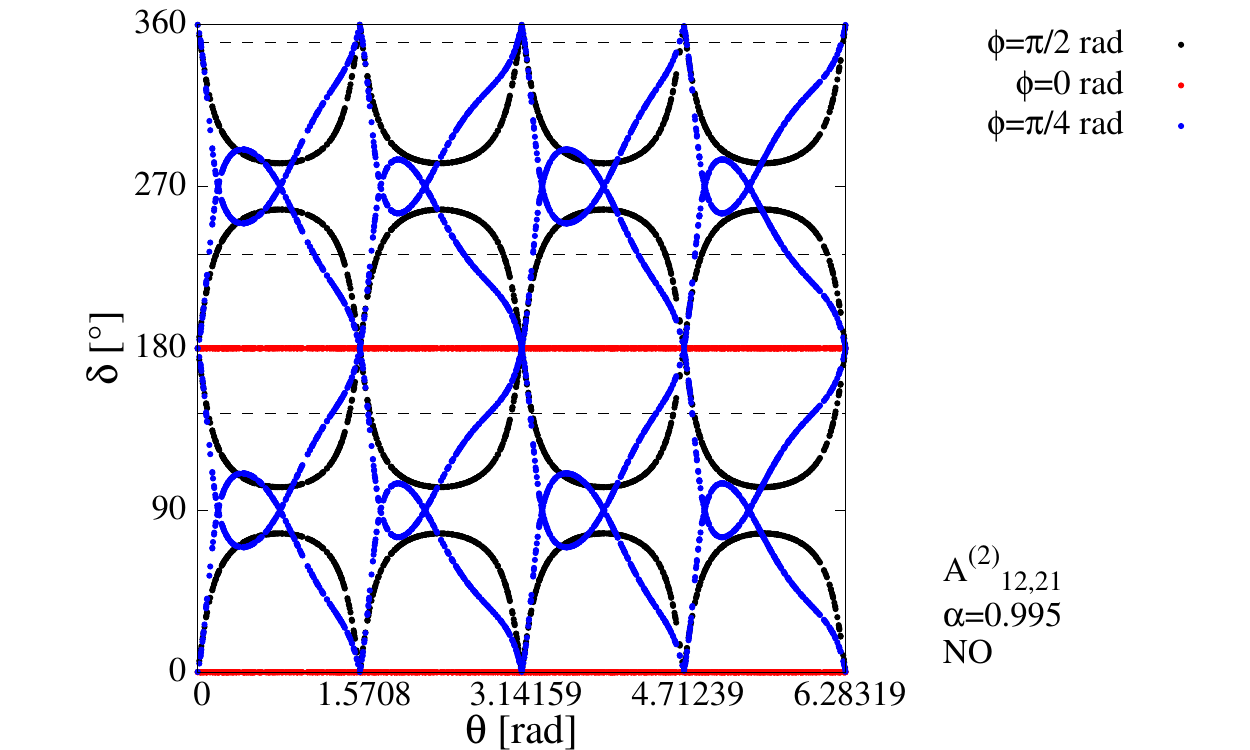}
\end{minipage}&
\begin{minipage}[t]{0.48\hsize}
\centering
\includegraphics[keepaspectratio, scale=0.5]{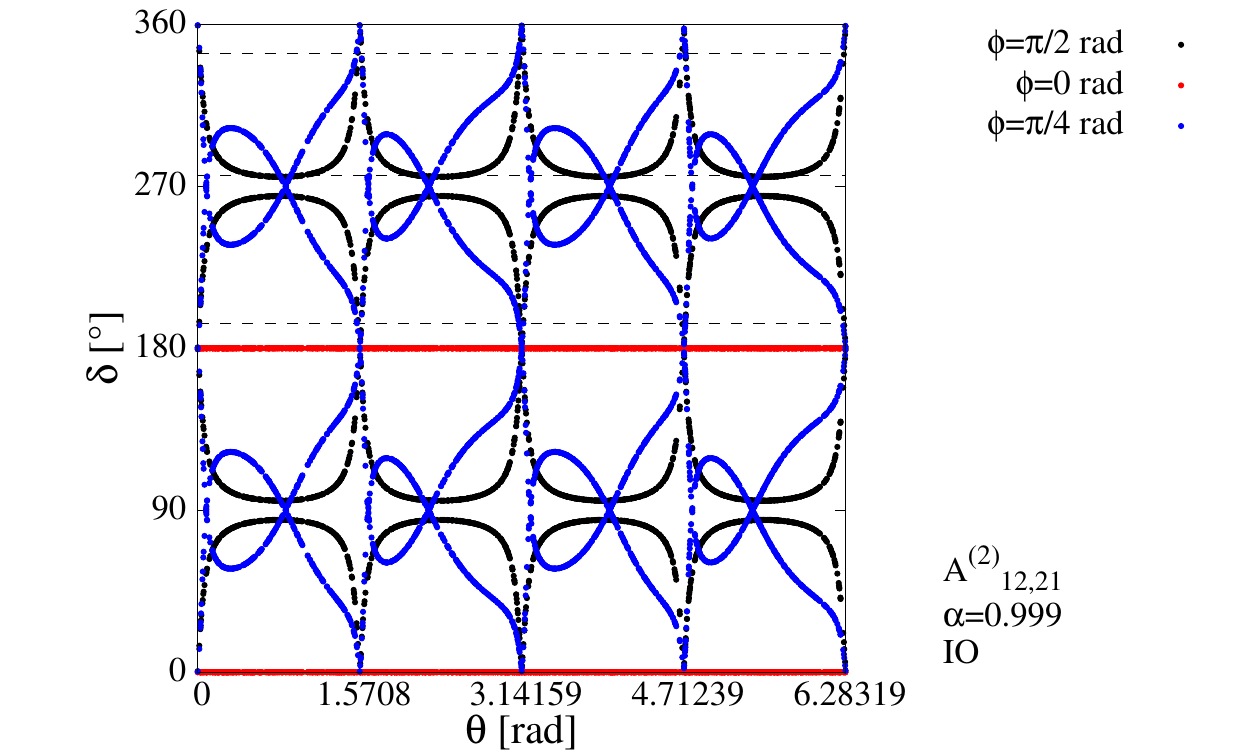}\\
\includegraphics[keepaspectratio, scale=0.5]{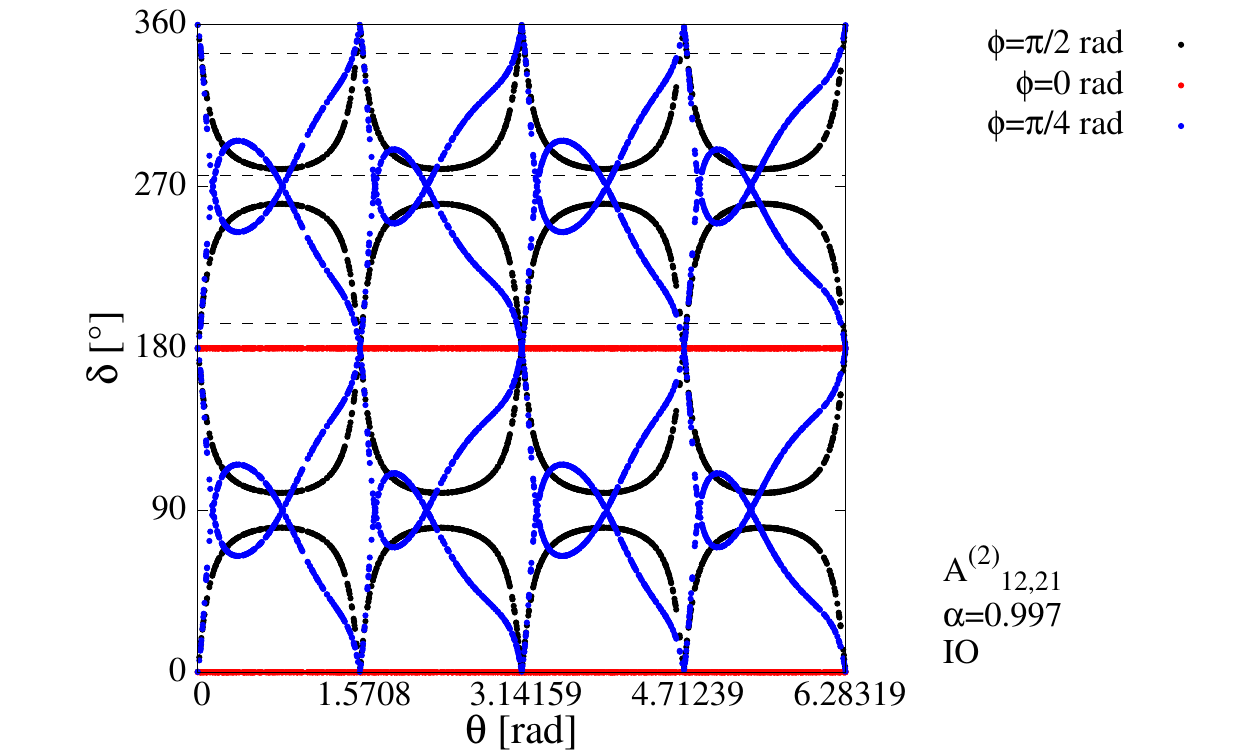}\\
\includegraphics[keepaspectratio, scale=0.5]{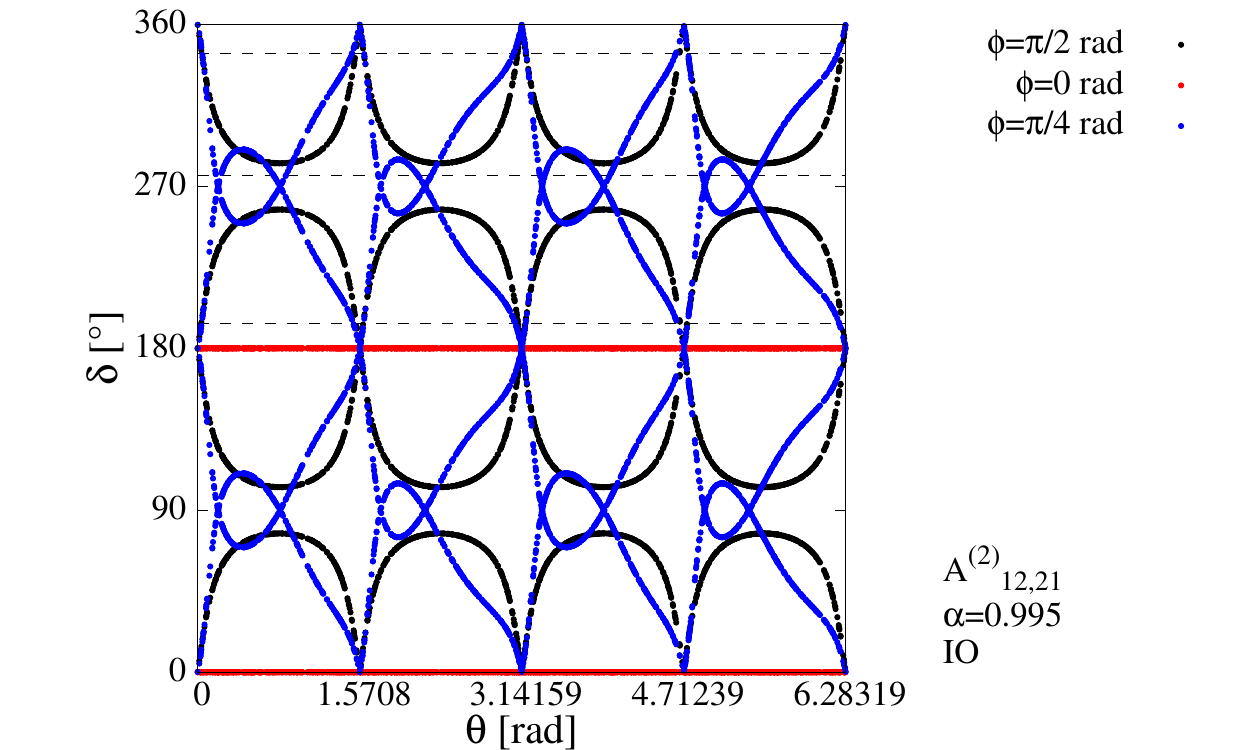}
 \end{minipage} \\
\end{tabular}
 \caption{$\delta$ vs $\theta$ for $\phi$ [rad] $=0, \frac{\pi}{4}, \frac{\pi}{2}$ in the case of MTM1($A_{12,21}^{(2)}$).}
 \label{Fig:MTM1_A1221_2_ap_d} 
  \end{figure}

\begin{figure}[t]
\begin{tabular}{cc}
\begin{minipage}[t]{0.48\hsize}
\centering
\includegraphics[keepaspectratio, scale=0.5]{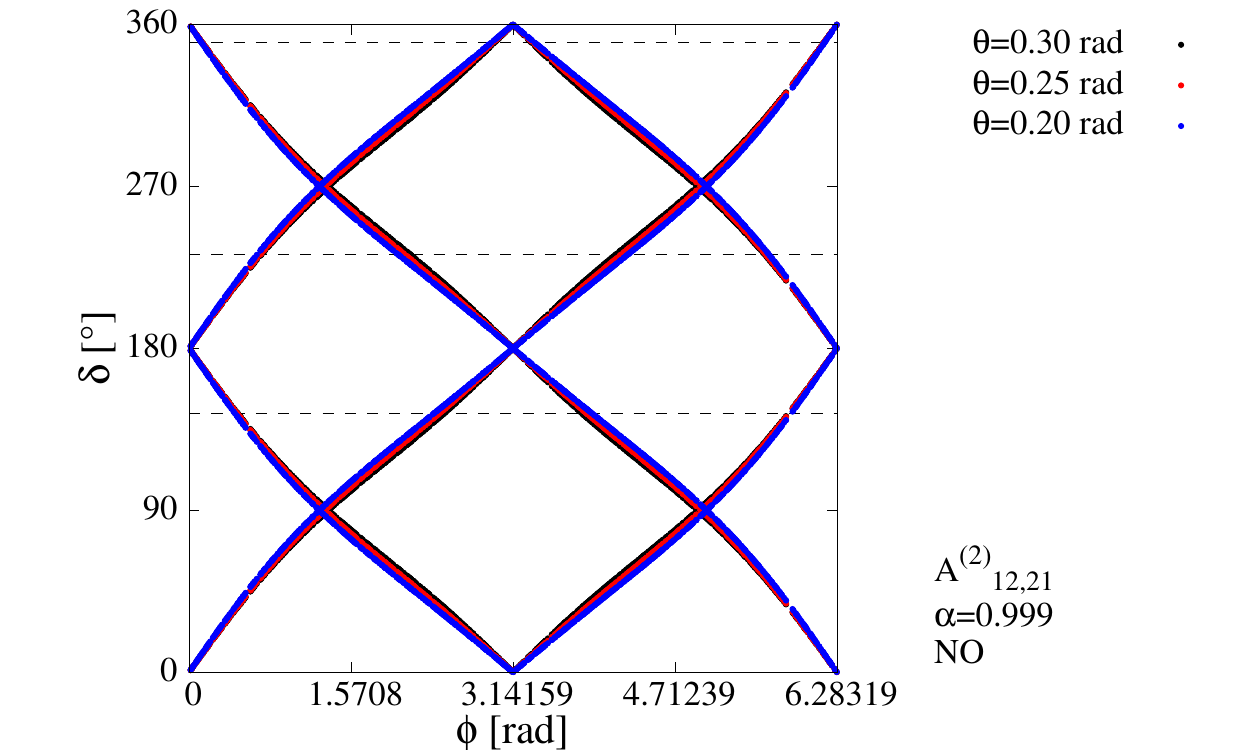}\\
\includegraphics[keepaspectratio, scale=0.5]{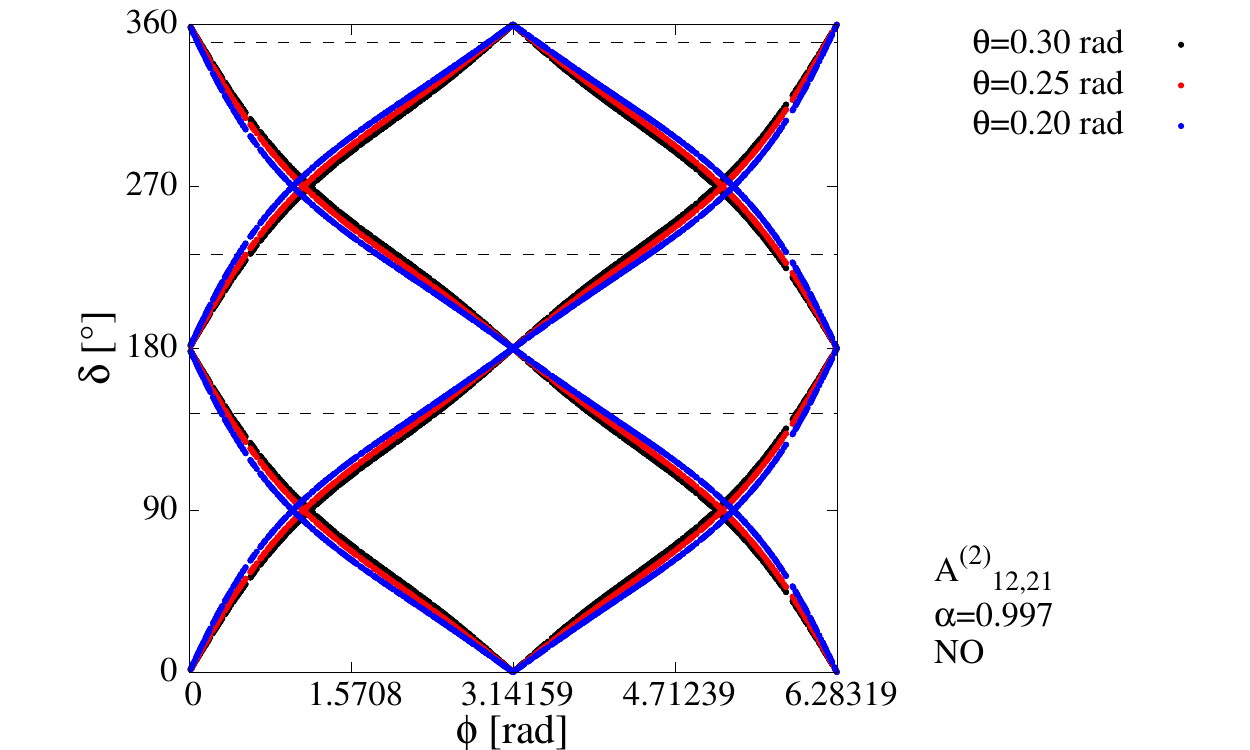}\\
\includegraphics[keepaspectratio, scale=0.5]{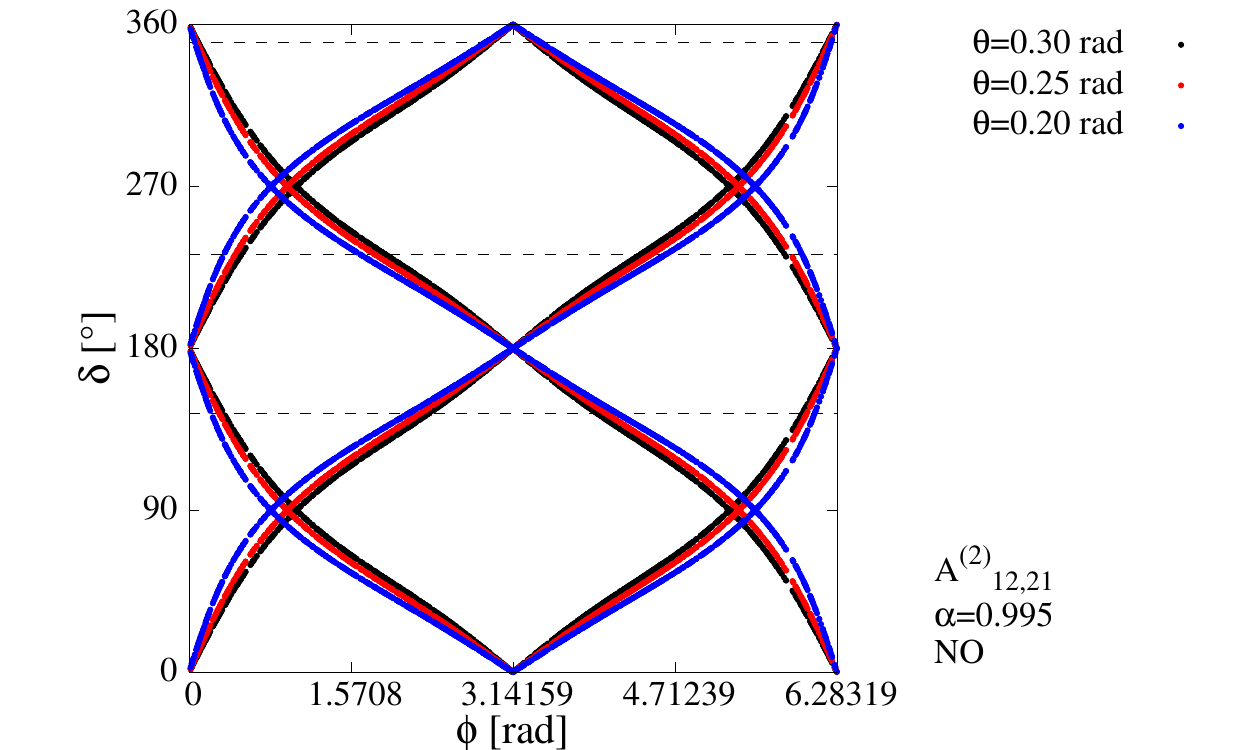}
\end{minipage}&
\begin{minipage}[t]{0.48\hsize}
\centering
\includegraphics[keepaspectratio, scale=0.5]{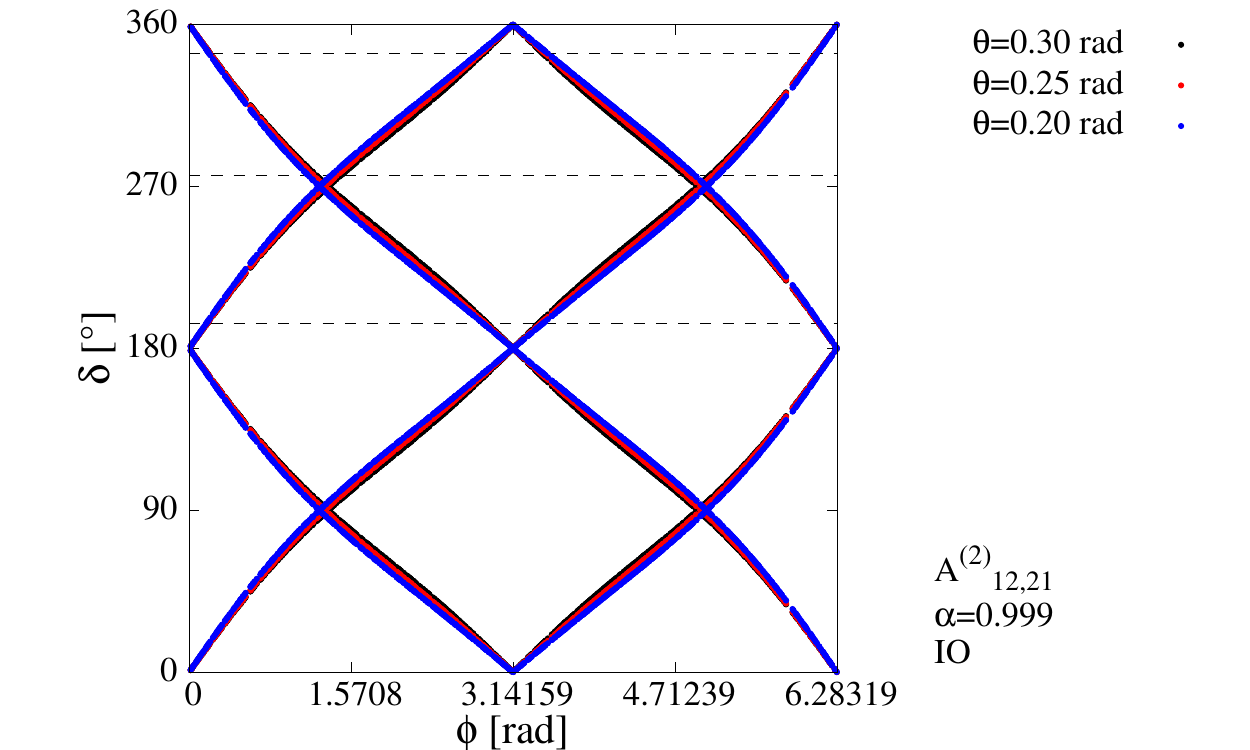}\\
\includegraphics[keepaspectratio, scale=0.5]{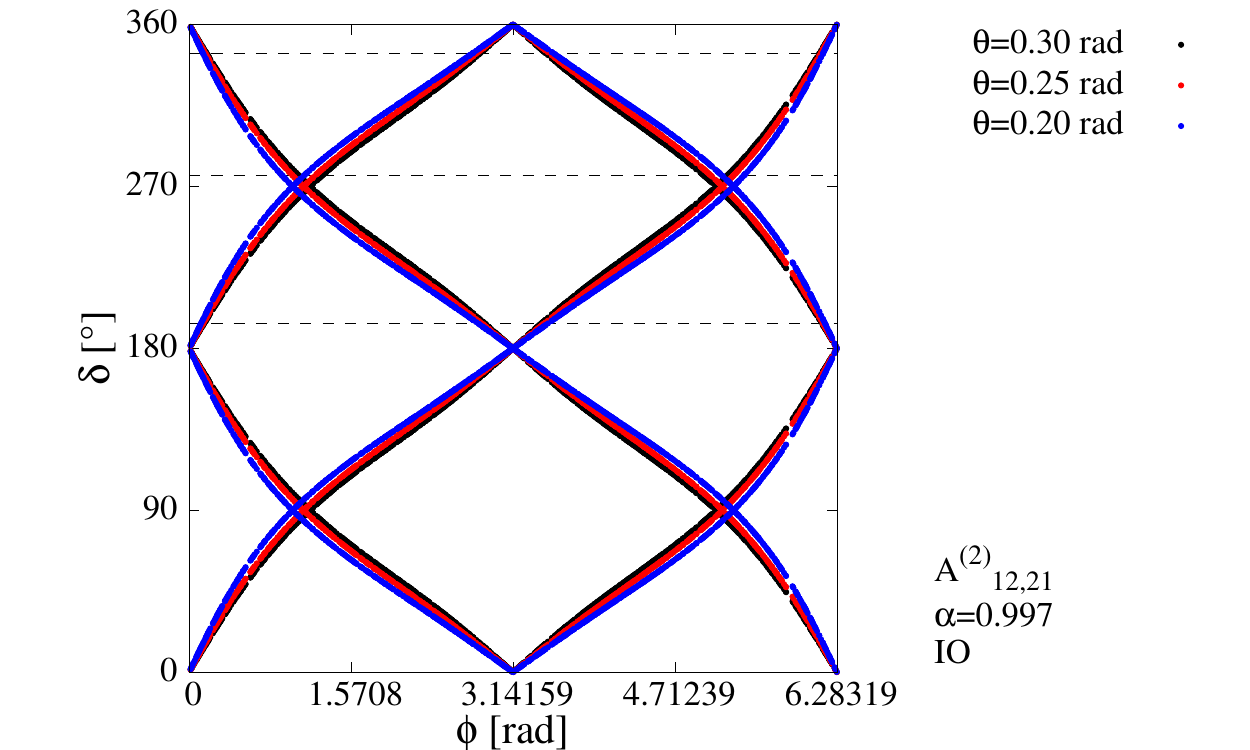}\\
\includegraphics[keepaspectratio, scale=0.5]{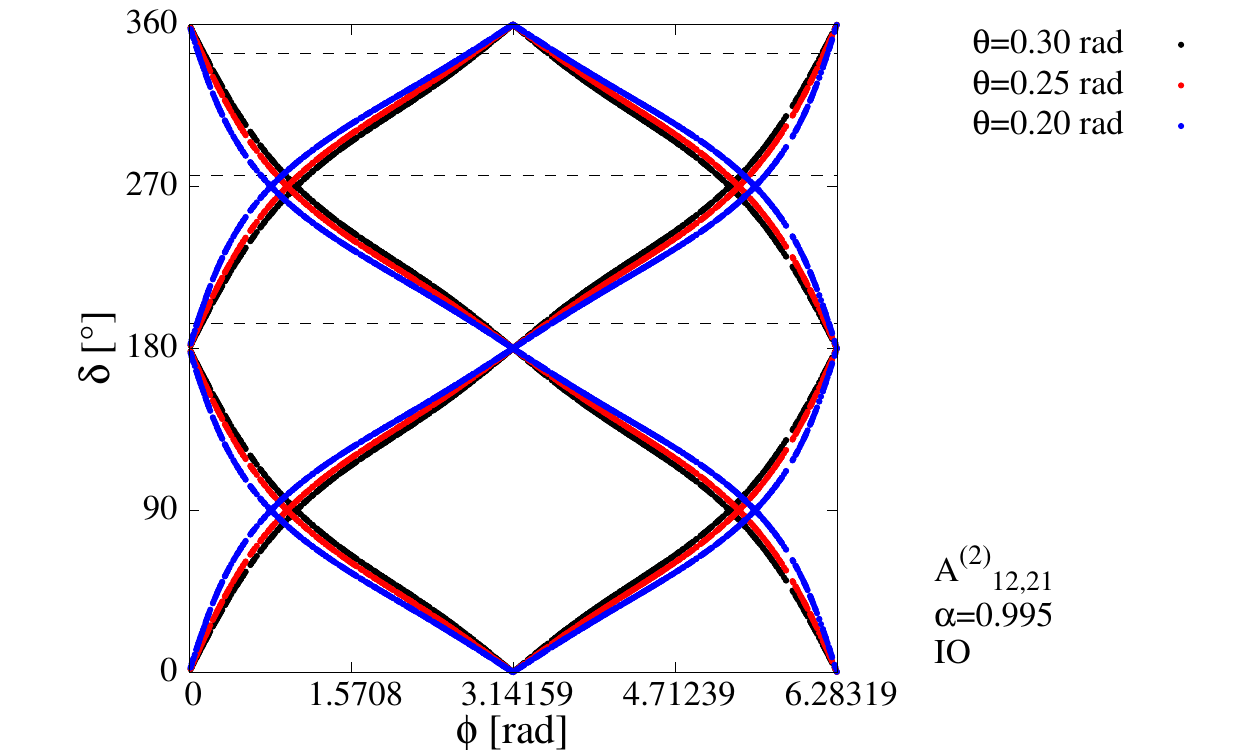}
 \end{minipage} \\
\end{tabular}
 \caption{$\delta$ vs $\phi$ for $\theta$ [rad] $=0.20, 0.25, 0.30$ in the case of MTM1($A_{12,21}^{(2)}$).}
 \label{Fig:MTM1_A1221_2_at_d} 
  \end{figure}

\begin{figure}[t]
\begin{tabular}{cc}
\begin{minipage}[t]{0.48\hsize}
\centering
\includegraphics[keepaspectratio, scale=0.5]{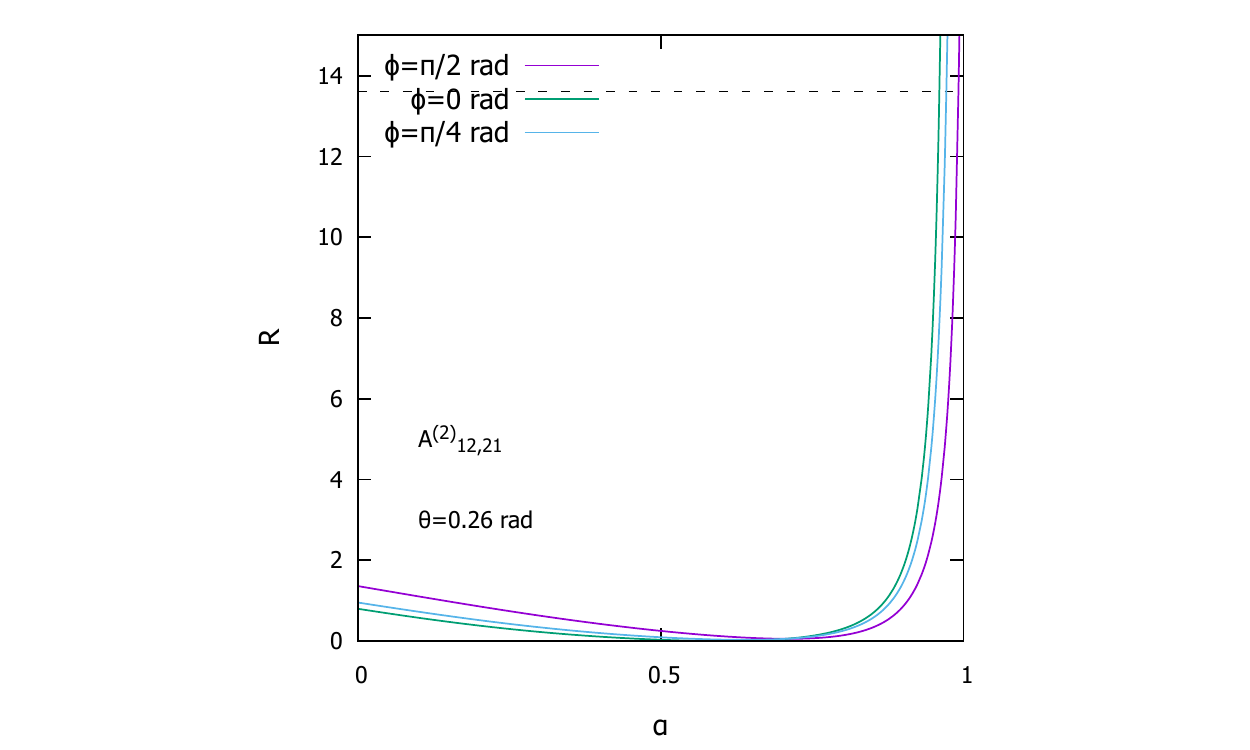}\\
\includegraphics[keepaspectratio, scale=0.5]{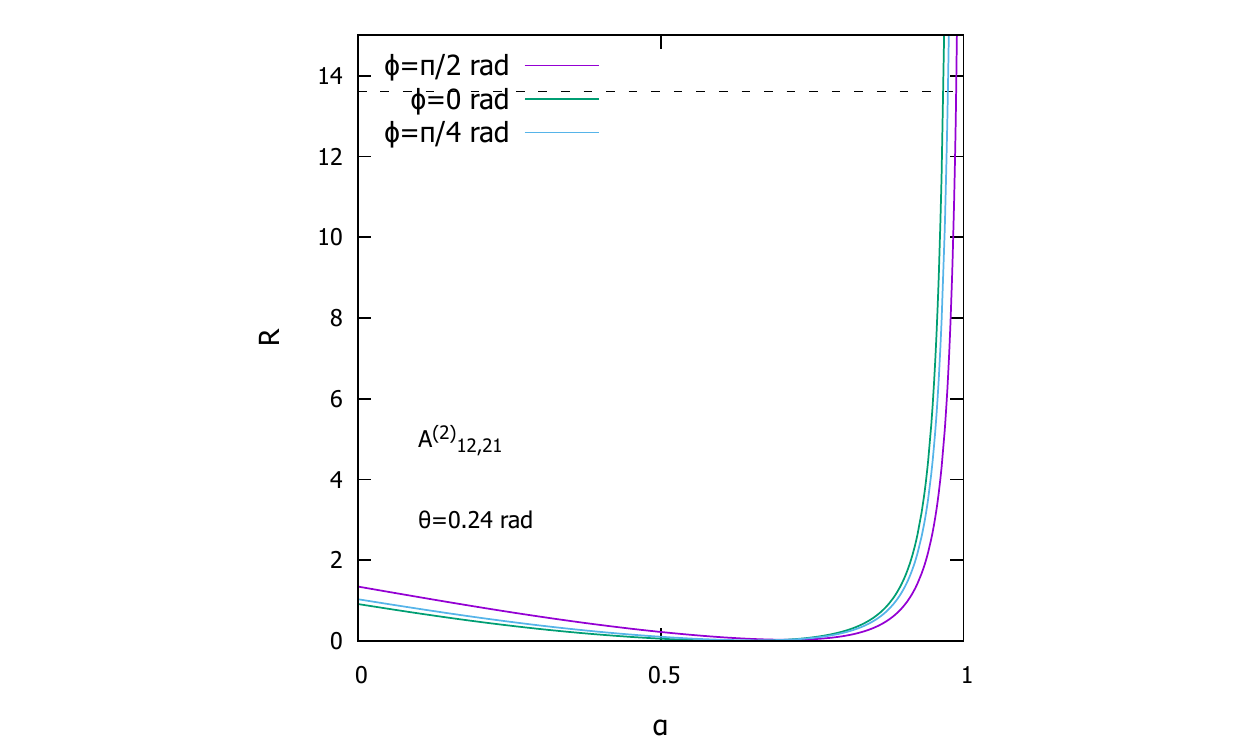}\\
\includegraphics[keepaspectratio, scale=0.5]{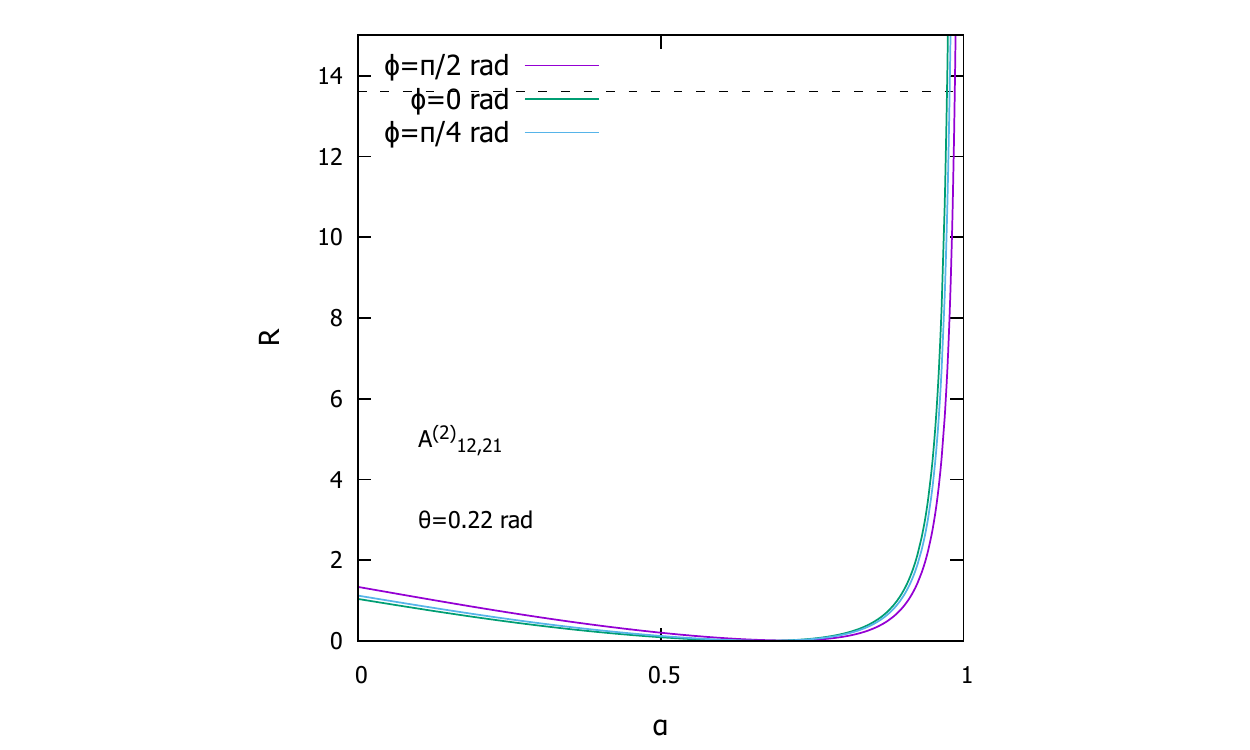}
\end{minipage}&
\begin{minipage}[t]{0.48\hsize}
\centering
\includegraphics[keepaspectratio, scale=0.5]{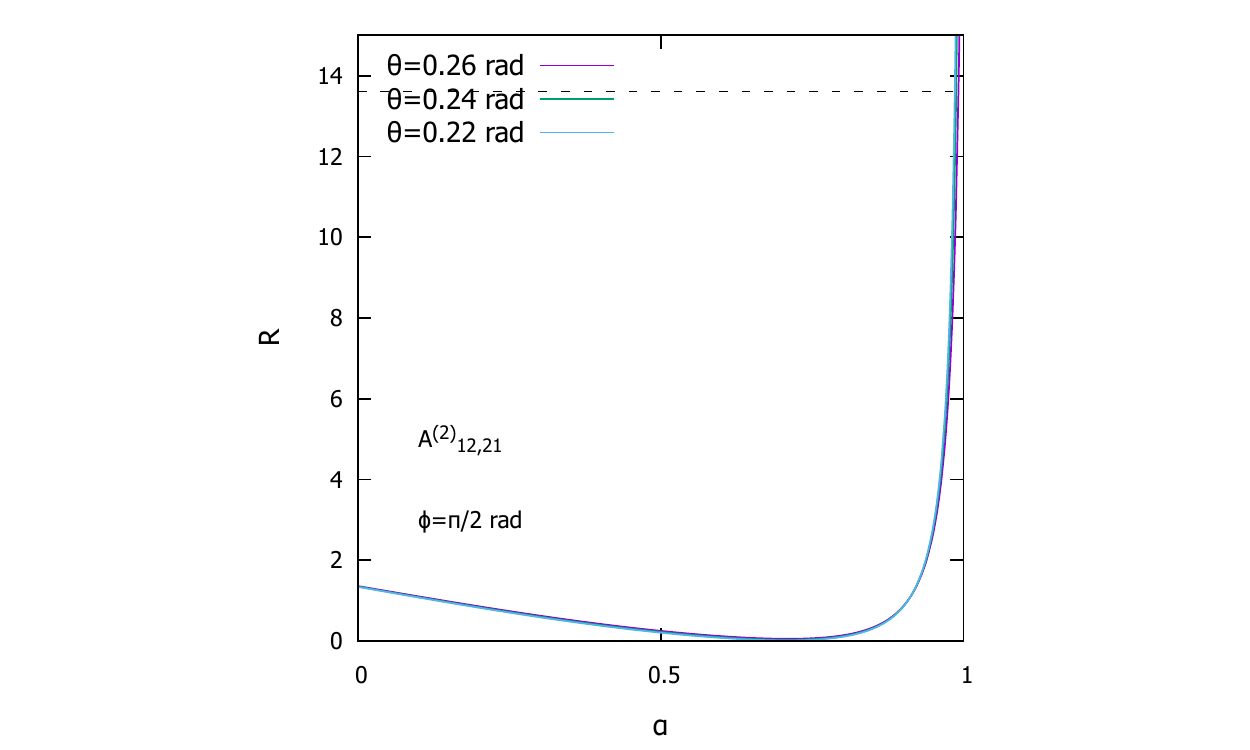}\\
\includegraphics[keepaspectratio, scale=0.5]{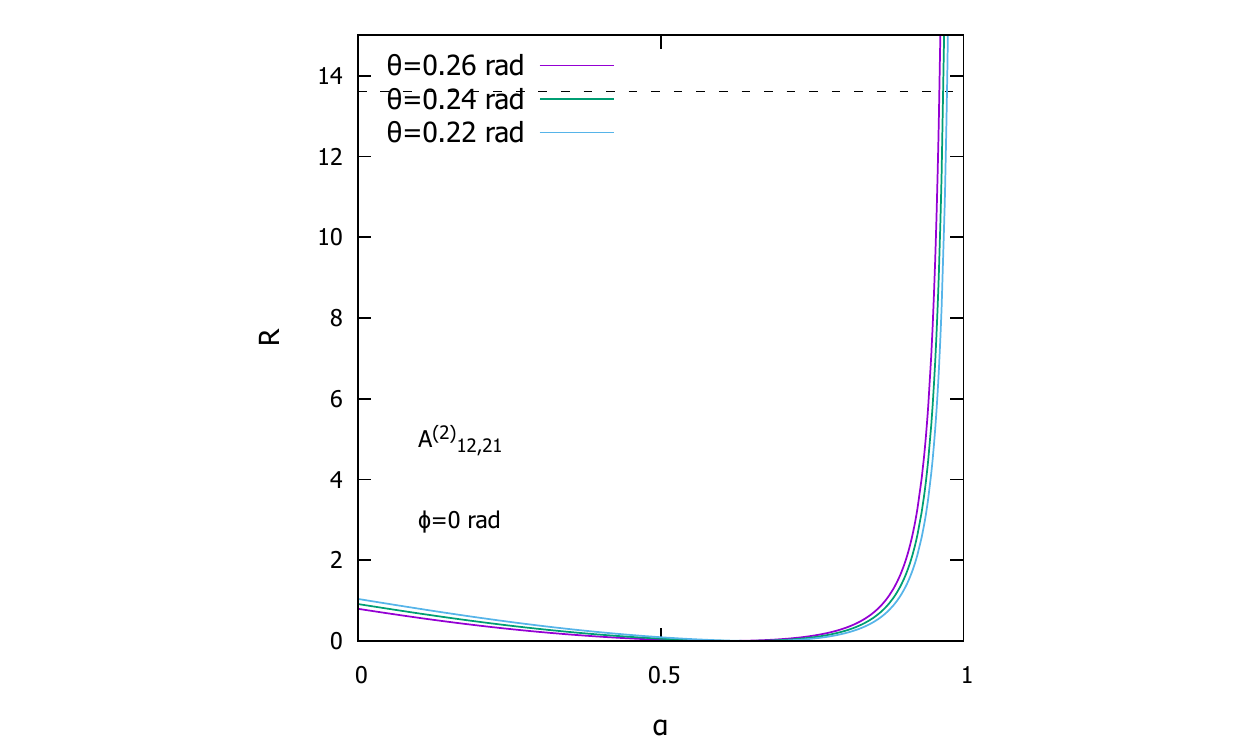}\\
\includegraphics[keepaspectratio, scale=0.5]{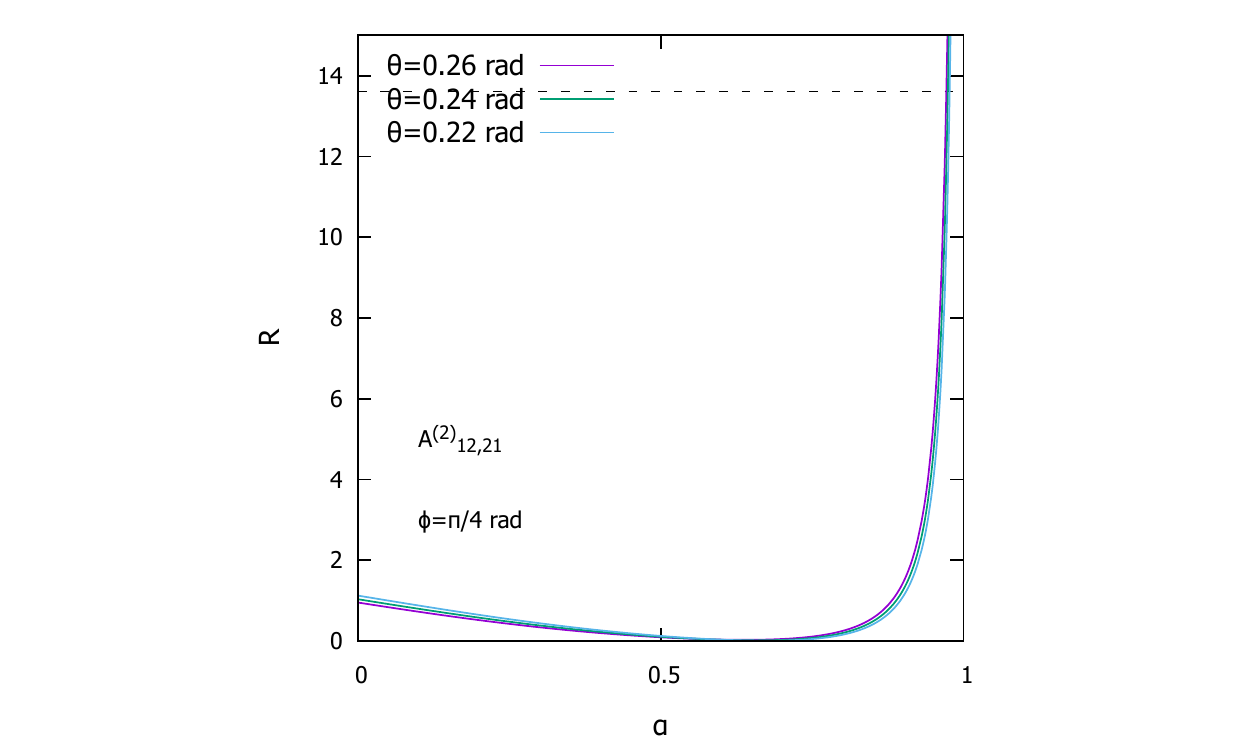}
 \end{minipage} \\
\end{tabular}
 \caption{Relation between $R$ and $\alpha$ in the case of MTM1($A_{12,21}^{(2)}$) for $\theta$ [rad] $=0.22, 0.24, 0.26$ (left-side panels) and  for $\phi$ [rad] $=0, \frac{\pi}{4}, \frac{\pi}{2}$ (right-side panels).}
 \label{Fig:MTM1_A1221_2_R} 
  \end{figure}

We modify TM1 mixing scheme using the rotation matrices $A_{12,21}^{(1)}$ and $A_{12,21}^{(2)}$ as follows: 
\begin{eqnarray}
\tilde{U}_1(A_{12,21}^{(1)}) = A_{12,21}^{(1)} U_1, \quad \tilde{U}_1(A_{12,21}^{(2)}) = A_{12,21}^{(2)} U_1.
\end{eqnarray}
We term $\tilde{U}_1(A_{12,21}^{(1)})$ and $\tilde{U}_1(A_{12,21}^{(2)})$ as MTM1($A_{12,21}^{(1)}$) mixing and MTM1($A_{12,21}^{(2)}$) mixing, respectively. The mixing angles are 
\begin{eqnarray}
s^2_{12} &=& \frac{5 \pm 4\alpha \sqrt{1-\alpha^2} - 3\alpha^2 - 6s^2_{13}}{6(1-s^2_{13})},  \label{Eq:MTM1_A1221_s12ss13s} \\
s^2_{23} &=& \frac{1}{6(1-s^2_{13})} \label{Eq:MTM1_A1221_s23ss13s} \left\{ 3\alpha^2 + (2-3\alpha^2 \mp 4\alpha \sqrt{1-\alpha^2})\sin^2\theta \right. \nonumber \\
  && \quad \left. + \sqrt{6}\alpha(\alpha \mp \sqrt{1-\alpha^2}) \sin 2\theta \cos\phi \right\}, \nonumber \\
s^2_{13} &=& \frac{1}{6} ( 3-3\alpha^2  -  B_1 + B_2 ), \label{Eq:MTM1_A1221_s13s}
\end{eqnarray}
where
\begin{eqnarray}
B_1 &=&  (1 - 3\alpha^2 \mp 4\alpha \sqrt{1-\alpha^2})\sin^2\theta,  \nonumber \\
B_2 &=& \sqrt{6}(1-\alpha^2 \pm \alpha \sqrt{1-\alpha^2}) \sin 2\theta \cos\phi.
\label{Eq:B1_B2}
\end{eqnarray}
The upper sign of $\pm$ and $\mp$ must be considered in the case of MTM1($A_{12,21}^{(1)}$), and the lower sign of $\pm$ and $\mp$ must be considered in the case of MTM1($A_{12,21}^{(2)}$).

Figure \ref{Fig:MTM1_A1221_12_13} illustrates the predicted values of $\theta_{13}$ and $\theta_{12}$ based on the MTM1($A_{12,21}^{(1)}$) and MTM1($A_{12,21}^{(2)}$). The horizontal and vertical dotted lines in each panel represent the same parameters as those represented in Fig. \ref{Fig:TM_13_12}. The upper panel indicates that the simultaneous reproducibility of $\theta_{12}$ and $\theta_{13}$ is diminished using the MTM1($A_{12,21}^{(1)}$). By contrast, the lower panel demonstrates that the simultaneous reproducibility of $\theta_{12}$ and $\theta_{13}$ is substantially enhanced using the MTM1($A_{12,21}^{(2)}$). For example, the best-fit values of $\theta_{12}$ and $\theta_{13}$ can be obtained simultaneously by introducing a small correction ($\alpha \simeq 0.9997$) in MTM1($A_{12,21}^{(2)}$). Therefore, we only consider the MTM1($A_{12,21}^{(2)}$) hereafter in this study.

Figure \ref{Fig:MTM1_A1221_cosphi_13_23} presents the predicted values of $\theta_{13}$ (upper panel) and $\theta_{23}$ (lower panel) as a function of $\phi$ in the MTM1($A_{12,21}^{(2)}$). The upper, middle, and lower horizontal dotted lines represent the $3\sigma$ upper limit, the best-fit value, and the $3\sigma$ lower limit of $\theta_{13}$ ($\theta_{23}$) in the NO case, respectively, in the upper (lower) panel. Fig. \ref{Fig:MTM1_A1221_cosphi_13_23} indicates that by appropriately selecting $\phi$ and $\theta$, we can obtain the values of $\theta_{13}$ and $\theta_{23}$ that are consistent with the observed data. Moreover, we confirmed that $s^2_{13}$ and $s^2_{23}$ can be obtained in the $3 \sigma$ region based on our numerical parameter search.

A benchmark point 
\begin{eqnarray}
(\alpha, \theta, \phi) = (0.99973, \ 15.50^\circ,\  82.56^\circ),
\label{Eq:MTM1_A1221_benchmark}
\end{eqnarray}
yields the best-fit values of $s_{12}^2$ and $s_{13}^2$ and the allowed value of $s_{23}^2$ as follows:
\begin{eqnarray}
(s_{12}^2, s_{23}^2, s_{13}^2, \delta) = (0.303, \ 0.529, \ 0.02225, \ 269.5^\circ). 
\end{eqnarray}

The reader may naturally ask the following issues:
\begin{description}
\item[Q1:] What values should be assigned to the parameters $\theta$, $\phi$ and $\alpha$, in order to obtain a reasonable leptonic mixing matrix? Can they be varied in a wide range? Are they correlated? 
\item[Q2:] What if the values of $\theta_{23}$ and the CP phase $\delta$ are finally pinned down? 
\end{description}
To answer these questions, we show Figures \ref{Fig:MTM1_A1221_2_a1p1_12_13} - \ref{Fig:MTM1_A1221_2_at_d}. These figures present the detailed behaviors of the parameters and the predicted values for $\alpha=0.995, 0.997, 0.999$ in the case of MTM1($A_{12,21}^{(2)}$). The dotted lines in these figures represent the best-fit value (center dotted line) and $3 \sigma$ region (upper and lower dotted lines) of the physical quantity which is shown in vertical axis. The left-side (right-side) panels in these figures excepted with Figure \ref{Fig:MTM1_A1221_2_a1p1_12_13} show the allowed region of the parameters, etc., for NO (IO).  The physical quantities compared in these figures are 
\begin{description}
\item[Figure \ref{Fig:MTM1_A1221_2_a1p1_12_13}:] $s_{12}^2$ (left-side panels) or $s_{13}^2$ (right-side panels) vs $\theta$ for $\phi$ [rad] $=0, \frac{\pi}{4}, \frac{\pi}{2}$,
\item[Figure \ref{Fig:MTM1_A1221_2_ap_23}:] $s_{23}^2$ vs $\theta$ for $\phi$ [rad] $=0, \frac{\pi}{4}, \frac{\pi}{2}$,
\item[Figure \ref{Fig:MTM1_A1221_2_at_23}:] $s_{23}^2$ vs $\phi$ for $\theta$ [rad] $=0.20, 0.25, 0.30$,
\item[Figure \ref{Fig:MTM1_A1221_2_ap_d}:] $\delta$ vs $\theta$ for $\phi$ [rad] $=0, \frac{\pi}{4}, \frac{\pi}{2}$,
\item[Figure \ref{Fig:MTM1_A1221_2_at_d}:] $\delta$ vs $\phi$ for $\theta$ [rad] $=0.20, 0.25, 0.30$.
\end{description}
From these figures, we can conclude that
\begin{description}
\item[A1:] The wide range of parameters $\theta$, $\phi$ and $\alpha$ are consistent with observation (see Figures \ref{Fig:MTM1_A1221_2_a1p1_12_13}, \ref{Fig:MTM1_A1221_2_ap_23} and \ref{Fig:MTM1_A1221_2_at_23}). The wide range of Dirac CP phase is also consistent with observation (see Figures \ref{Fig:MTM1_A1221_2_ap_d} and \ref{Fig:MTM1_A1221_2_at_d}).
\item[A2:] If the values of $\theta_{23}$ and the CP phase $\delta$ are finally pinned down, we can reproduce these fixed values with appropriate values of $\alpha$, $\theta$ and $\phi$.
\end{description}

Moreover, the reader may also naturally ask the following question:
\begin{description}
\item[Q3:] What if the best-fit values change in the future?
\end{description}
To answer this critical question, first, we define the ratio
\begin{eqnarray}
R=\frac{s_{12}^2}{s_{13}^2}.
\label{Eq:R_1213}
\end{eqnarray}
The value of $R$ for the current best-fit values is calculated as
\begin{eqnarray}
R_{\rm{best-fit}}=13.618.
\label{Eq:R_1213}
\end{eqnarray}
Then, we show Figure \ref{Fig:MTM1_A1221_2_R}. Figure \ref{Fig:MTM1_A1221_2_R} presents the relation between $R$ and $\alpha$ in the case of MTM1($A_{12,21}^{(2)}$) for $\theta$ [rad] $=0.22, 0.24, 0.26$ (left-side panels) and  for $\phi$ [rad] $=0, \frac{\pi}{4}, \frac{\pi}{2}$ (right-side panels). The dotted lines in the panels represent $R_{\rm{best-fit}}$. From Figure \ref{Fig:MTM1_A1221_2_R}, it turned out that 
\begin{description}
\item[A3:] if the best-fit values (or equivalently the ratio $R_{\rm{best-fit}}$) change in the future, the new best-fit values can be reproduced with appropriate selection of the values of $\alpha$, $\theta$, $\phi$.
\end{description}
%

\subsection{$A_{23,32}^{(1)}$ and $A_{23,32}^{(2)}$}
We modify TM1 mixing using the rotation matrix $A_{23,32}^{(1)}$ and $A_{23,32}^{(2)}$ as follows:
\begin{eqnarray}
\tilde{U}_1(A_{23,32}^{(1)}) = A_{23,32}^{(1)} U_1, \quad \tilde{U}_1(A_{23,32}^{(2)}) = A_{23,32}^{(2)} U_1, 
\end{eqnarray}
and we call these mixings MTM1($A_{23,32}^{(1)}$) and MTM1($A_{23,32}^{(2)}$), respectively. The mixing angles are 
\begin{eqnarray}
s^2_{12} &=& \frac{1-3s^2_{13}}{3(1-s^2_{13})}, \label{Eq:MTM1_A2332_s12ss13s} \\
s^2_{23} &=& \frac{1}{12(1-s^2_{13})} \left\{5 \mp  2 \alpha \sqrt{1-\alpha^2}  + (1 \mp 10\alpha \sqrt{1-\alpha^2})\cos 2\theta \right. \nonumber \\
  && \quad \left. - 2\sqrt{6}(1- 2\alpha^2) \sin 2\theta \cos\phi \right\}, \label{Eq:MTM1_A2332_s23ss13s}\nonumber \\
s^2_{13} &=& \frac{1}{3}\sin^2\theta. \label{Eq:MTM1_A2332_s13s}
\end{eqnarray}
The negative sign of $\mp$ should be taken for MTM1$(A_{23,32}^{(1)})$, and the positive sign should be taken for MTM1$(A_{23,32}^{(2)})$.

Eq.(\ref{Eq:MTM1_A2332_s12ss13s}) is equal to Eq.(\ref{Eq:s12ss13sUTM1}). Moreover, Eq.(\ref{Eq:MTM1_A2332_s13s}) is equal to Eq.(\ref{Eq:s13sUTM1}). Therefore, the simultaneous reproducibility of $\theta_{12}$ and $\theta_{13}$ is not improved in MTM1($A_{23,32}^{(1)}$) and MTM1($A_{23,32}^{(2)}$). For this reason, no further analysis is performed for MTM1($A_{23,32}^{(1)}$) and MTM1($A_{23,32}^{(2)}$).

\subsection{$A_{13,31}^{(1)}$ and $A_{13,31}^{(2)}$}

\begin{figure}[t]
\begin{center}
\includegraphics[scale=1.0]{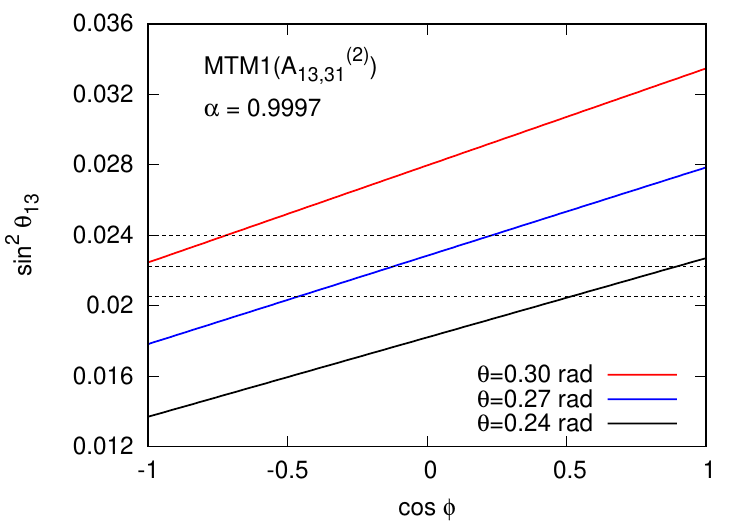}
\includegraphics[scale=1.0]{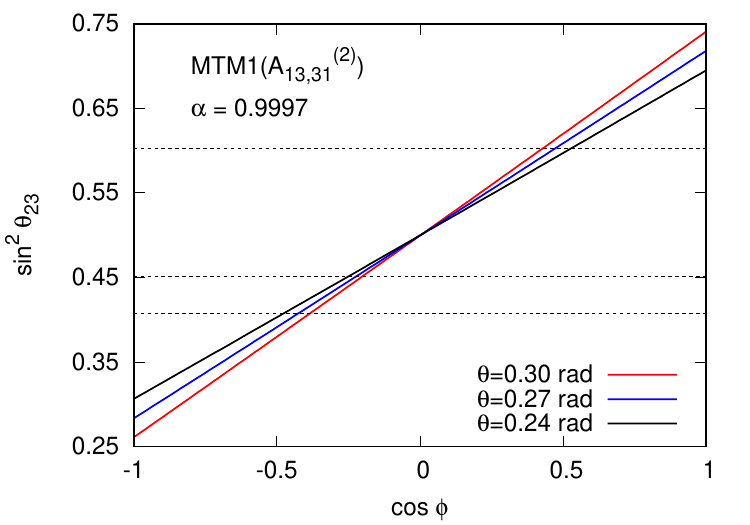}
\caption{Same as Fig. \ref{Fig:MTM1_A1221_cosphi_13_23} but for MTM1($A_{13,31}^{(2)}$).}
\label{Fig:MTM1_A1331_cosphi_13_23} 
\end{center}
\end{figure}

\begin{figure}[t]
\begin{tabular}{cc}
\begin{minipage}[t]{0.3\hsize}
\centering
\includegraphics[keepaspectratio, scale=0.5]{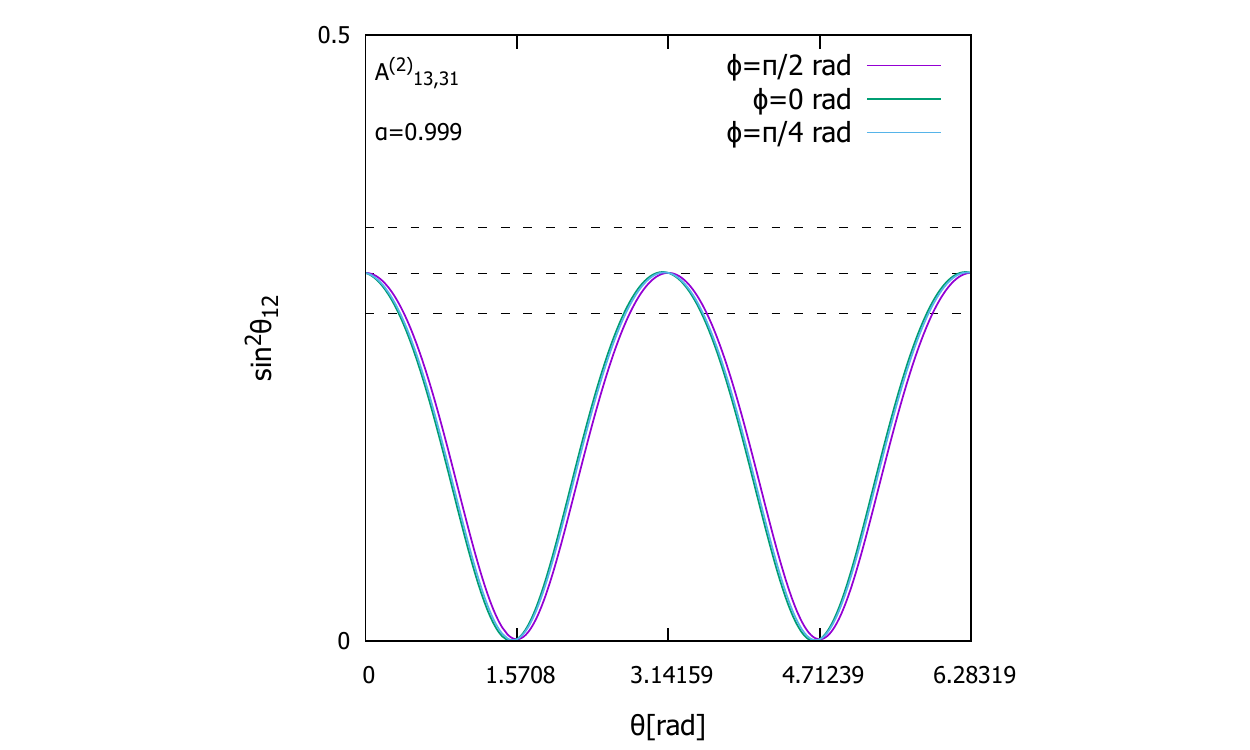}\\
\includegraphics[keepaspectratio, scale=0.5]{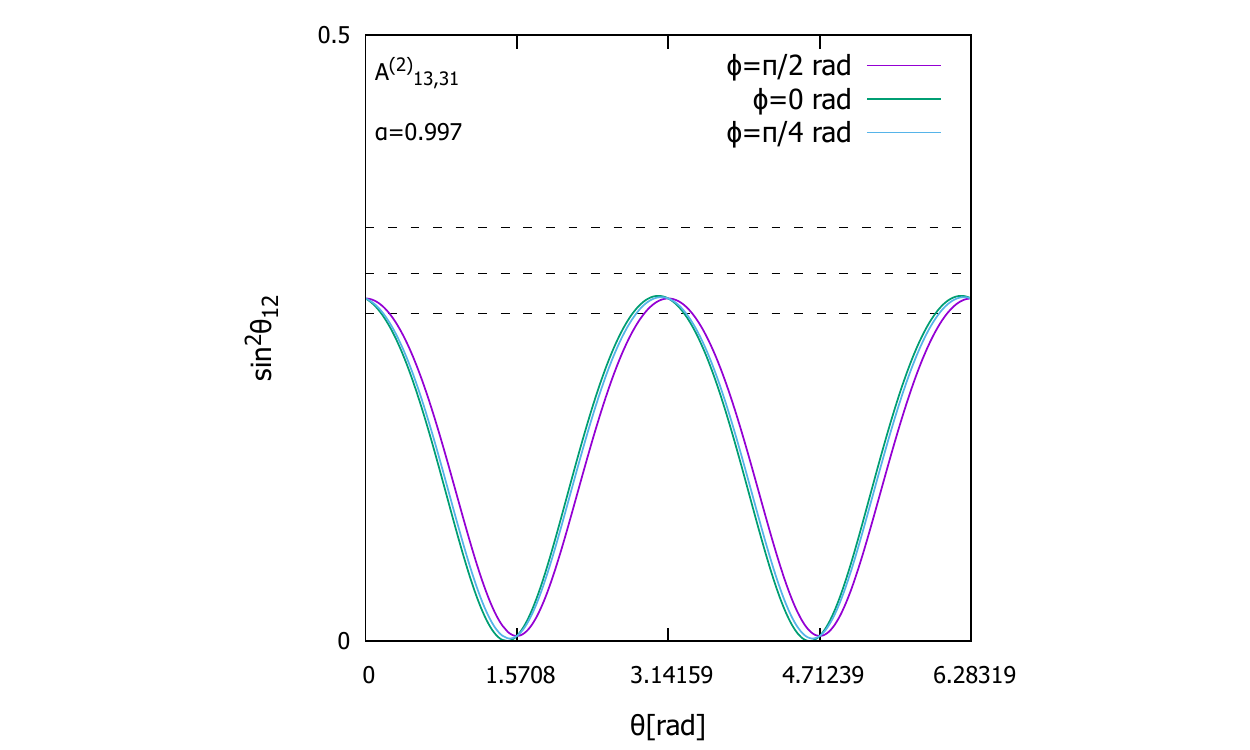}\\
\includegraphics[keepaspectratio, scale=0.5]{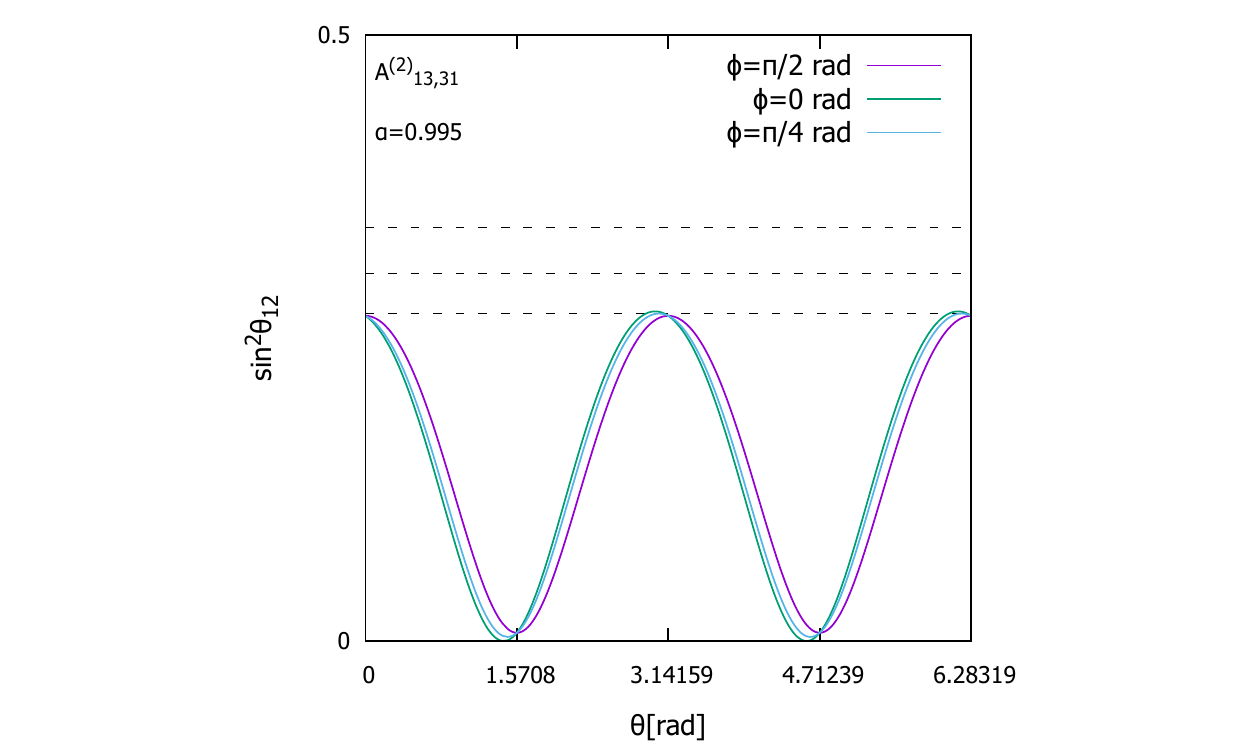}
\end{minipage}&
\begin{minipage}[t]{0.3\hsize}
\centering
\includegraphics[keepaspectratio, scale=0.5]{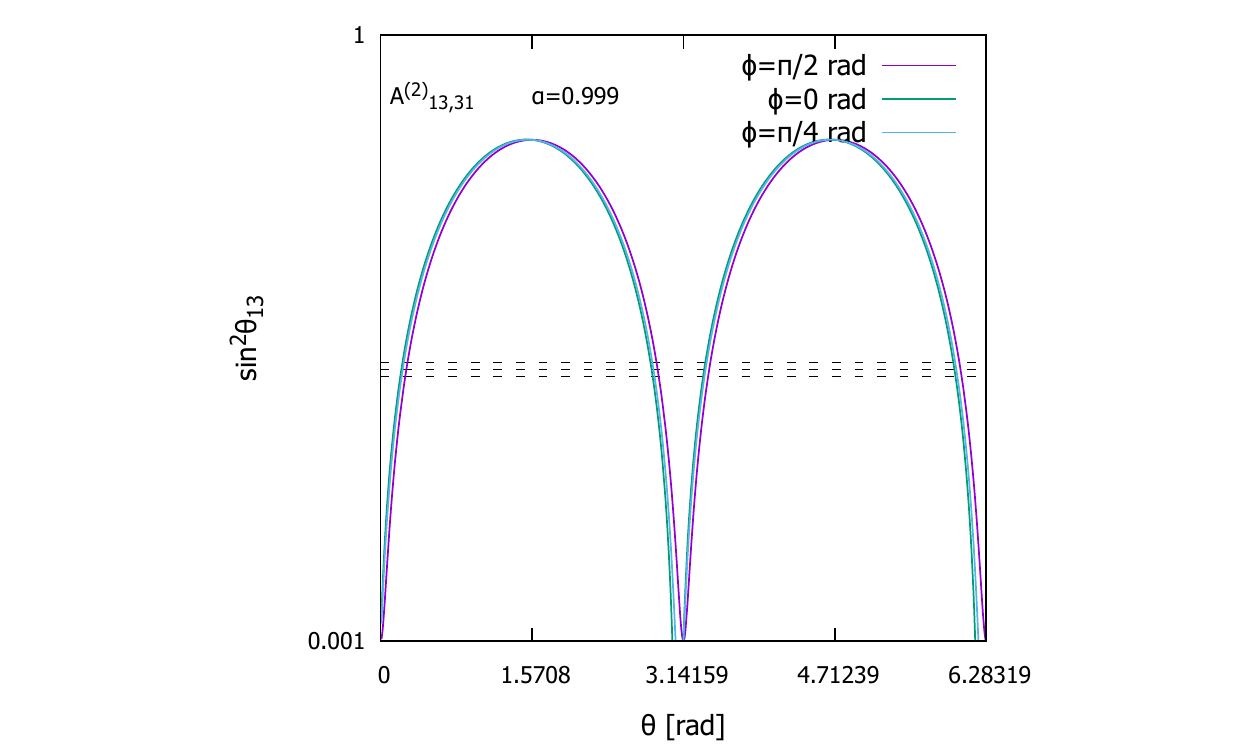}\\
\includegraphics[keepaspectratio, scale=0.5]{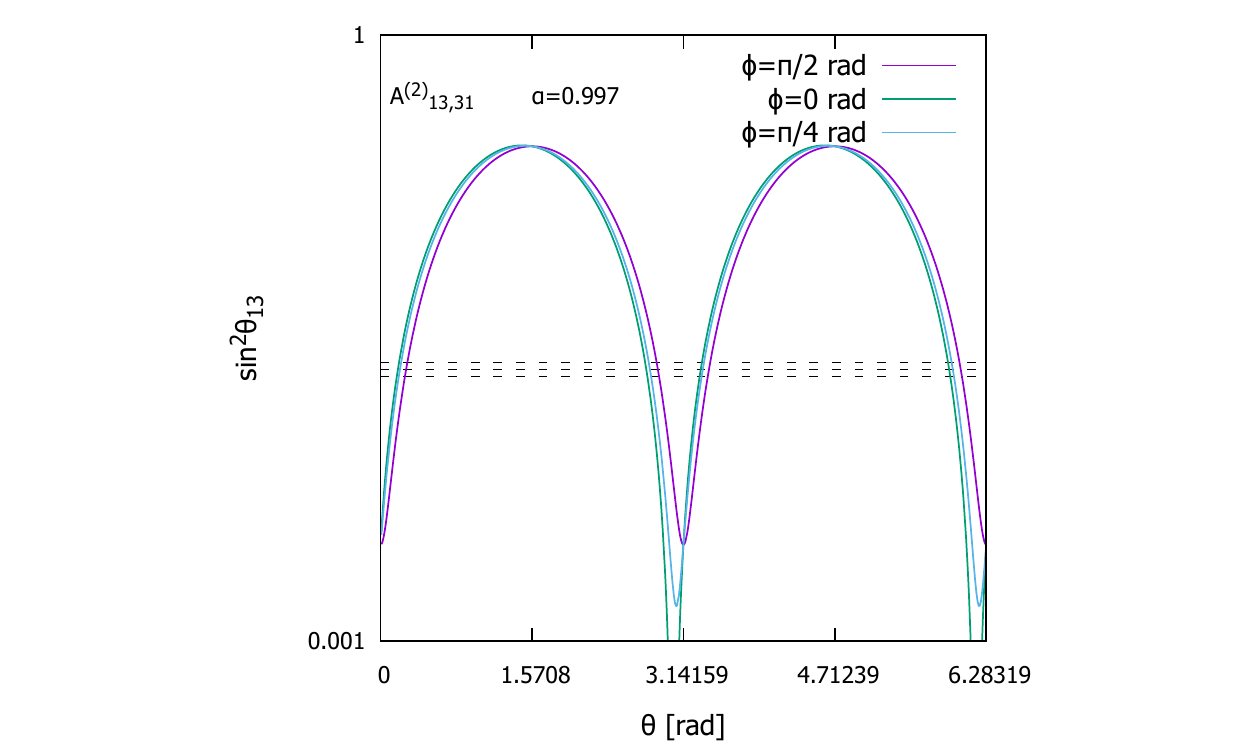}\\
\includegraphics[keepaspectratio, scale=0.5]{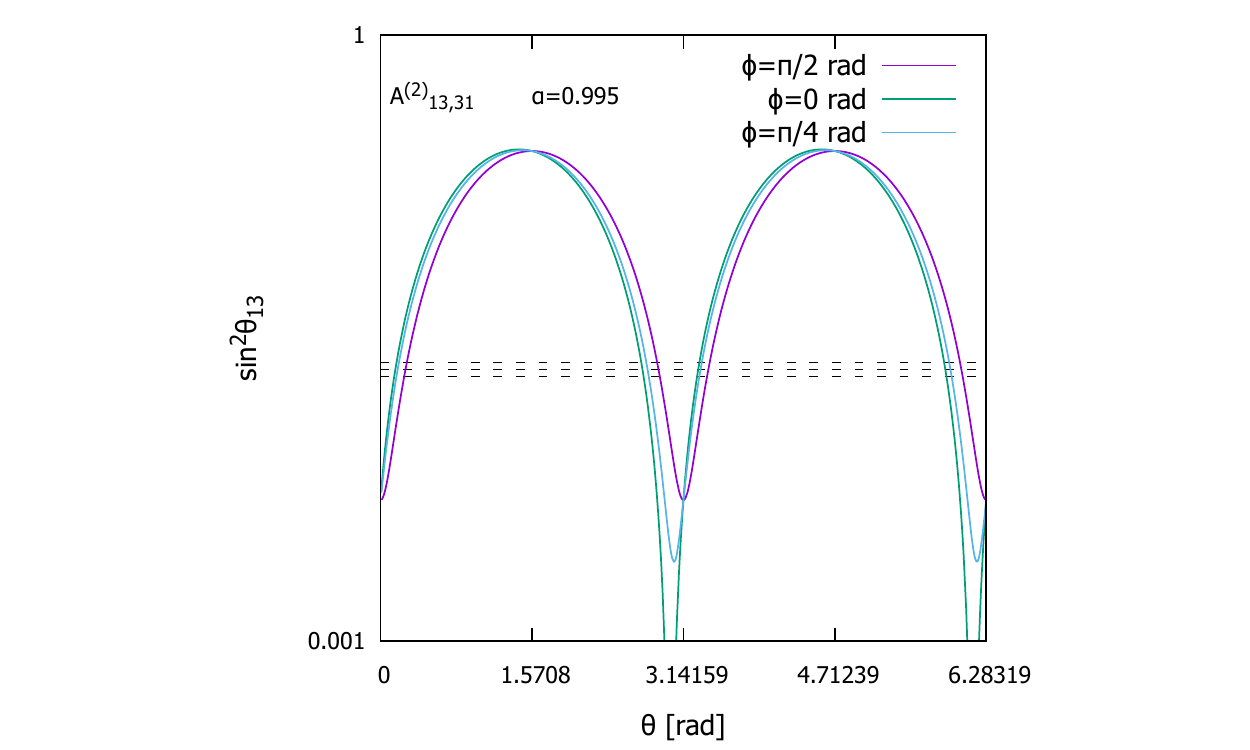}
 \end{minipage} \\
\end{tabular}
 \caption{Same as Fig. \ref{Fig:MTM1_A1221_2_a1p1_12_13} but for MTM1($A_{13,31}^{(2)}$).}
 \label{Fig:MTM1_A1331_2_a1p1_12_13} 
  \end{figure}

\begin{figure}[t]
\begin{tabular}{cc}
\begin{minipage}[t]{0.48\hsize}
\centering
\includegraphics[keepaspectratio, scale=0.5]{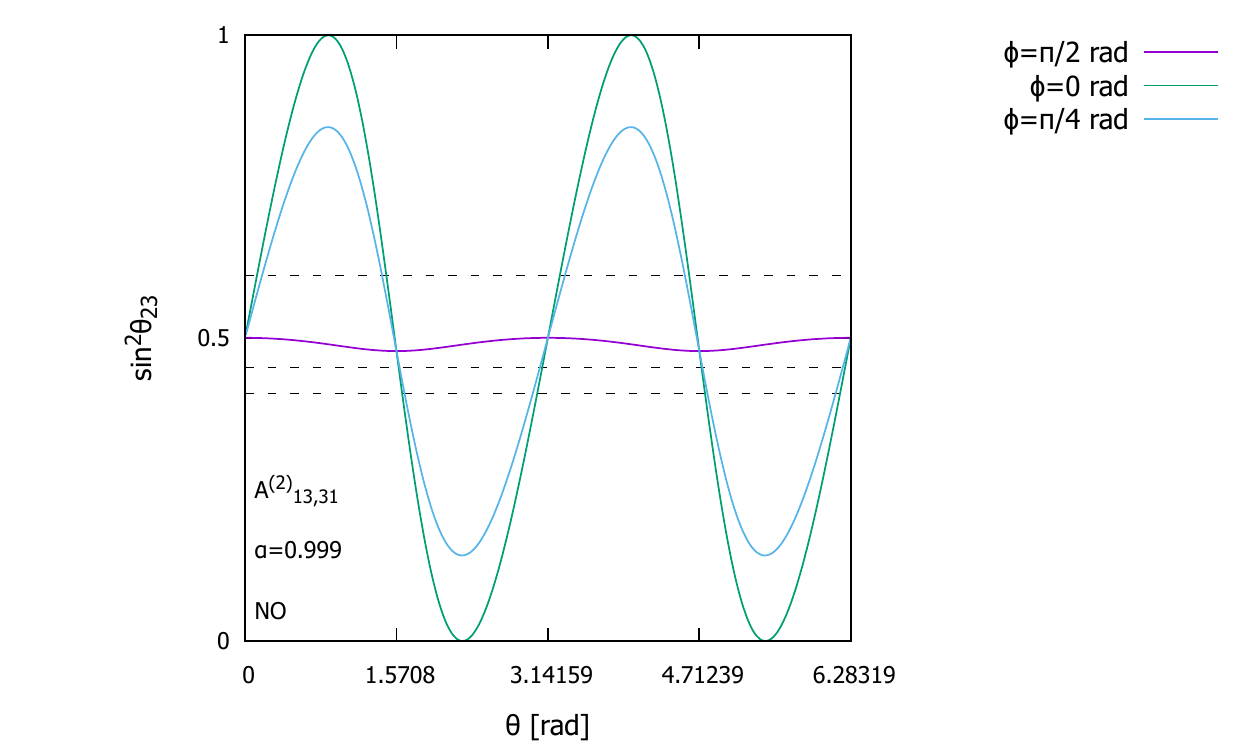}\\
\includegraphics[keepaspectratio, scale=0.5]{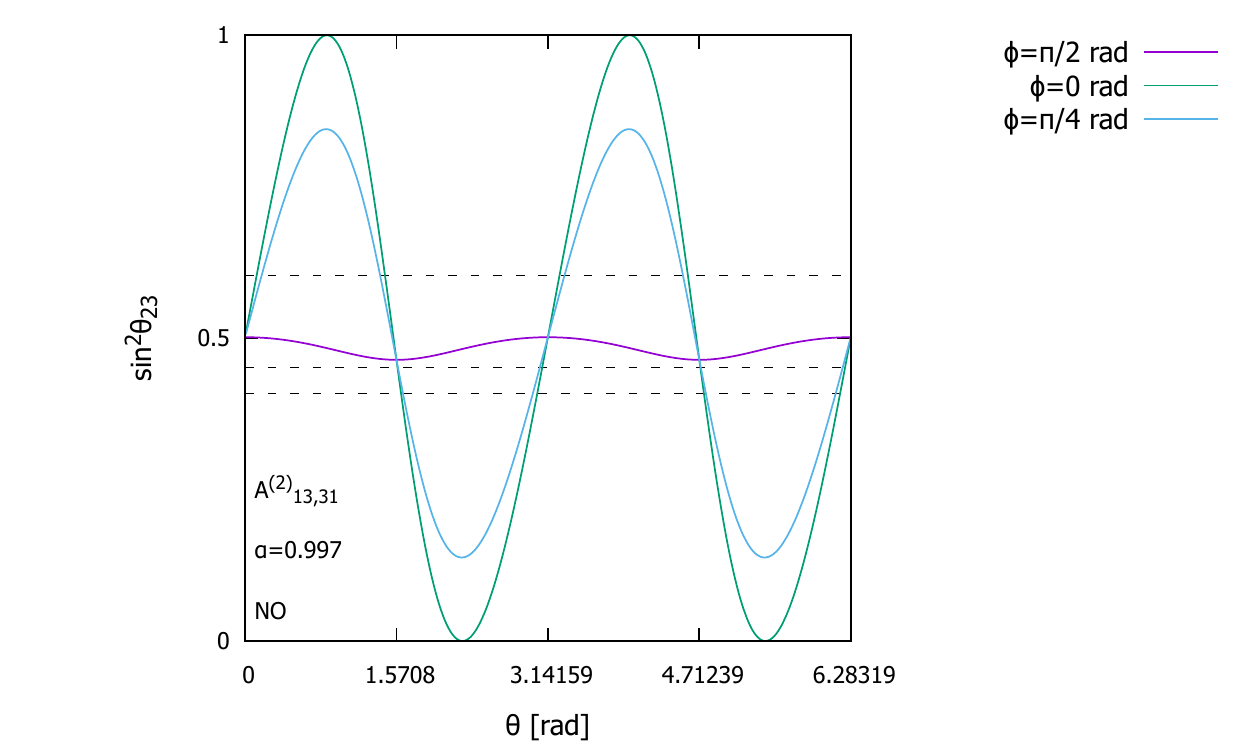}\\
\includegraphics[keepaspectratio, scale=0.5]{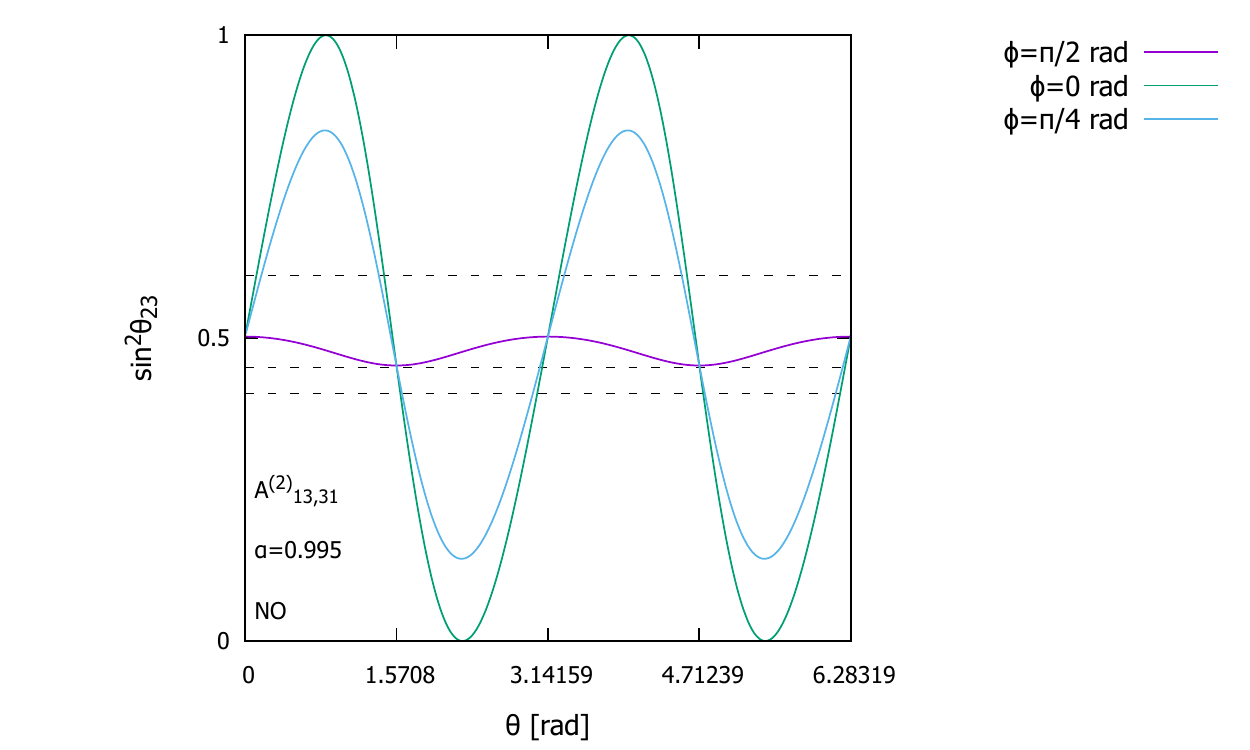}
\end{minipage}&
\begin{minipage}[t]{0.48\hsize}
\centering
\includegraphics[keepaspectratio, scale=0.5]{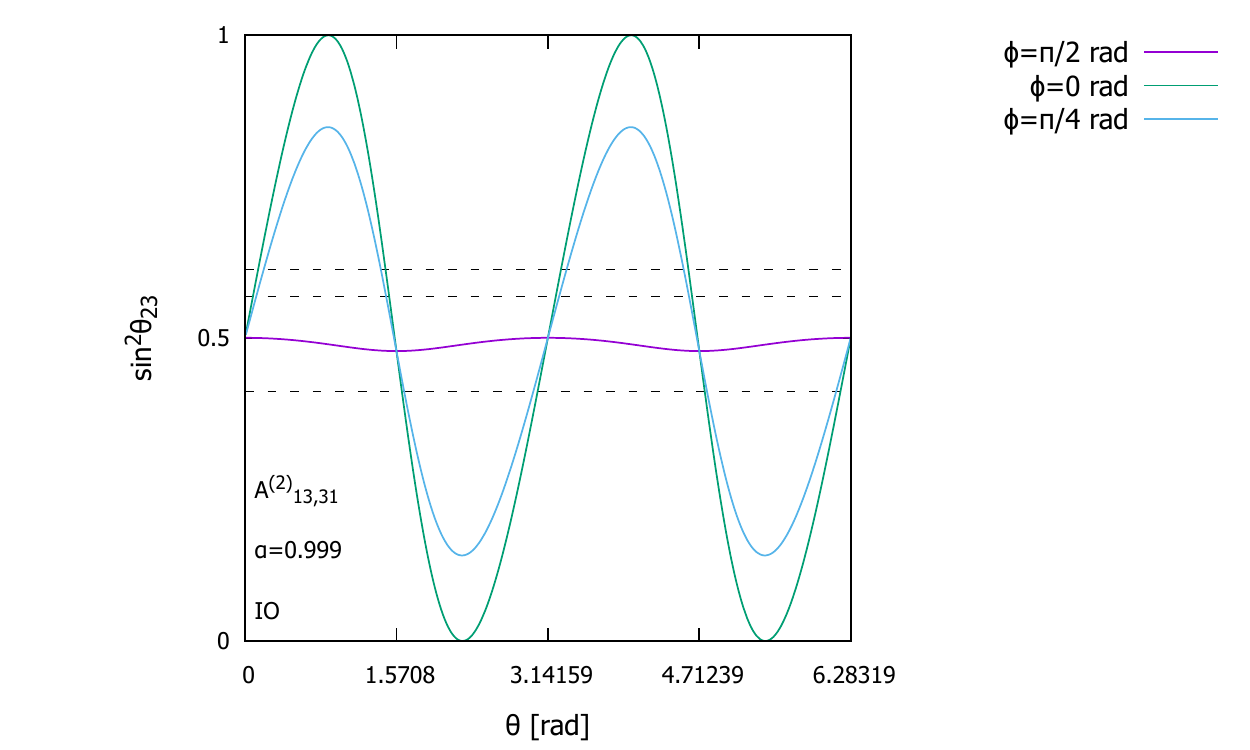}\\
\includegraphics[keepaspectratio, scale=0.5]{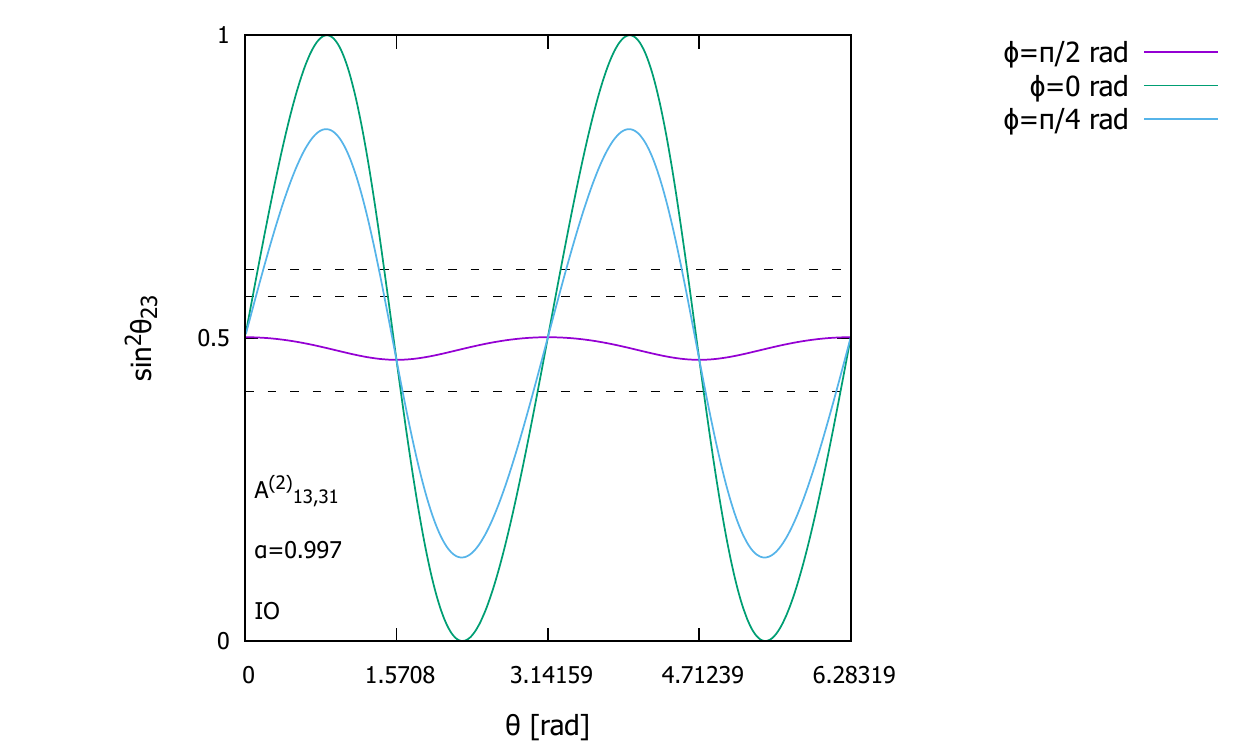}\\
\includegraphics[keepaspectratio, scale=0.5]{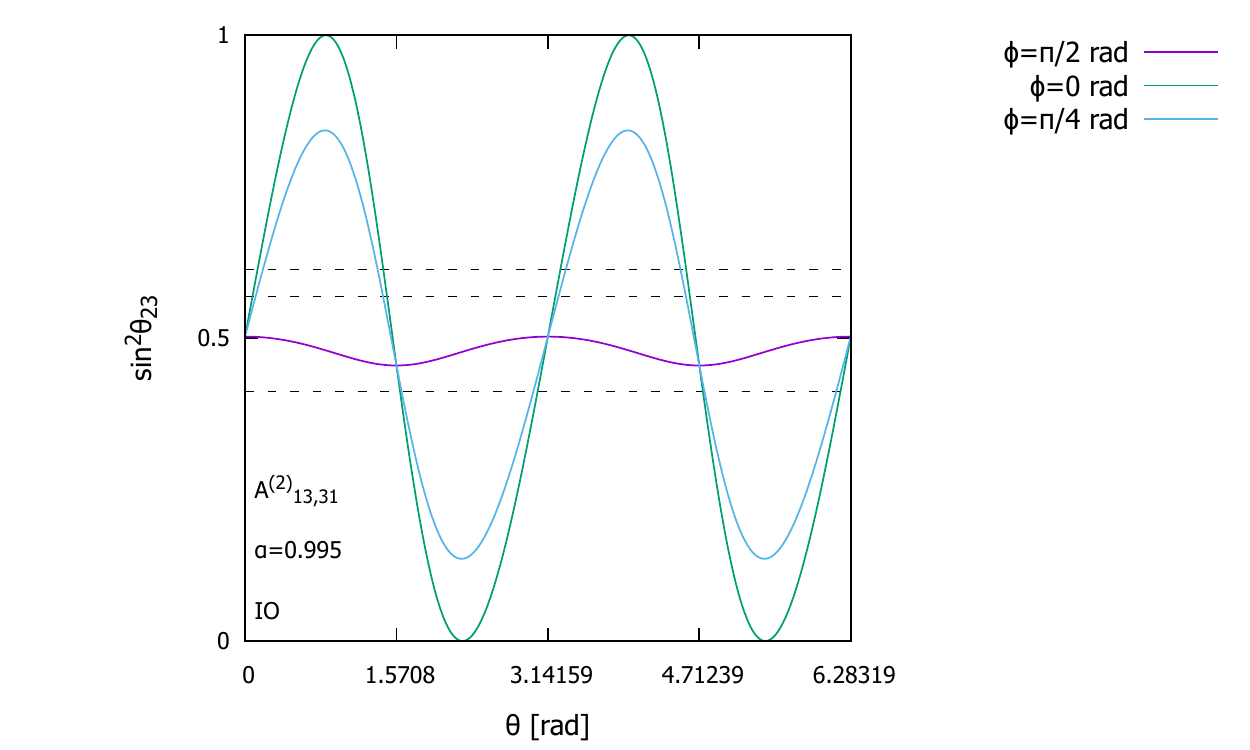}
 \end{minipage} \\
\end{tabular}
 \caption{Same as Fig. \ref{Fig:MTM1_A1221_2_ap_23} but for MTM1($A_{13,31}^{(2)}$)}
 \label{Fig:MTM1_A1331_2_ap_23} 
  \end{figure}

\begin{figure}[t]
\begin{tabular}{cc}
\begin{minipage}[t]{0.48\hsize}
\centering
\includegraphics[keepaspectratio, scale=0.5]{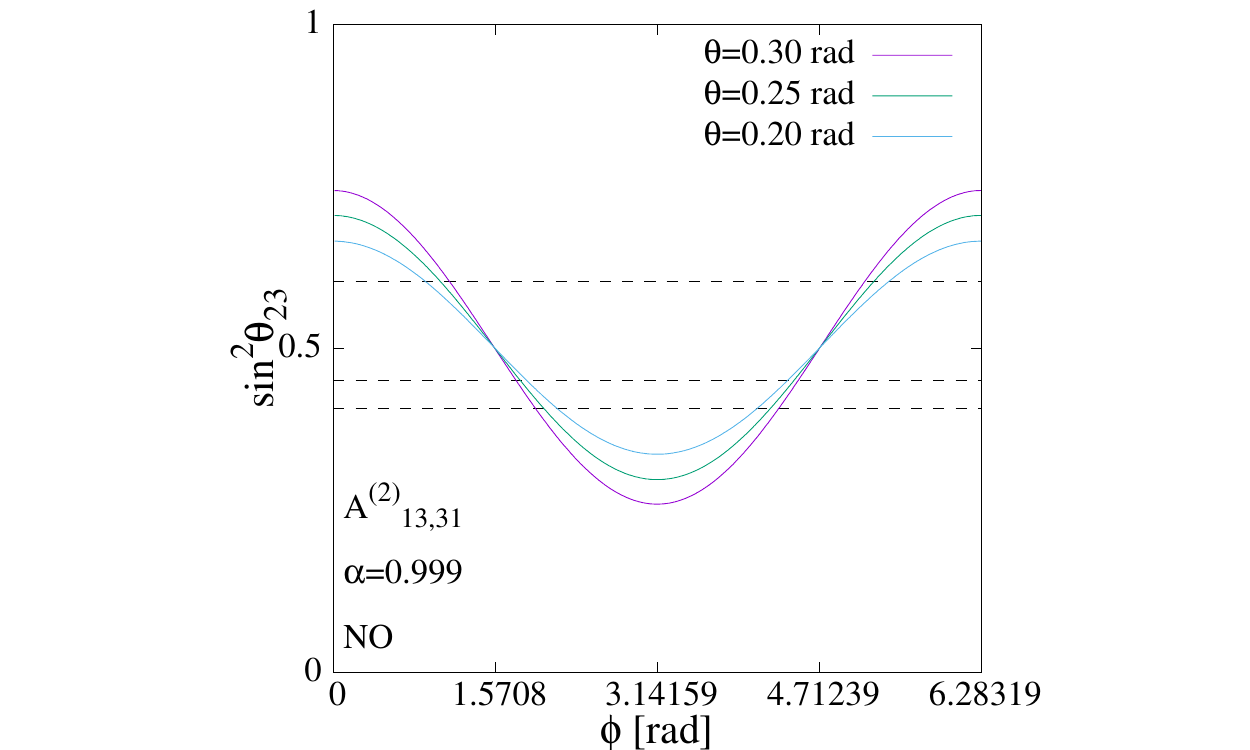}\\
\includegraphics[keepaspectratio, scale=0.5]{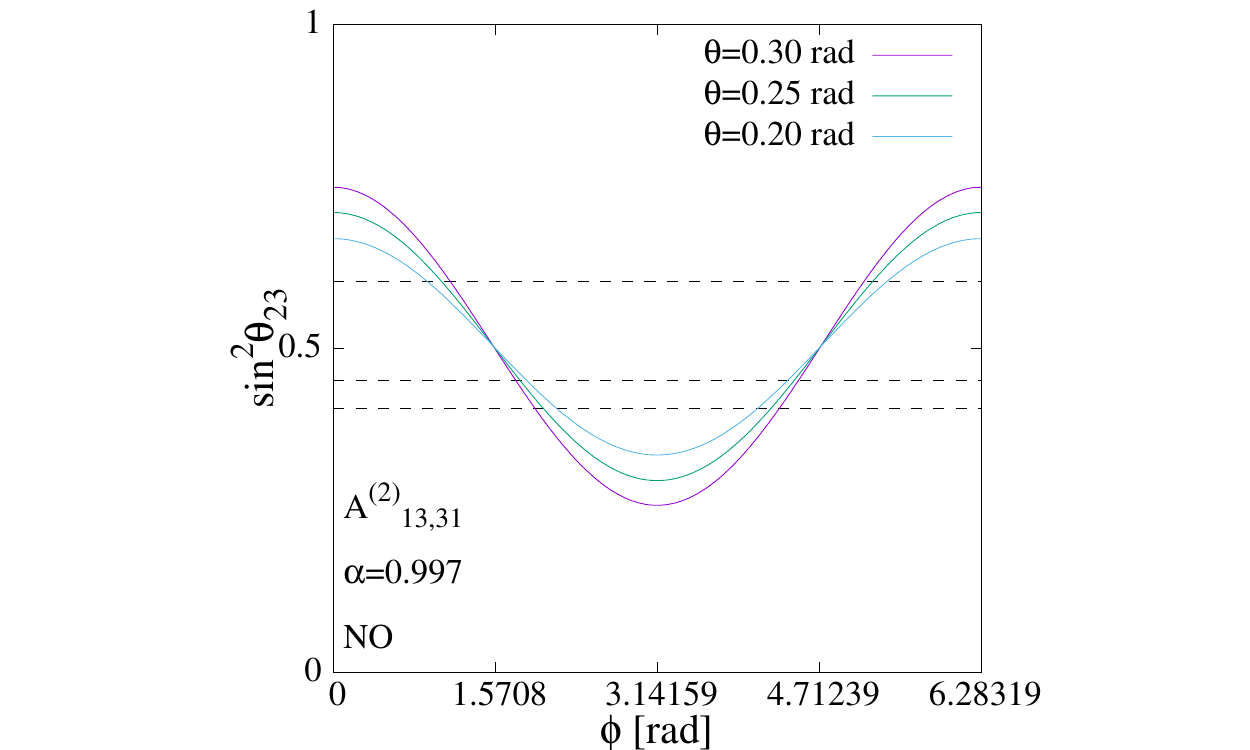}\\
\includegraphics[keepaspectratio, scale=0.5]{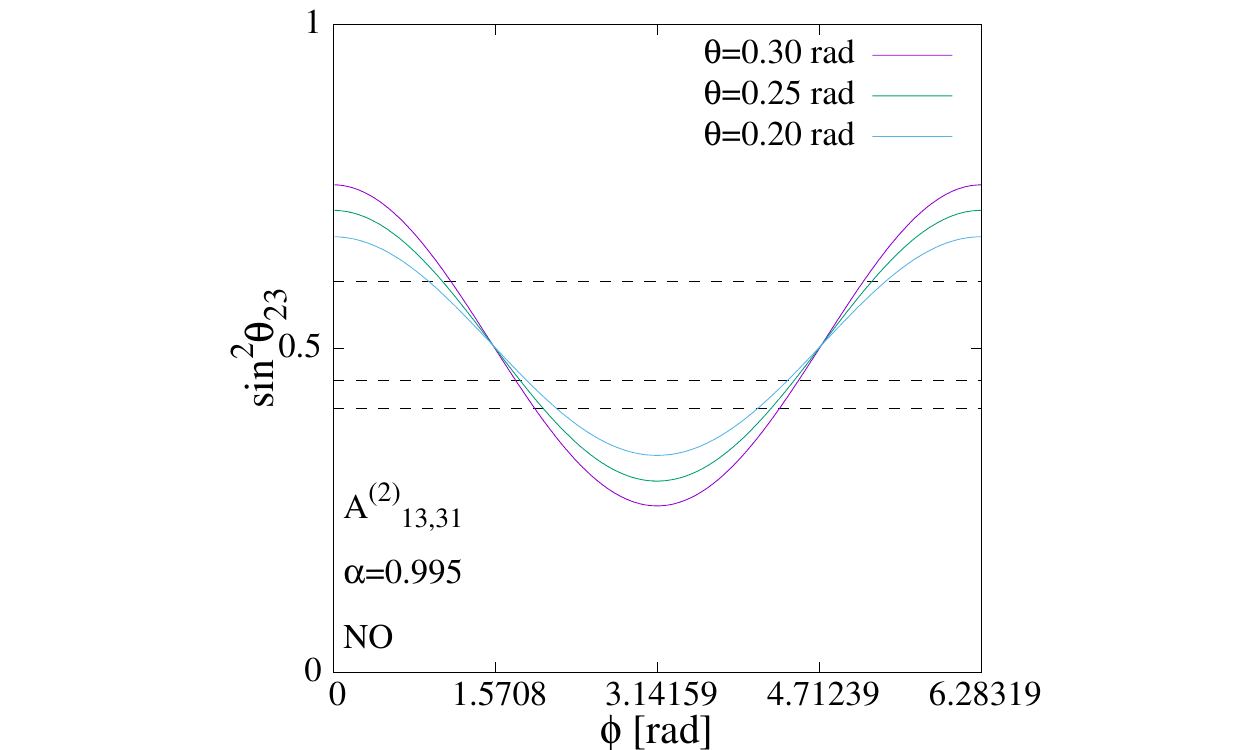}
\end{minipage}&
\begin{minipage}[t]{0.48\hsize}
\centering
\includegraphics[keepaspectratio, scale=0.5]{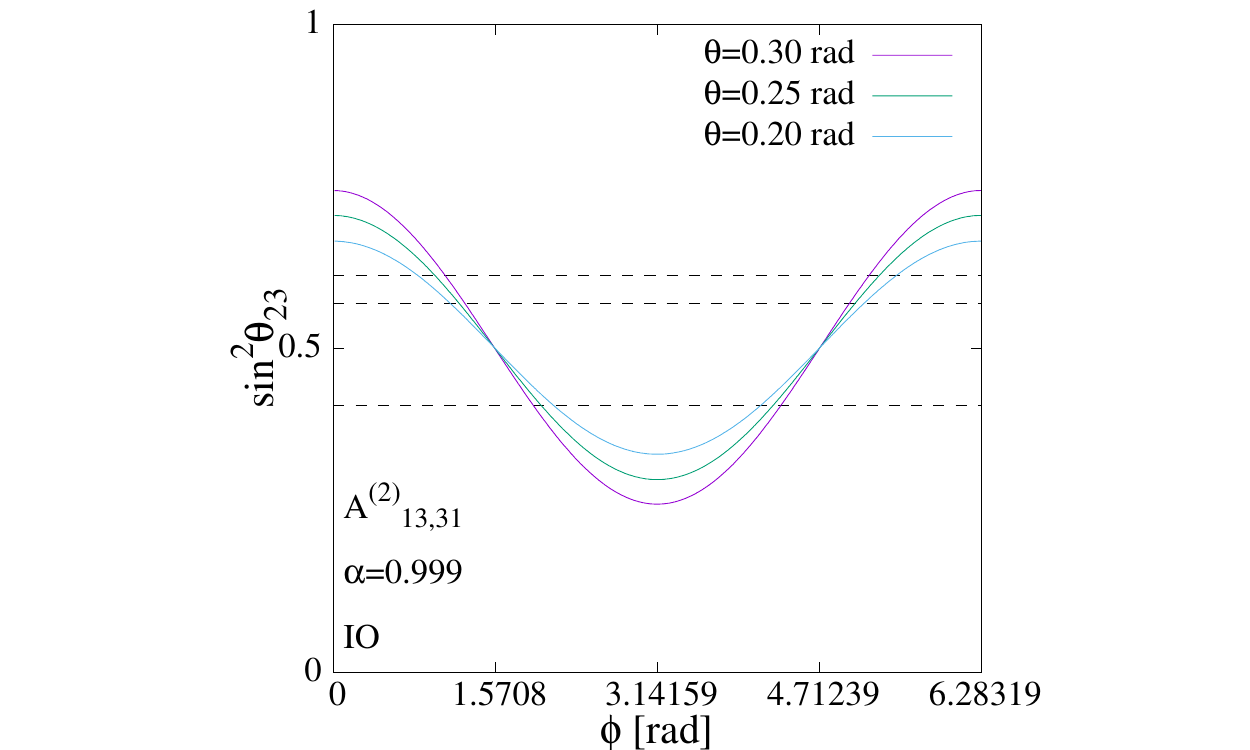}\\
\includegraphics[keepaspectratio, scale=0.5]{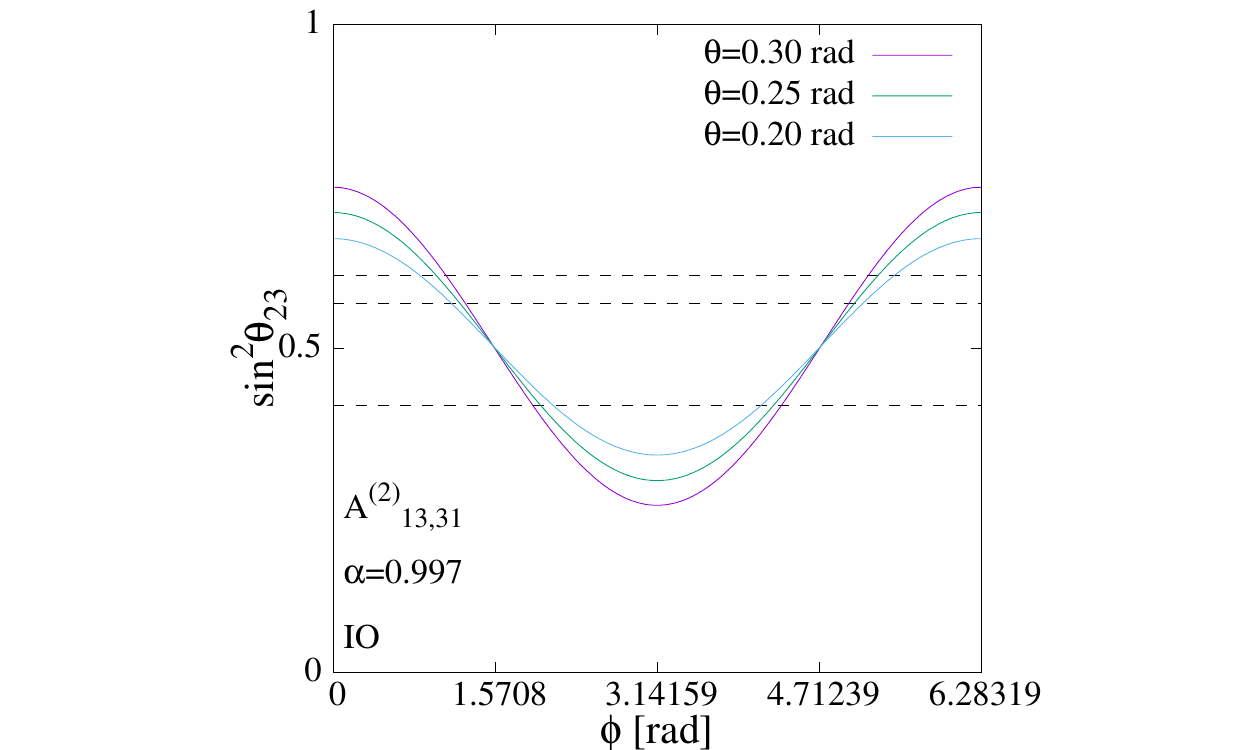}\\
\includegraphics[keepaspectratio, scale=0.5]{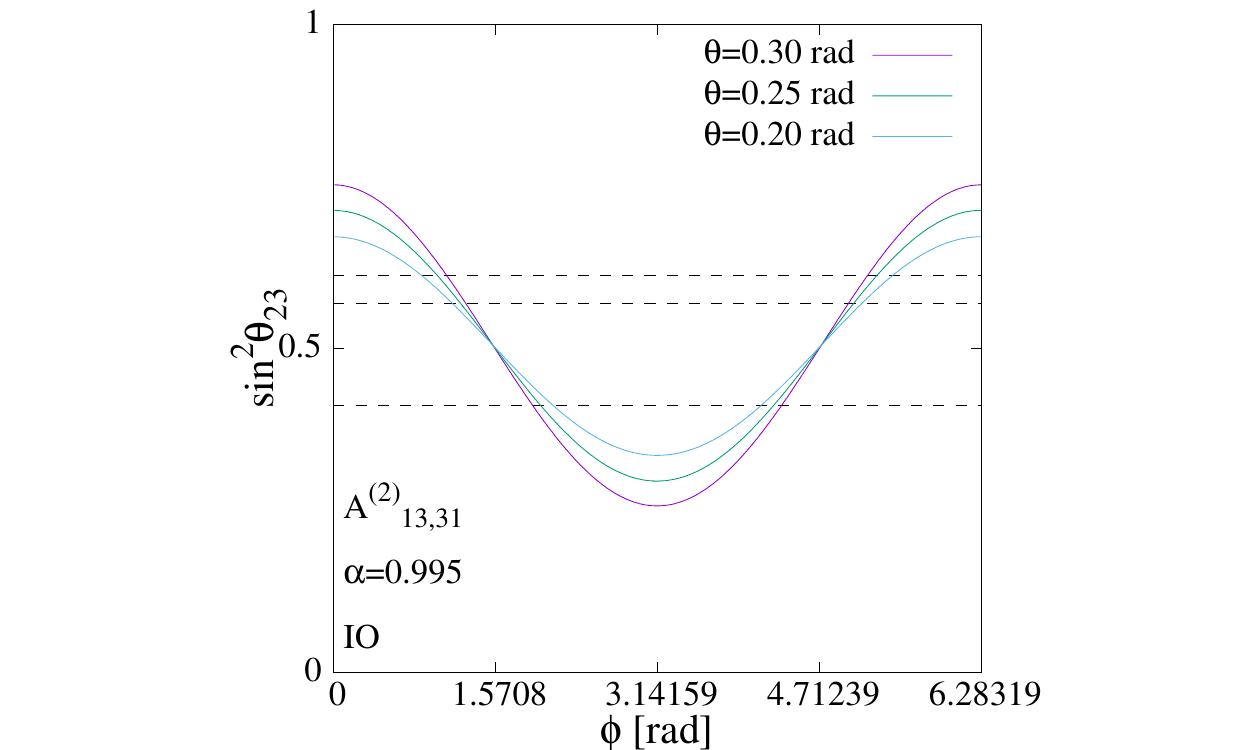}
 \end{minipage} \\
\end{tabular}
 \caption{Same as Fig. \ref{Fig:MTM1_A1221_2_at_23} but for MTM1($A_{13,31}^{(2)}$)}
 \label{Fig:MTM1_A1331_2_at_23} 
  \end{figure}

\begin{figure}[t]
\begin{tabular}{cc}
\begin{minipage}[t]{0.48\hsize}
\centering
\includegraphics[keepaspectratio, scale=0.5]{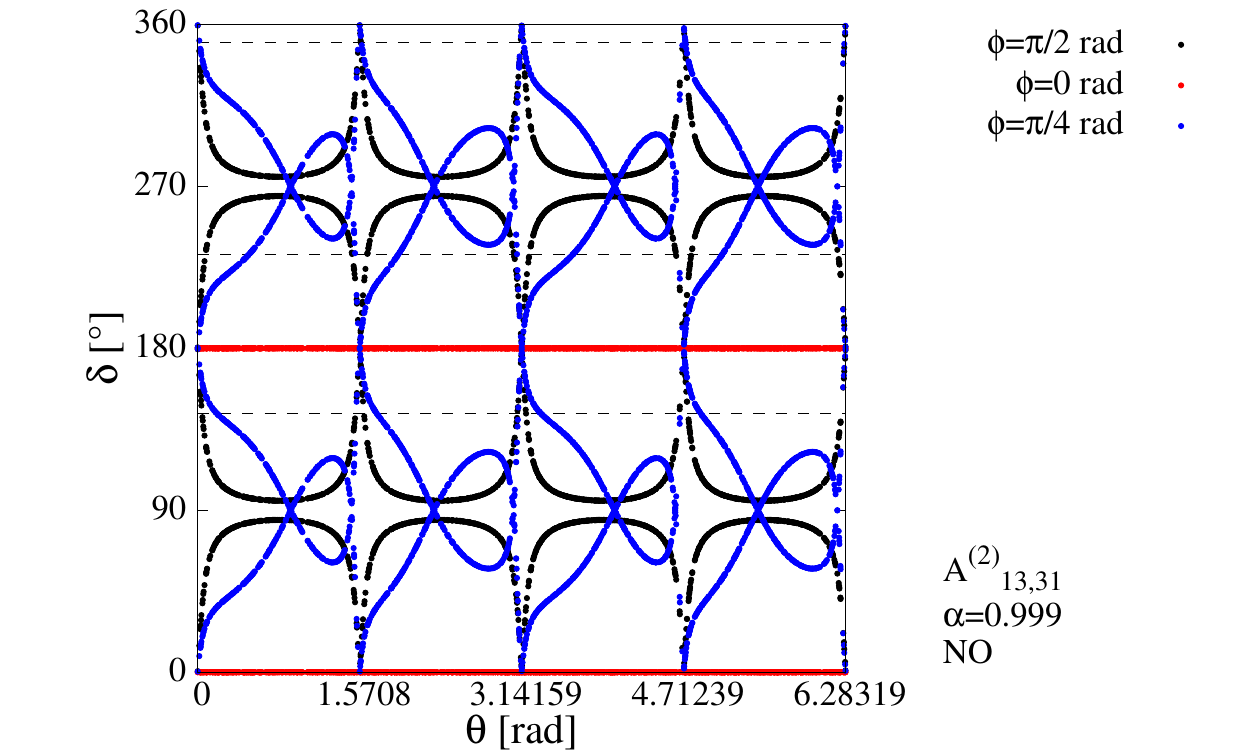}\\
\includegraphics[keepaspectratio, scale=0.5]{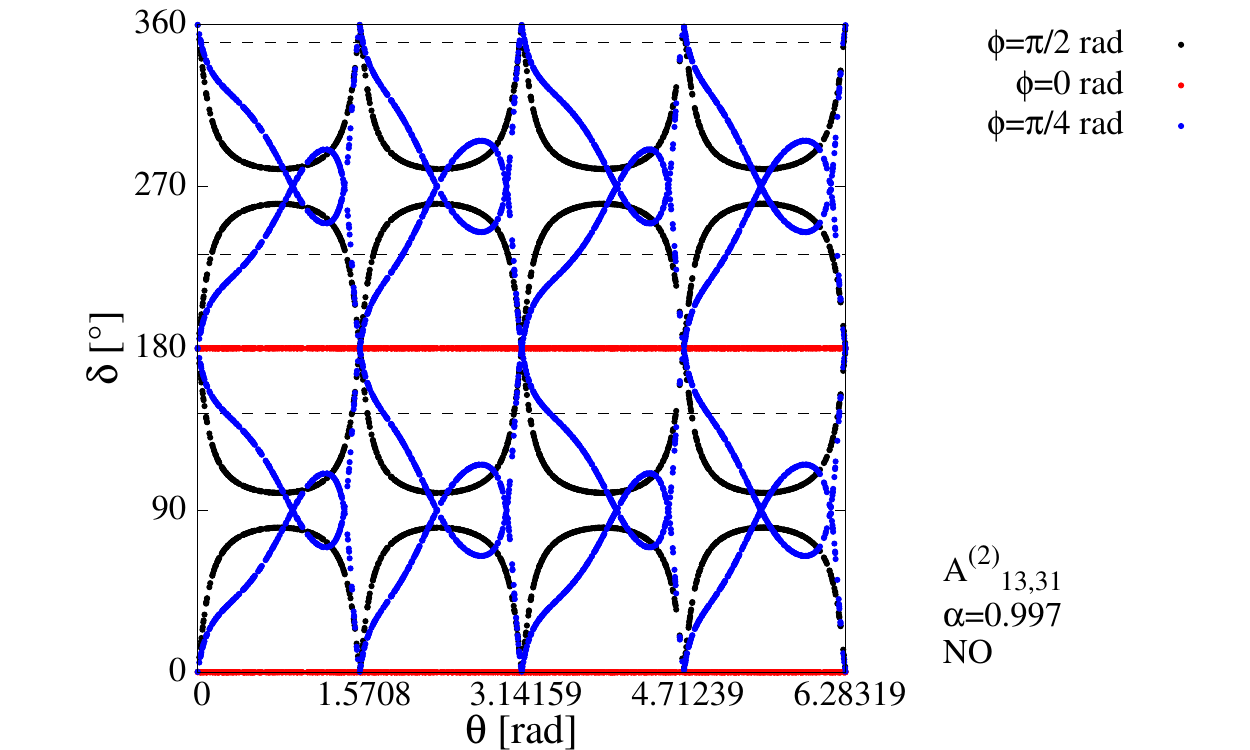}\\
\includegraphics[keepaspectratio, scale=0.5]{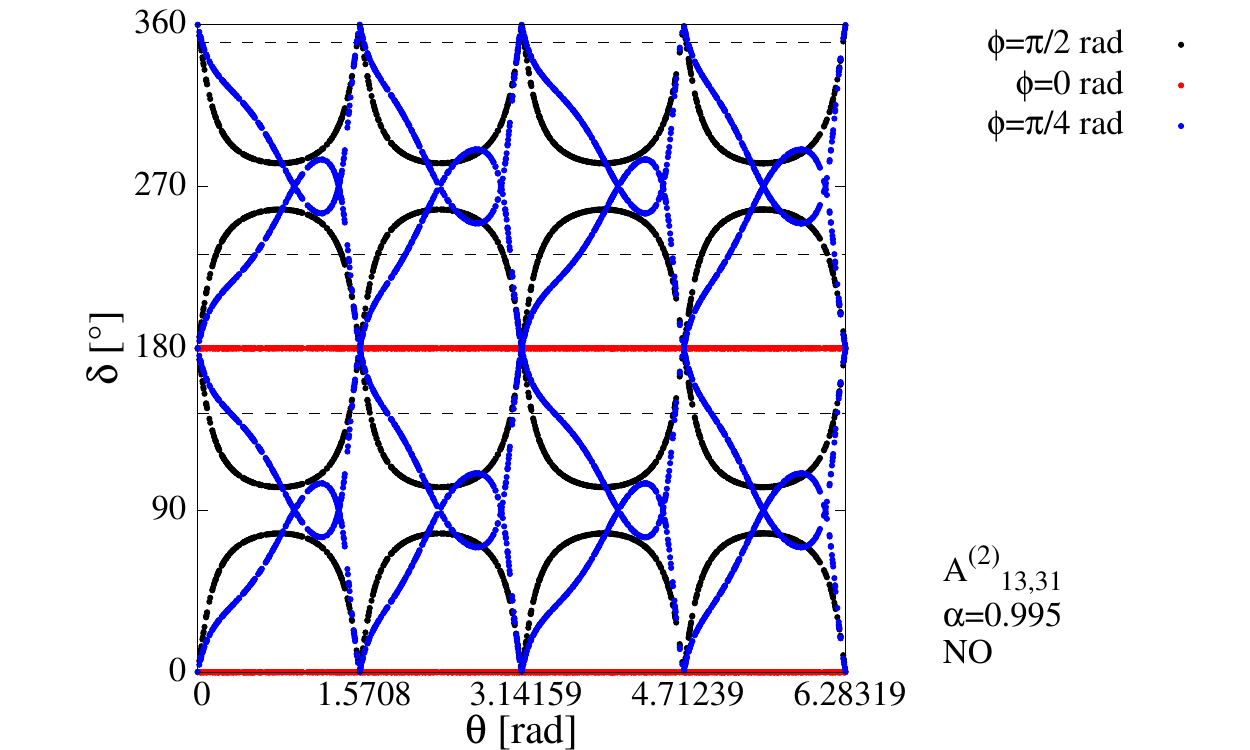}
\end{minipage}&
\begin{minipage}[t]{0.48\hsize}
\centering
\includegraphics[keepaspectratio, scale=0.5]{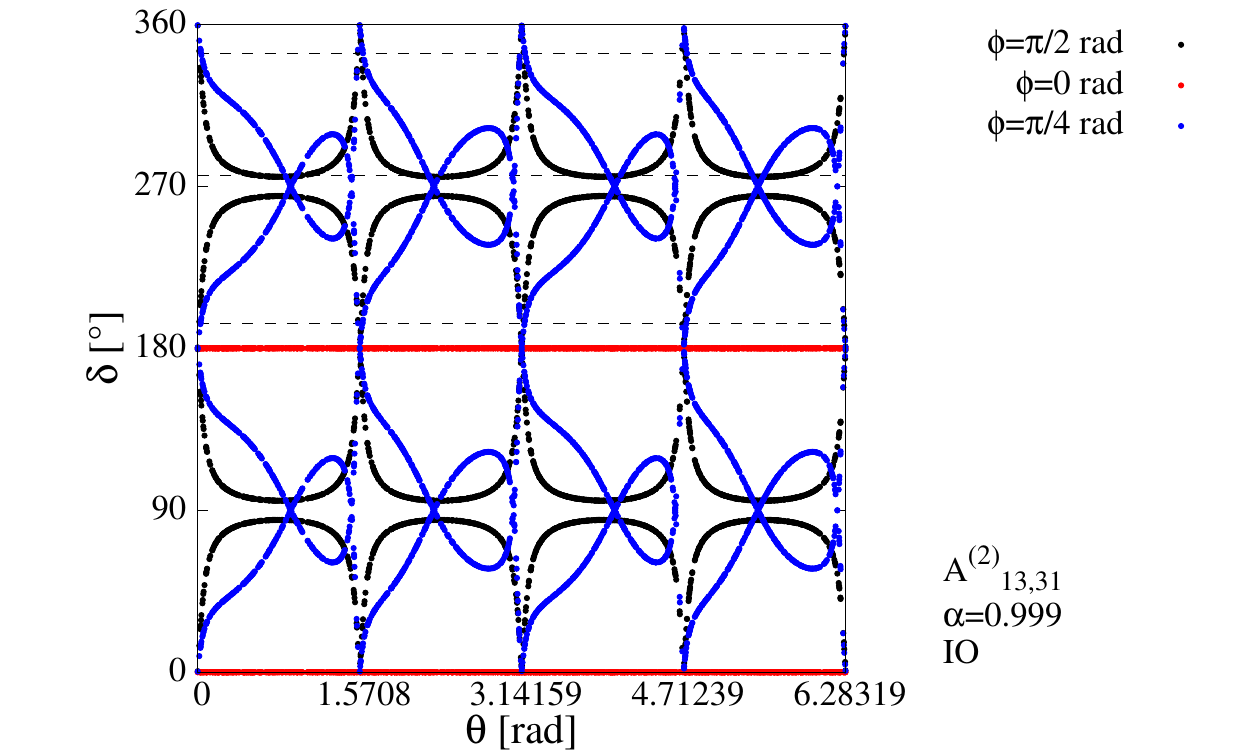}\\
\includegraphics[keepaspectratio, scale=0.5]{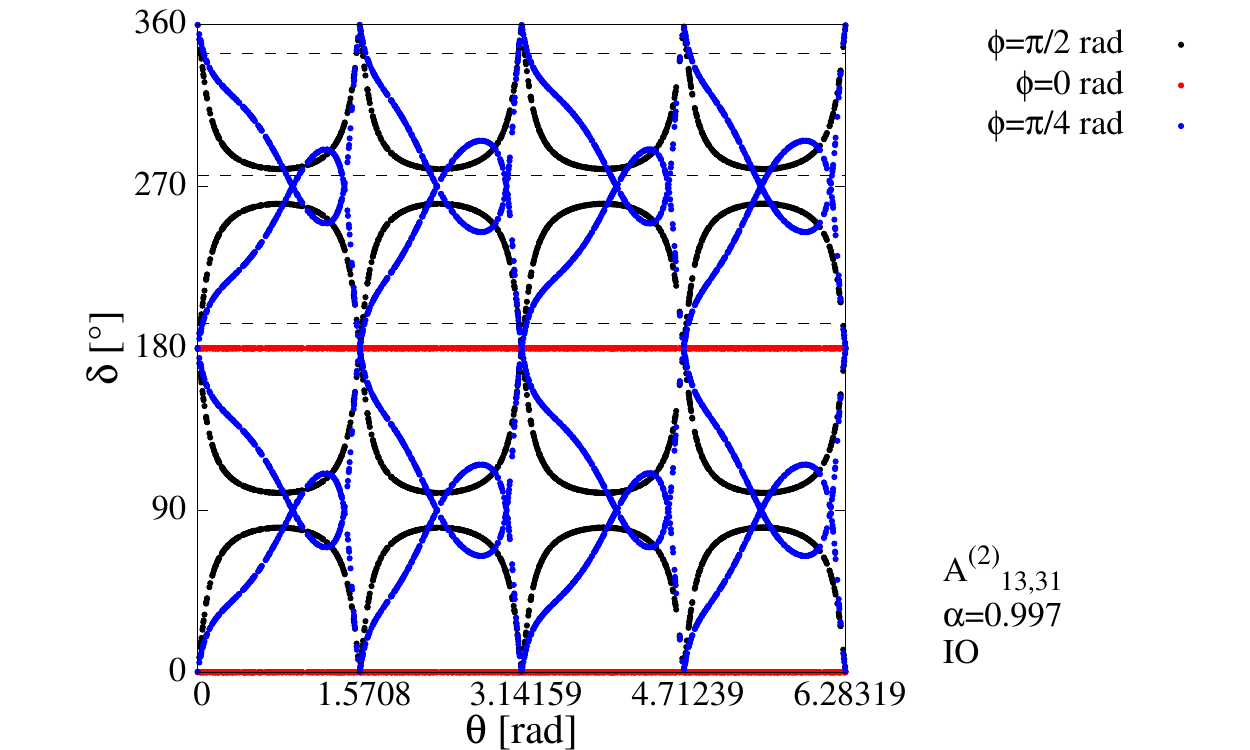}\\
\includegraphics[keepaspectratio, scale=0.5]{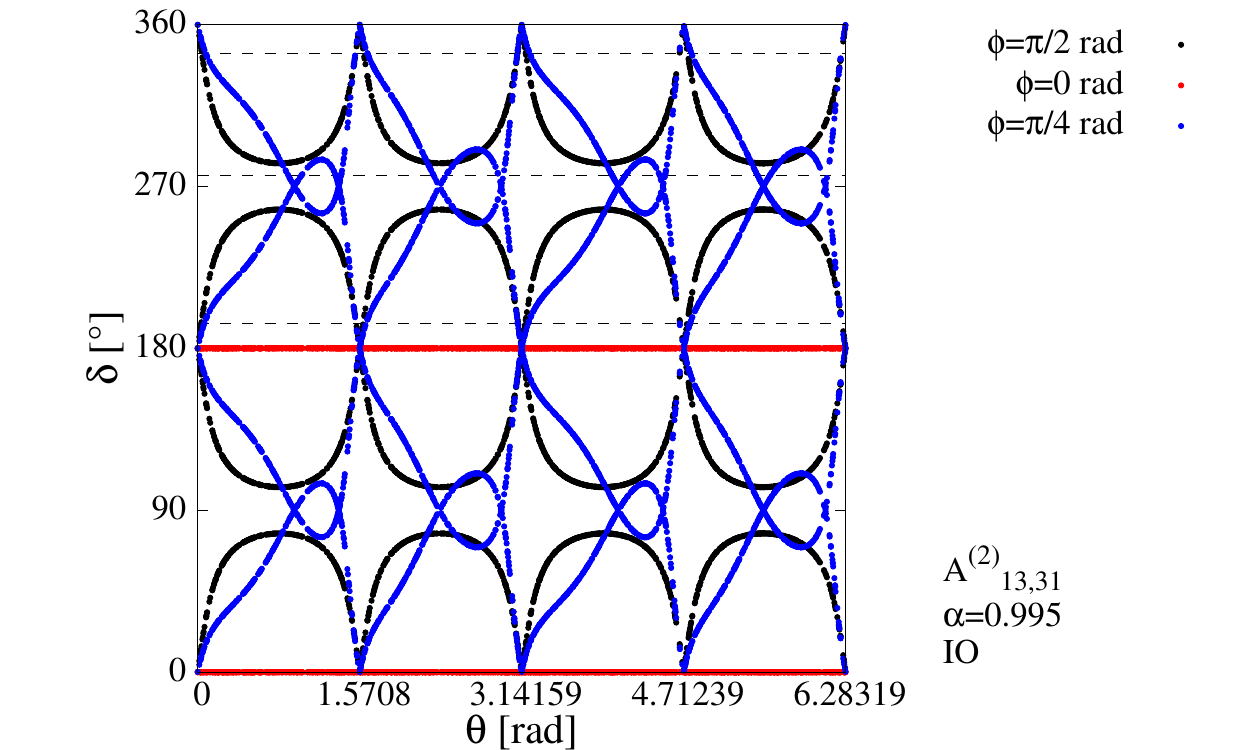}
 \end{minipage} \\
\end{tabular}
 \caption{Same as Fig. \ref{Fig:MTM1_A1221_2_ap_d} but for MTM1($A_{13,31}^{(2)}$)}
 \label{Fig:MTM1_A1331_2_ap_d} 
  \end{figure}

\begin{figure}[t]
\begin{tabular}{cc}
\begin{minipage}[t]{0.48\hsize}
\centering
\includegraphics[keepaspectratio, scale=0.5]{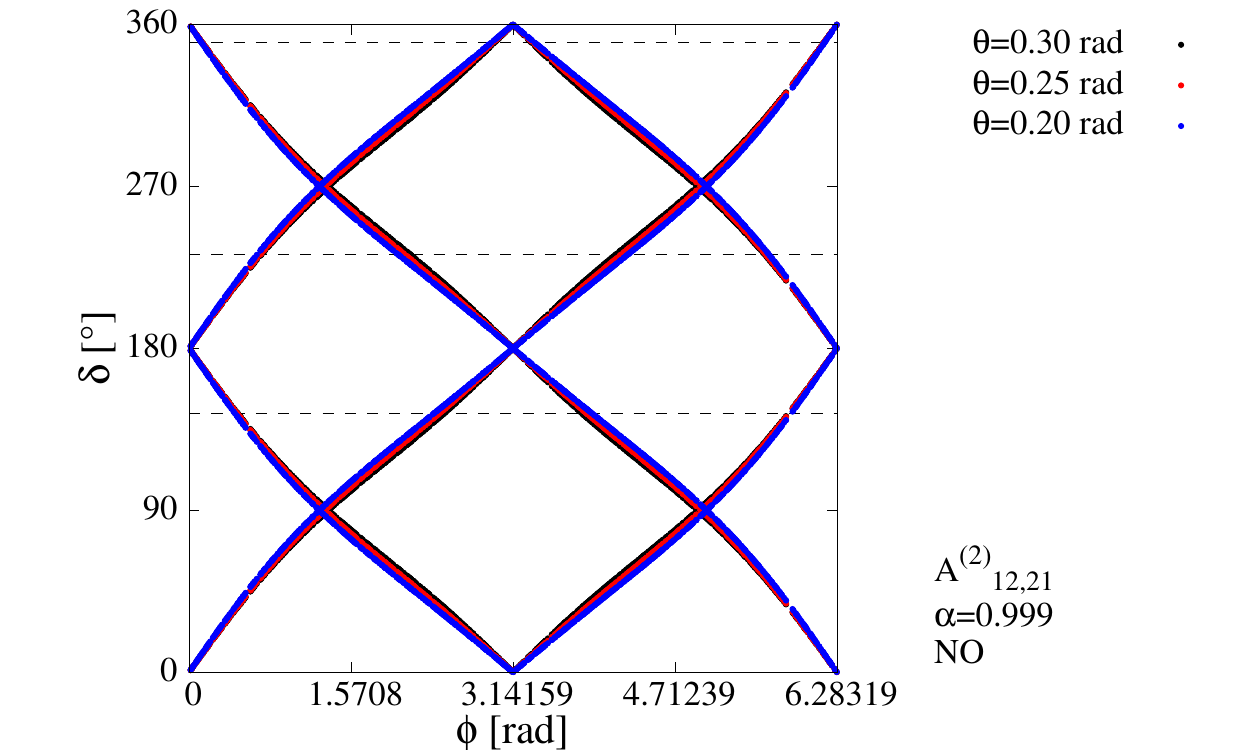}\\
\includegraphics[keepaspectratio, scale=0.5]{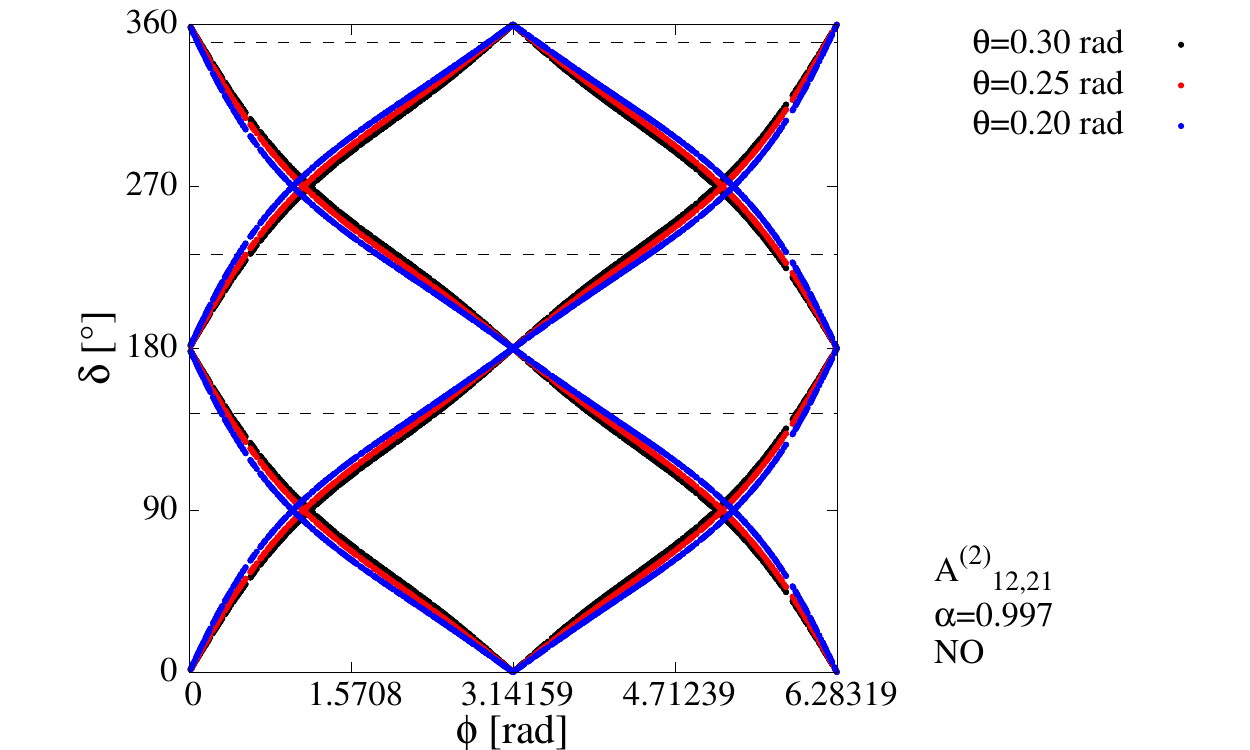}\\
\includegraphics[keepaspectratio, scale=0.5]{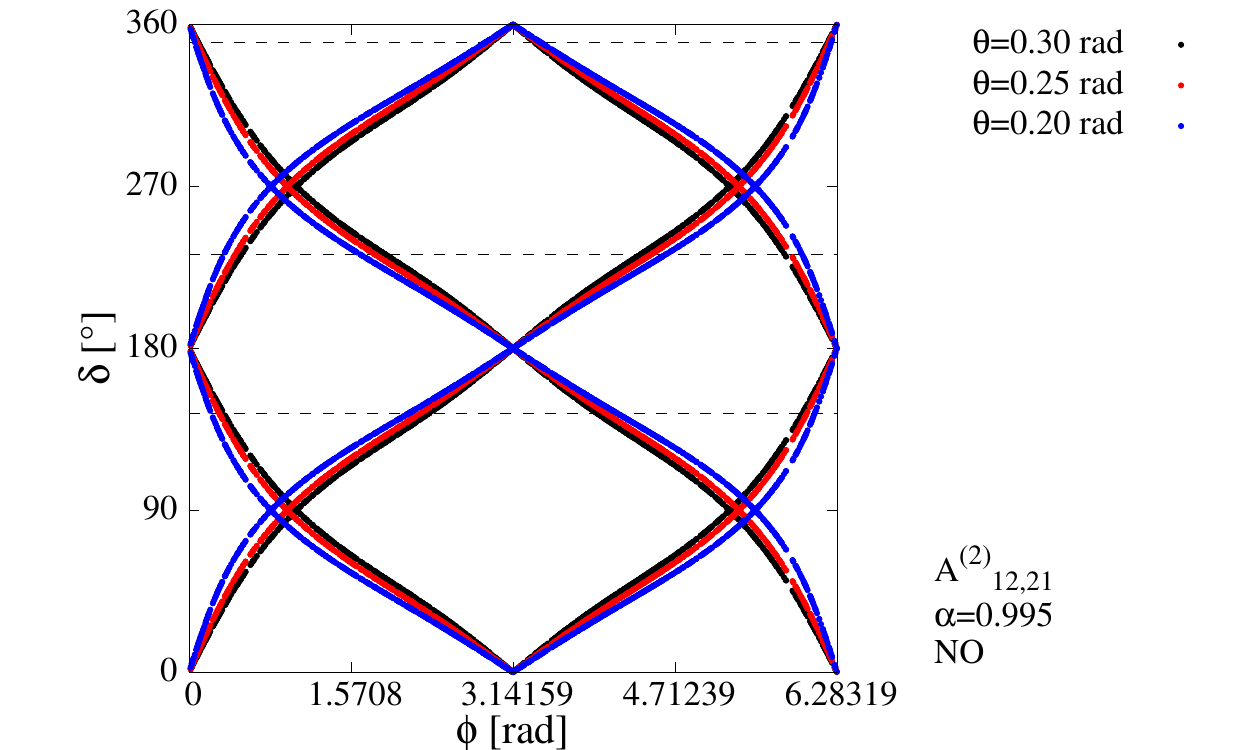}
\end{minipage}&
\begin{minipage}[t]{0.48\hsize}
\centering
\includegraphics[keepaspectratio, scale=0.5]{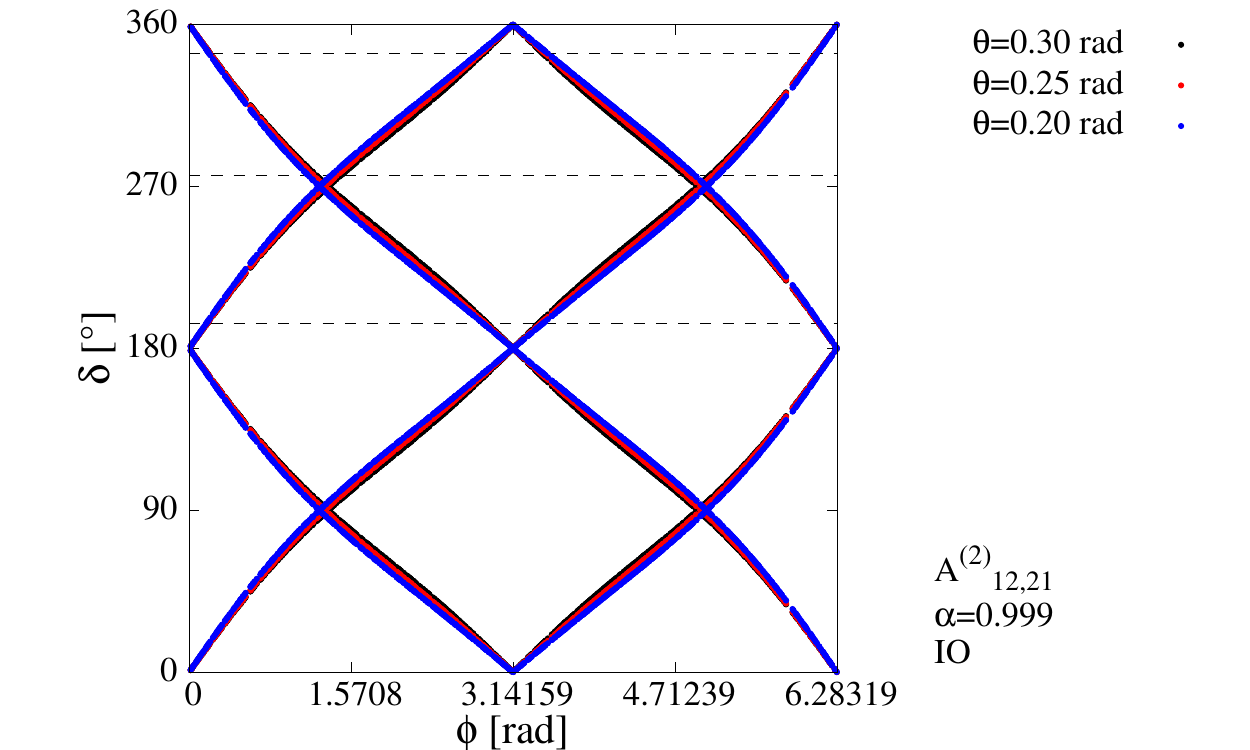}\\
\includegraphics[keepaspectratio, scale=0.5]{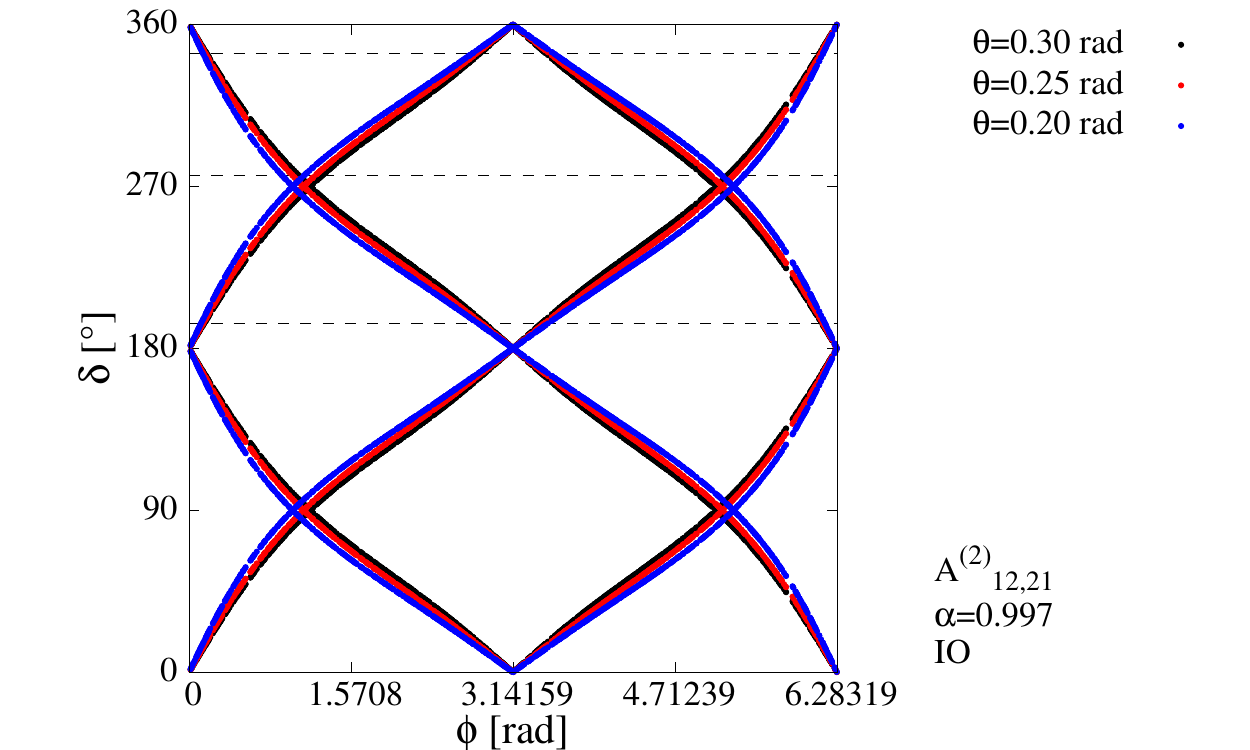}\\
\includegraphics[keepaspectratio, scale=0.5]{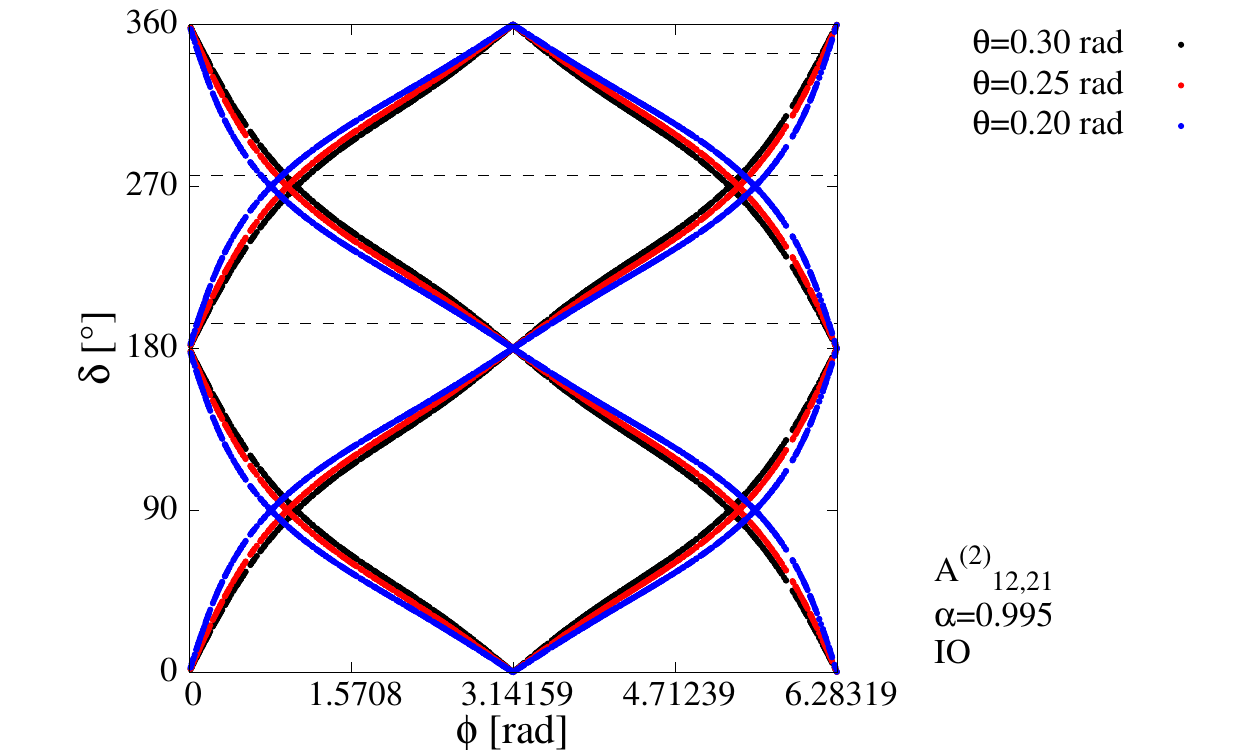}
 \end{minipage} \\
\end{tabular}
 \caption{Same as Fig. \ref{Fig:MTM1_A1221_2_at_d} but for MTM1($A_{13,31}^{(2)}$)}
 \label{Fig:MTM1_A1331_2_at_d} 
  \end{figure}

\begin{figure}[t]
\begin{tabular}{cc}
\begin{minipage}[t]{0.48\hsize}
\centering
\includegraphics[keepaspectratio, scale=0.5]{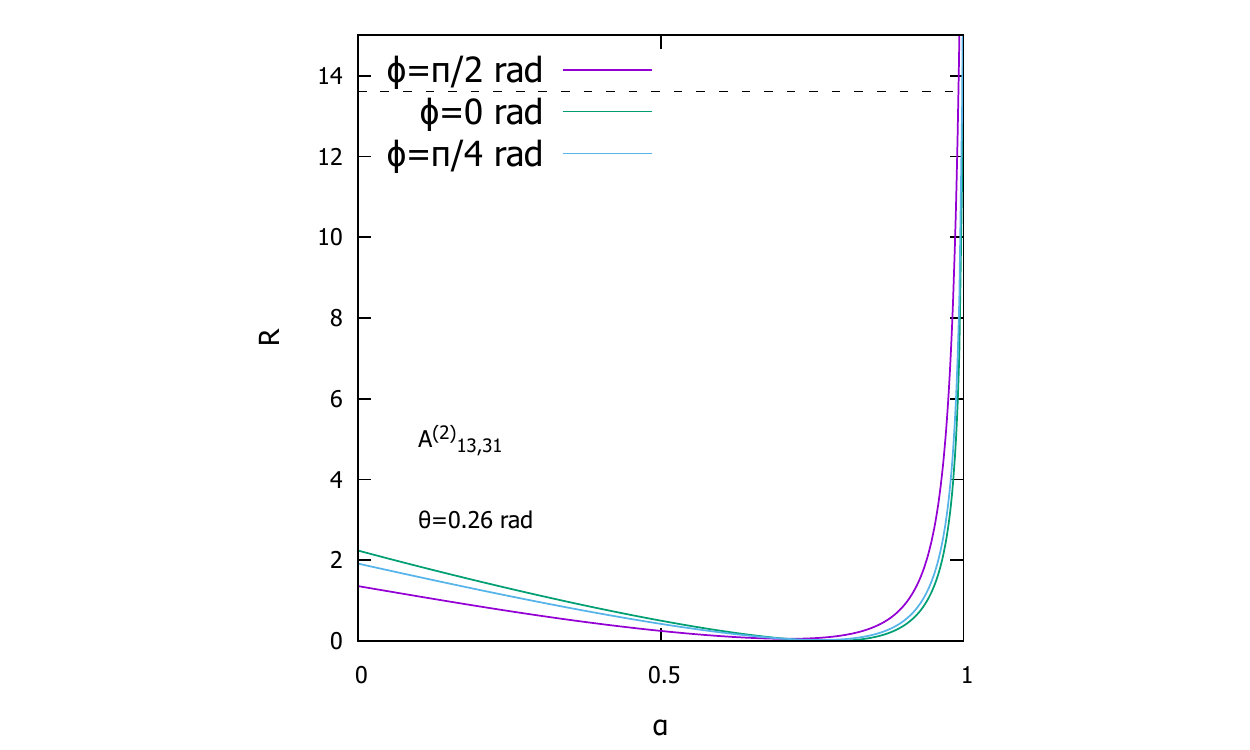}\\
\includegraphics[keepaspectratio, scale=0.5]{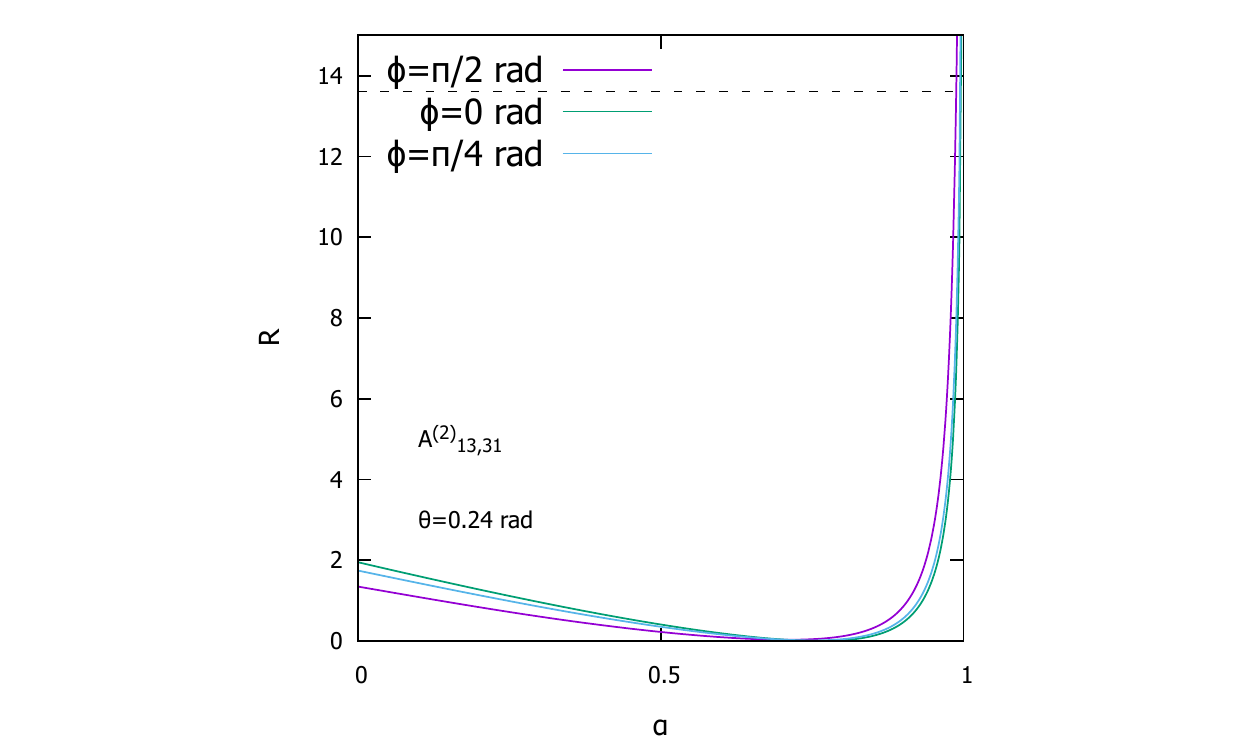}\\
\includegraphics[keepaspectratio, scale=0.5]{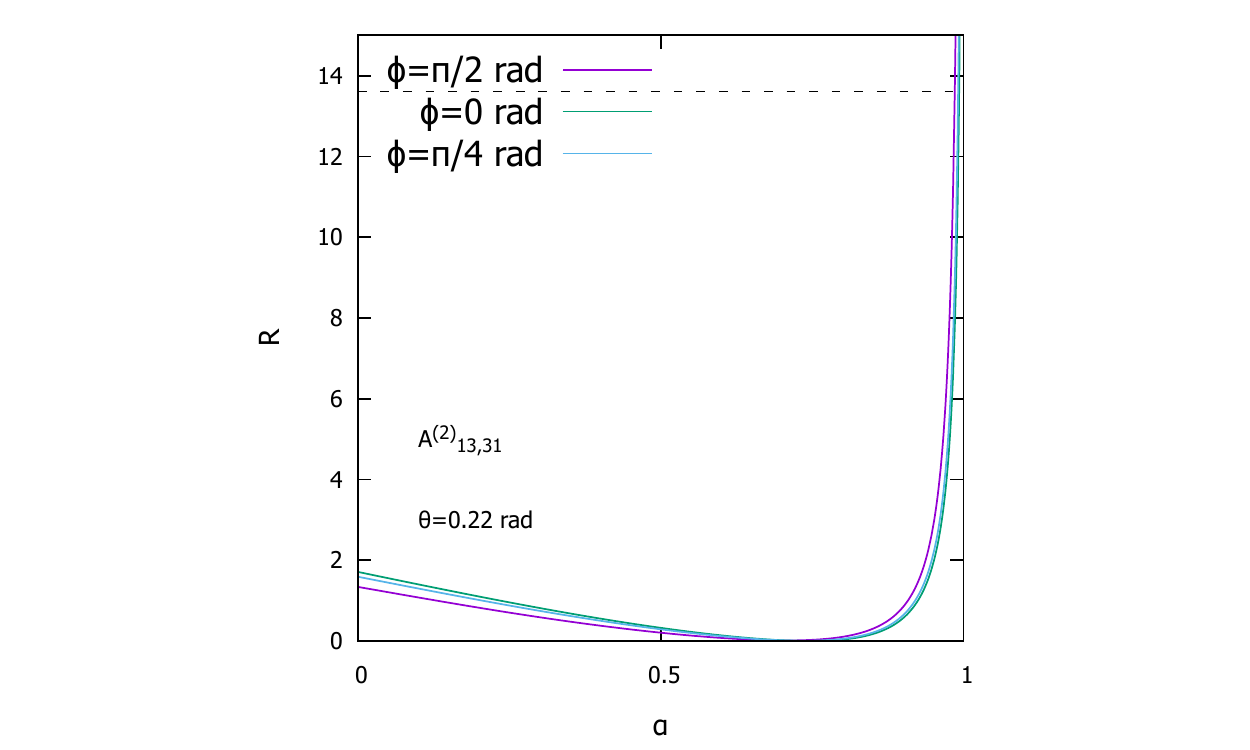}
\end{minipage}&
\begin{minipage}[t]{0.48\hsize}
\centering
\includegraphics[keepaspectratio, scale=0.5]{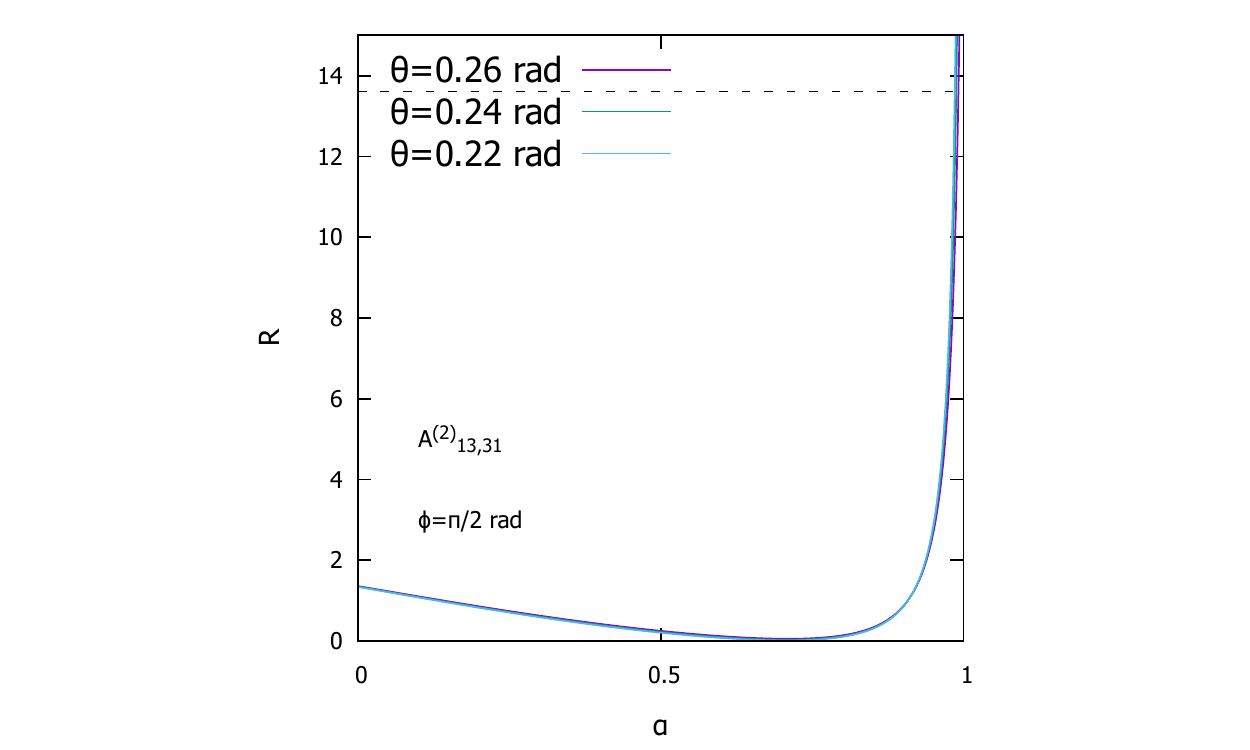}\\
\includegraphics[keepaspectratio, scale=0.5]{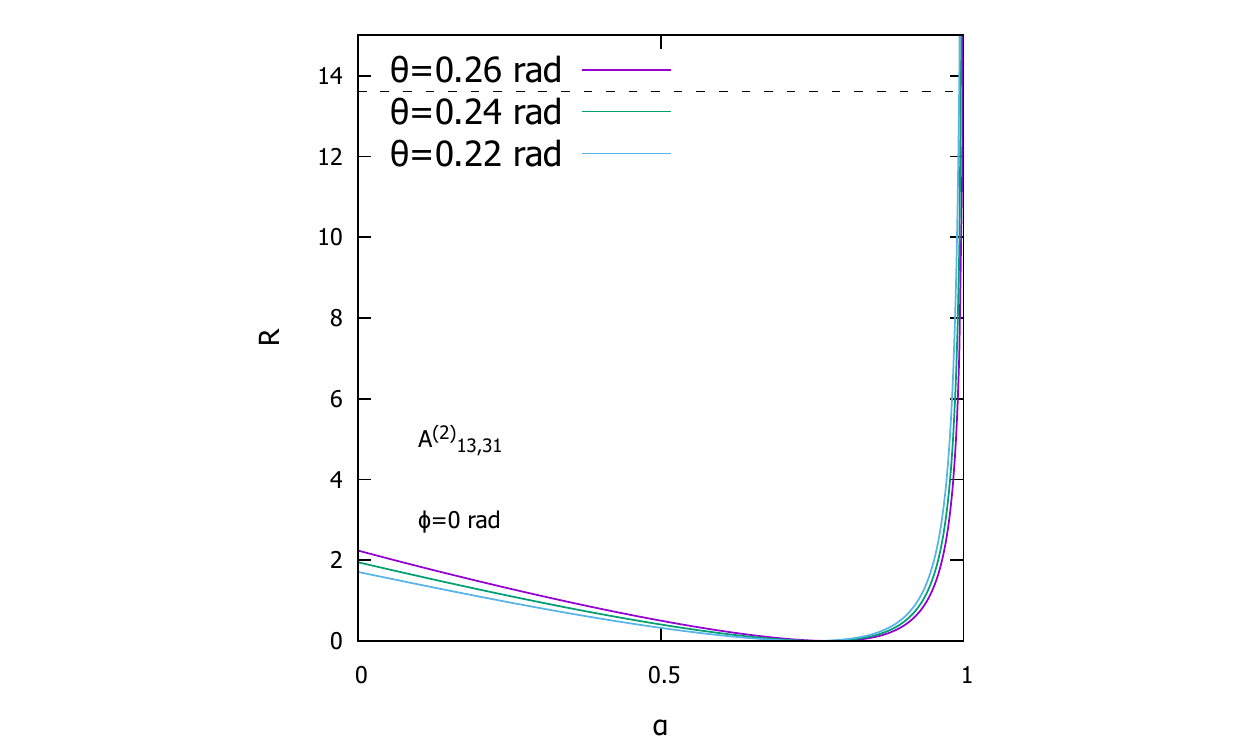}\\
\includegraphics[keepaspectratio, scale=0.5]{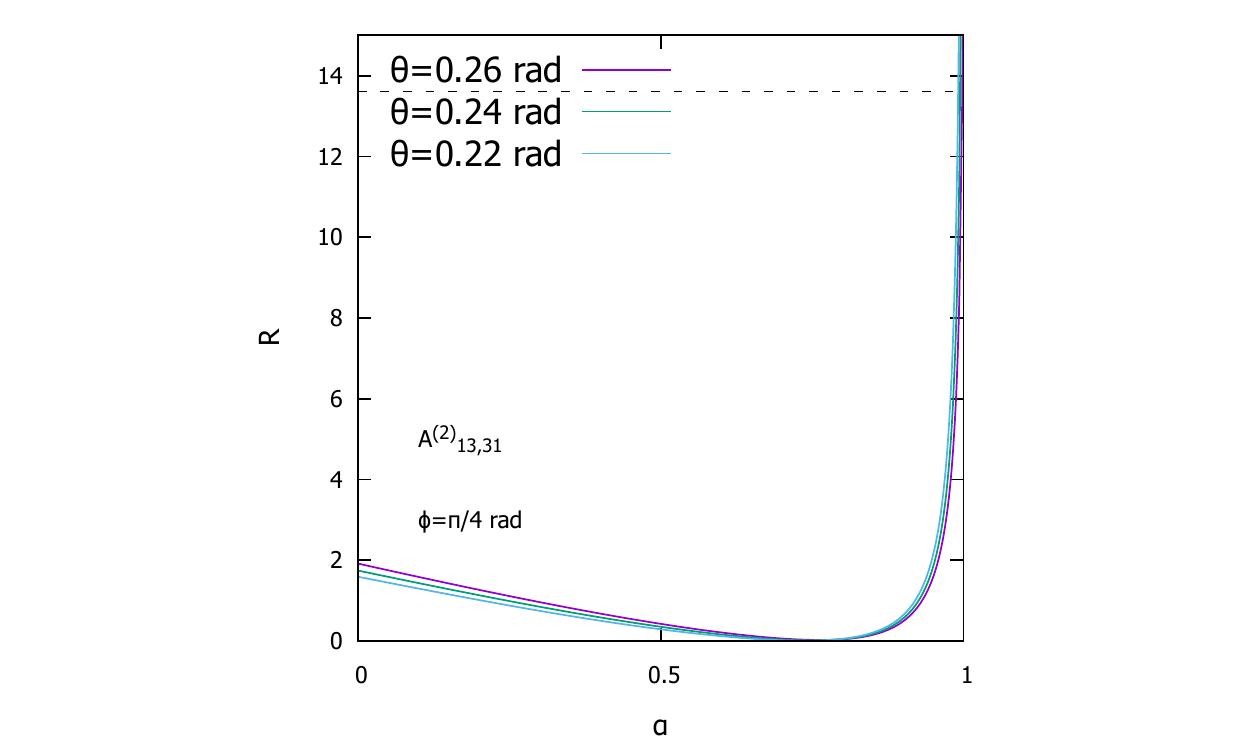}
 \end{minipage} \\
\end{tabular}
 \caption{Same as Fig. \ref{Fig:MTM1_A1221_2_R} but for MTM1($A_{13,31}^{(2)}$)}
 \label{Fig:MTM1_A1331_2_R} 
  \end{figure}

We modify the TM1 mixing scheme using the rotation matrix $A_{13,31}^{(1)}$ and $A_{13,31}^{(2)}$ as follows:
\begin{eqnarray}
\tilde{U}_1(A_{13,31}^{(1)}) = A_{13,31}^{(1)} U_1, \quad \tilde{U}_1(A_{13,31}^{(2)}) = A_{13,31}^{(2)} U_1, 
\end{eqnarray}
and we call these mixings MTM1($A_{13,31}^{(1)}$) and MTM1($A_{13,31}^{(2)}$), respectively. The mixing angles are
\begin{eqnarray}
s^2_{12} &=& \frac{5 \pm 4\alpha \sqrt{1-\alpha^2} - 3\alpha^2 - 6s^2_{13}}{6(1-s^2_{13})},   \label{Eq:MTM1_A1331_s12ss13s} \\
s^2_{23} &=& \frac{5+\cos 2\theta+2\sqrt{6}\sin 2\theta\cos\phi}{12(1-s^2_{13})}, \label{Eq:MTM1_A1331_s23ss13s}\\
s^2_{13} &=& \frac{1}{6} ( 3-3\alpha^2  -  B_1 - B_2 ). \label{Eq:MTM1_A1331_s13s}
 \end{eqnarray}
The upper sign of $\pm$ and $\mp$ should be taken for MTM1$(A_{13,31}^{(1)})$. The lower sign of these should be taken for MTM1$(A_{13,31}^{(2)})$.

Eq.(\ref{Eq:MTM1_A1331_s12ss13s}) is equal to Eq.(\ref{Eq:MTM1_A1221_s12ss13s}). Therefore, the simultaneous reproducibility of $\theta_{12}$ and $\theta_{13}$ in MTM1($A_{13,31}^{(1)}$) and MTM1($A_{13,31}^{(2)}$) is same as MTM1($A_{12,21}^{(1)}$) and MTM1($A_{12,21}^{(2)}$). Thus, the simultaneous reproducibility of $\theta_{12}$ and $\theta_{13}$ is reduced in MTM1($A_{13,31}^{(1)}$). In contrast, this reproducibility is substantially improved in MTM1($A_{13,31}^{(2)}$). Hereafter, we consider only MTM1($A_{13,31}^{(2)}$).

Figure \ref{Fig:MTM1_A1331_cosphi_13_23} shows the prediction of $\theta_{13}$ (upper panel) and $\theta_{23}$ (lower panel) as a function of $\phi$ in MTM1($A_{13,31}^{(2)}$). Similar to MTM1($A_{12,21}^{(2)}$), Fig. \ref{Fig:MTM1_A1331_cosphi_13_23} suggests that we can obtain values of $\theta_{13}$ and $\theta_{23}$ that are consistent with the observations by choosing $\phi$ and $\theta$. In fact, we have confirmed that $s^2_{13}$ and $s^2_{23}$ can be obtained in the $3 \sigma$ region by numerical parameter search. 

A benchmark point 
\begin{eqnarray}
(\alpha, \theta, \phi) = (0.99973, \ 15.50^\circ,\  97.40^\circ),
\label{Eq:MTM1_A1331_benchmark}
\end{eqnarray}
yields the best-fit values of $s_{12}^2$ and $s_{13}^2$ simultaneous reproducibility and the allowed value of $s_{23}^2$ as follows:
\begin{eqnarray}
(s_{12}^2, s_{23}^2, s_{13}^2,\delta) = (0.303, \ 0.471,\  0.02225, \ 268.1^\circ). 
\end{eqnarray}

As same as the case of MTM1($A_{12,21}^{(2)}$), the reader may wish to know the ballpark figures of the parameter space and the possible ranges of those mixing parameters in the case of MTM1($A_{13,31}^{(2)}$). Moreover, the reader may wonder what would happen if the best-fit values were changed slightly., or the possible ranges of $\theta_{23}$ and $\delta$ are narrowed down in the near future. 

To answer these questions, we show Figures \ref{Fig:MTM1_A1331_2_a1p1_12_13} - \ref{Fig:MTM1_A1331_2_R}. These figures are same as Figures \ref{Fig:MTM1_A1221_2_a1p1_12_13} - \ref{Fig:MTM1_A1221_2_R} but for MTM1($A_{13,31}^{(2)}$). From these figures, we have the following similar conclusions for  MTM1($A_{13,31}^{(2)}$) as for  MTM1($A_{12,21}^{(2)}$),
\begin{itemize}
\item The wide range of parameters $\theta$, $\phi$ and $\alpha$ are consistent with observation (Figures \ref{Fig:MTM1_A1331_2_a1p1_12_13}, \ref{Fig:MTM1_A1331_2_ap_23} and \ref{Fig:MTM1_A1331_2_at_23}). The wide range of Dirac CP phase is also consistent with observation (Figures \ref{Fig:MTM1_A1331_2_ap_d} and \ref{Fig:MTM1_A1331_2_at_d}).
\item If the values of $\theta_{23}$ and the CP phase $\delta$ are finally pinned down, we can reproduce these fixed values with appropriate values of $\alpha$, $\theta$ and $\phi$.
\item If the best-fit values change in the future, the new best-fit values can be reproduced with appropriate selection of the values of $\alpha$, $\theta$, $\phi$ (Figure \ref{Fig:MTM1_A1331_2_R}).
\end{itemize}
%

\subsection{$Z_2$ symmetry breaking}
Thus far, it has been found that the modifications of the TM1 mixing using $A_{12,21}^{(2)}$ and $A_{13,31}^{(2)}$ can improve the simultaneous reproducibility of $\theta_{12}$ and $\theta_{13}$. However, as a consequence of this improvement, the mass matrix in the modified mixing scheme breaks the $Z_2$ symmetry, which was strictly preserved in the original TM1 mixing. 

The flavor neutrino mass matrix for the modified TM1 mixing with $A=\{A_{12,21}^{(2)},A_{13,31}^{(2)}\}$ is obtained 
\begin{eqnarray}
\tilde{M}_1 = (A U_1)^\ast D(A U_1)^\dag = A^\ast M_1 A^\dag.
\end{eqnarray}
The $Z_2$ transformation of this mass matrix is performed
\begin{eqnarray}
S_1 \tilde{M}_1 S_1^{\rm T}  &=& S_1 (A^\ast M_1 A^\dag) S_1^{\rm T}\nonumber \\
&=& S_1A^\ast (S_1^{\rm T}  S_1) M_1 (S_1^{\rm T}  S_1)  A^\dag S_1^{\rm T}\nonumber \\
&=& (S_1A^\ast S_1^{\rm T})  M_1 (S_1  A^\dag S_1^{\rm T}).
\end{eqnarray}
Using the following definition of $\delta A$ 
\begin{eqnarray}
S_1A^\ast S_1^{\rm T} = A^\ast + \delta A^\ast, \quad
S_1A^\dag S_1^{\rm T} = A^\dag + \delta A^\dag, 
\end{eqnarray}
we can write the $Z_2$ transformation as 
\begin{eqnarray}
S_1 \tilde{M}_1 S_1^{\rm T}  = \tilde{M}_1 + \delta \tilde{M}_1,
\end{eqnarray}
where 
\begin{eqnarray}
\delta \tilde{M}_1 &=& A^\ast M_1 \delta A^\dag + \delta A^\ast M_1 A^\dag  + \delta A^\ast M_1 \delta A^\dag \nonumber \\ 
 &=& A^\ast  M_1 (A^\dag A) \delta A^\dag + \delta A^\ast (A^{\rm T} A^\ast) M_1 A^\dag  + \delta A^\ast  (A^{\rm T} A^\ast)  M_1  (A^\dag A) \delta A^\dag  \\ 
 &=& \tilde{M}_1 A \delta A^\dag + (A \delta A^\dag)^{\rm T} \tilde{M}_1  +(A \delta A^\dag)^{\rm T} \tilde{M}_1 A \delta A^\dag. \nonumber 
\end{eqnarray}

The magnitude of the symmetry breaking of $Z_2$ can be evaluated by the following matrix:
\begin{eqnarray}
 \delta_{\rm Z2}^{\rm MTM1}= \left(
\begin{matrix}
\delta_{11} & \delta_{12} &   \delta_{13}  \\
*&  \delta_{22}  & \delta_{23} \\
 *  & *   &  \delta_{33}   \\
\end{matrix}
\right),
\label{Eq:Z2breaking}
\end{eqnarray}
where
\begin{eqnarray}
\delta_{ij} =  \frac{\left| (\delta \tilde{M}_1)_{ij} \right| }{\left| (\tilde{M}_1)_{ij} \right|},
\end{eqnarray}
and $*$ indicates the symmetric elements.

For MTM1($A_{12,21}^{(2)}$), the magnitude of $Z_2$ symmetry breaking at the benchmark point shown in Eq.(\ref{Eq:MTM1_A1221_benchmark}) is obtained as follows: 
\begin{eqnarray}
 \delta_{\rm Z2}^{\rm MTM1}(A_{12,21}^{(2)})= \left(
\begin{matrix}
0.132 & 0.116 &   0.169 \\
*&  0.0304  & 0.00878 \\
 *  & *   & 0.0452   \\
\end{matrix}
\right),
\end{eqnarray}
where $(m_1, m_2, m_3) = (0, \Delta m_{21}^2, \Delta m_{31}^2)$ with the best-fit values of $\Delta m_{ij}^2$ in NO. The $Z_2$ symmetry breaking in the first row (maximum $16.9 \%$) is larger than that in the second and third rows (maximum $4.5\%$). Unfavorably, the magnitude of the symmetry breaking of $Z_2$ is not small in the electron flavor sector.

Similarly, for MTM1($A_{13,31}^{(2)}$), the magnitude of the symmetry breaking of $Z_2$ at the benchmark point shown in Eq.(\ref{Eq:MTM1_A1331_benchmark}) is obtained as follows:
\begin{eqnarray}
 \delta_{\rm Z2}^{\rm MTM1}(A_{13,31}^{(2)})= \left(
\begin{matrix}
0.133 & 0.171 &   0.117 \\
*&  0.0456  & 0.00880 \\
 *  & *   & 0.0307   \\
\end{matrix}
\right).
\end{eqnarray}
The symmetry breaking of $Z_2$ in the first row (maximum $17.1 \%$) is larger than that in the second and third rows (maximum $4.6\%$). Again, the magnitude of the symmetry breaking of $Z_2$ is not small in the electron flavor sector.

While the theoretical origin of the $Z_2$ symmetry breaking in the MTM1($A_{12,21}^{(2)}$) and MTM1($A_{13,31}^{(2)}$) mixings is a critical consideration, it is acknowledged that the primary focus of this study is to improve the simultaneous reproducibility of $\theta_{12}$ and $\theta_{13}$. Therefore, for the purposes of this study, the problem of $Z_2$ symmetry breaking will be disregarded. 

\section{Modified TM2 mixing \label{sec:MTM2}}

\subsection{$A_{12,21}^{(1)}$ and $A_{12,21}^{(2)}$}

\begin{figure}[t]
\begin{center}
\includegraphics[scale=1.0]{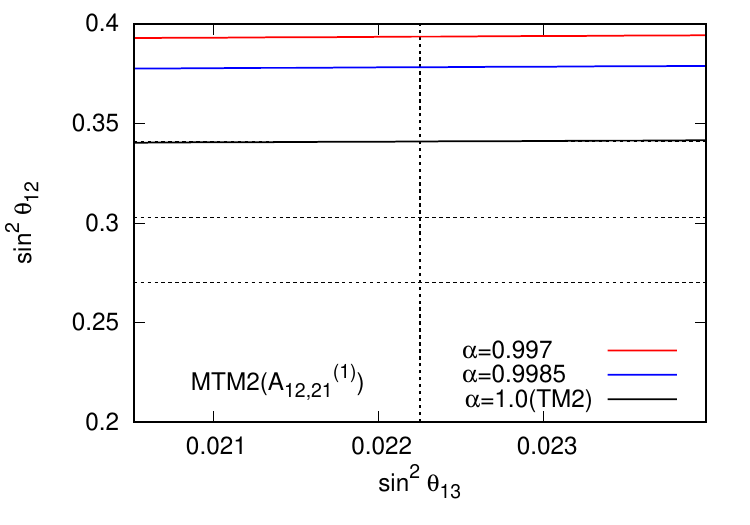}
\includegraphics[scale=1.0]{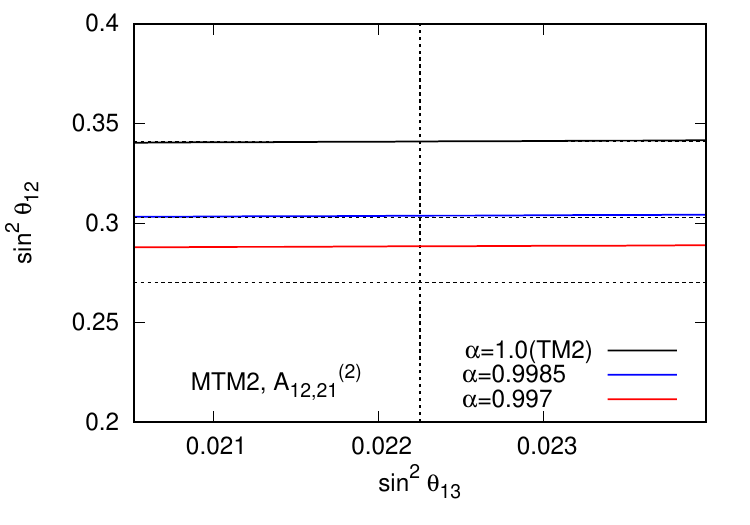}
\caption{Same as Fig. \ref{Fig:MTM1_A1221_12_13} but for MTM2.}
\label{Fig:MTM2_A1221_12_13} 
\end{center}
\end{figure}

\begin{figure}[t]
\begin{center}
\includegraphics[scale=1.0]{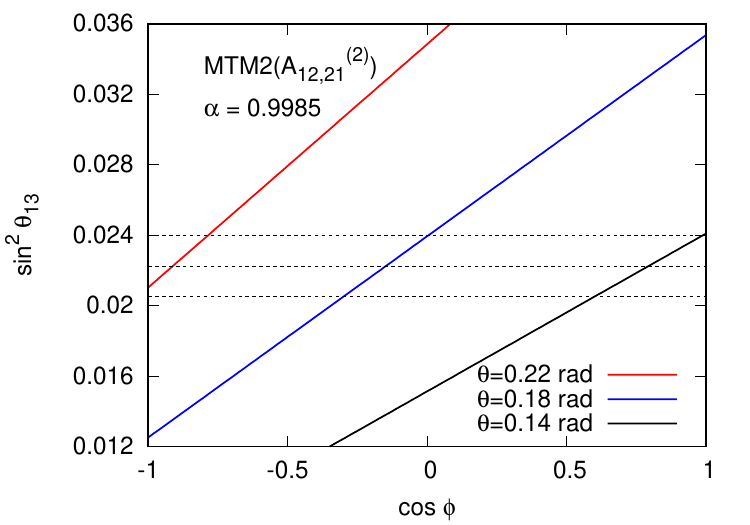}
\includegraphics[scale=1.0]{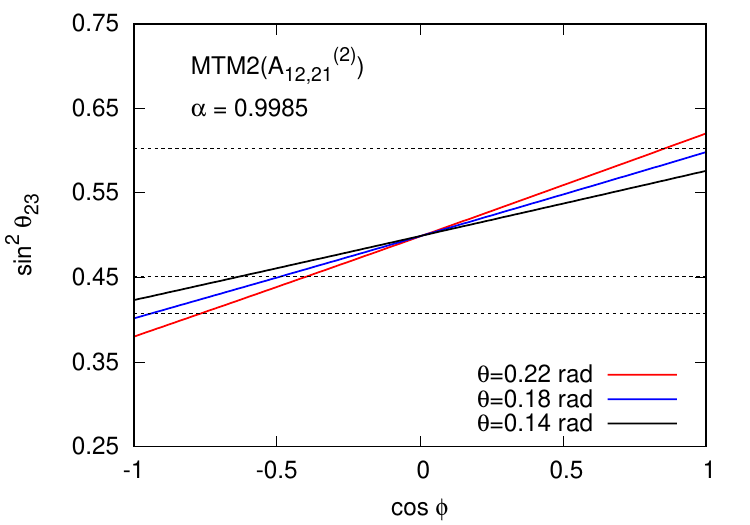}
\caption{Same as Fig. \ref{Fig:MTM1_A1221_cosphi_13_23} but for MTM2.}
\label{Fig:MTM2_A1221_cosphi_13_23} 
\end{center}
\end{figure}

\begin{figure}[t]
\begin{tabular}{cc}
\begin{minipage}[t]{0.3\hsize}
\centering
\includegraphics[keepaspectratio, scale=0.5]{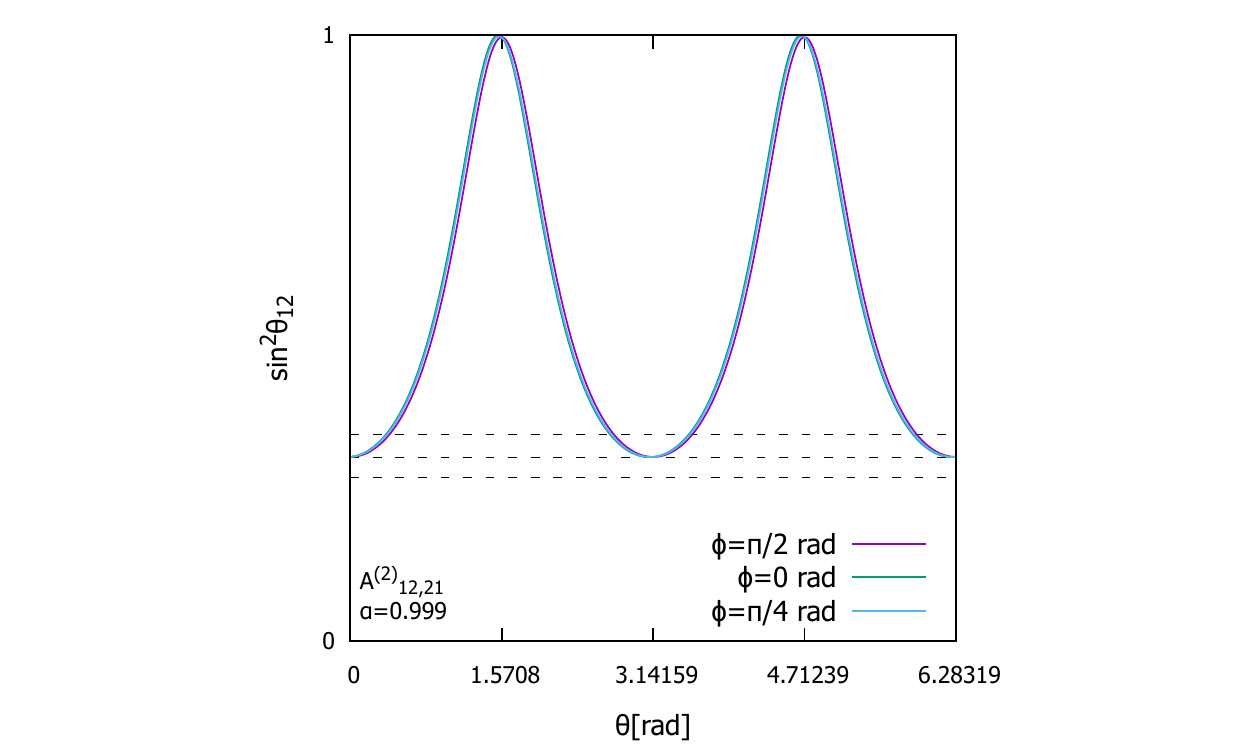}\\
\includegraphics[keepaspectratio, scale=0.5]{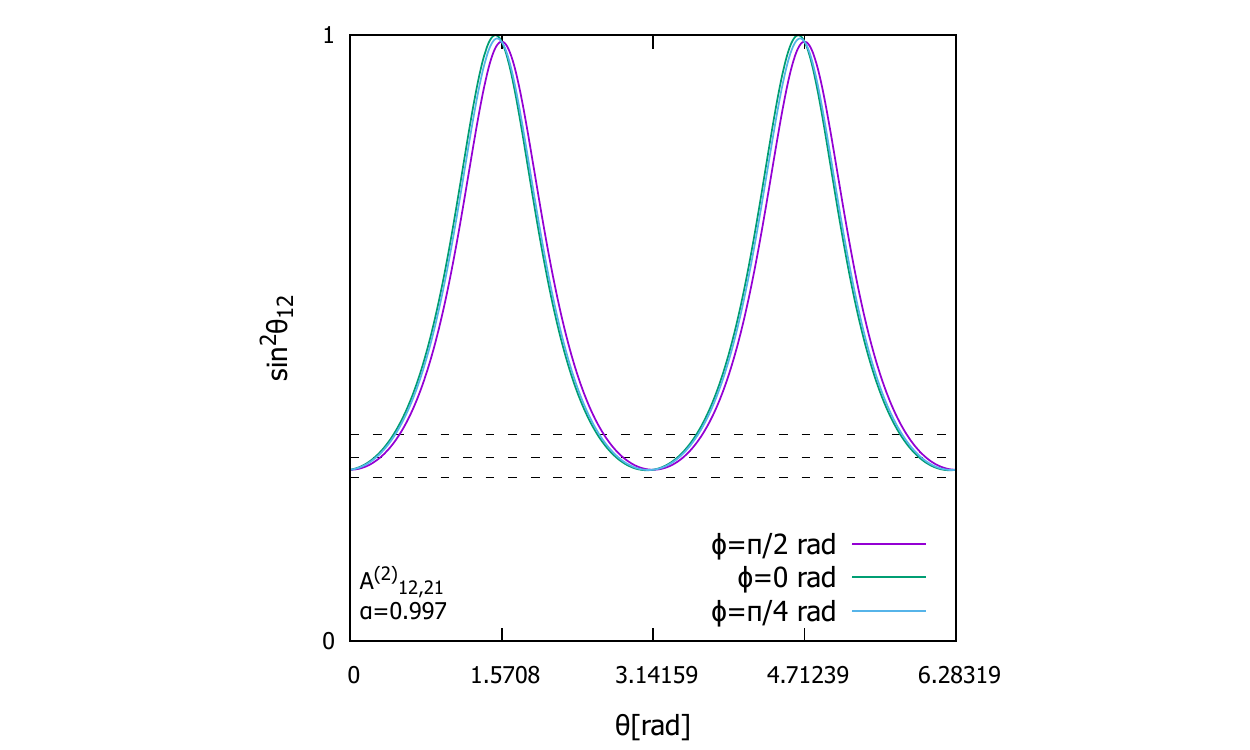}\\
\includegraphics[keepaspectratio, scale=0.5]{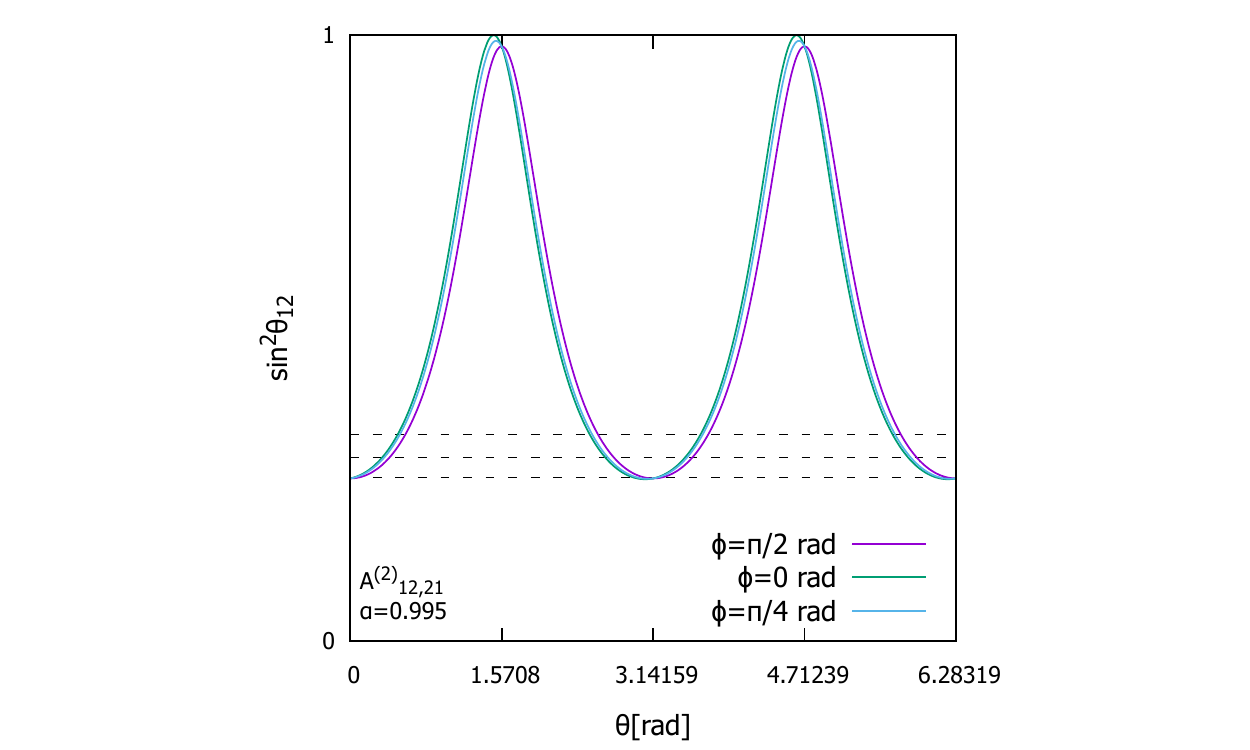}
\end{minipage}&
\begin{minipage}[t]{0.3\hsize}
\centering
\includegraphics[keepaspectratio, scale=0.5]{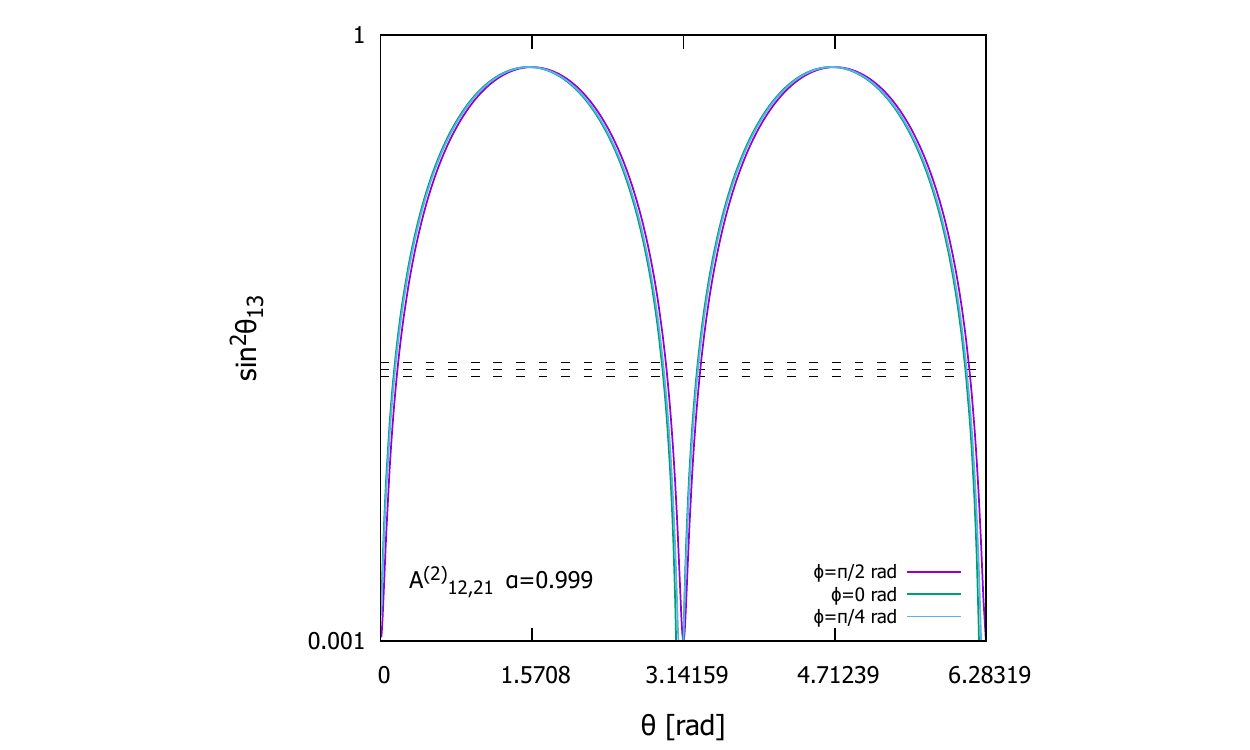}\\
\includegraphics[keepaspectratio, scale=0.5]{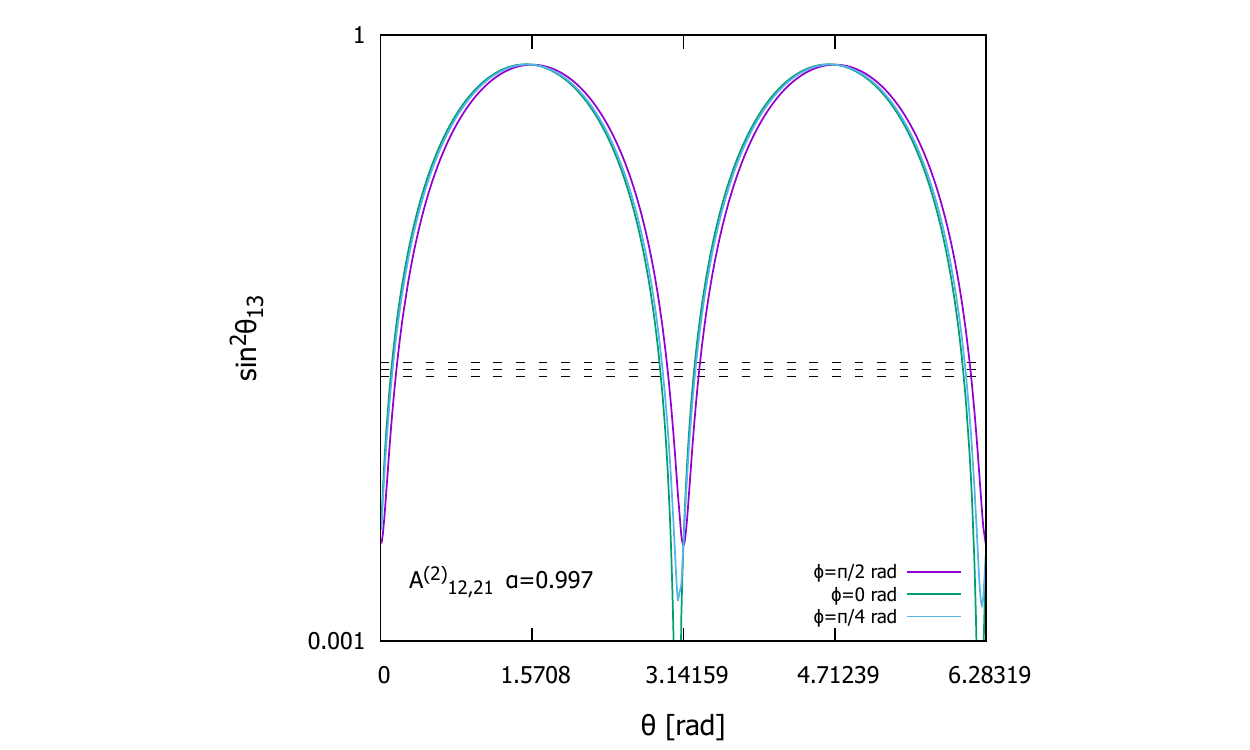}\\
\includegraphics[keepaspectratio, scale=0.5]{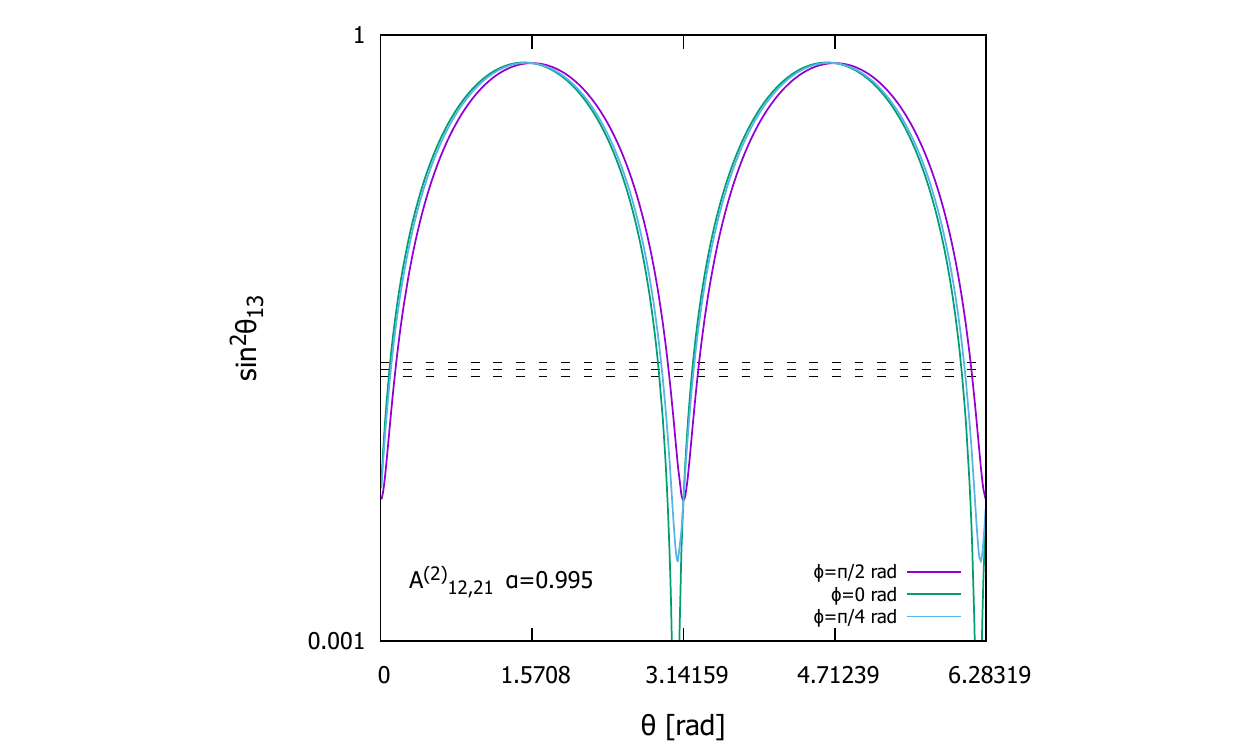}
 \end{minipage} \\
\end{tabular}
 \caption{Same as Fig. \ref{Fig:MTM1_A1221_2_a1p1_12_13} but for MTM2($A_{12,21}^{(2)}$).}
 \label{Fig:MTM2_A1221_2_a1p1_12_13} 
  \end{figure}
\begin{figure}[t]
\begin{tabular}{cc}
\begin{minipage}[t]{0.48\hsize}
\centering
\includegraphics[keepaspectratio, scale=0.5]{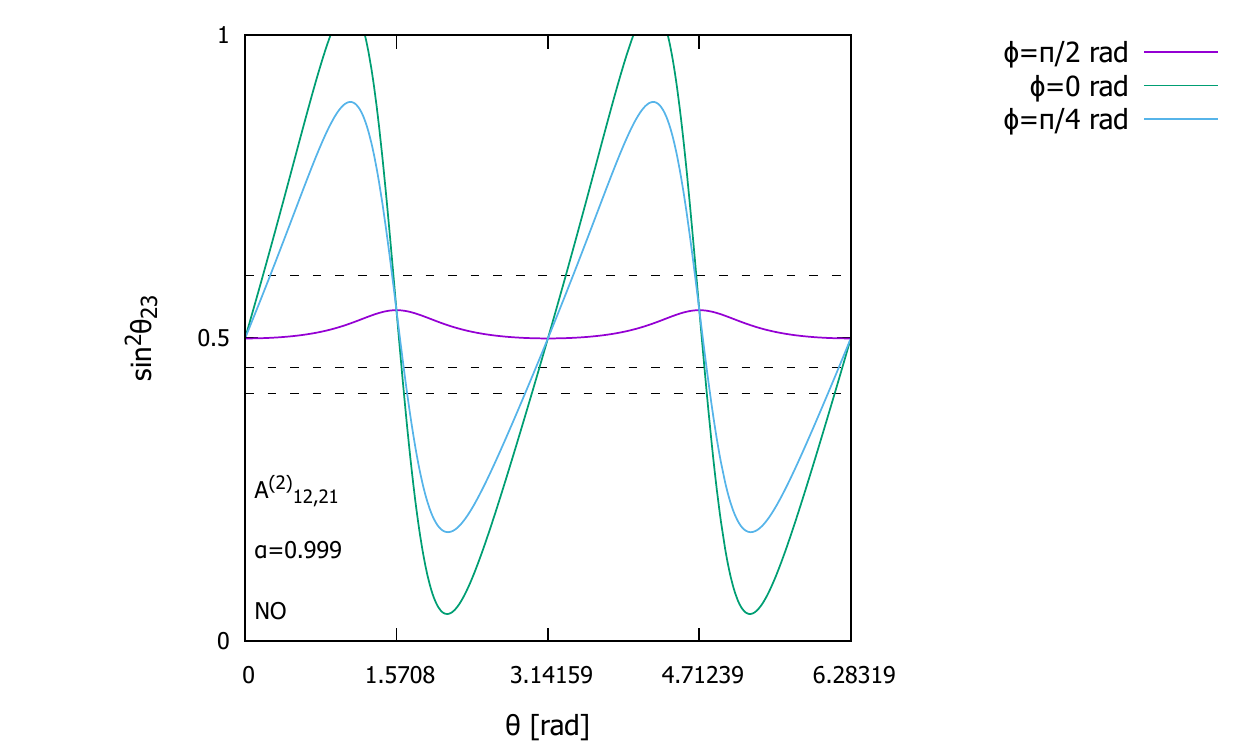}\\
\includegraphics[keepaspectratio, scale=0.5]{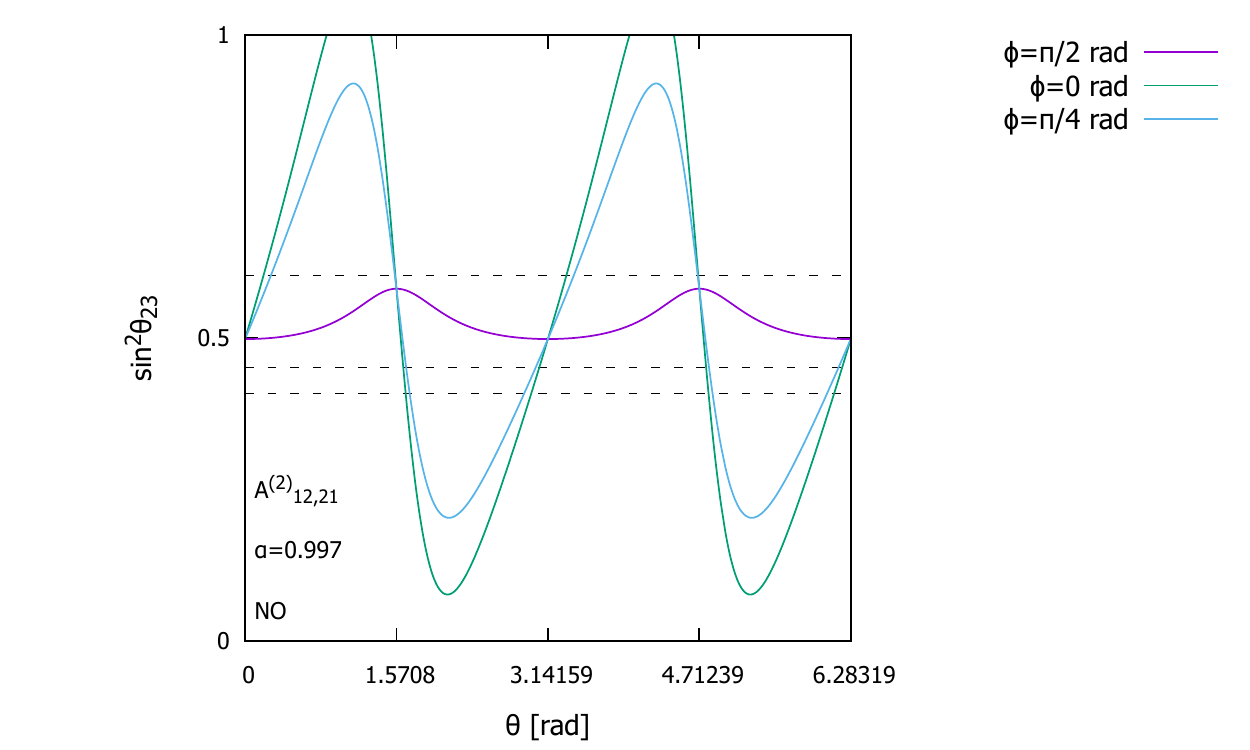}\\
\includegraphics[keepaspectratio, scale=0.5]{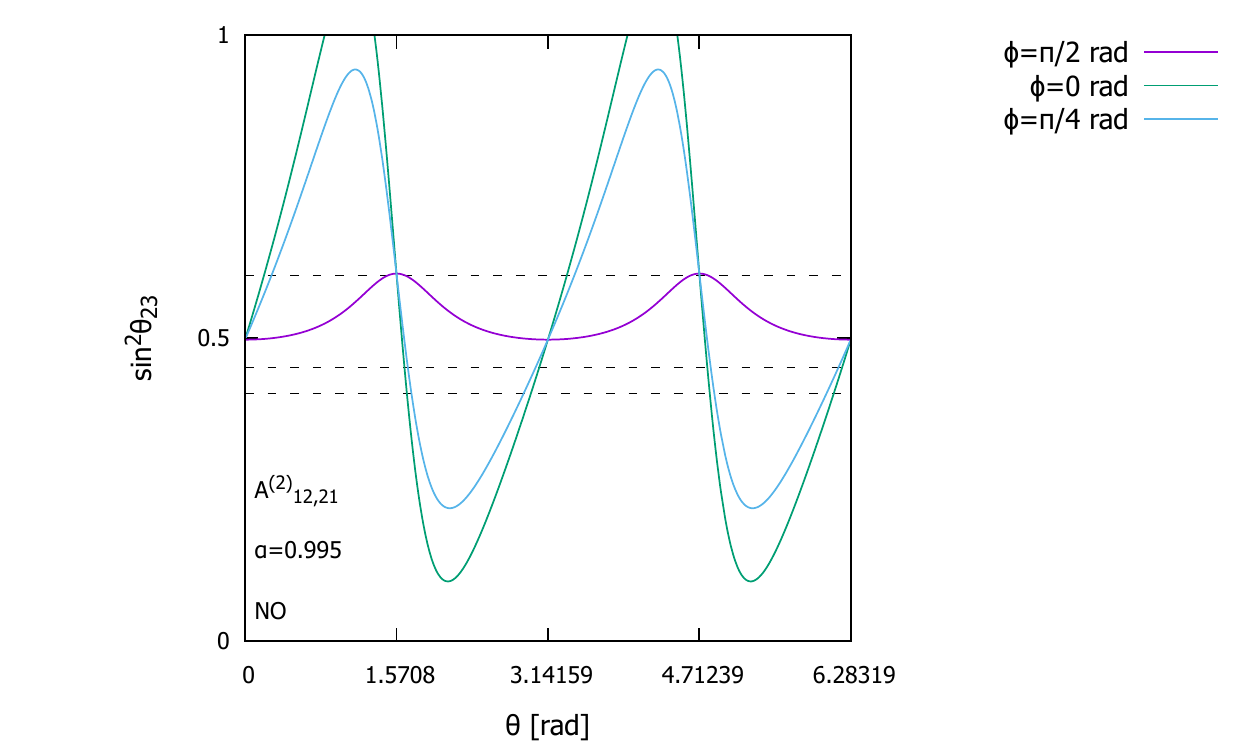}
\end{minipage}&
\begin{minipage}[t]{0.48\hsize}
\centering
\includegraphics[keepaspectratio, scale=0.5]{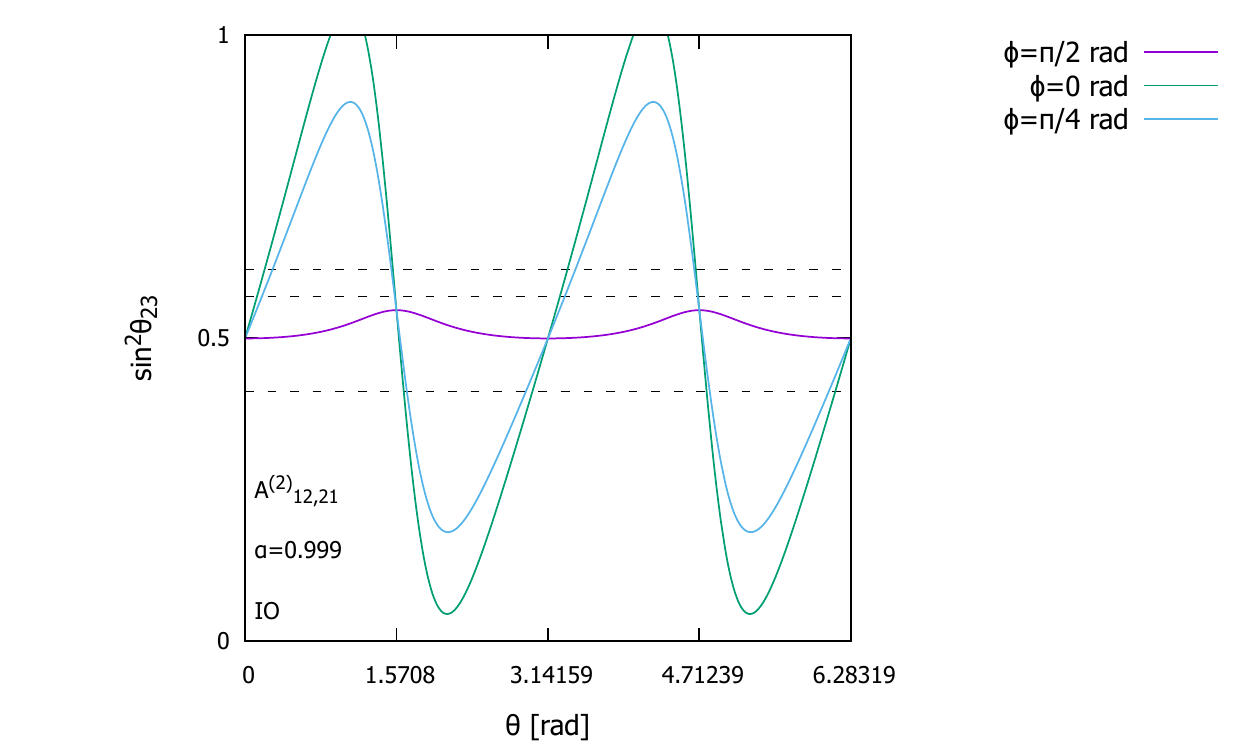}\\
\includegraphics[keepaspectratio, scale=0.5]{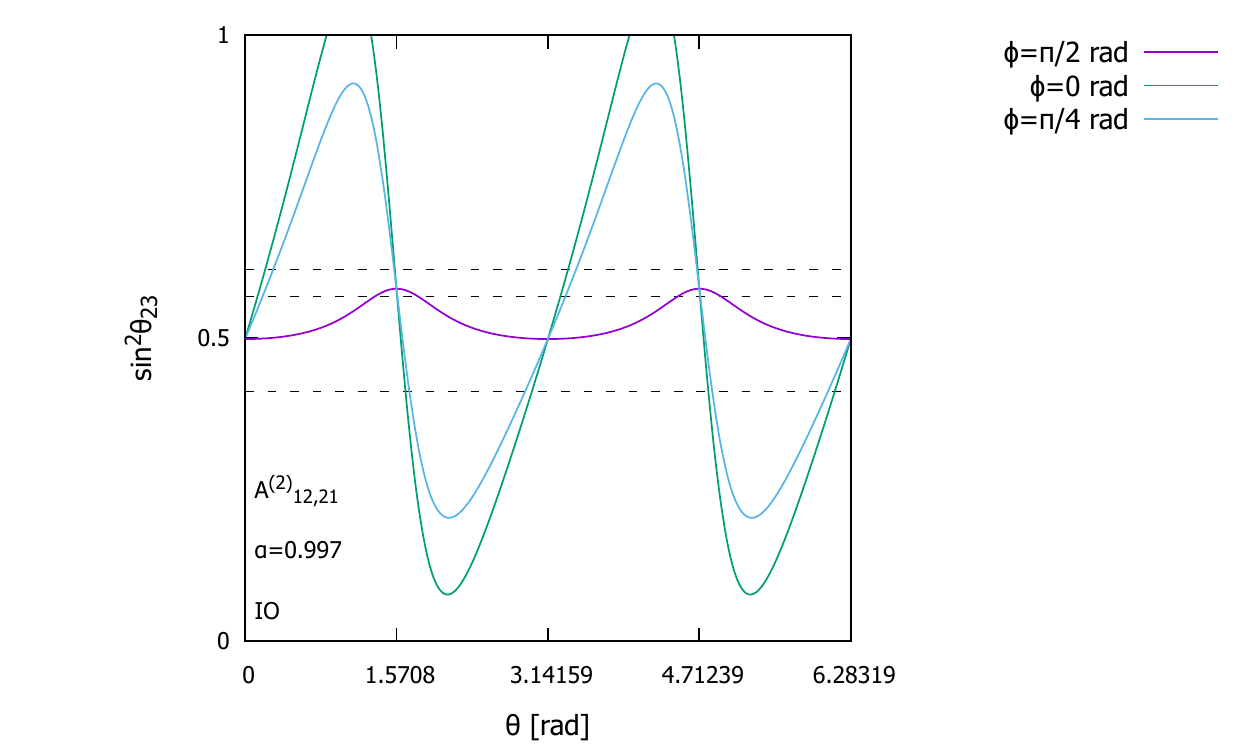}\\
\includegraphics[keepaspectratio, scale=0.5]{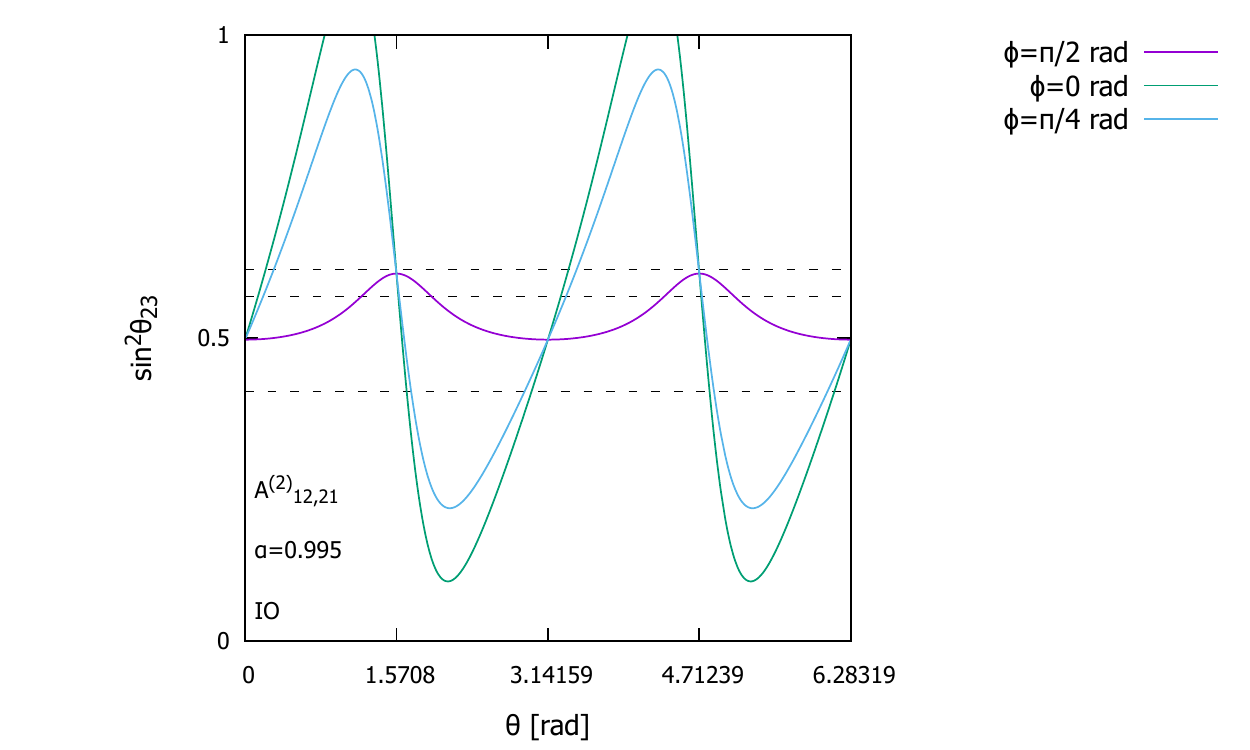}
 \end{minipage} \\
\end{tabular}
 \caption{Same as Fig. \ref{Fig:MTM1_A1221_2_ap_23} but for MTM2($A_{12,21}^{(2)}$)}
 \label{Fig:MTM2_A1221_2_ap_23} 
  \end{figure}

\begin{figure}[t]
\begin{tabular}{cc}
\begin{minipage}[t]{0.48\hsize}
\centering
\includegraphics[keepaspectratio, scale=0.5]{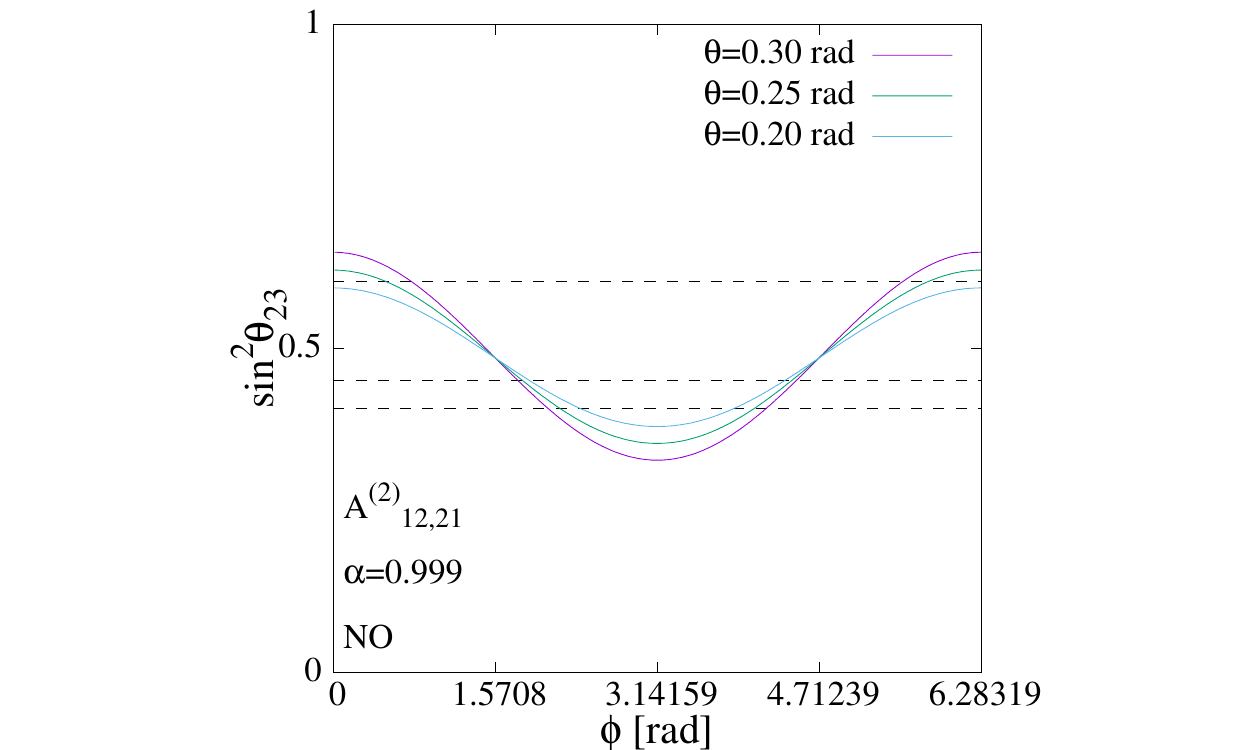}\\
\includegraphics[keepaspectratio, scale=0.5]{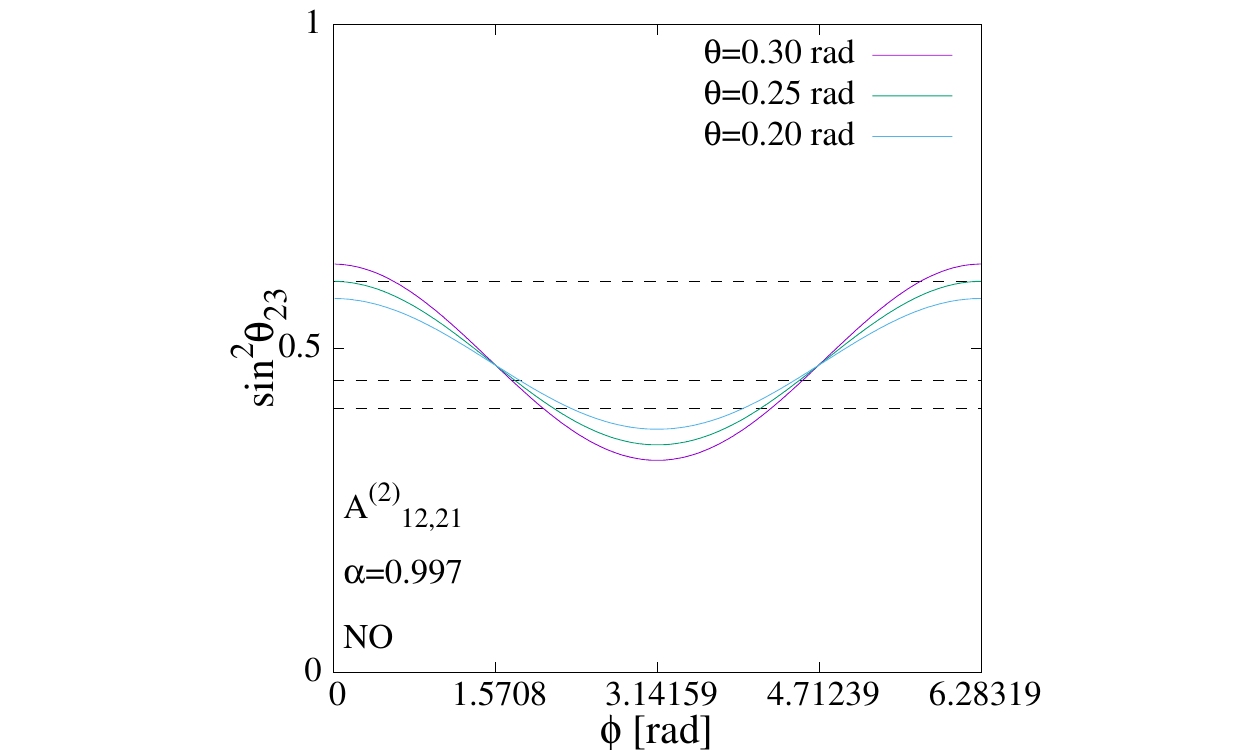}\\
\includegraphics[keepaspectratio, scale=0.5]{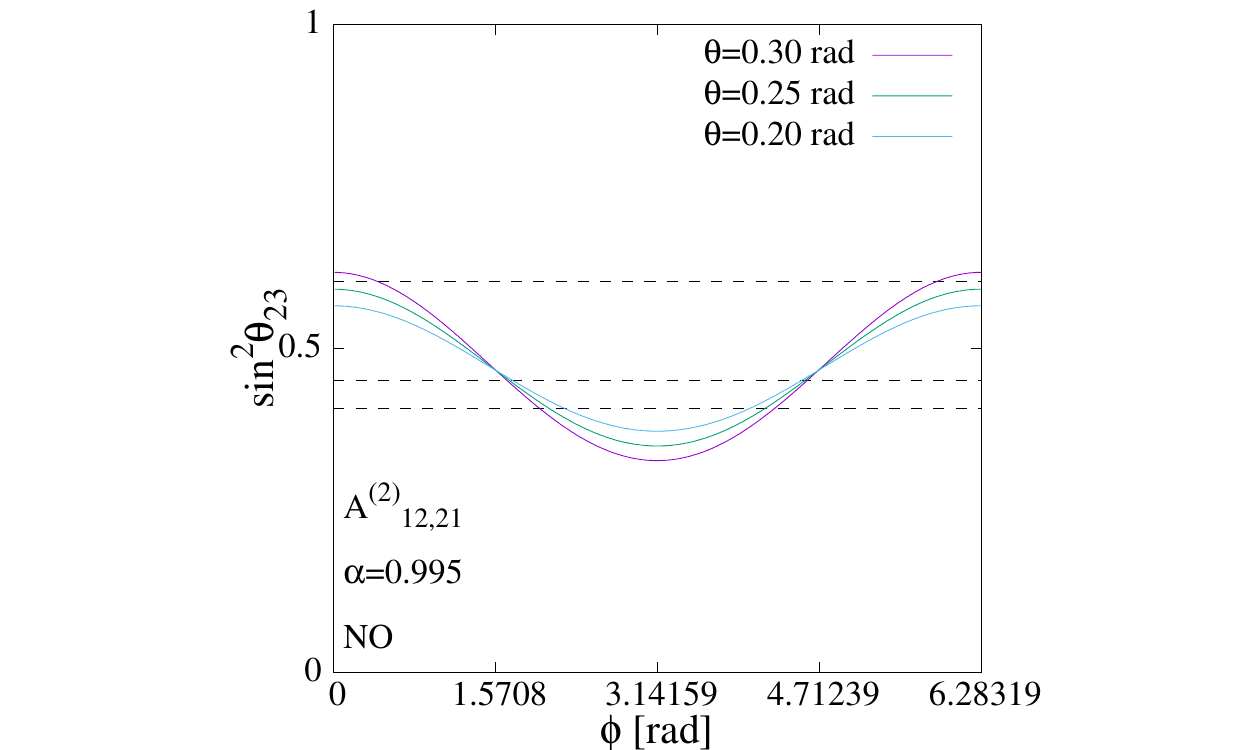}
\end{minipage}&
\begin{minipage}[t]{0.48\hsize}
\centering
\includegraphics[keepaspectratio, scale=0.5]{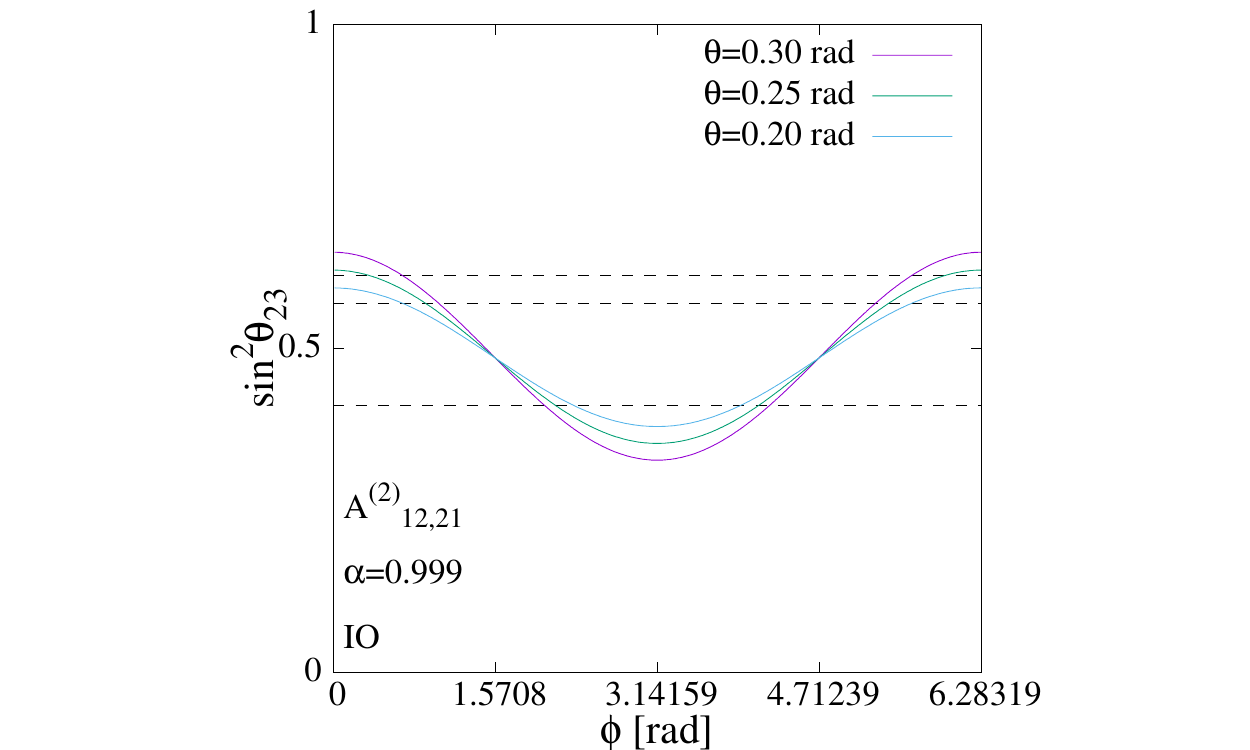}\\
\includegraphics[keepaspectratio, scale=0.5]{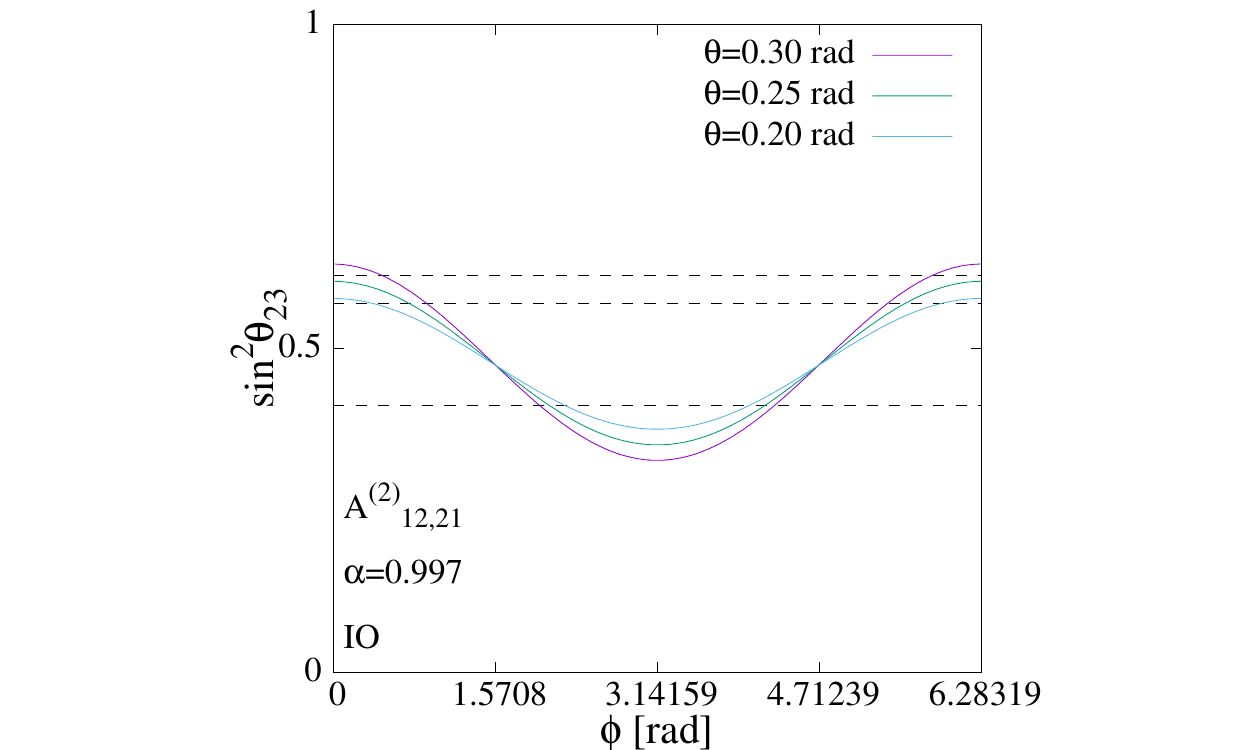}\\
\includegraphics[keepaspectratio, scale=0.5]{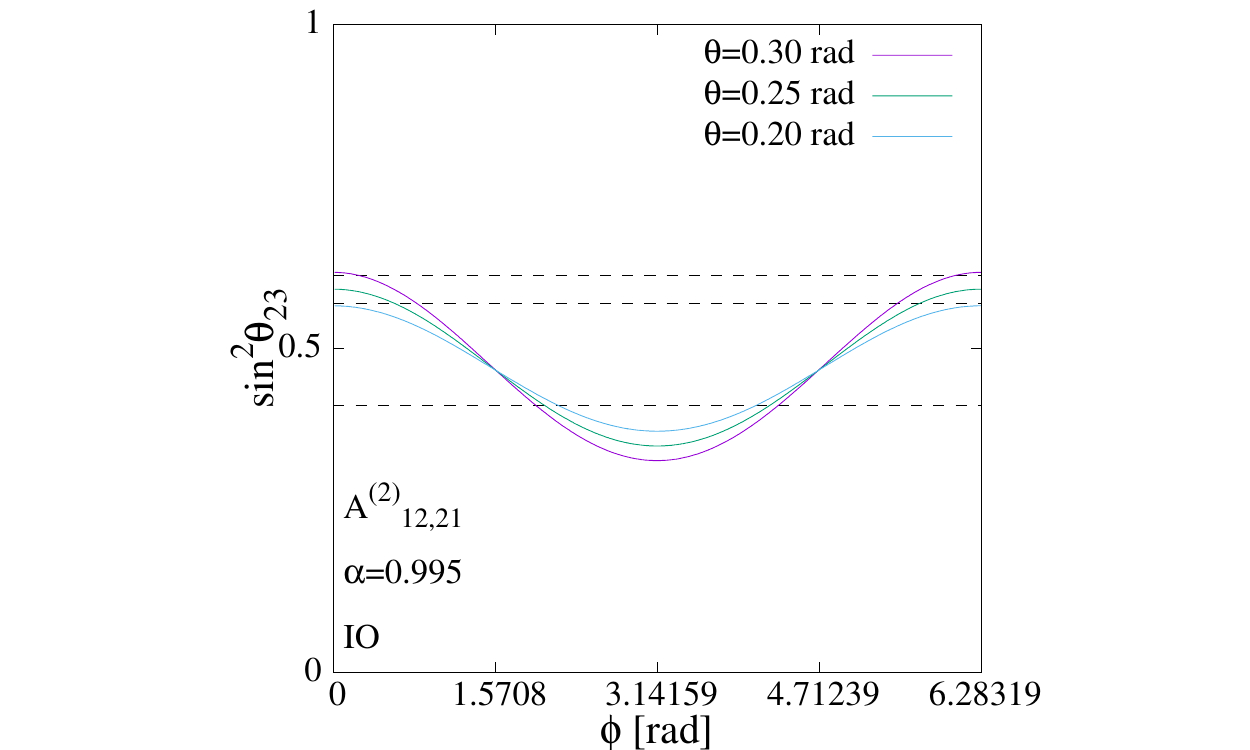}
 \end{minipage} \\
\end{tabular}
 \caption{Same as Fig. \ref{Fig:MTM1_A1221_2_at_23} but for MTM2($A_{12,21}^{(2)}$)}
 \label{Fig:MTM2_A1221_2_at_23} 
  \end{figure}

\begin{figure}[t]
\begin{tabular}{cc}
\begin{minipage}[t]{0.48\hsize}
\centering
\includegraphics[keepaspectratio, scale=0.5]{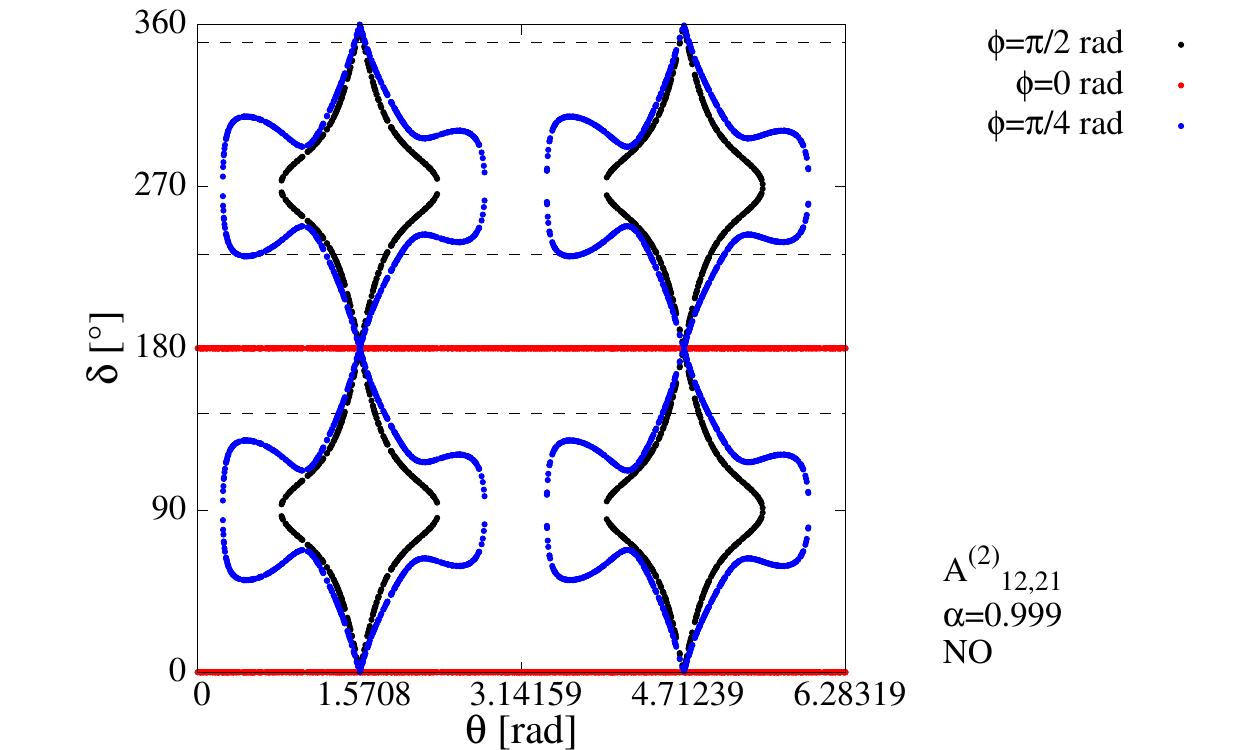}\\
\includegraphics[keepaspectratio, scale=0.5]{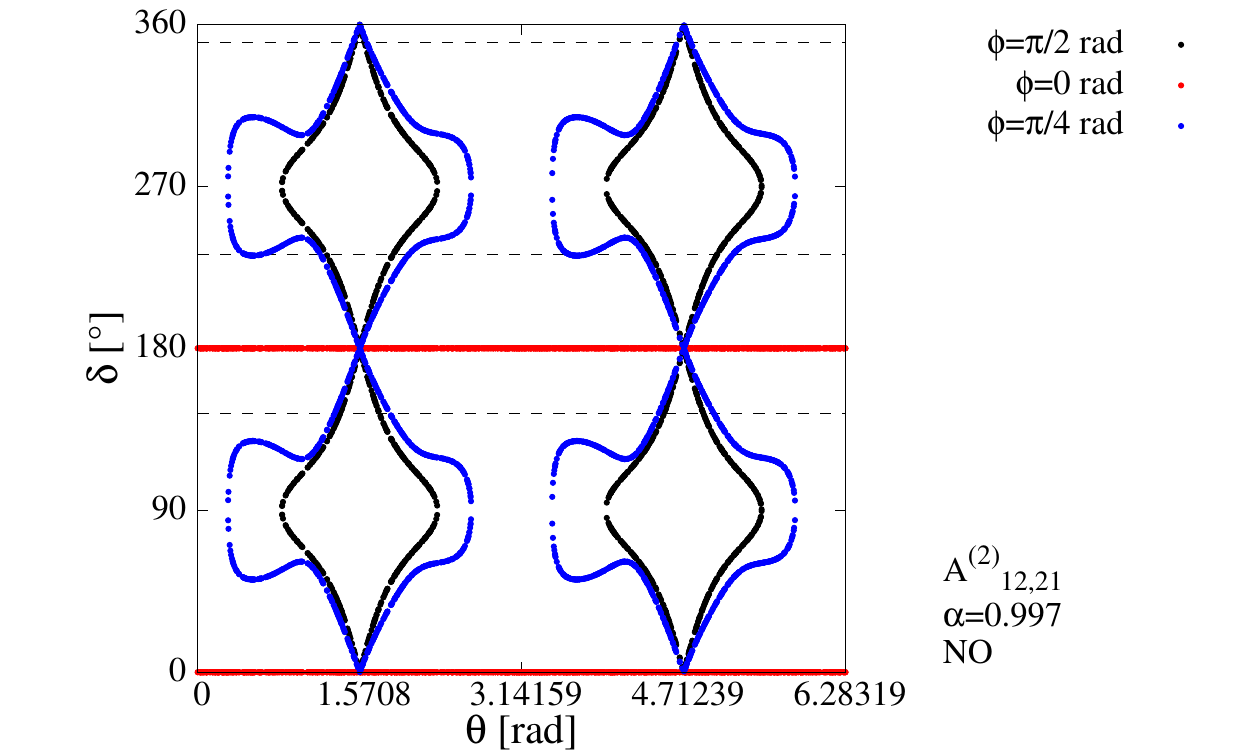}\\
\includegraphics[keepaspectratio, scale=0.5]{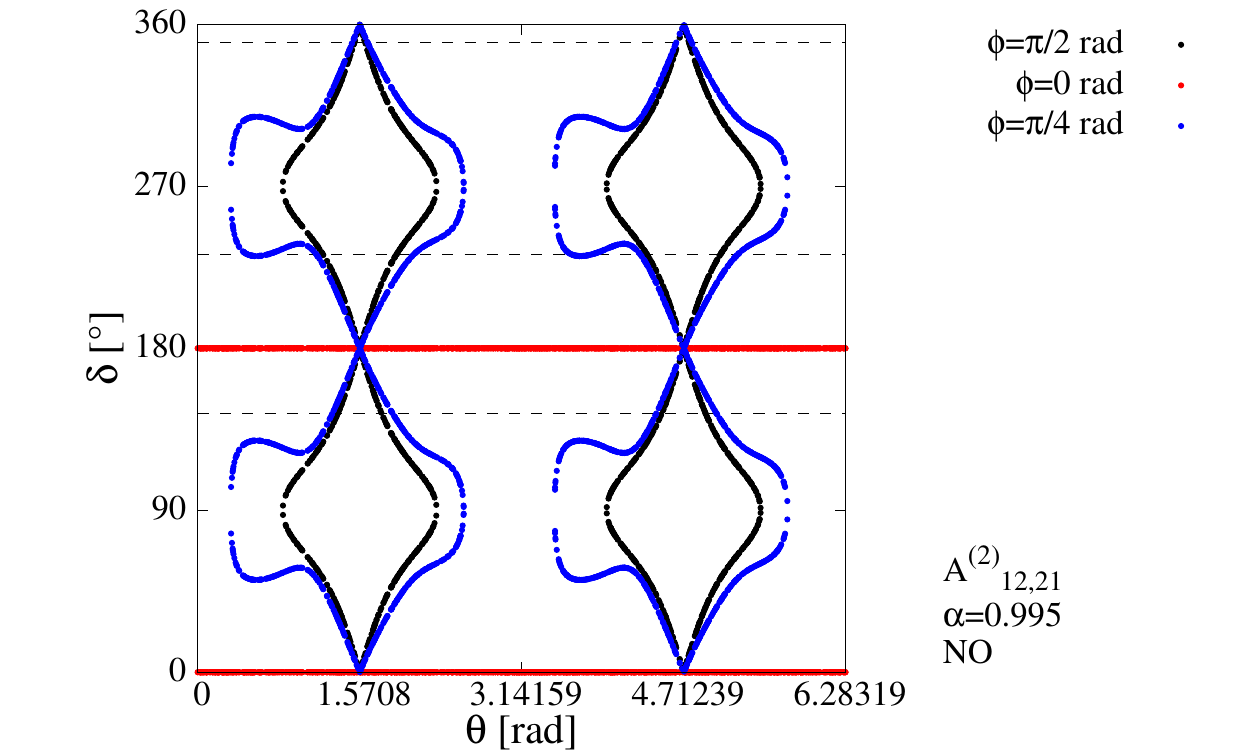}
\end{minipage}&
\begin{minipage}[t]{0.48\hsize}
\centering
\includegraphics[keepaspectratio, scale=0.5]{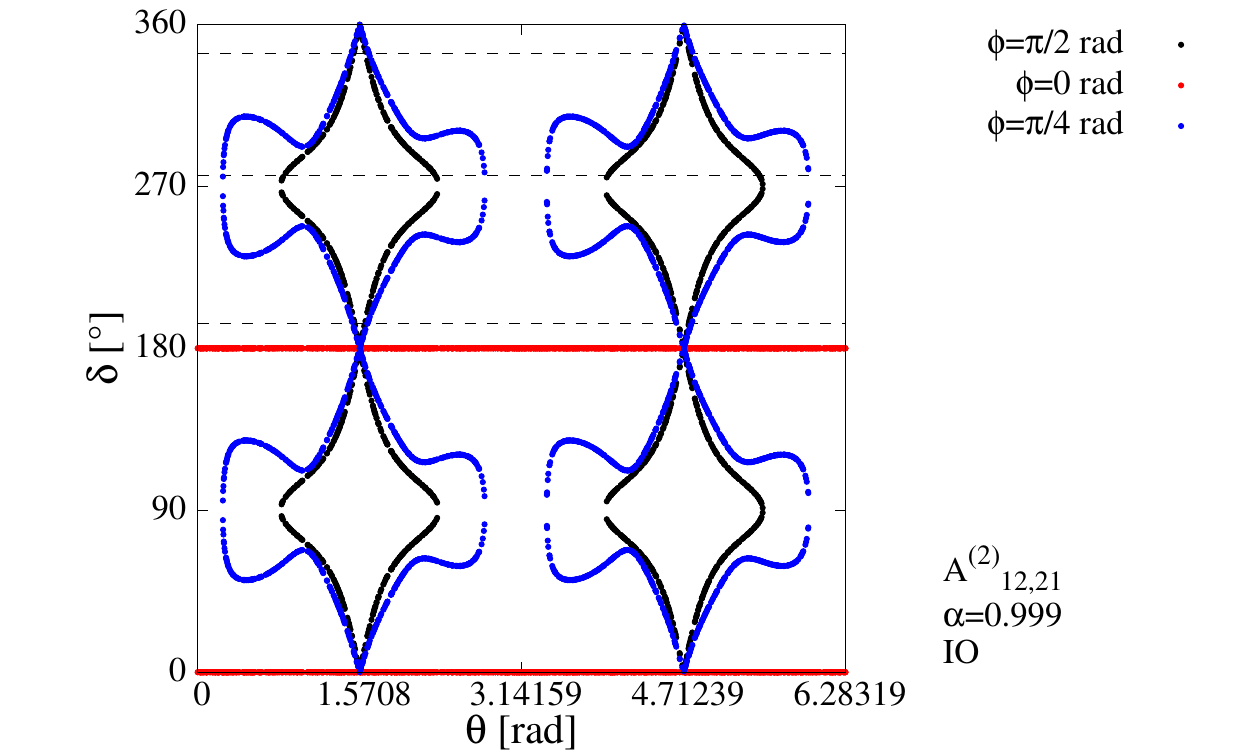}\\
\includegraphics[keepaspectratio, scale=0.5]{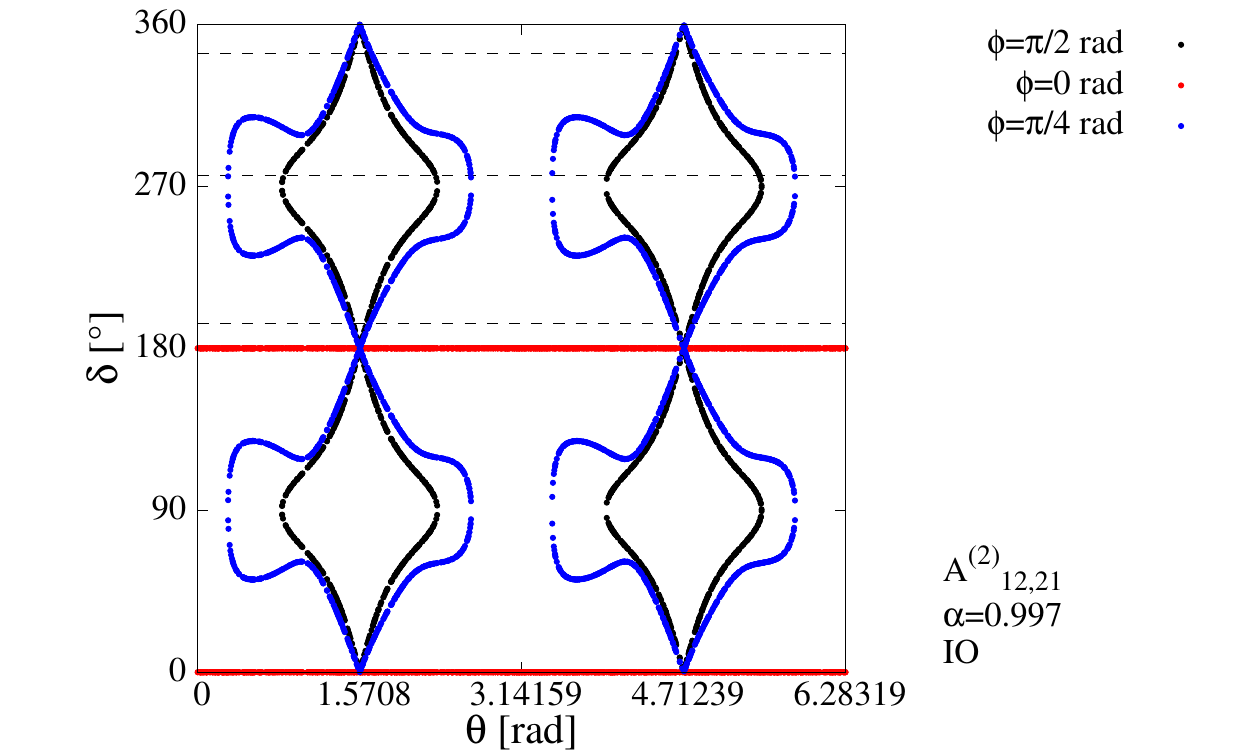}\\
\includegraphics[keepaspectratio, scale=0.5]{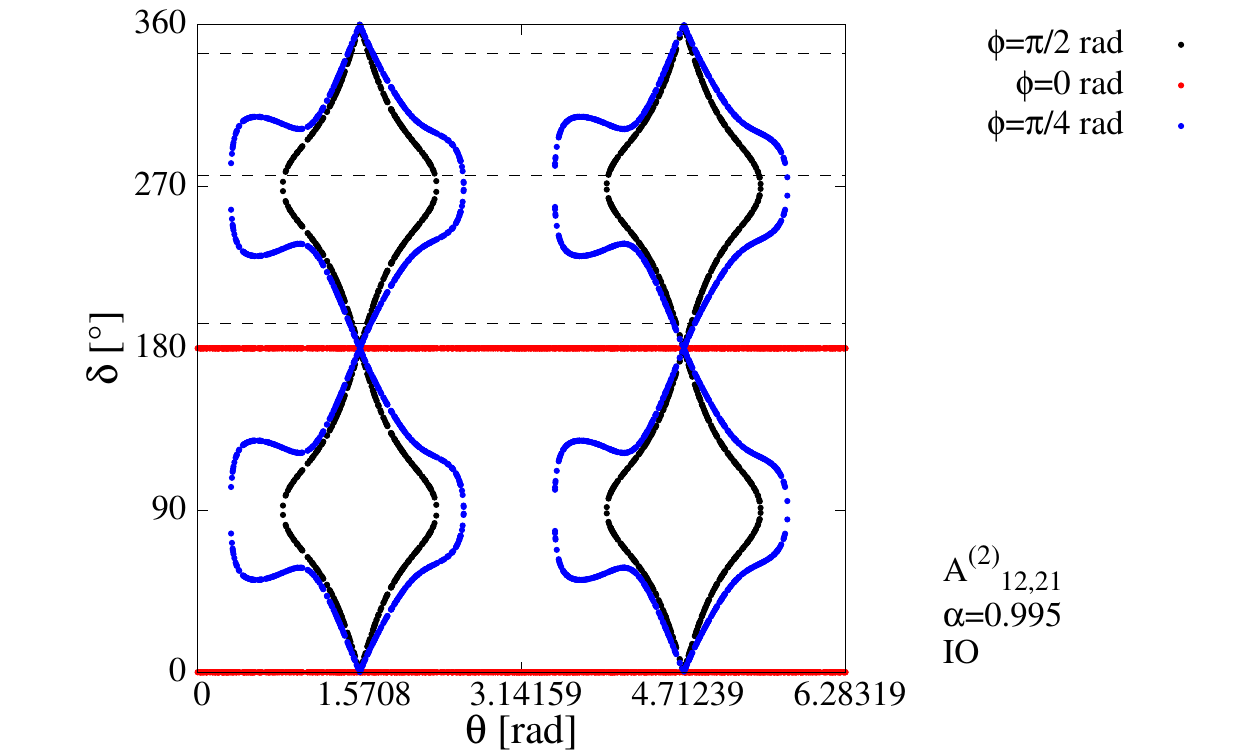}
 \end{minipage} \\
\end{tabular}
 \caption{Same as Fig. \ref{Fig:MTM1_A1221_2_ap_d} but for MTM2($A_{12,21}^{(2)}$)}
 \label{Fig:MTM2_A1221_2_ap_d} 
  \end{figure}

\begin{figure}[t]
\begin{tabular}{cc}
\begin{minipage}[t]{0.48\hsize}
\centering
\includegraphics[keepaspectratio, scale=0.5]{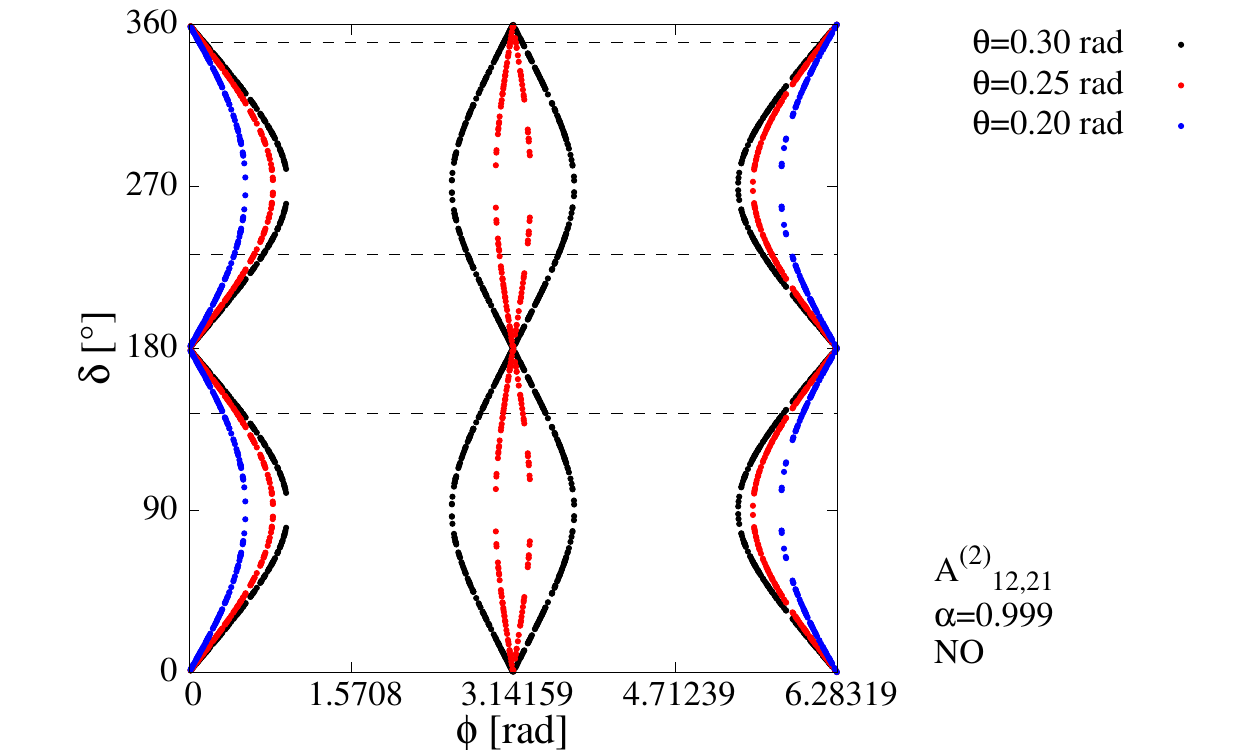}\\
\includegraphics[keepaspectratio, scale=0.5]{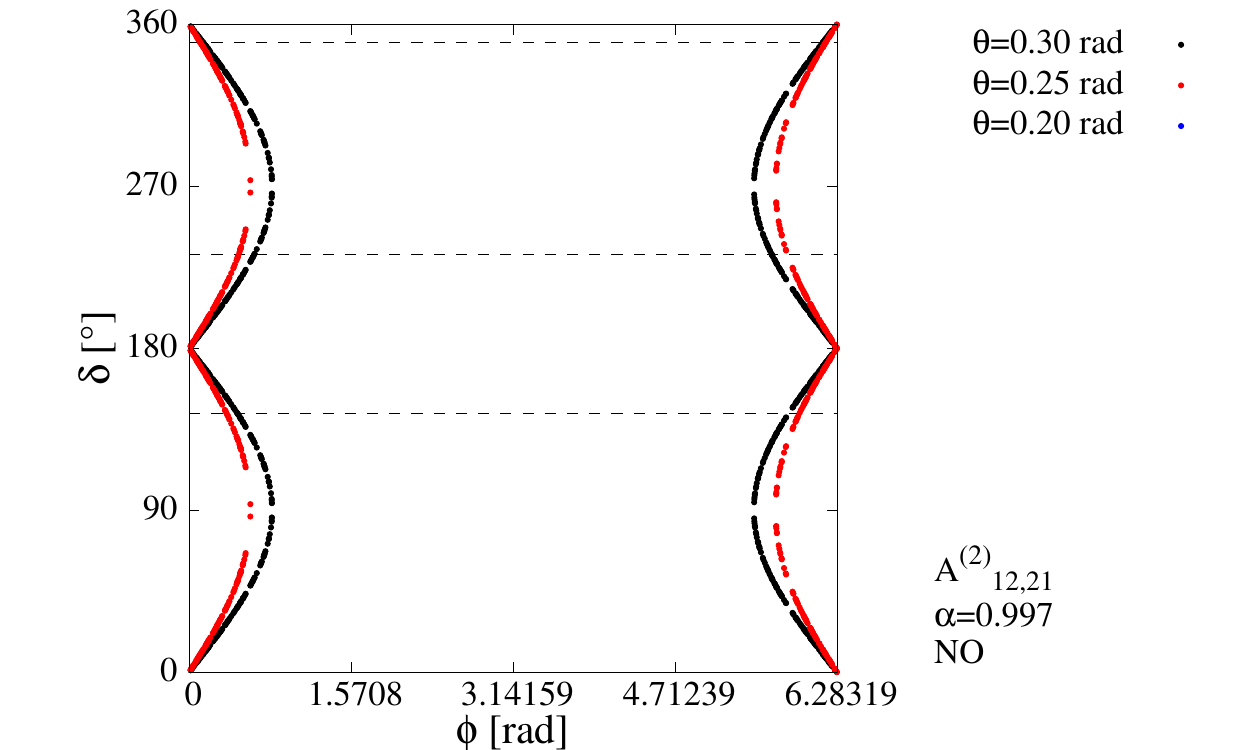}\\
\includegraphics[keepaspectratio, scale=0.5]{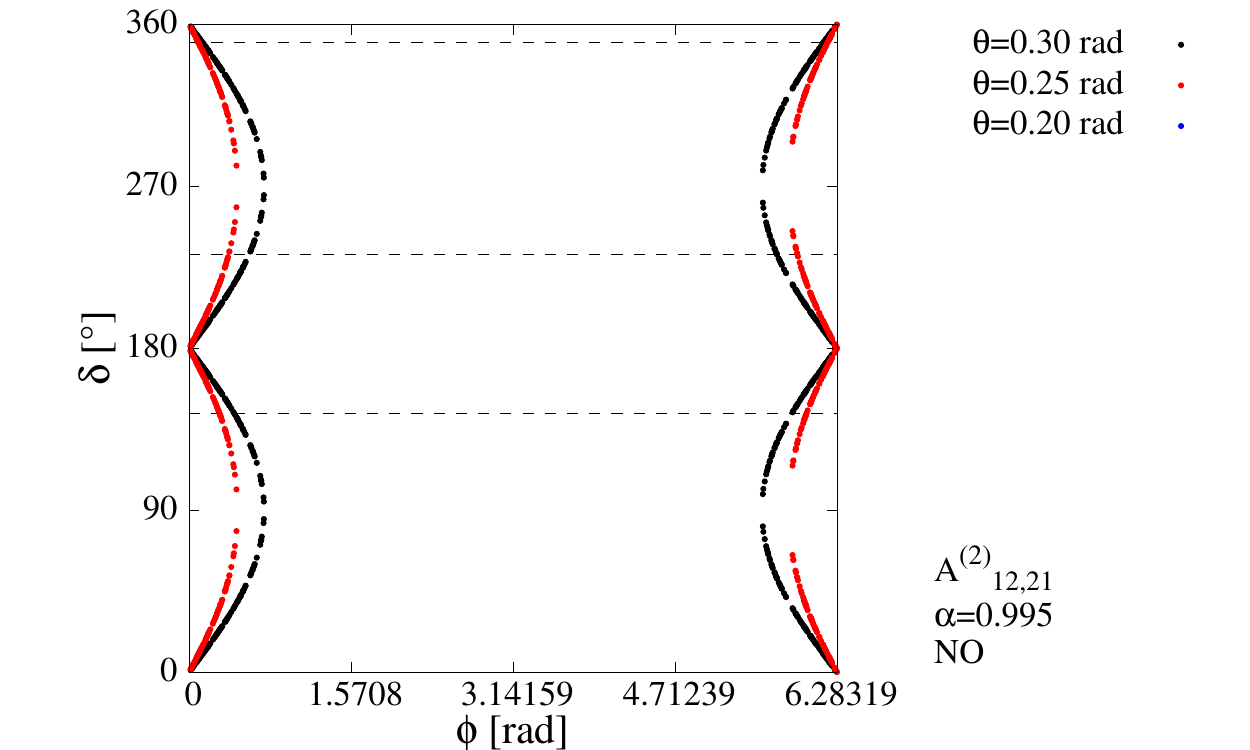}
\end{minipage}&
\begin{minipage}[t]{0.48\hsize}
\centering
\includegraphics[keepaspectratio, scale=0.5]{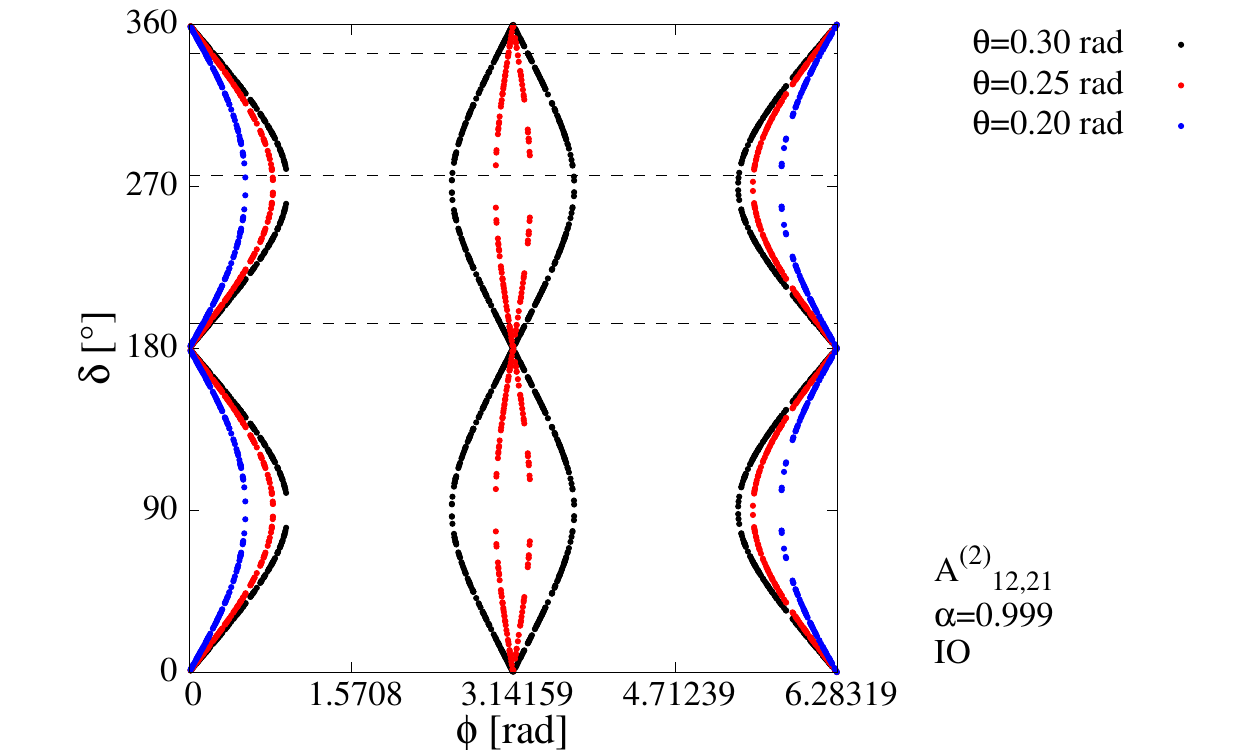}\\
\includegraphics[keepaspectratio, scale=0.5]{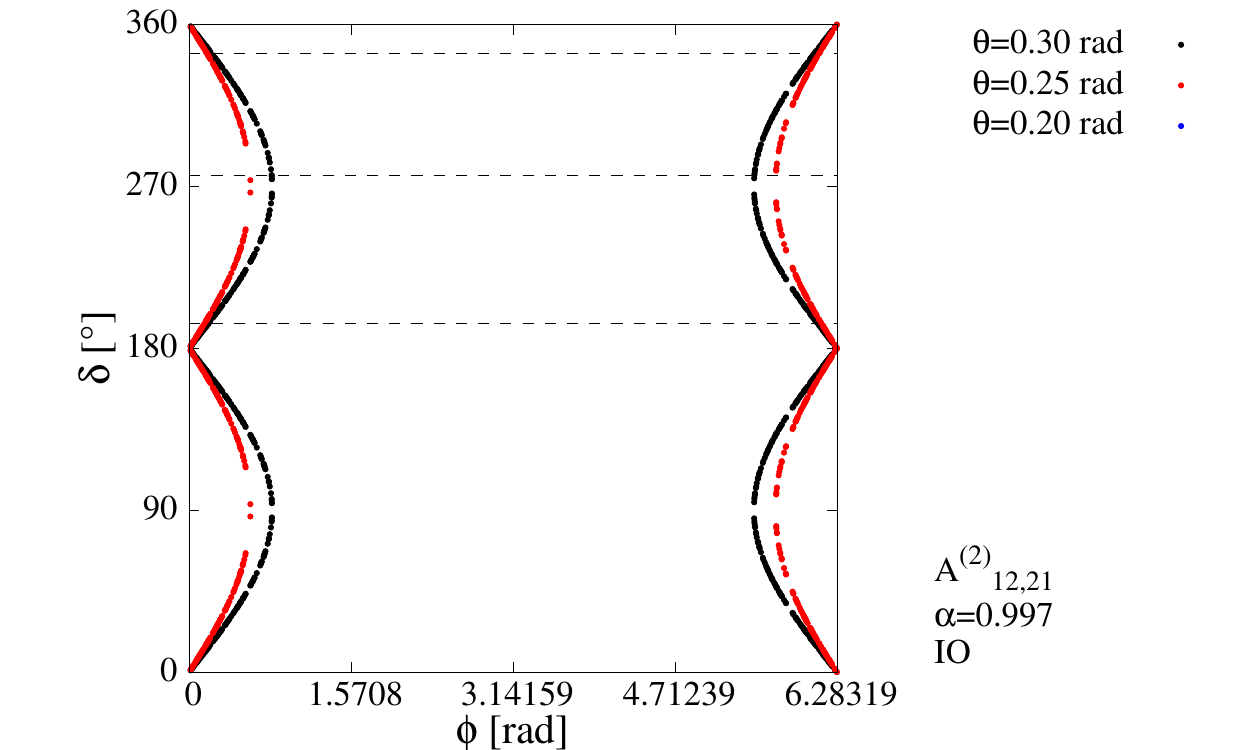}\\
\includegraphics[keepaspectratio, scale=0.5]{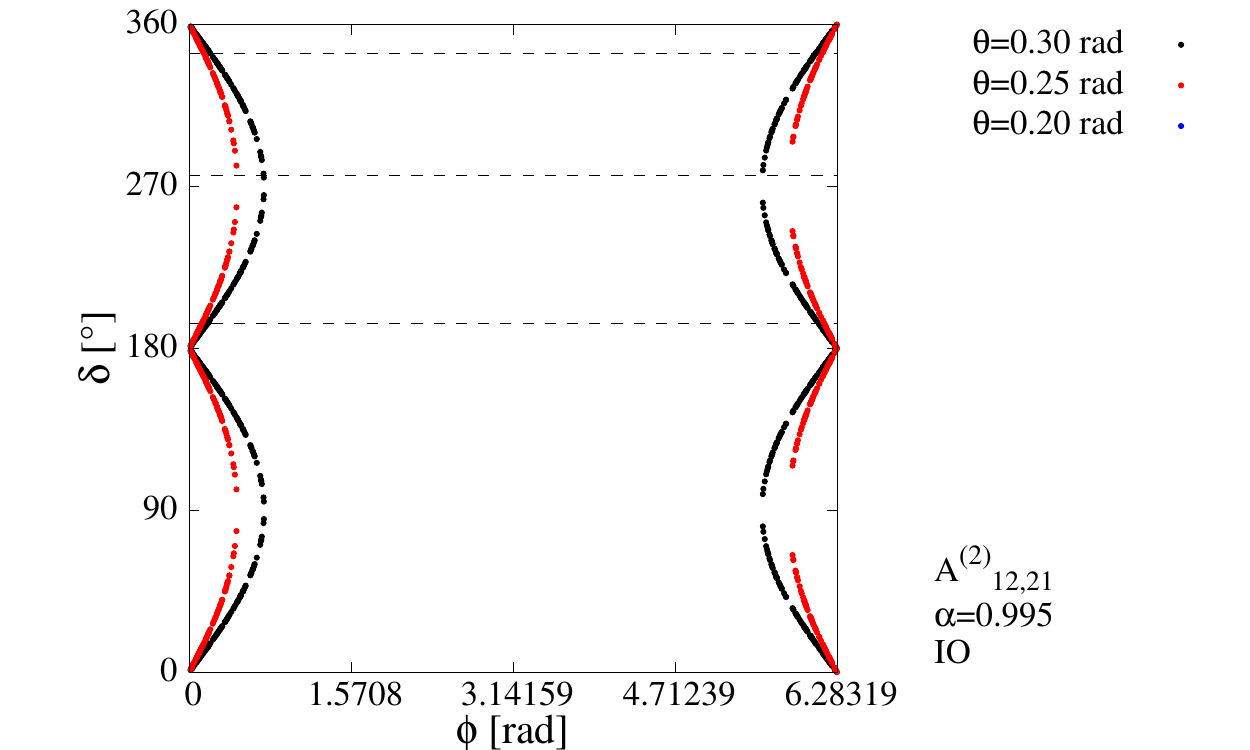}
 \end{minipage} \\
\end{tabular}
 \caption{Same as Fig. \ref{Fig:MTM1_A1221_2_at_d} but for MTM2($A_{12,21}^{(2)}$).}
 \label{Fig:MTM2_A1221_2_at_d} 
  \end{figure}
\begin{figure}[t]
\begin{tabular}{cc}
\begin{minipage}[t]{0.48\hsize}
\centering
\includegraphics[keepaspectratio, scale=0.5]{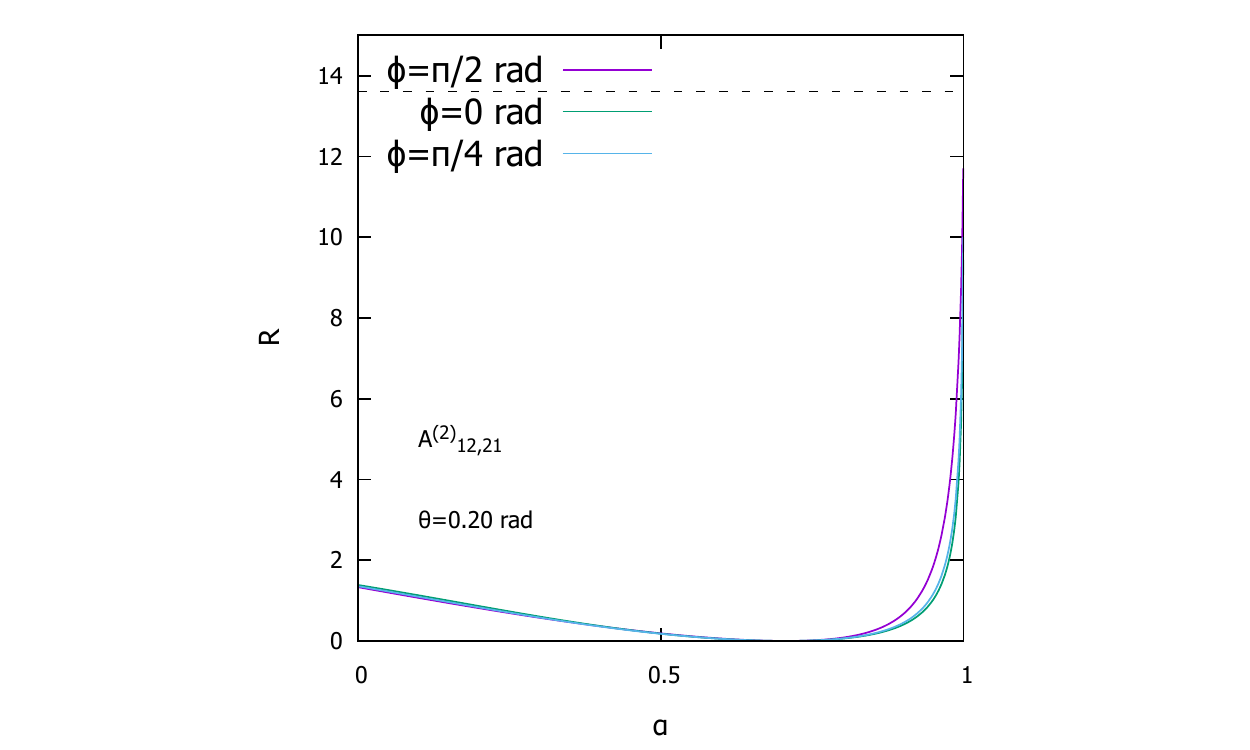}\\
\includegraphics[keepaspectratio, scale=0.5]{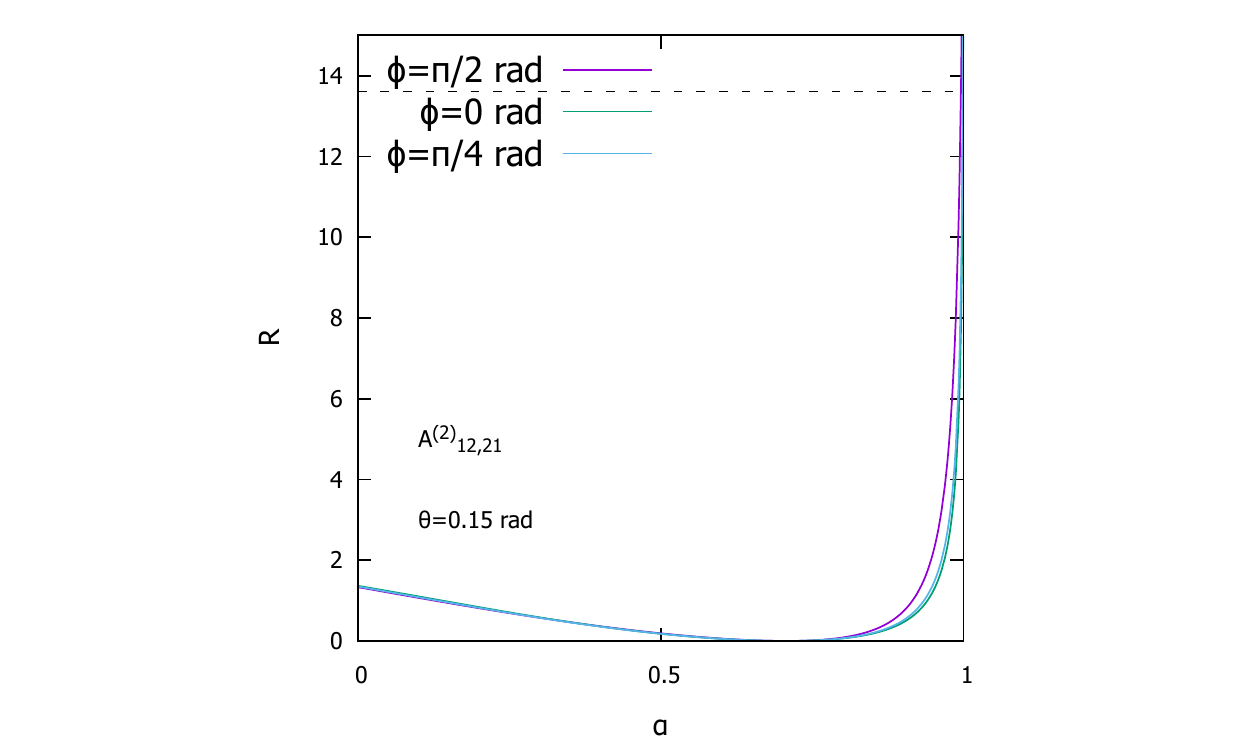}\\
\includegraphics[keepaspectratio, scale=0.5]{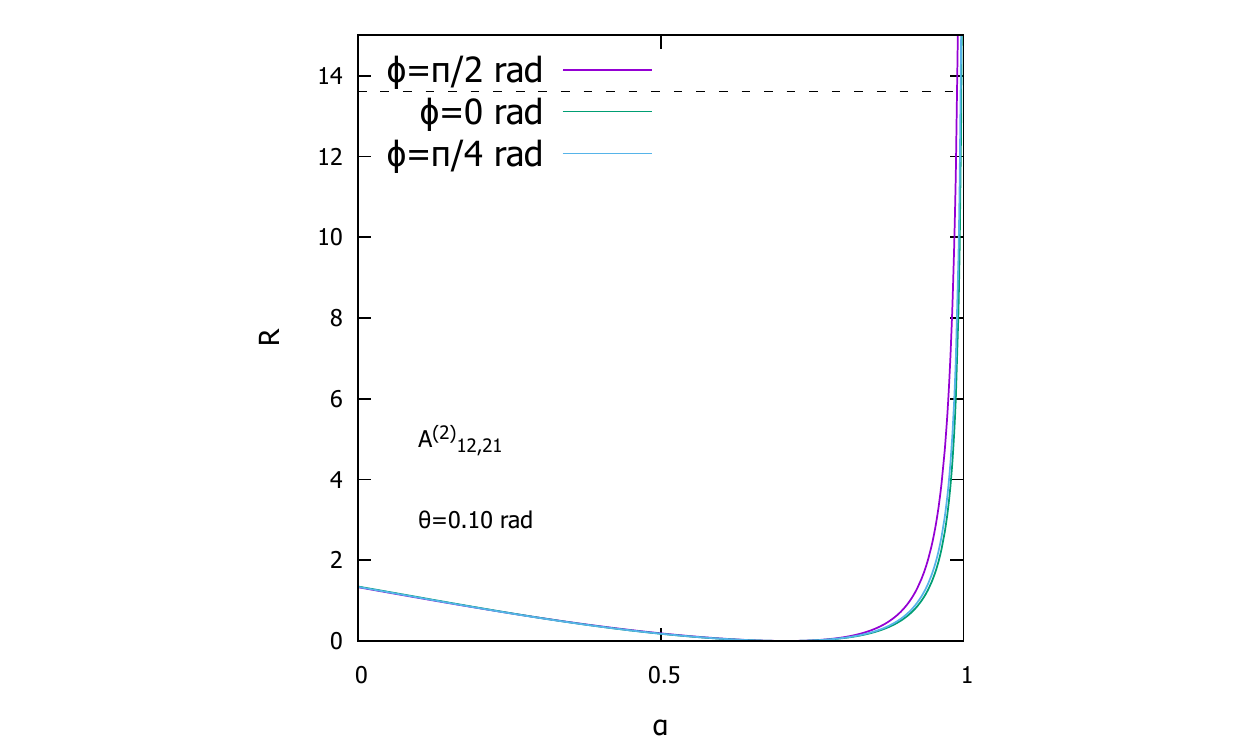}
\end{minipage}&
\begin{minipage}[t]{0.48\hsize}
\centering
\includegraphics[keepaspectratio, scale=0.5]{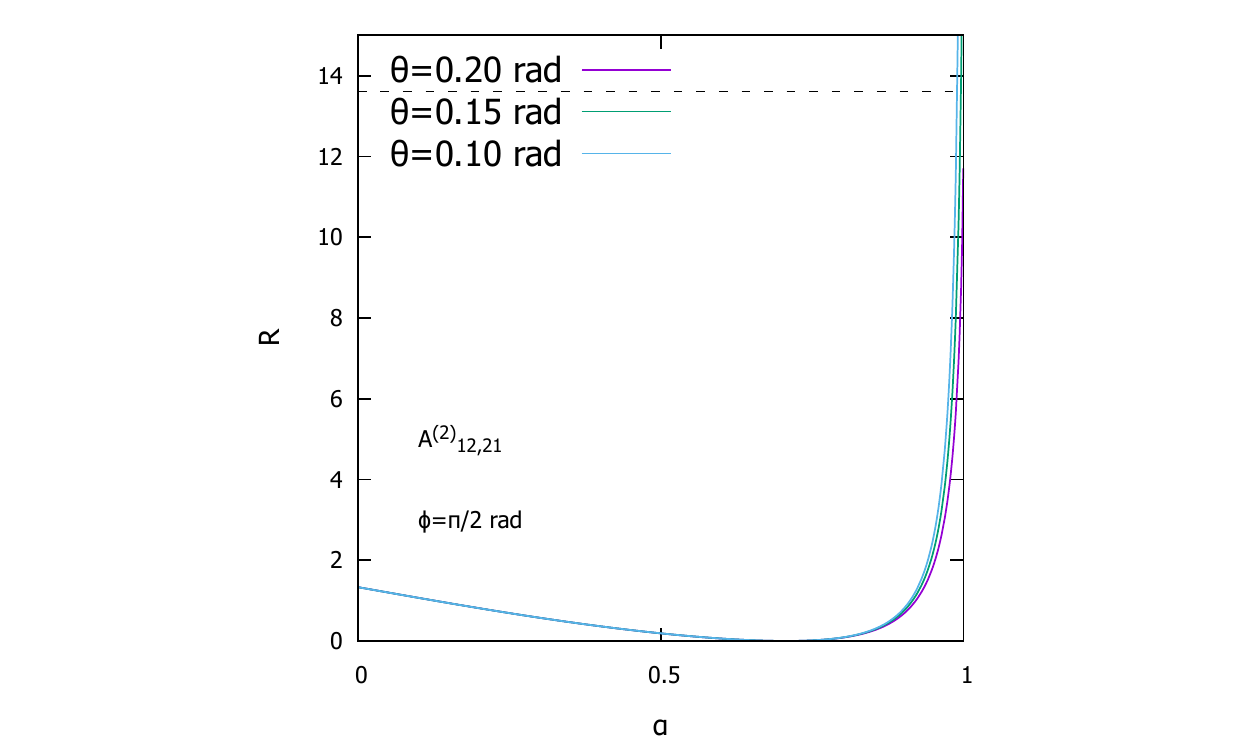}\\
\includegraphics[keepaspectratio, scale=0.5]{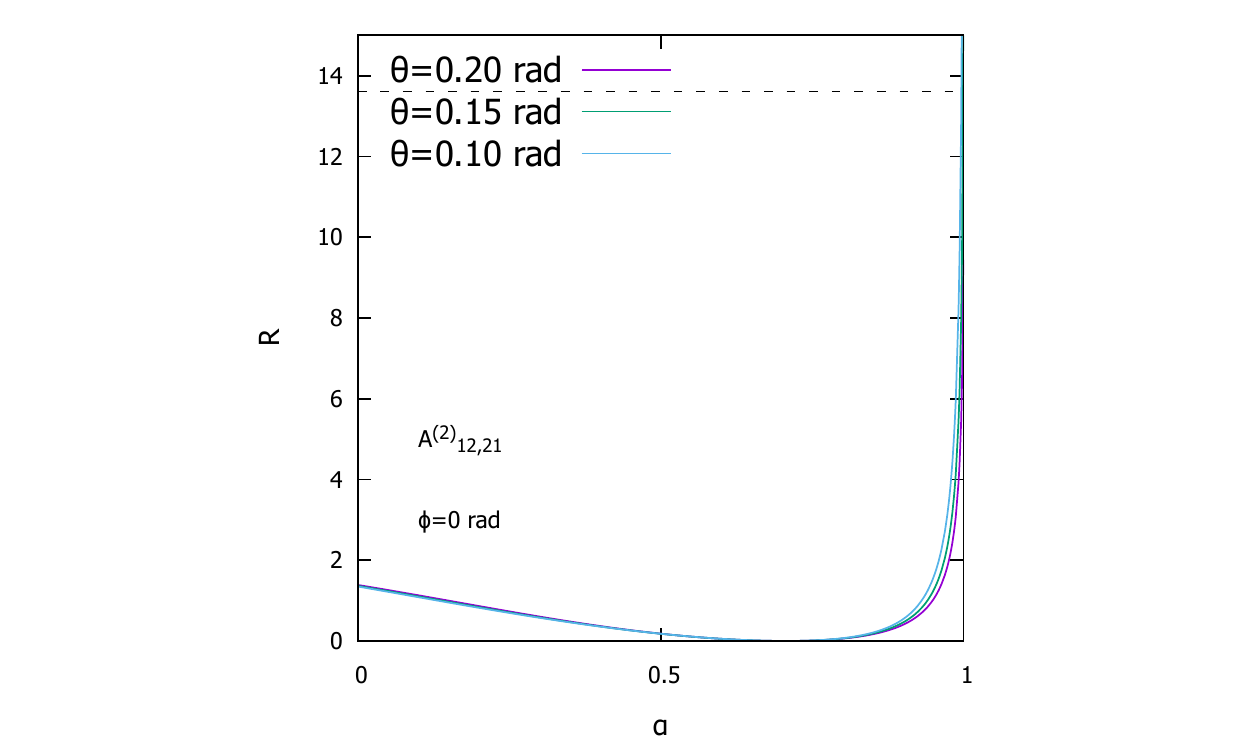}\\
\includegraphics[keepaspectratio, scale=0.5]{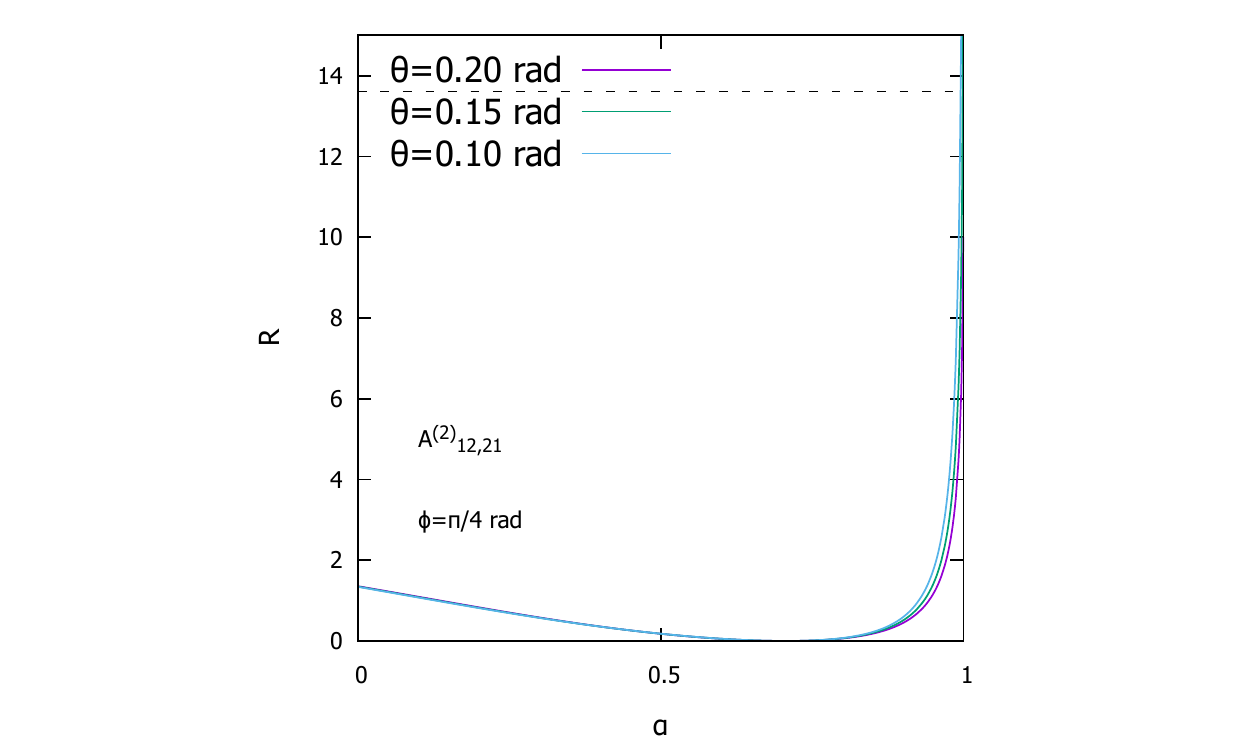}
 \end{minipage} \\
\end{tabular}
 \caption{Same as Fig. \ref{Fig:MTM1_A1221_2_R} but for MTM2($A_{12,21}^{(2)}$).}
 \label{Fig:MTM2_A1221_2_R} 
  \end{figure}

The TM2 mixing can also be subject to the same modification method employed for the TM1 mixing. The modified TM2 mixing, utilizing the rotation matrices $A_{12,21}^{(1)}$ and $A_{12,21}^{(2)}$, can be expressed as follows:
\begin{eqnarray}
\tilde{U}_2(A_{12,21}^{(1)}) = A_{12,21}^{(1)} U_2, \quad \tilde{U}_2(A_{12,21}^{(2)}) = A_{12,21}^{(2)} U_2.
\end{eqnarray}
We call these mixings MTM2($A_{12,21}^{(1)}$) and MTM2($A_{12,21}^{(2)}$), respectively. The mixing angles are 
\begin{eqnarray}
s^2_{12} &=& \frac{1 \pm 2\alpha \sqrt{1-\alpha^2}}{3(1-s^2_{13})},  \label{Eq:MTM2_A1221_s12ss13s} \\
s^2_{23} &=& \frac{1}{6(1-s^2_{13})} \label{Eq:MTM2_A1221_s23ss13s} \left\{ 3\alpha^2 + (4-6\alpha^2 \pm 4\alpha \sqrt{1-\alpha^2})\sin^2\theta \right. \nonumber \\
  && \quad \left. + \sqrt{3}\alpha(\alpha \pm 2\sqrt{1-\alpha^2}) \sin 2\theta \cos\phi \right\}, \nonumber \\
s^2_{13} &=& \frac{1}{6} (2 \mp  2 \alpha \sqrt{1-\alpha^2} + B_3 + B_4),   \label{Eq:MTM2_A1221_13s}
\end{eqnarray}
where
\begin{eqnarray}
B_3 &=&  (1 - 3\alpha^2 \pm 2\alpha \sqrt{1-\alpha^2})\cos 2\theta,  \nonumber \\
B_4 &=& \sqrt{3}(1-\alpha^2 \mp 2\alpha \sqrt{1-\alpha^2}) \sin 2\theta \cos\phi.
\label{Eq:B3_B4}
\end{eqnarray}
The upper sign of $\pm$ and $\mp$ should be taken for MTM2($A_{12,21}^{(1)}$). The lower sign of these should be taken for MTM2($A_{12,21}^{(2)}$).

Figure \ref{Fig:MTM2_A1221_12_13} presents the predictions of $\theta_{13}$ and $\theta_{12}$ in the MTM2($A_{12,21}^{(1)}$) and MTM2($A_{12,21}^{(2)}$) scenarios. The horizontal and vertical dotted lines in each panel correspond to those in Figure \ref{Fig:MTM1_A1221_12_13}. Similar to MTM1($A_{12,21}^{(1)}$) and MTM1($A_{12,21}^{(2)}$), the simultaneous reproducibility of $\theta_{12}$ and $\theta_{13}$ is reduced in MTM2($A_{12,21}^{(1)}$). Conversely, this reproducibility is significantly improved in MTM2($A_{12,21}^{(2)}$). For instance, the best-fit values of $\theta_{12}$ and $\theta_{13}$ can be obtained simultaneously with a small correction ($\alpha \simeq 0.9985$). Moving forward, we will solely focus on MTM2($A_{12,21}^{(2)}$).

Figure \ref{Fig:MTM2_A1221_cosphi_13_23} illustrates the prediction of $\theta_{13}$ (upper panel) and $\theta_{23}$ (lower panel) as a function of $\phi$ in the MTM2($A_{12,21}^{(2)}$) scenario. Similar to MTM1($A_{12,21}^{(2)}$), the figure suggests that by selecting appropriate values of $\phi$ and $\theta$, we can obtain $\theta_{13}$ and $\theta_{23}$ values that are consistent with the observations. Through our numerical parameter search, we have confirmed that $s^2_{13}$ and $s^2_{23}$ can be obtained within the $3\sigma$ region. 

A benchmark point 
\begin{eqnarray}
(\alpha, \theta, \phi) = (0.99847, \ 10.31^\circ,\  98.61^\circ),
\label{Eq:MTM2_A1221_benchmark}
\end{eqnarray}
yields the best-fit values of $s_{12}^2$ and $s_{13}^2$ and the allowed value of $s_{23}^2$ as follows:
\begin{eqnarray}
(s_{12}^2, s_{23}^2, s_{13}^2, \delta) = (0.303, \ 0.484,\ 0.02225,\ 263.8^\circ). 
\end{eqnarray}

Figures \ref{Fig:MTM2_A1221_2_a1p1_12_13} - \ref{Fig:MTM2_A1221_2_R} show that the ballpark figures of the parameter space and the possible ranges of those mixing parameters. From these figures, we have the similar conclusions for MTM2($A_{12,21}^{(2)}$) as for  MTM1($A_{12,21}^{(2)}$) as follows,
\begin{itemize}
\item The wide range of parameters $\theta$, $\phi$ and $\alpha$ are consistent with observatoins (Figures \ref{Fig:MTM2_A1221_2_a1p1_12_13}, \ref{Fig:MTM2_A1221_2_ap_23} and \ref{Fig:MTM2_A1221_2_at_23}). The wide range of Dirac CP phase is also consistent with observation (Figures \ref{Fig:MTM2_A1221_2_ap_d} and \ref{Fig:MTM2_A1221_2_at_d}).
\item If the values of $\theta_{23}$ and the CP phase $\delta$ are finally pinned down, we can reproduce these fixed values with appropriate values of $\alpha$, $\theta$ and $\phi$.
\item If the best-fit values change in the future, the new best-fit values can be reproduced with appropriate selection of the values of $\alpha$, $\theta$, $\phi$ (Figure \ref{Fig:MTM2_A1221_2_R}).
\end{itemize}
%

\subsection{$A_{23,32}^{(1)}$ and $A_{23,32}^{(2)}$}
We modify TM2 mixing using the rotation matrix $A_{23,32}^{(1)}$ and $A_{23,32}^{(2)}$ as follows: 
\begin{eqnarray}
\tilde{U}_2(A_{23,32}^{(1)}) = A_{23,32}^{(1)} U_2, \quad \tilde{U}_2(A_{23,32}^{(2)}) = A_{23,32}^{(2)} U_2, 
\end{eqnarray}
and we call these mixings MTM2($A_{23,32}^{(1)}$) and MTM2($A_{23,32}^{(2)}$), respectively. The mixing angles are 
\begin{eqnarray}
s^2_{12} &=& \frac{1}{3(1-s^2_{13})},  \label{Eq:MTM2_A2332_s12ss13s} \\
s^2_{23} &=& \frac{1}{6(1-s^2_{13})} \left\{ 2 \mp 2\alpha\sqrt{1-\alpha^2}  +(1 \mp 4\alpha \sqrt{1-\alpha^2}\cos 2\theta \right.  \label{Eq:MTM2_A2332_s23ss13s} \nonumber \\
  &&  \left. + \sqrt{3}(2\alpha^2-1) \sin 2\theta \cos\phi)\right\}. \nonumber \\
s^2_{13} &=& \frac{2}{3} \sin^2\theta. \label{Eq:MTM2_A2332_s13s} 
\end{eqnarray}
The negative sign of $\mp$ should be taken for MTM2$(A_{23,32}^{(1)})$, and the positive sign should be taken for MTM2$(A_{23,32}^{(2)})$.

After comparing Eq. (\ref{Eq:MTM2_A2332_s12ss13s}) and Eq. (\ref{Eq:s12ss13sUTM2}), as well as Eq. (\ref{Eq:MTM2_A2332_s13s}) and Eq. (\ref{Eq:s13sUTM2}), it is evident that the simultaneous reproducibility of $\theta_{12}$ and $\theta_{13}$ is not improved in MTM2($A_{23,32}^{(1)}$) and MTM2($A_{23,32}^{(2)}$). Therefore, no further analysis is conducted for MTM2($A_{23,32}^{(1)}$) and MTM2($A_{23,32}^{(2)}$).

\subsection{$A_{13,31}^{(1)}$ and $A_{13,31}^{(2)}$}

\begin{figure}[t]
\begin{center}
\includegraphics[scale=1.0]{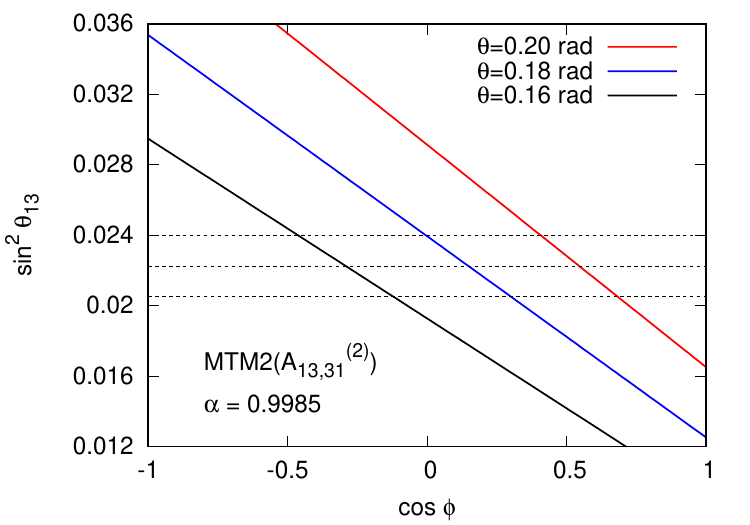}
\includegraphics[scale=1.0]{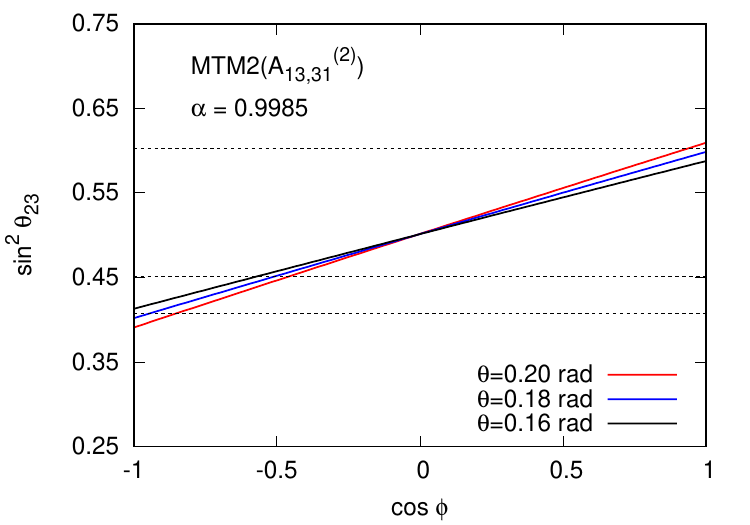}
\caption{Same as Fig. \ref{Fig:MTM1_A1221_cosphi_13_23} but for MTM2.}
\label{Fig:MTM2_A1331_cosphi_13_23} 
\end{center}
\end{figure}

\begin{figure}[t]
\begin{tabular}{cc}
\begin{minipage}[t]{0.3\hsize}
\centering
\includegraphics[keepaspectratio, scale=0.5]{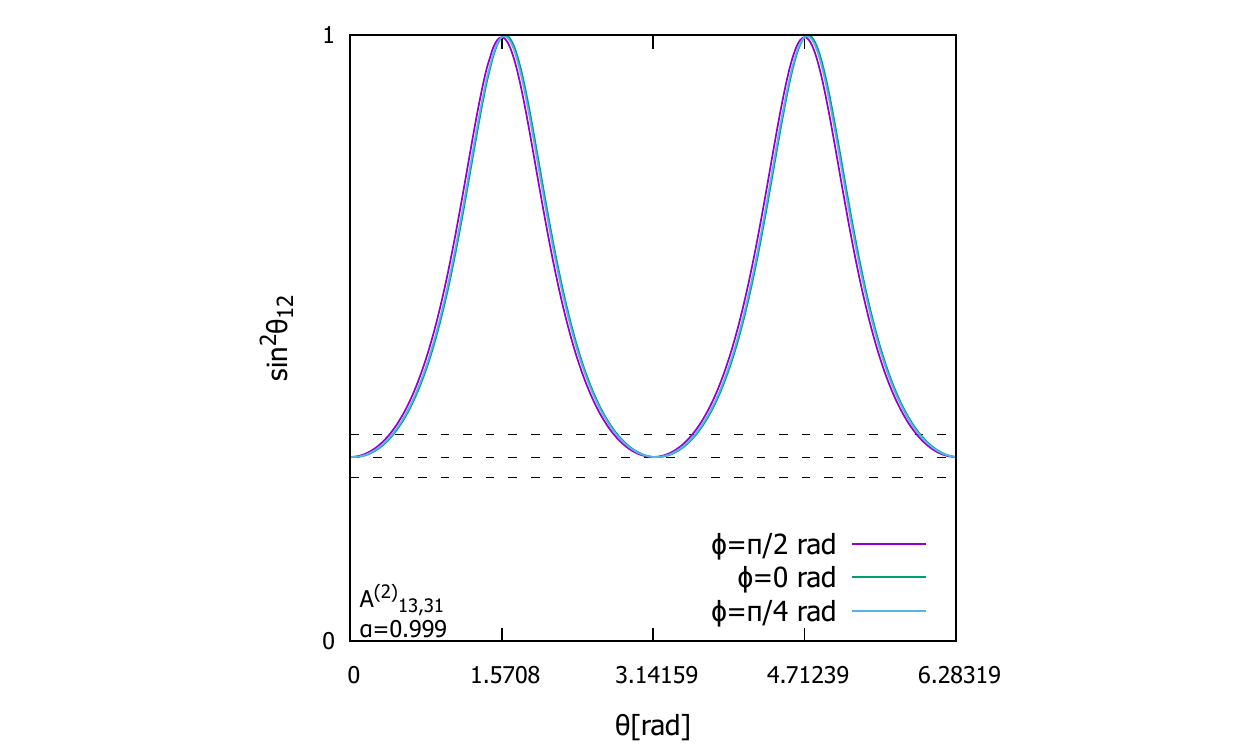}\\
\includegraphics[keepaspectratio, scale=0.5]{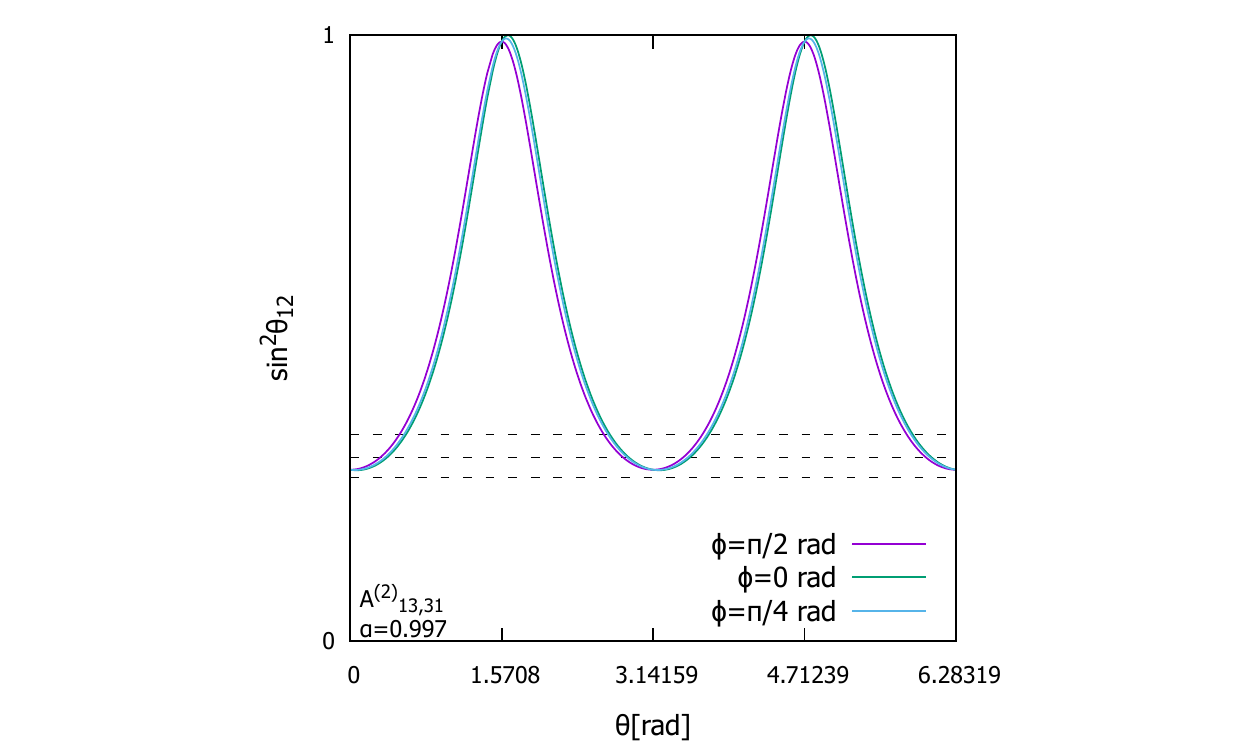}\\
\includegraphics[keepaspectratio, scale=0.5]{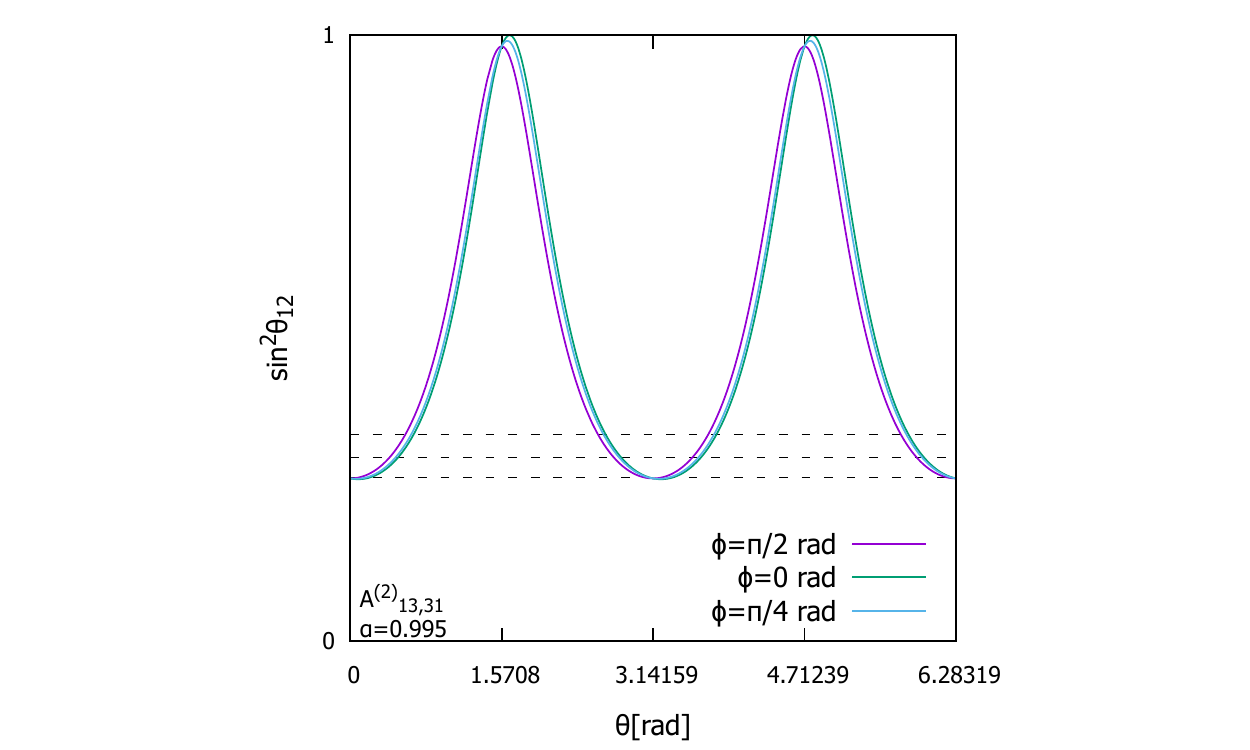}
\end{minipage}&
\begin{minipage}[t]{0.3\hsize}
\centering
\includegraphics[keepaspectratio, scale=0.5]{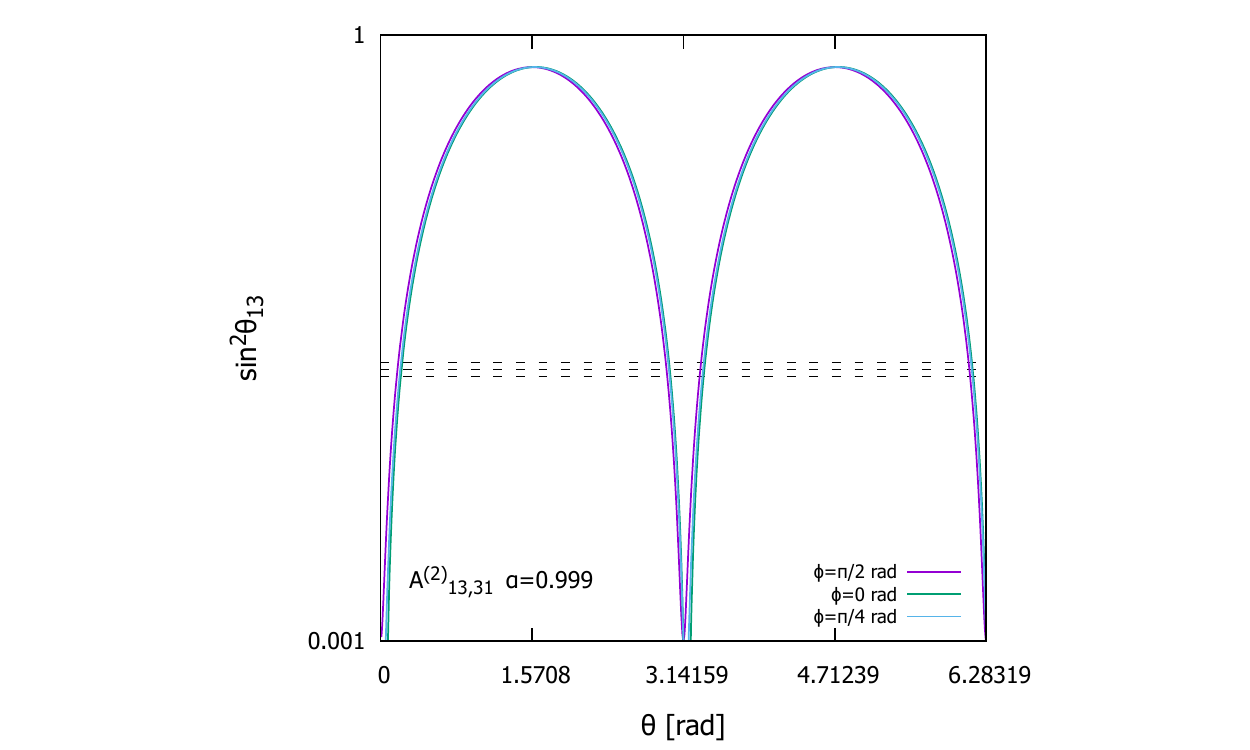}\\
\includegraphics[keepaspectratio, scale=0.5]{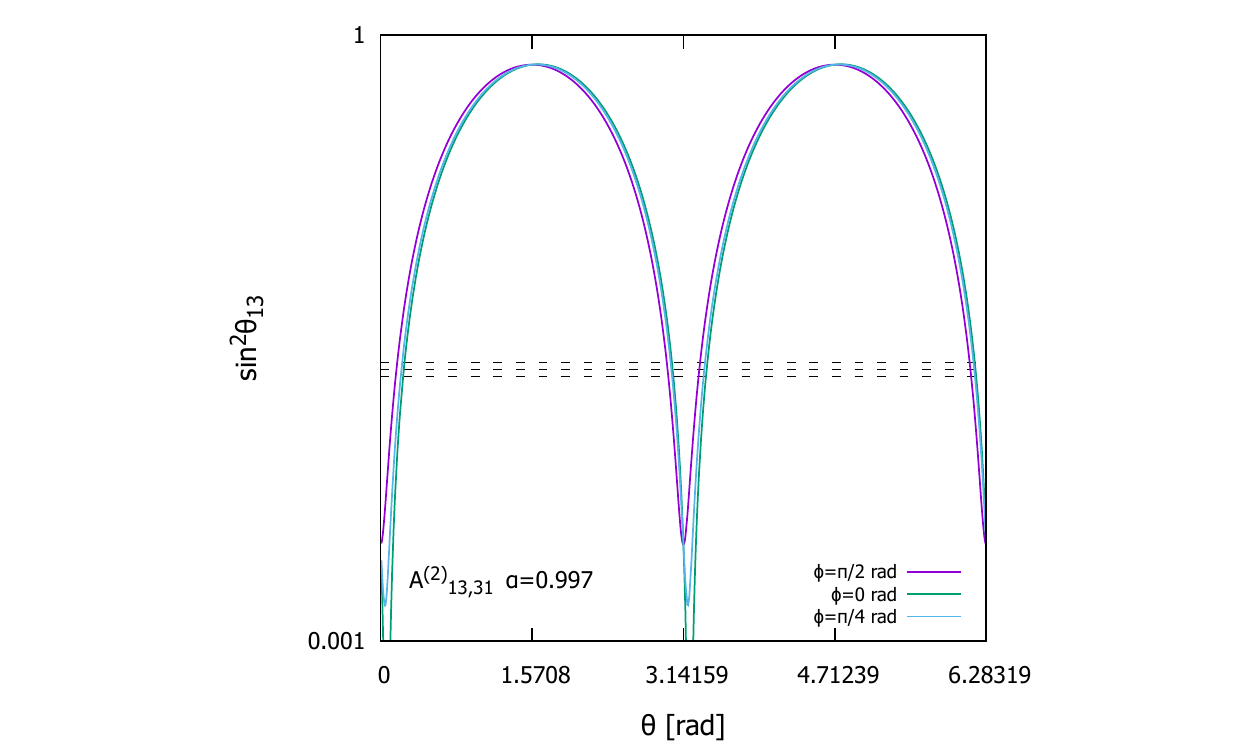}\\
\includegraphics[keepaspectratio, scale=0.5]{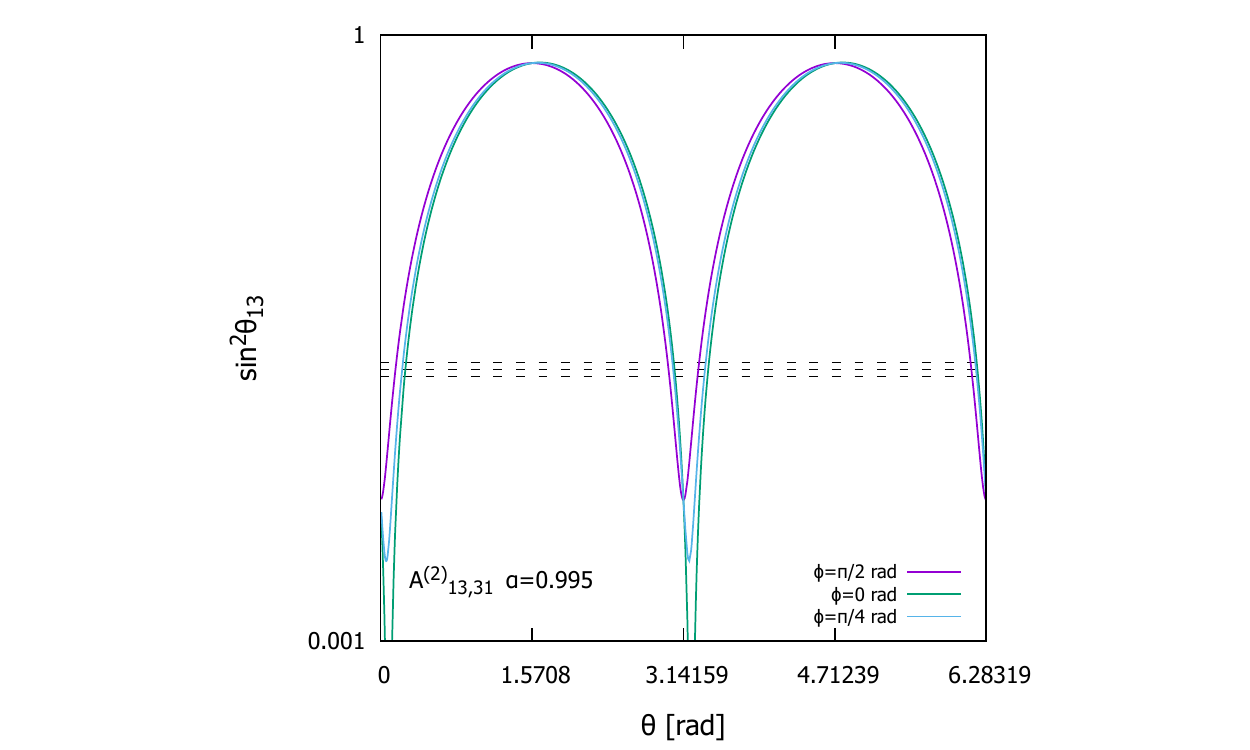}
 \end{minipage} \\
\end{tabular}
 \caption{Same as Fig. \ref{Fig:MTM1_A1221_2_a1p1_12_13} but for MTM2($A_{13,31}^{(2)}$).}
 \label{Fig:MTM2_A1331_2_a1p1_12_13} 
  \end{figure}
\begin{figure}[t]
\begin{tabular}{cc}
\begin{minipage}[t]{0.48\hsize}
\centering
\includegraphics[keepaspectratio, scale=0.5]{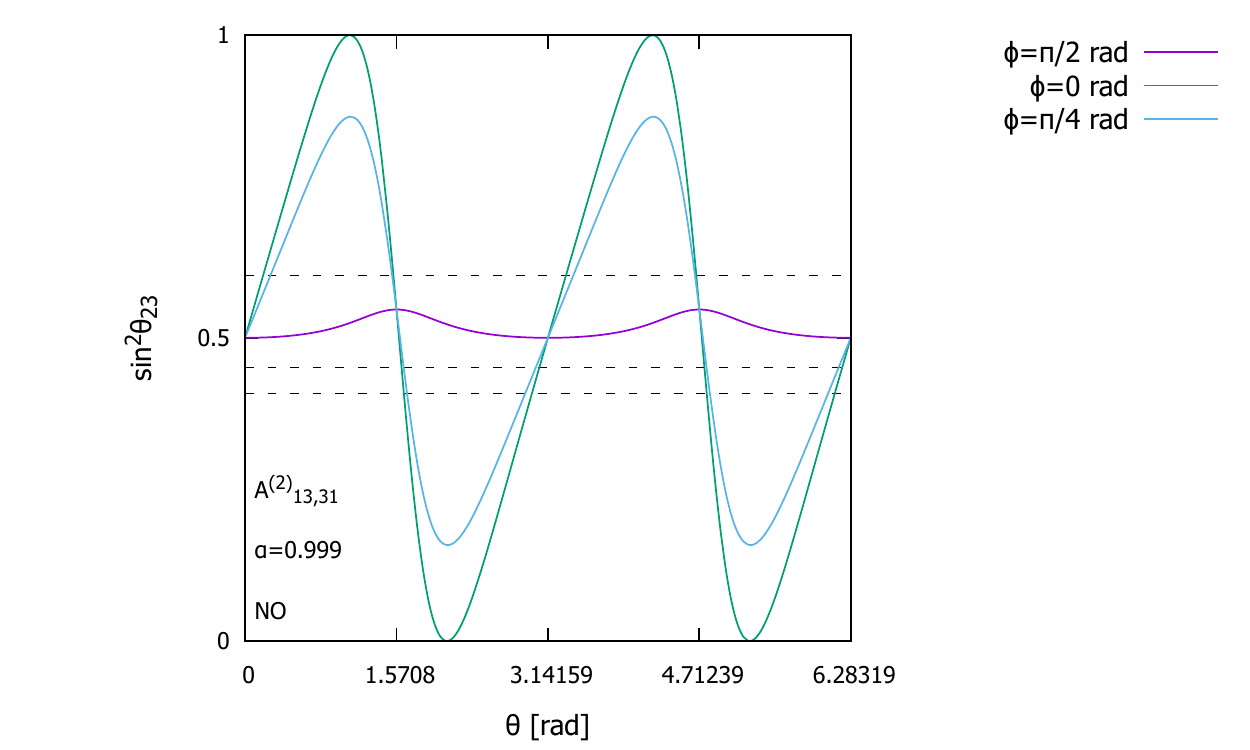}\\
\includegraphics[keepaspectratio, scale=0.5]{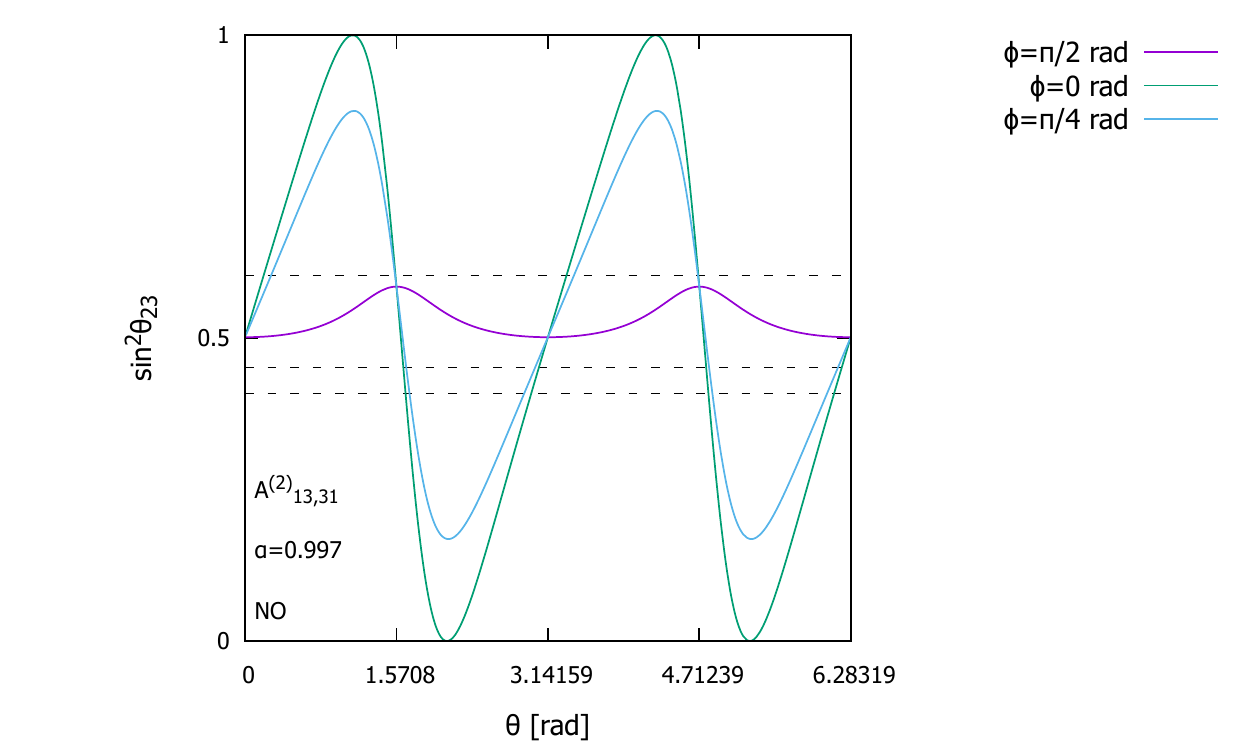}\\
\includegraphics[keepaspectratio, scale=0.5]{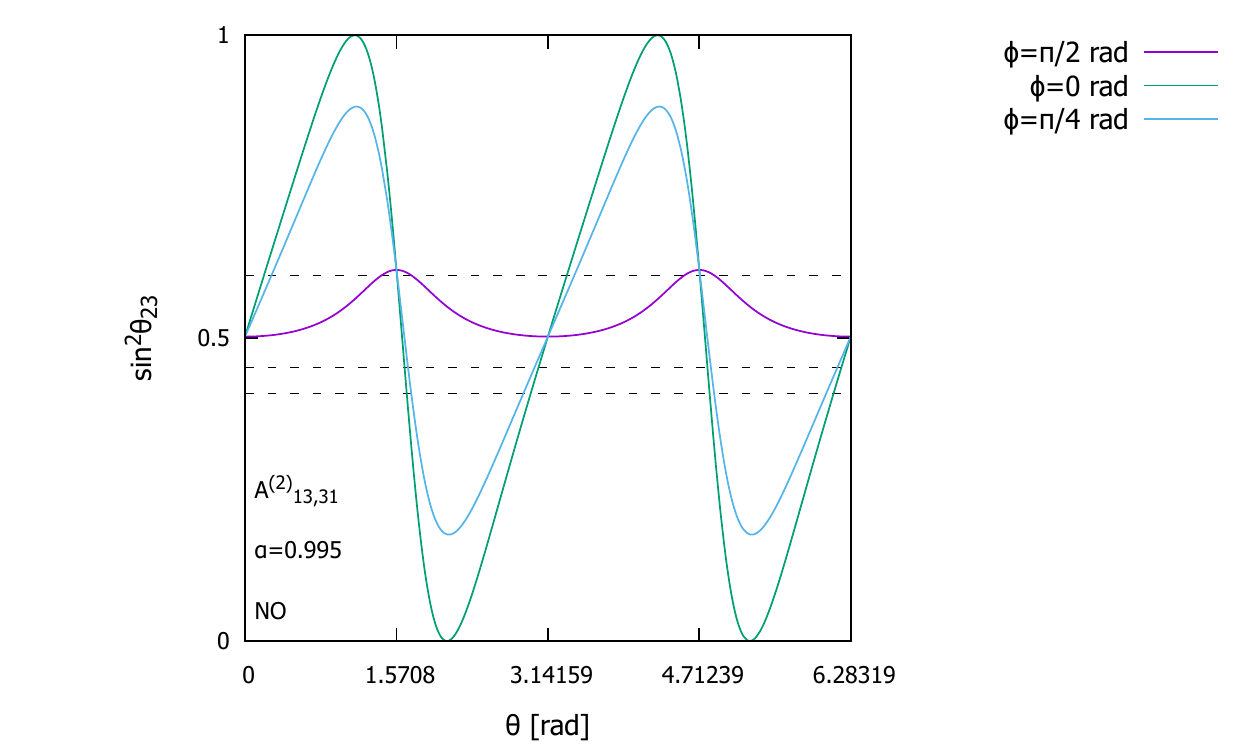}
\end{minipage}&
\begin{minipage}[t]{0.48\hsize}
\centering
\includegraphics[keepaspectratio, scale=0.5]{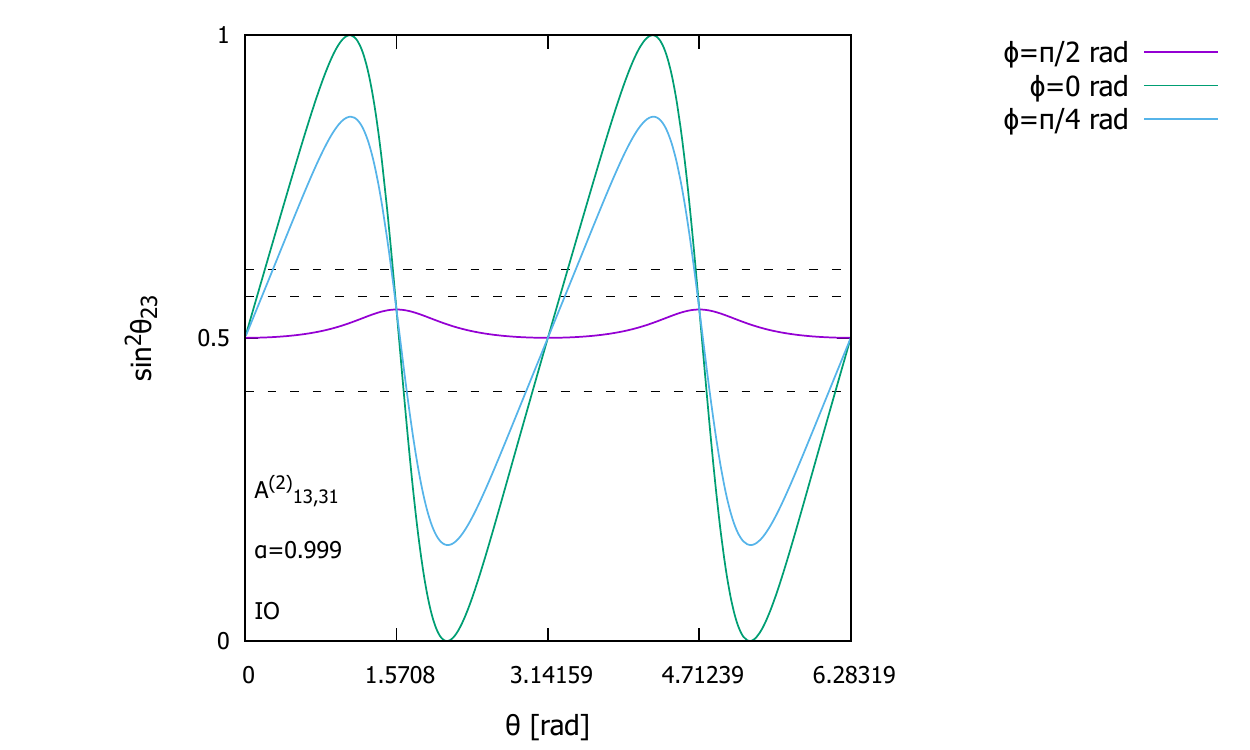}\\
\includegraphics[keepaspectratio, scale=0.5]{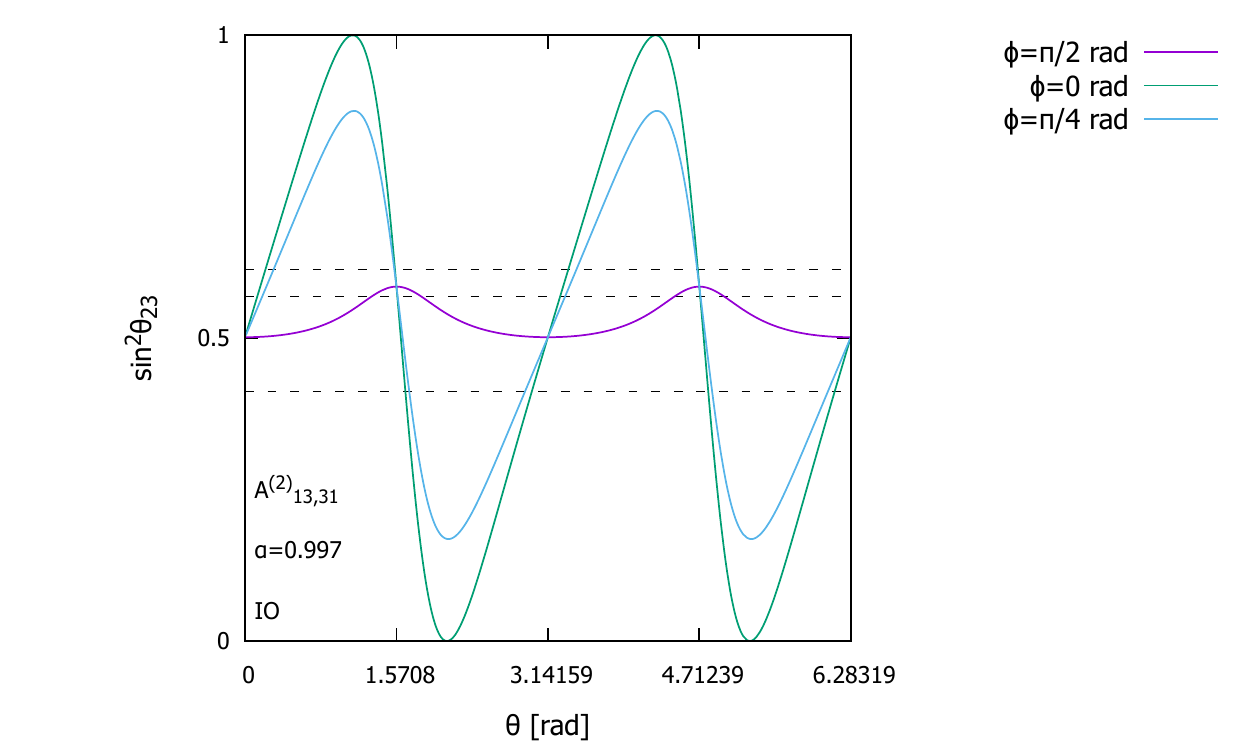}\\
\includegraphics[keepaspectratio, scale=0.5]{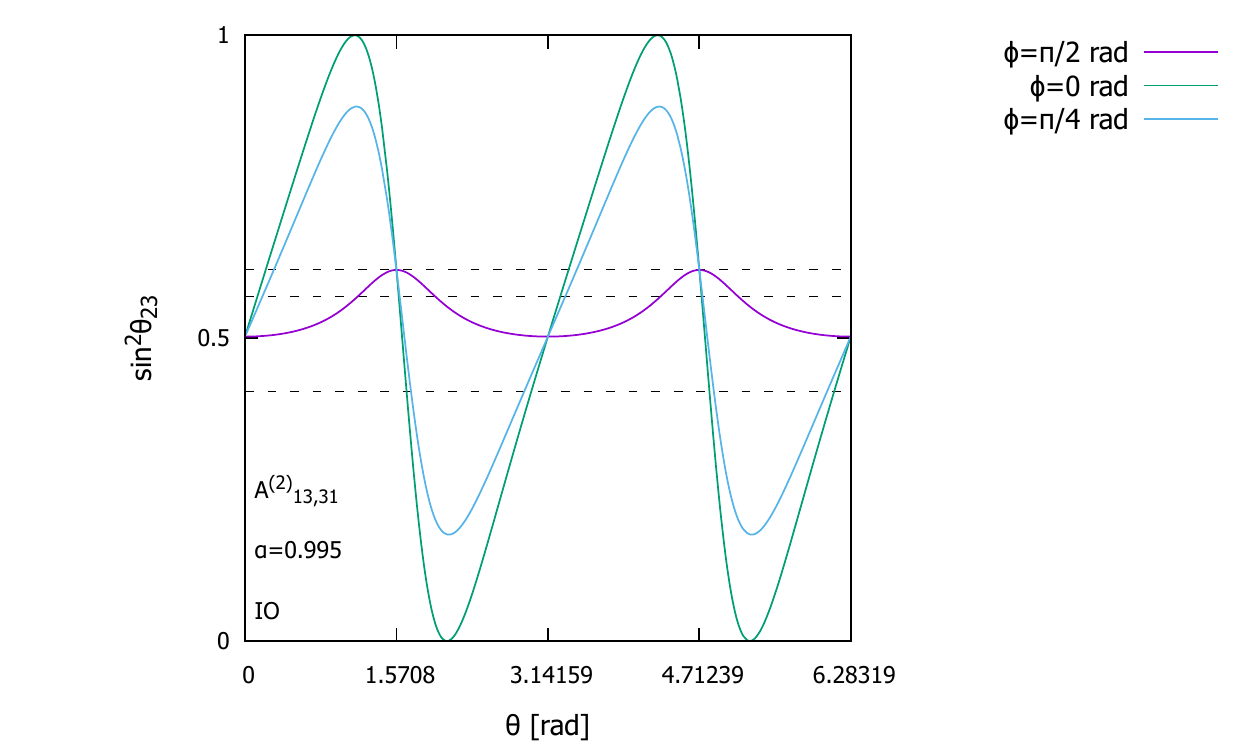}
 \end{minipage} \\
\end{tabular}
 \caption{Same as Fig. \ref{Fig:MTM1_A1221_2_ap_23} but for MTM2($A_{13,31}^{(2)}$)}
 \label{Fig:MTM2_A1331_2_ap_23} 
  \end{figure}

\begin{figure}[t]
\begin{tabular}{cc}
\begin{minipage}[t]{0.48\hsize}
\centering
\includegraphics[keepaspectratio, scale=0.5]{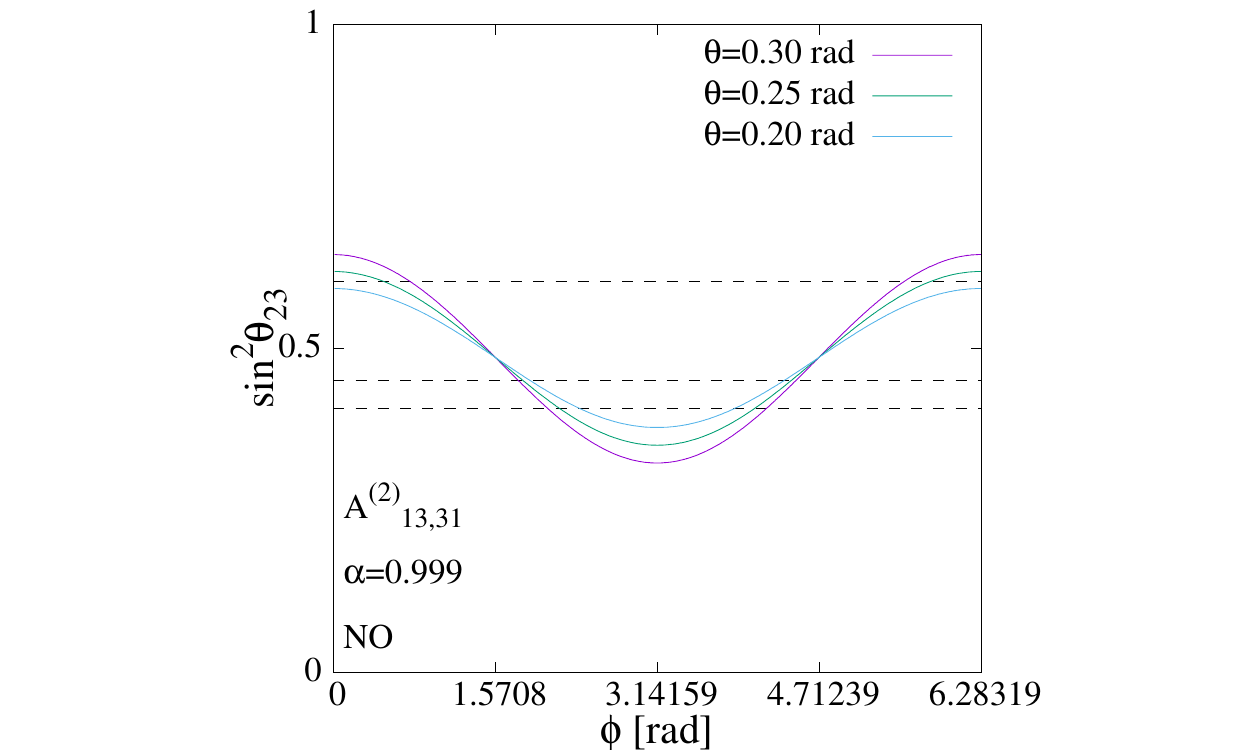}\\
\includegraphics[keepaspectratio, scale=0.5]{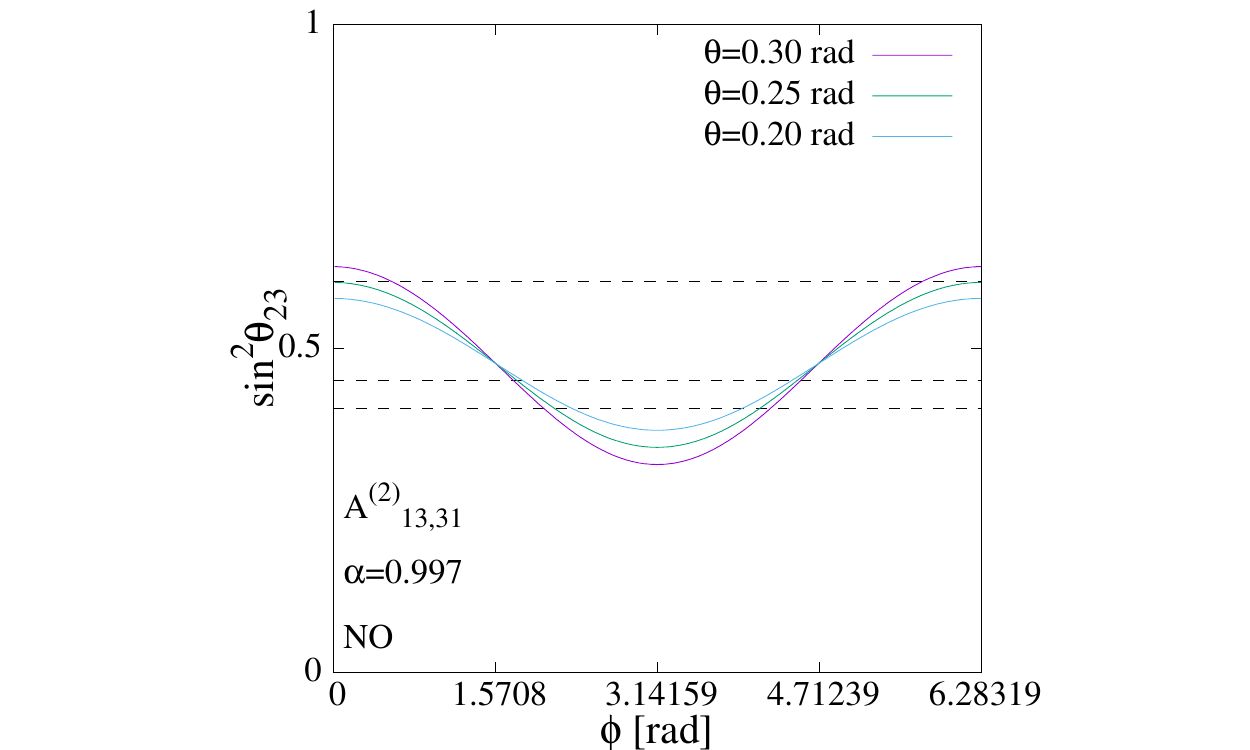}\\
\includegraphics[keepaspectratio, scale=0.5]{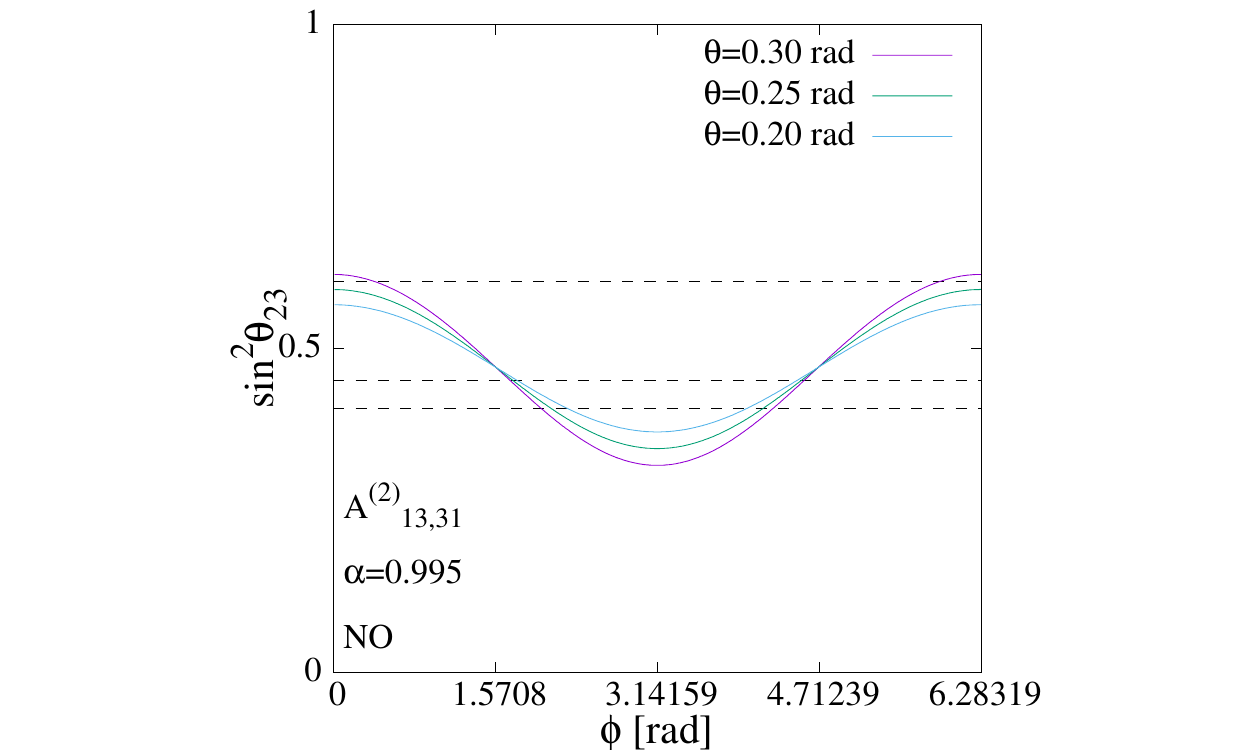}
\end{minipage}&
\begin{minipage}[t]{0.48\hsize}
\centering
\includegraphics[keepaspectratio, scale=0.5]{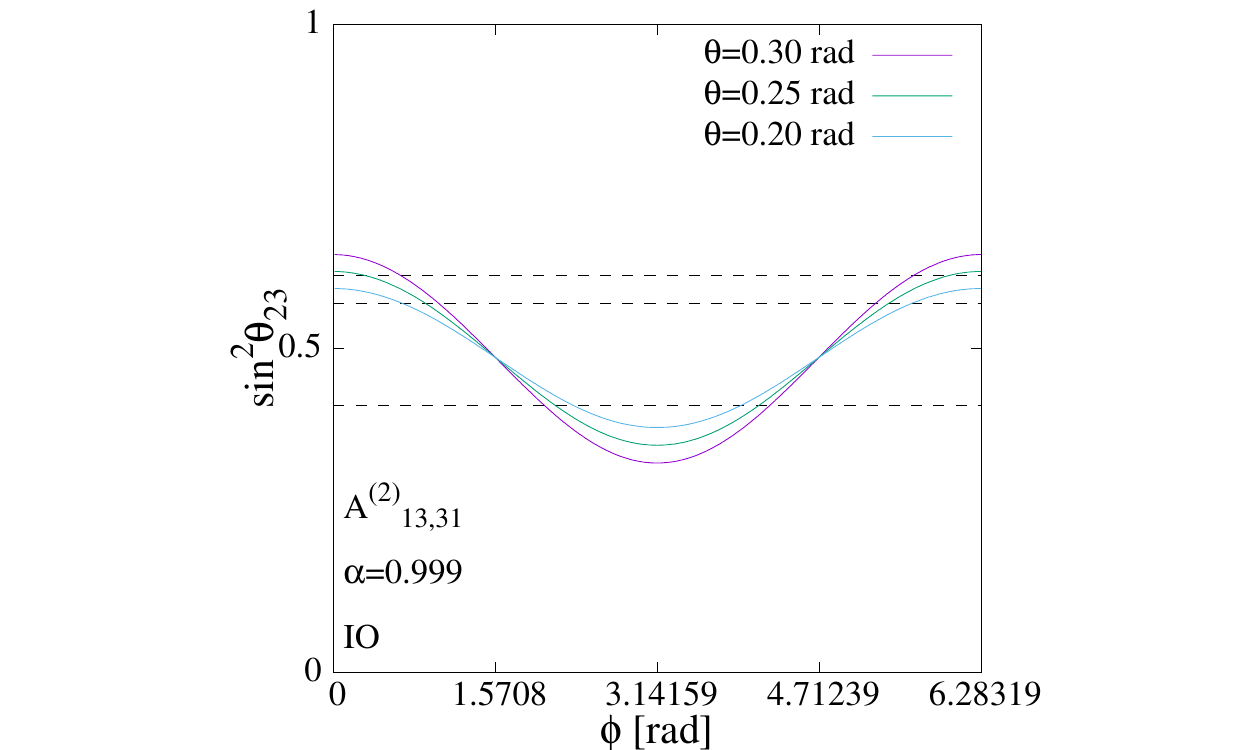}\\
\includegraphics[keepaspectratio, scale=0.5]{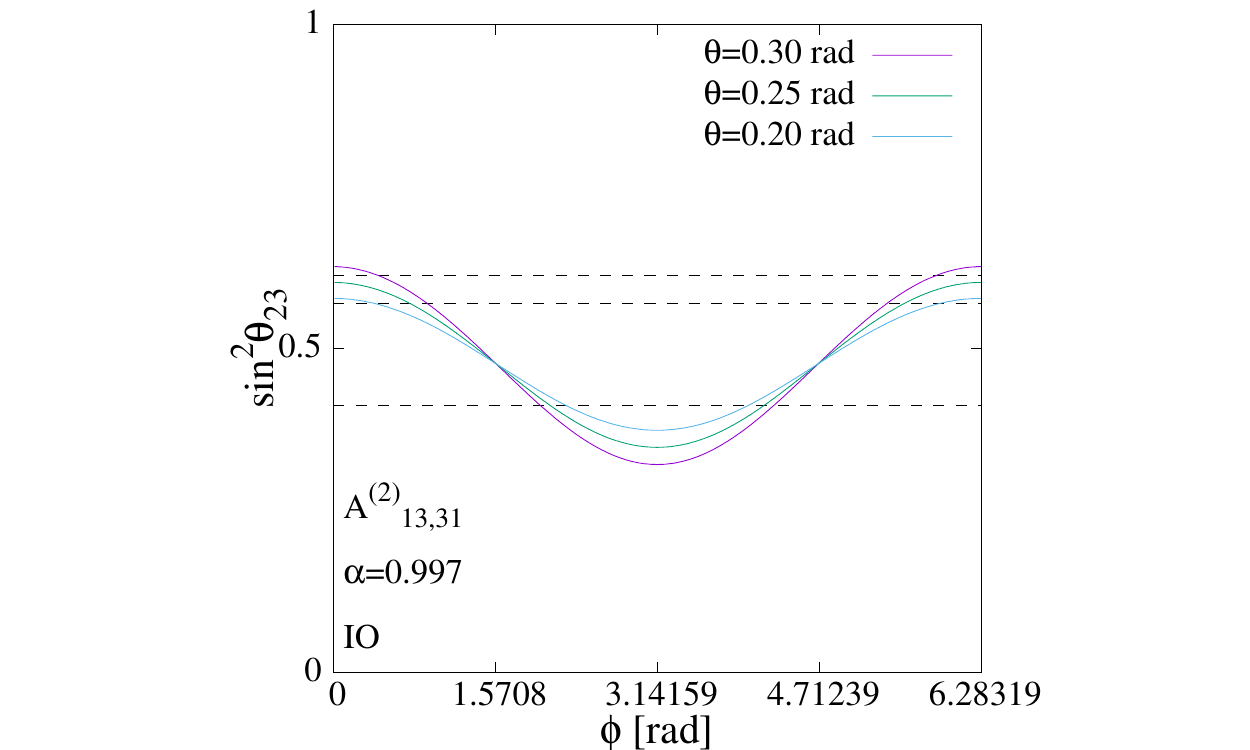}\\
\includegraphics[keepaspectratio, scale=0.5]{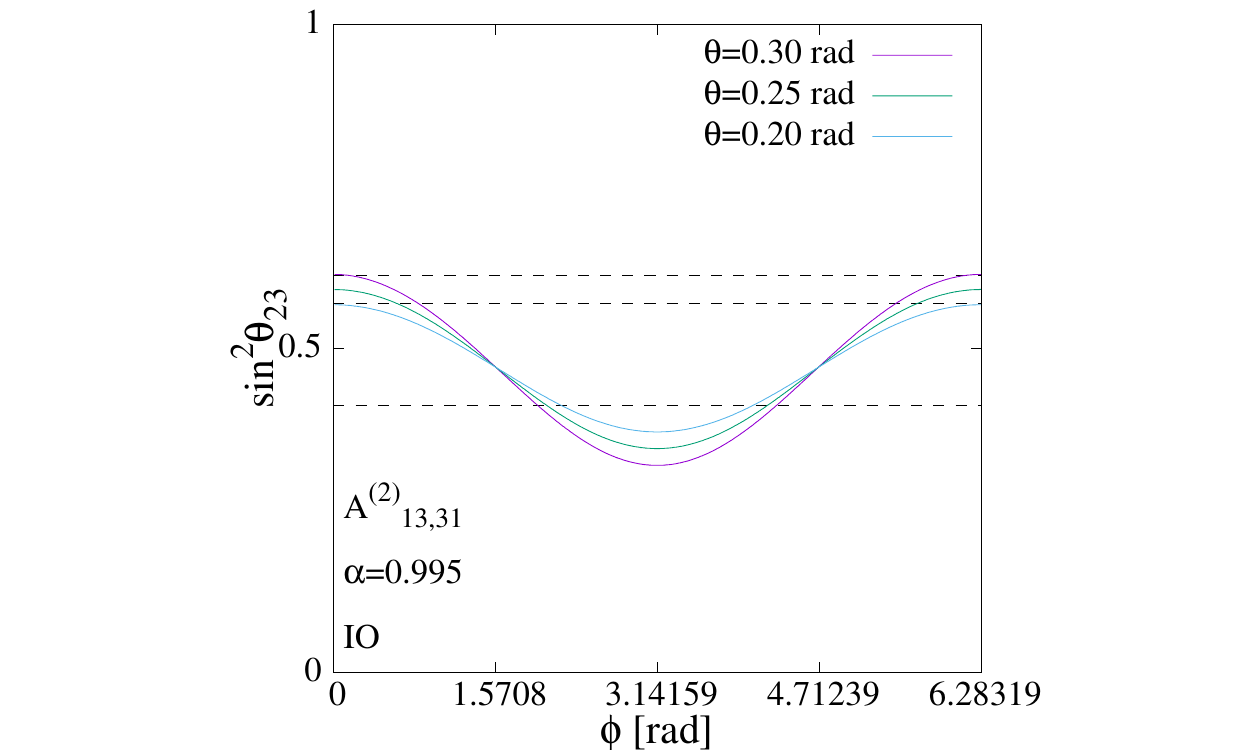}
 \end{minipage} \\
\end{tabular}
 \caption{Same as Fig. \ref{Fig:MTM1_A1221_2_at_23} but for MTM2($A_{13,31}^{(2)}$)}
 \label{Fig:MTM2_A1331_2_at_23} 
  \end{figure}

\begin{figure}[t]
\begin{tabular}{cc}
\begin{minipage}[t]{0.48\hsize}
\centering
\includegraphics[keepaspectratio, scale=0.5]{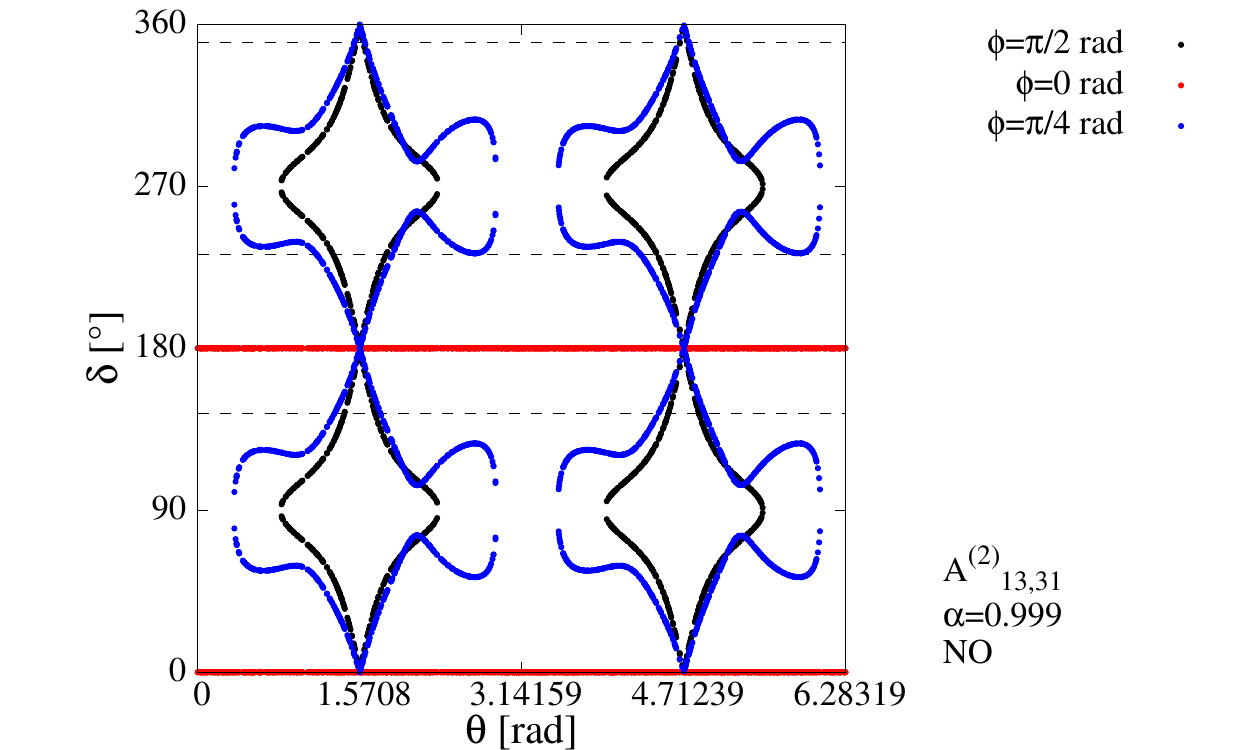}\\
\includegraphics[keepaspectratio, scale=0.5]{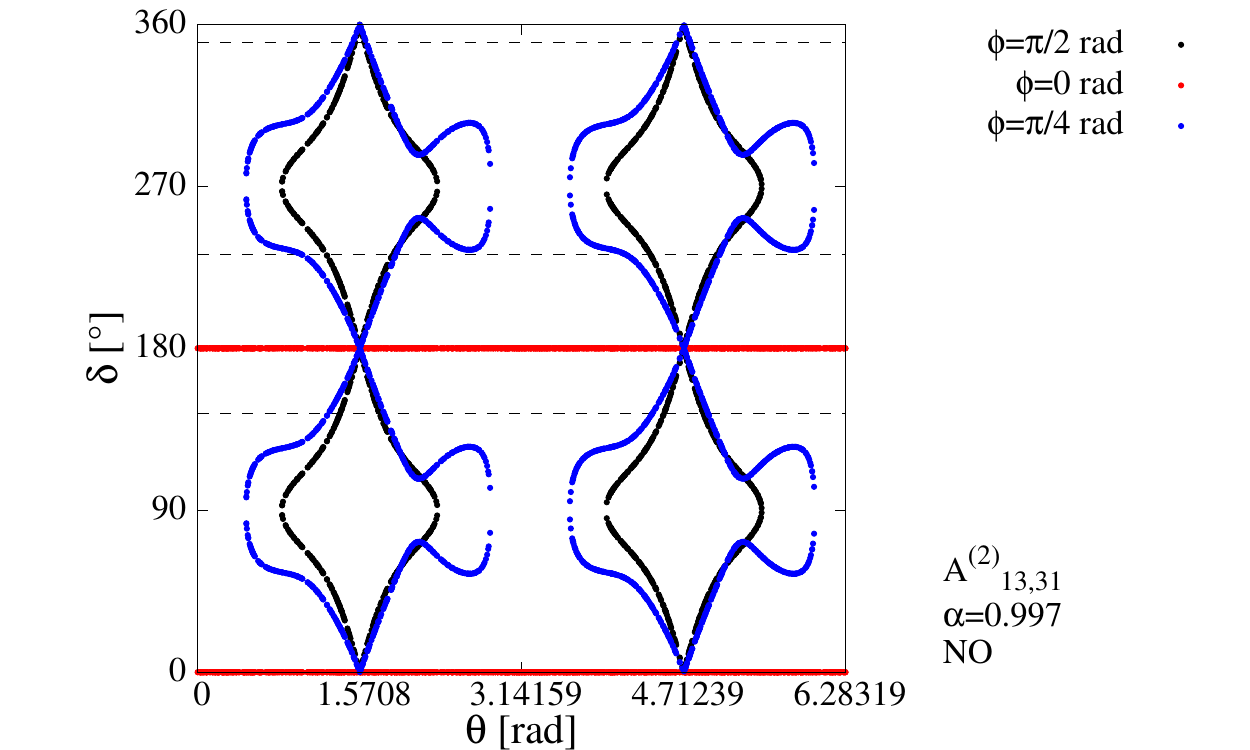}\\
\includegraphics[keepaspectratio, scale=0.5]{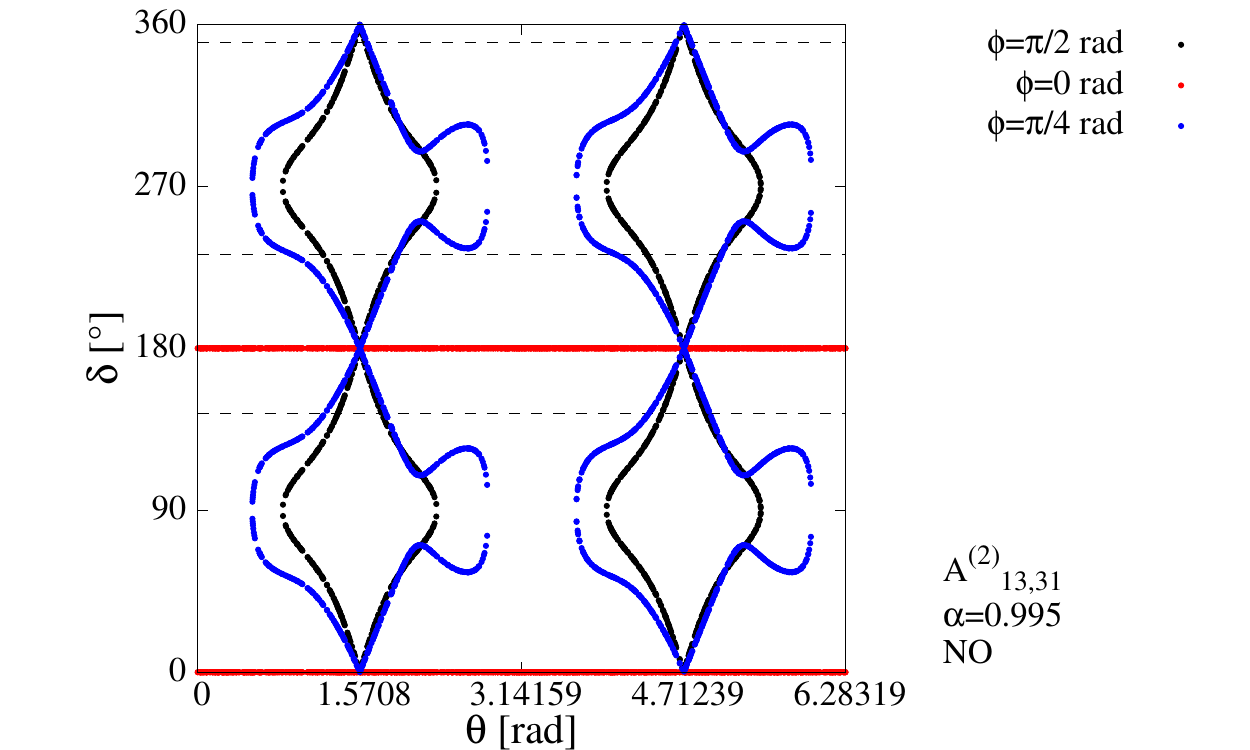}
\end{minipage}&
\begin{minipage}[t]{0.48\hsize}
\centering
\includegraphics[keepaspectratio, scale=0.5]{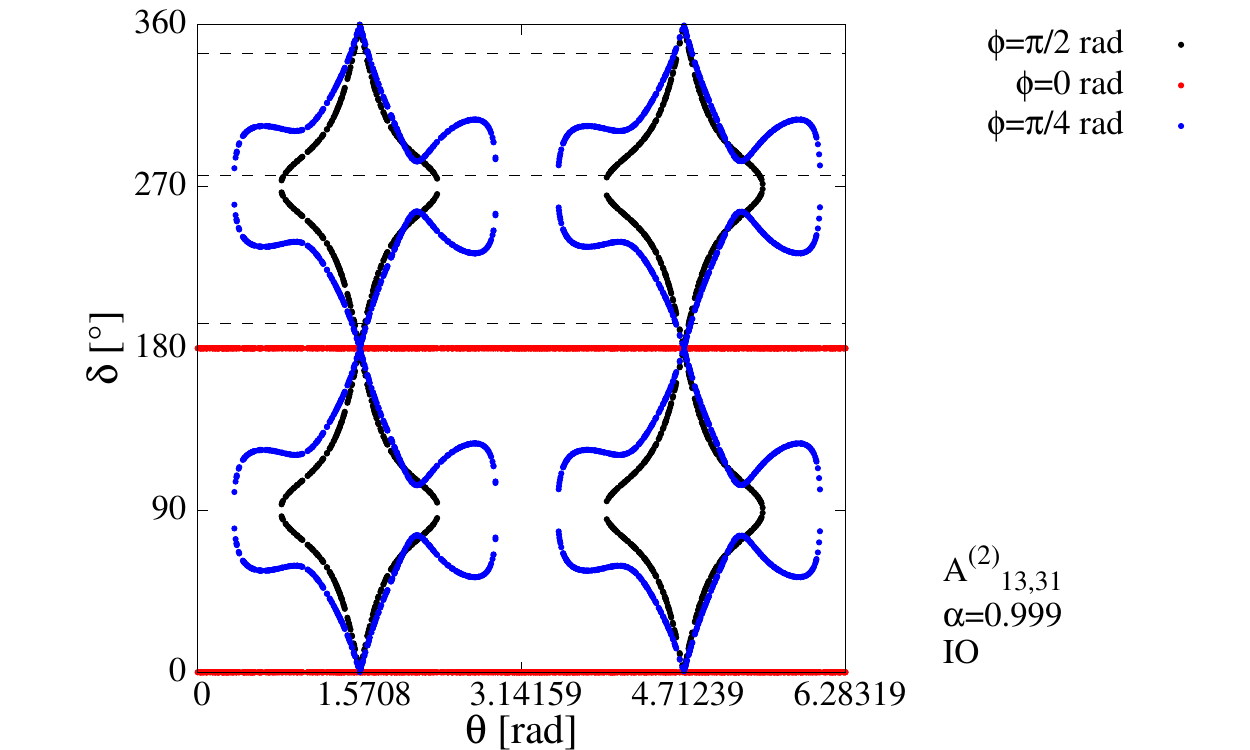}\\
\includegraphics[keepaspectratio, scale=0.5]{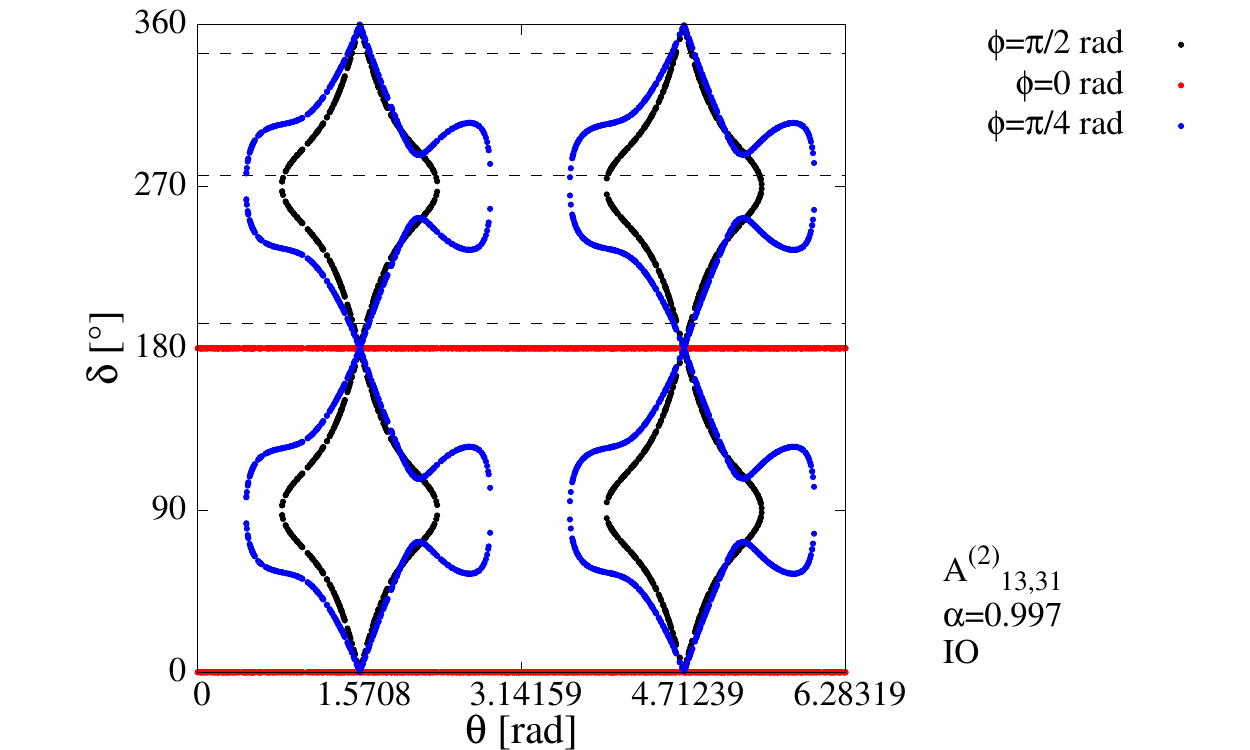}\\
\includegraphics[keepaspectratio, scale=0.5]{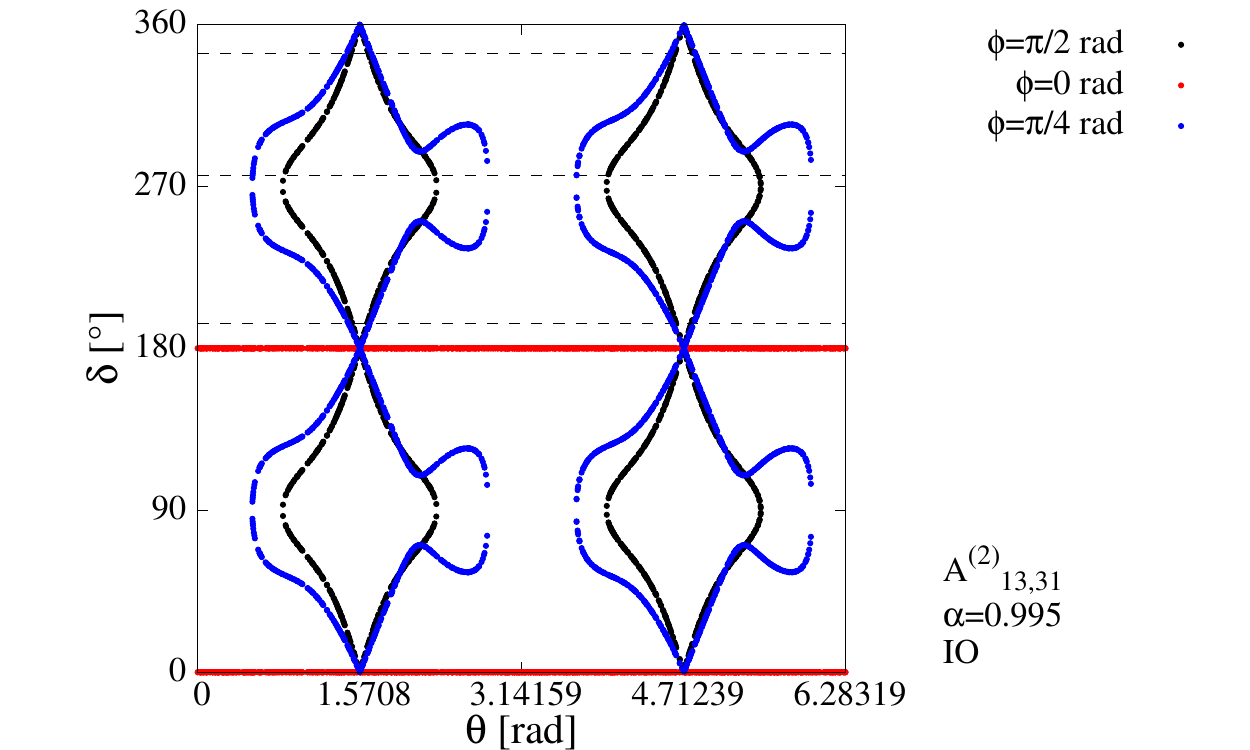}
 \end{minipage} \\
\end{tabular}
 \caption{Same as Fig. \ref{Fig:MTM1_A1221_2_ap_d} but for MTM2($A_{13,31}^{(2)}$).}
 \label{Fig:MTM2_A1331_2_ap_d} 
  \end{figure}

\begin{figure}[t]
\begin{tabular}{cc}
\begin{minipage}[t]{0.48\hsize}
\centering
\includegraphics[keepaspectratio, scale=0.5]{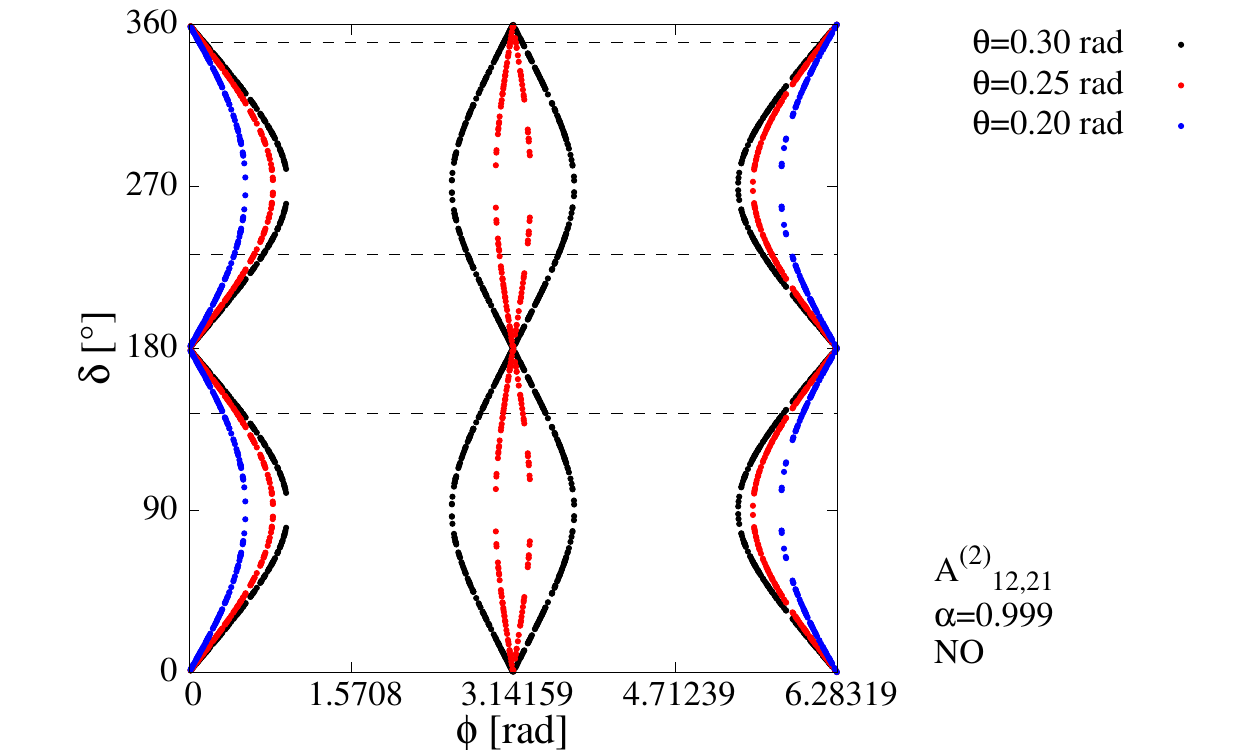}\\
\includegraphics[keepaspectratio, scale=0.5]{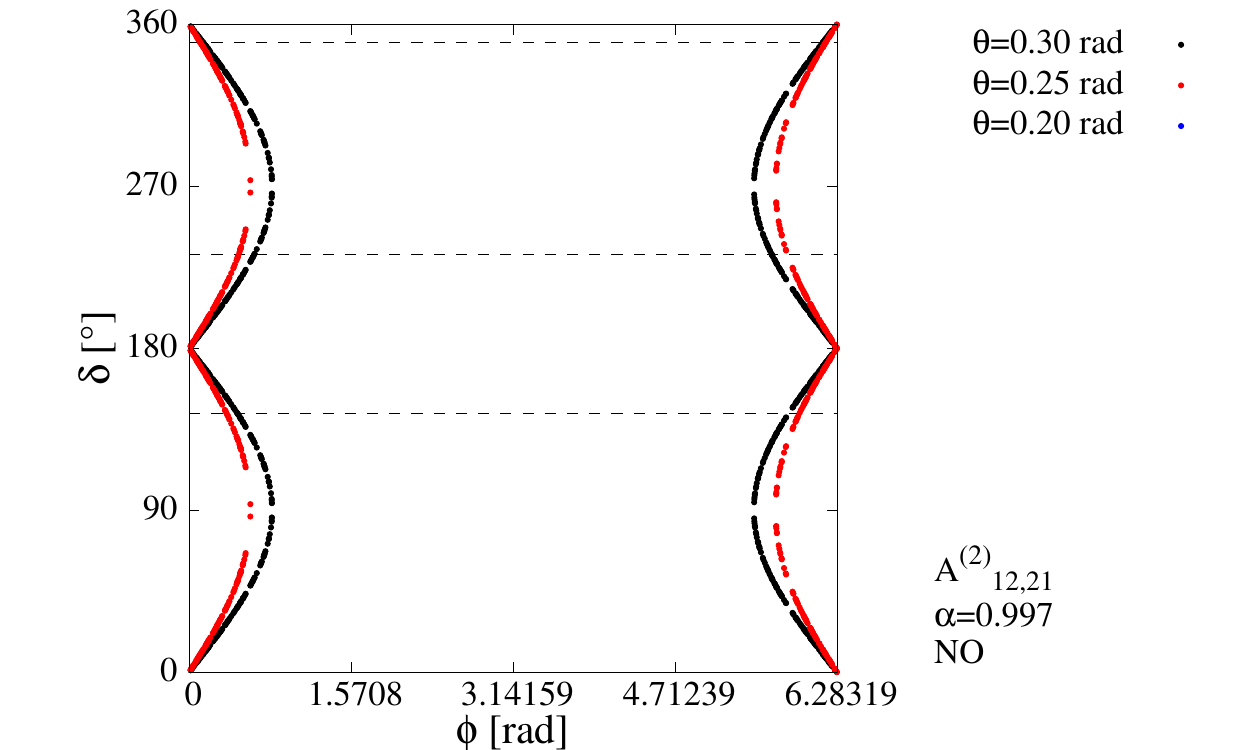}\\
\includegraphics[keepaspectratio, scale=0.5]{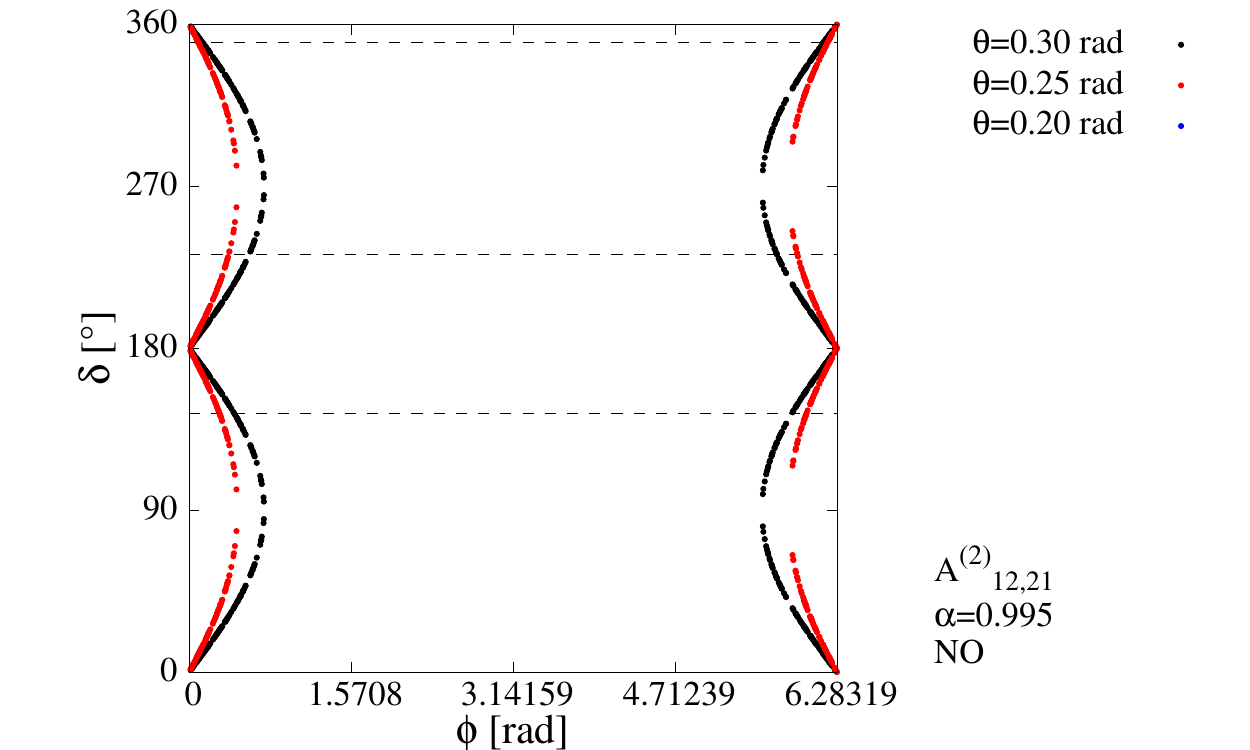}
\end{minipage}&
\begin{minipage}[t]{0.48\hsize}
\centering
\includegraphics[keepaspectratio, scale=0.5]{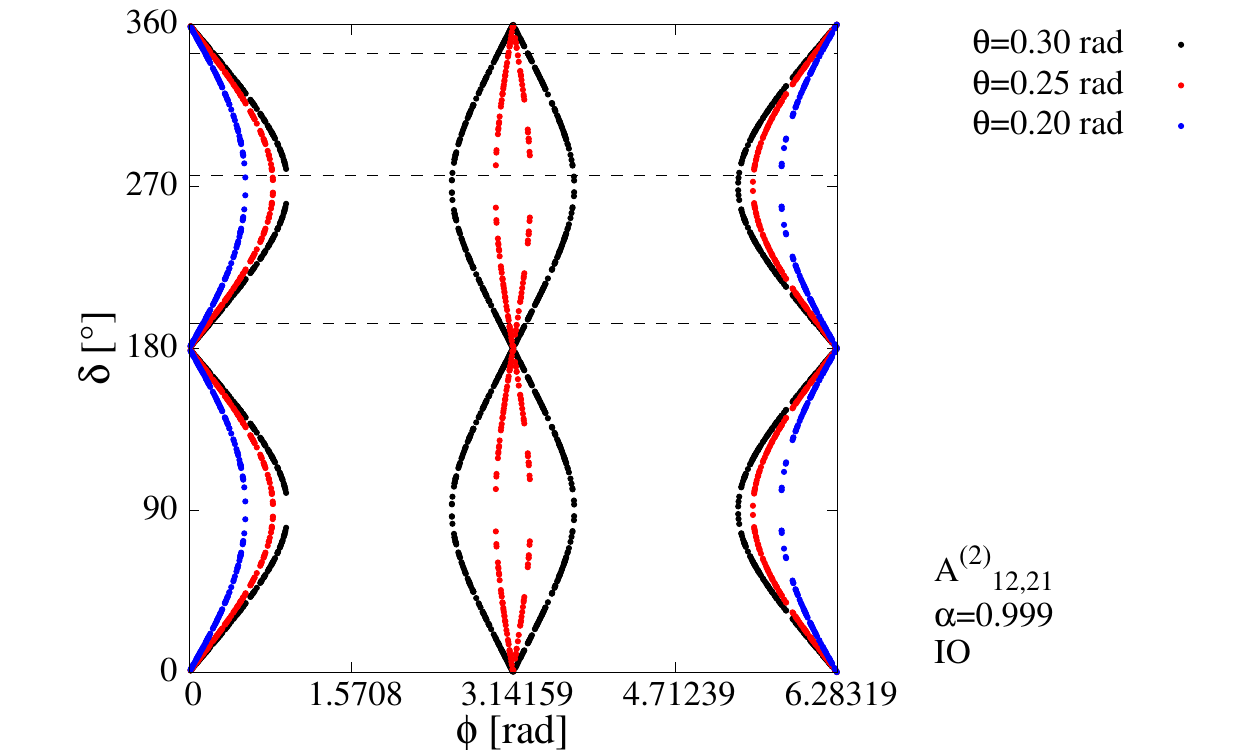}\\
\includegraphics[keepaspectratio, scale=0.5]{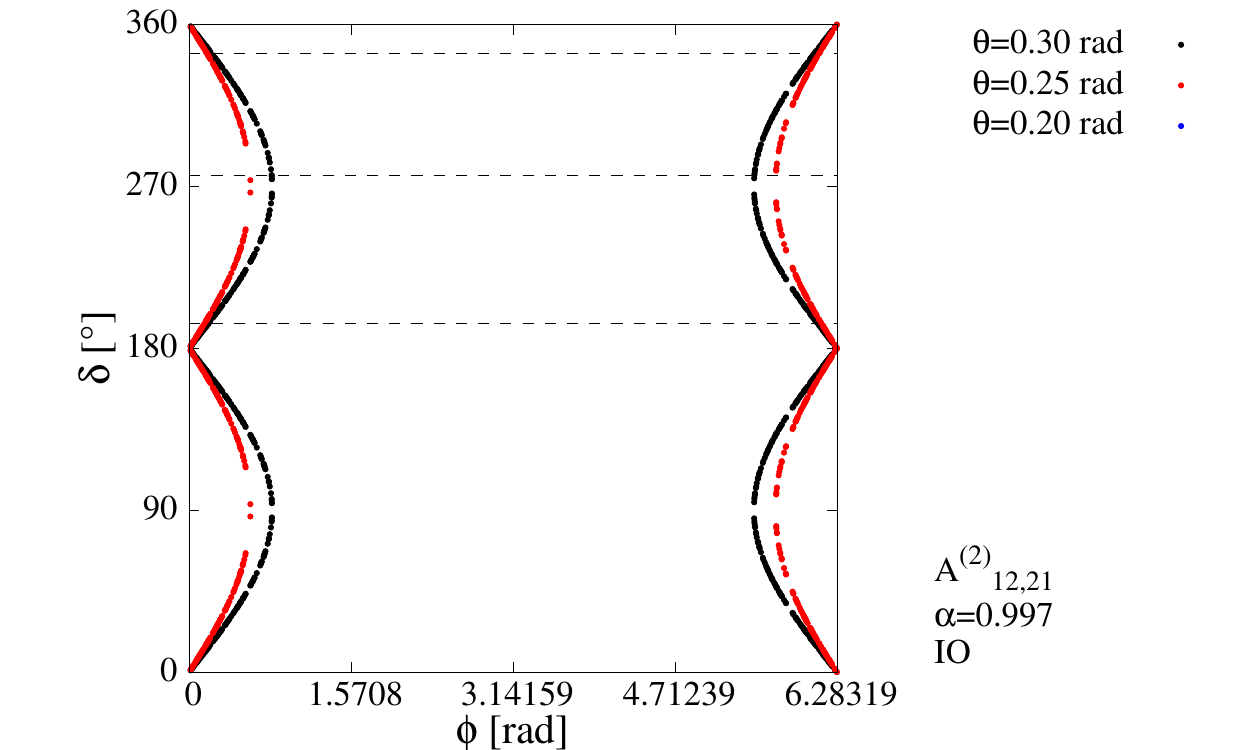}\\
\includegraphics[keepaspectratio, scale=0.5]{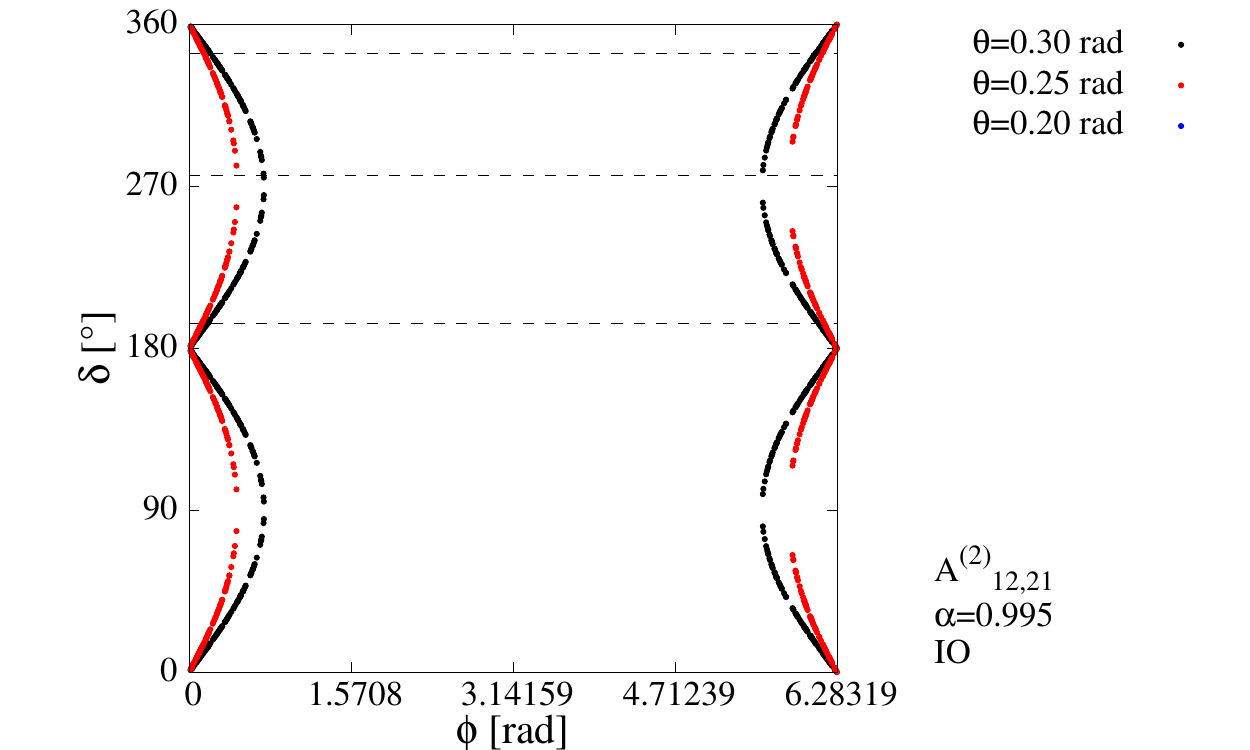}
 \end{minipage} \\
\end{tabular}
 \caption{Same as Fig. \ref{Fig:MTM1_A1221_2_at_d} but for MTM2($A_{13,31}^{(2)}$).}
 \label{Fig:MTM2_A1331_2_at_d} 
  \end{figure}

\begin{figure}[t]
\begin{tabular}{cc}
\begin{minipage}[t]{0.48\hsize}
\centering
\includegraphics[keepaspectratio, scale=0.5]{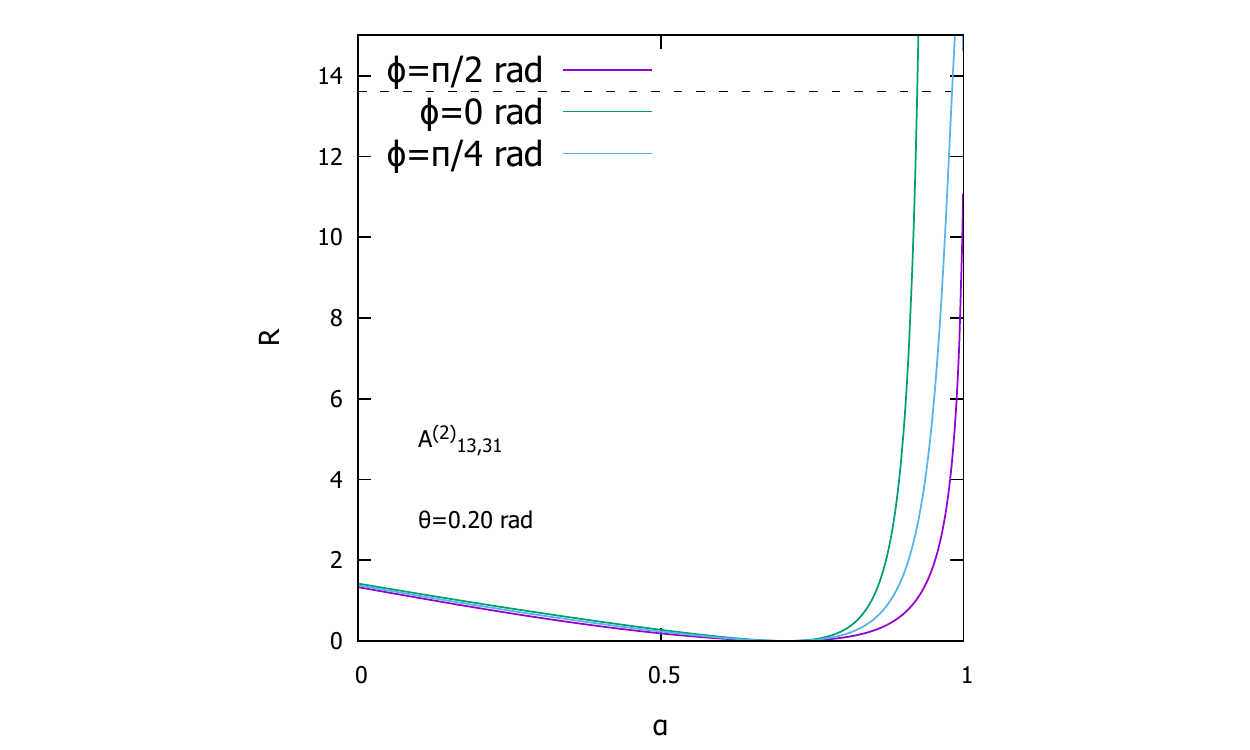}\\
\includegraphics[keepaspectratio, scale=0.5]{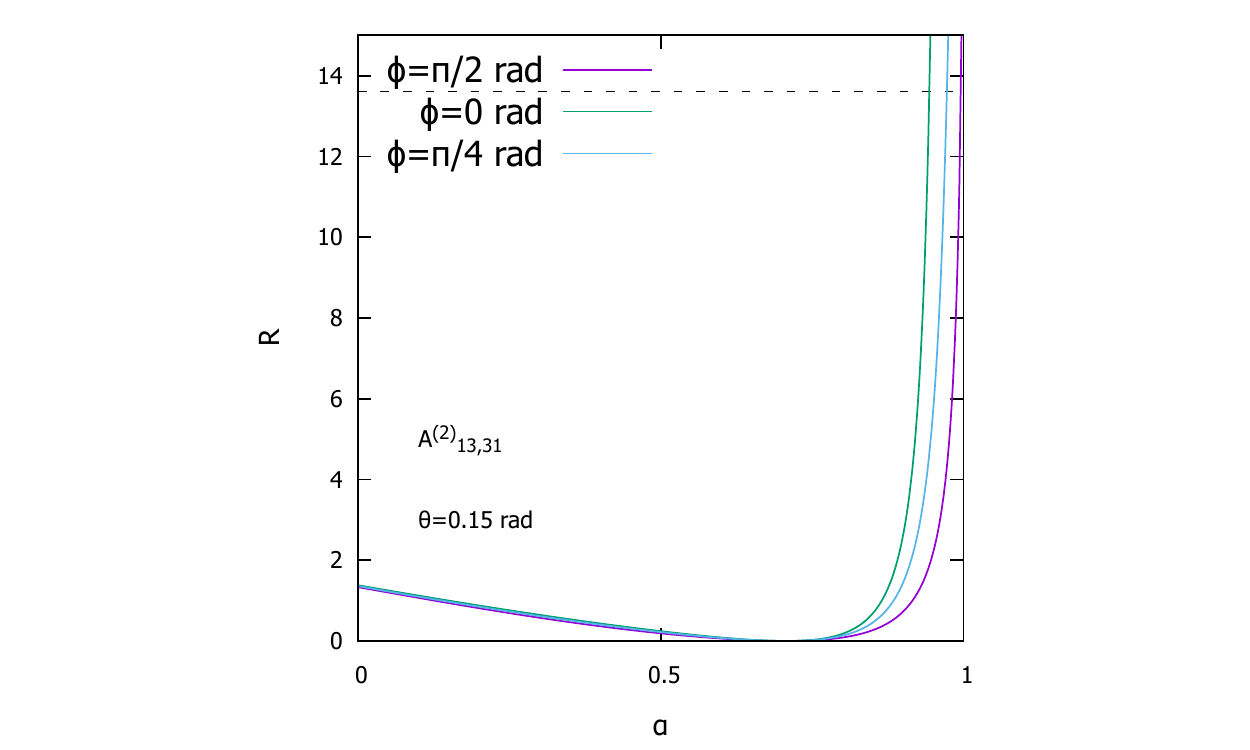}\\
\includegraphics[keepaspectratio, scale=0.5]{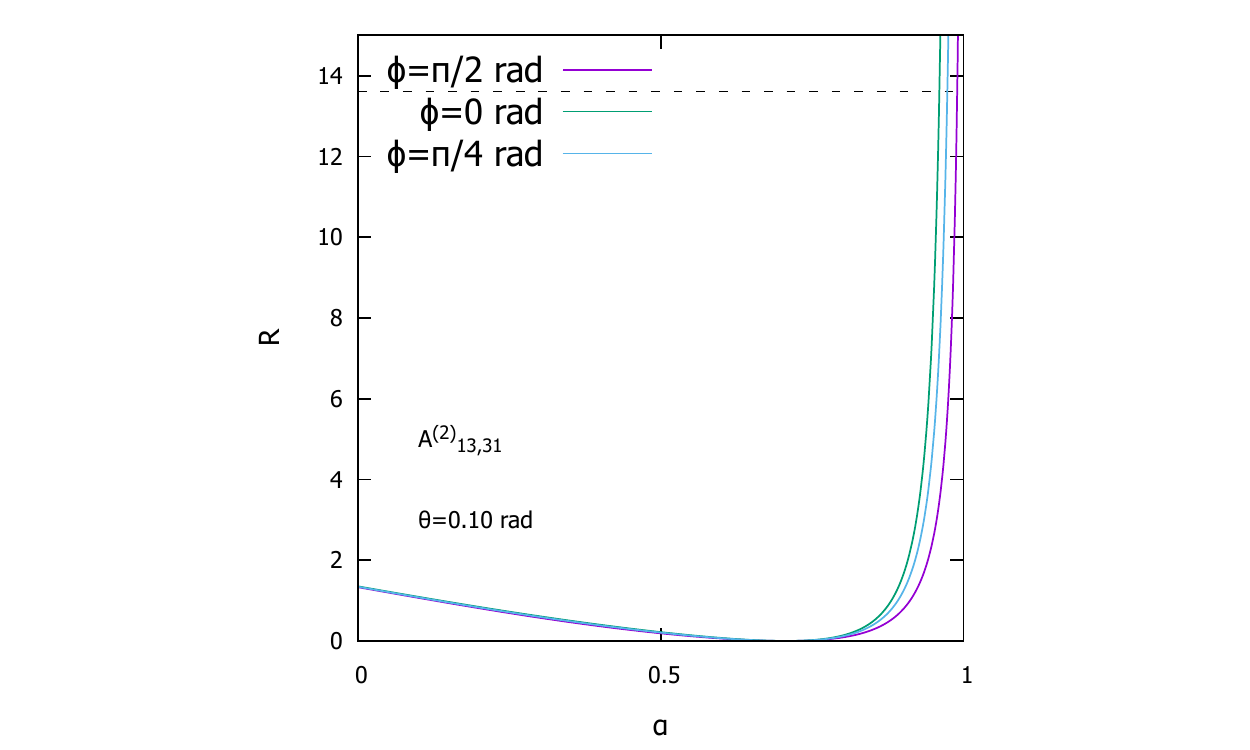}
\end{minipage}&
\begin{minipage}[t]{0.48\hsize}
\centering
\includegraphics[keepaspectratio, scale=0.5]{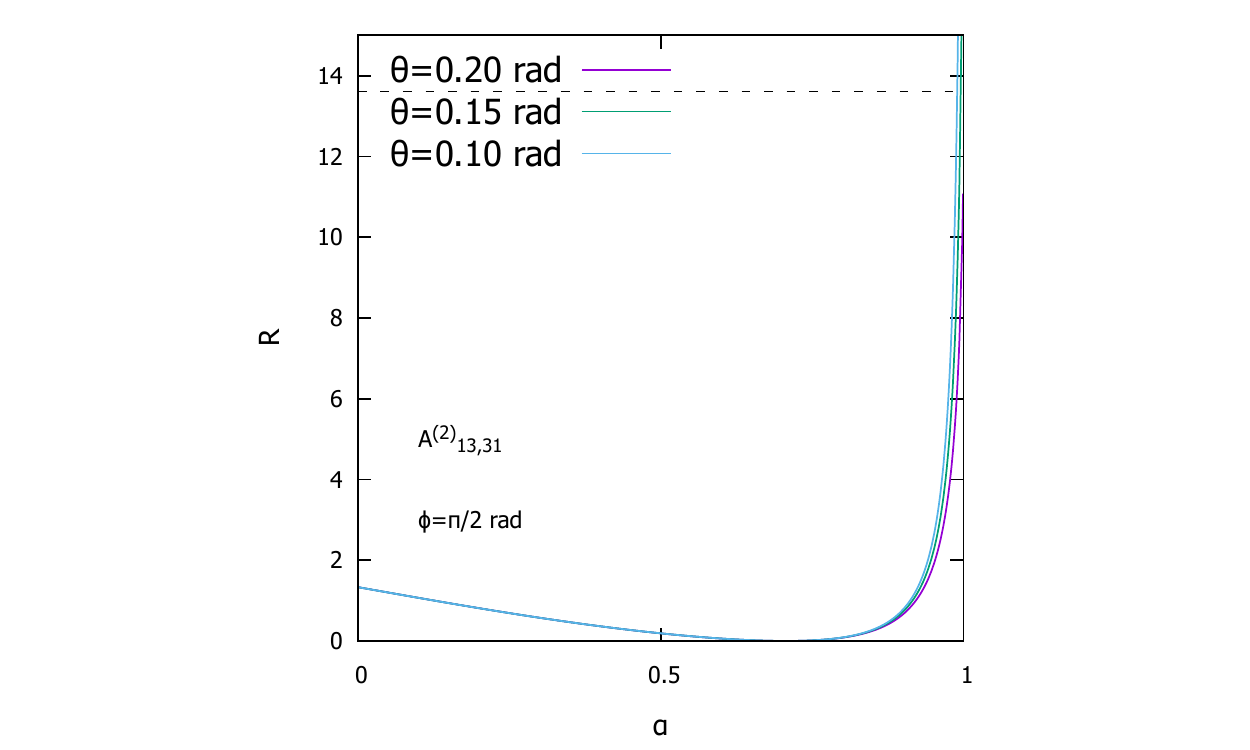}\\
\includegraphics[keepaspectratio, scale=0.5]{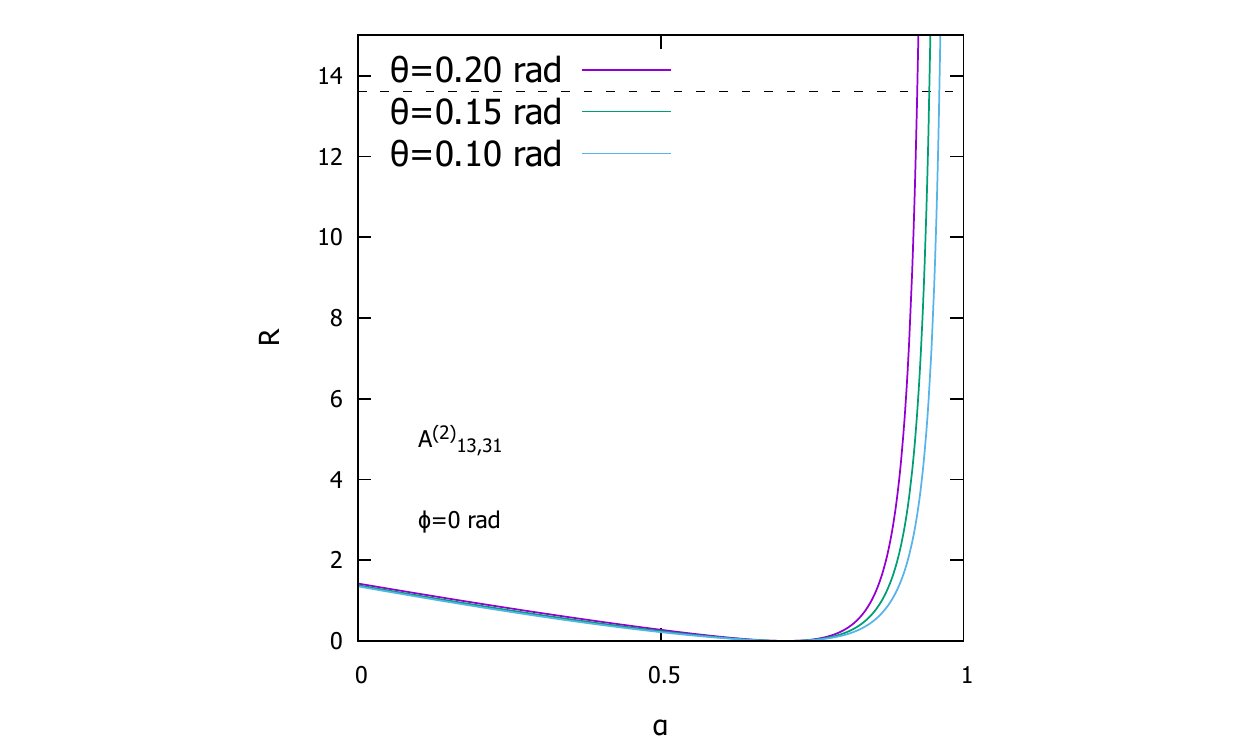}\\
\includegraphics[keepaspectratio, scale=0.5]{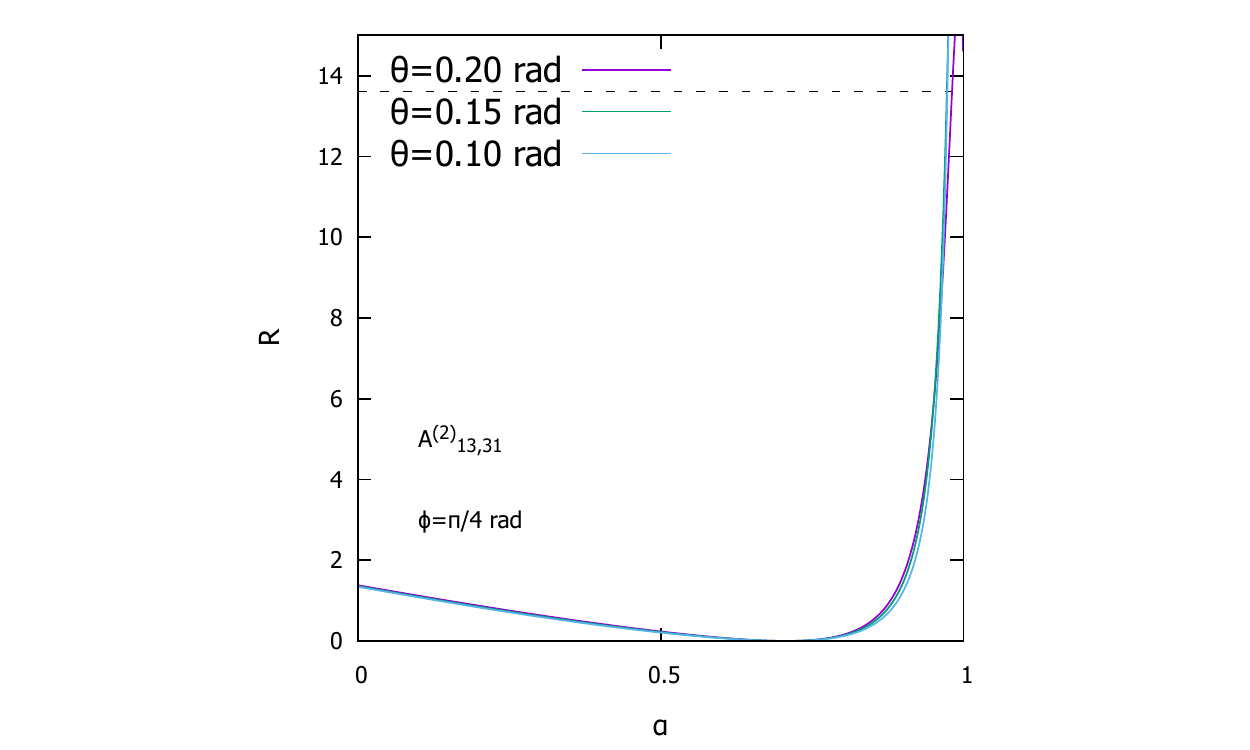}
 \end{minipage} \\
\end{tabular}
 \caption{Same as Fig. \ref{Fig:MTM1_A1221_2_R} but for MTM2($A_{13,31}^{(2)}$).}
 \label{Fig:MTM2_A1331_2_R} 
  \end{figure}

The modified TM2 mixing scheme involving the rotation matrices $A_{13,31}^{(1)}$ and $A_{13,31}^{(2)}$ is obtained as follows:
\begin{eqnarray}
\tilde{U}_2(A_{13,31}^{(1)}) = A_{13,31}^{(1)} U_2, \quad \tilde{U}_2(A_{13,31}^{(2)}) = A_{13,31}^{(2)} U_2, 
\end{eqnarray}
and we term these mixing schemes as MTM2($A_{13,31}^{(1)}$) and MTM2($A_{13,31}^{(2)}$), respectively. The mixing angles are
\begin{eqnarray}
s^2_{12} &=& \frac{1 \pm 2\alpha \sqrt{1-\alpha^2}}{3(1-s^2_{13})},  \label{Eq:MTM2_A1331_s12ss13s} \\
s^2_{23} &=& \frac{2+\cos 2\theta + \sqrt{3}\sin 2\theta \cos\phi}{6(1-s^2_{13})},  \label{Eq:MTM2_23sA13_1_8} \\
s^2_{13} &=& \frac{1}{6} (2 \mp  2 \alpha \sqrt{1-\alpha^2} + B_3 - B_4).  \label{Eq:MTM2_13sA13_1_8} 
\end{eqnarray}
The upper sign of $\pm$ and $\mp$ must be considered in case of the MTM2$(A_{13,31}^{(1)})$, and the lower sign of $\pm$ and $\mp$ must be considered in case of the MTM2$(A_{13,31}^{(2)})$.

Eq.(\ref{Eq:MTM2_A1331_s12ss13s}) is equal to Eq.(\ref{Eq:MTM2_A1221_s12ss13s}). Thus, the simultaneous reproducibility of $\theta_{12}$ and $\theta_{13}$ is diminished using the MTM2($A_{13,31}^{(1)}$). By contrast, the simultaneous reproducibility of $\theta_{12}$ and $\theta_{13}$ is substantially enhanced using the MTM2($A_{13,31}^{(2)}$). Therefore, we only consider the MTM2($A_{13,31}^{(2)}$) for analyses in this study.

Figure \ref{Fig:MTM2_A1331_cosphi_13_23} depicts the predicted values of $\theta_{13}$ (upper panel) and $\theta_{23}$ (lower panel) with respect to $\phi$ based on the MTM2($A_{13,31}^{(2)}$). Similar to the results obtained using the MTM2($A_{12,21}^{(2)}$), Fig. \ref{Fig:MTM2_A1331_cosphi_13_23} indicates that the values of $\theta_{13}$ and $\theta_{23}$ that are consistent with the observed data can be obtained by choosing the appropriate values of $\phi$ and $\theta$. We confirmed this result based on our numerical parameter search. 

\color{black}
A benchmark point 
\begin{eqnarray}
(\alpha, \theta, \phi) = (0.99847, \ 10.31^\circ\  81.41^\circ),
\label{Eq:MTM2_A1331_benchmark}
\end{eqnarray}
yields the best-fit values of $s_{12}^2$ and $s_{13}^2$ and the allowed value of $s_{23}^2$ as follows:
\begin{eqnarray}
(s_{12}^2, s_{23}^2, s_{13}^2,\delta) = (0.303,\  0.514,\  0.02225, \  262.3^\circ). 
\end{eqnarray}

Figures \ref{Fig:MTM2_A1331_2_a1p1_12_13} - \ref{Fig:MTM2_A1331_2_R} show that the ballpark figures of the parameter space and the possible ranges of those mixing parameters. From these figures, we have the similar conclusions for MTM2($A_{13,31}^{(2)}$) as for  MTM1($A_{13,31}^{(2)}$) as follows,
\begin{itemize}
\item The wide range of parameters $\theta$, $\phi$ and $\alpha$ are allowed with observations (Figures \ref{Fig:MTM2_A1331_2_a1p1_12_13}, \ref{Fig:MTM2_A1331_2_ap_23} and \ref{Fig:MTM2_A1331_2_at_23}). The wide range of Dirac CP phase is also consistent with observation (Figures \ref{Fig:MTM2_A1331_2_ap_d} and \ref{Fig:MTM2_A1331_2_at_d}).
\item If the values of $\theta_{23}$ and the CP phase $\delta$ are finally pinned down, we can reproduce these fixed values with appropriate values of $\alpha$, $\theta$ and $\phi$.
\item If the best-fit values change in the future, the new best-fit values can be reproduced with appropriate selection of the values of $\alpha$, $\theta$, $\phi$ (Figure \ref{Fig:MTM2_A1331_2_R}).
\end{itemize}
%

\subsection{$Z_2$ symmetry breaking}
Similar to the results of modification introduced in the TM1 mixing scheme, modifications to the TM2 mixing scheme realized via the introduction of $A_{12,21}^{(2)}$ and $A_{13,31}^{(2)}$ can improve the simultaneous reproducibility of $\theta_{12}$ and $\theta_{13}$. However, the $Z_2$ symmetry related to the mass matrices is broken owing to the modifications introduced the mixing scheme.

With respect to the MTM2($A_{12,21}^{(2)}$), the magnitude of $Z_2$ symmetry breaking at the benchmark point shown in Eq.(\ref{Eq:MTM2_A1221_benchmark}) can be determined as follows: 
\begin{eqnarray}
 \delta_{\rm Z2}^{\rm MTM2}(A_{12,21}^{(2)})= \left(
\begin{matrix}
0.211 & 0.139 &   0.267 \\
*&  0.1201  & 0.0137 \\
 *  & *   & 0.0804   \\
\end{matrix}
\right),
\end{eqnarray}
where $(m_1, m_2, m_3) = (0, \Delta m_{21}^2, \Delta m_{31}^2)$ with the best-fit values of $\Delta m_{ij}^2$ in the case of NO. 

Similarly, for MTM2($A_{13,31}^{(2)}$), the magnitude of $Z_2$ symmetry breaking at the benchmark point shown in Eq.(\ref{Eq:MTM2_A1331_benchmark}) can be determined as follows:
\begin{eqnarray}
 \delta_{\rm Z2}^{\rm MTM1}(A_{13,31}^{(2)})= \left(
\begin{matrix}
0.208 & 0.261 &   0.134 \\
*&  0.0775  & 0.0136 \\
 *  & *   & 0.116   \\
\end{matrix}
\right).
\end{eqnarray}

The magnitude of maximum $Z_2$ symmetry breaking with respect to the MTM1($A_{12,21}^{(2)}$) and MTM1($A_{13,31}^{(2)}$) is ~$17\%$, while the magnitude of maximum $Z_2$ symmetry breaking with respect to the MTM2($A_{12,21}^{(2)}$) and MTM2($A_{13,31}^{(2)}$) is ~$27\%$. This is because the simultaneous reproducibility of $\theta_{12}$ and $\theta_{13}$ in the original TM1 scheme is larger than that in the original TM2 scheme. Hence, substantial modifications are necessary in the TM2 scheme to achieve the desired level of reproducibility.
\section{Summary\label{sec:summary}}
The two types of trimaximal mixing schemes, TM1 and TM2, are widely studied neutrino mixing schemes. The values of neutrino mixing angles predicted using TM1 and TM2 are consistent with the experimentally observed data. However, TM1 and TM2 cannot simultaneously predict the best-fit values of $\theta_{12}$ and $\theta_{13}$. Specifically, in the case of TM2, the predicted value of $\theta_{12}$ approaches the upper limit of the $3\sigma$ allowed region. Hence, the TM2 mixing scheme may soon be excluded from neutrino oscillation experiments.

Although simultaneous reproducibility of $\theta_{12}$ and $\theta_{13}$ cannot be realized using the TM1 and TM2 mixing schemes, these schemes are still widely studied because the flavor neutrino mass matrices in the TM1 and TM2 mixing schemes are invariant under the exact $Z_2$ symmetry. We have attempted to develop modified versions of the TM1 and TM2 mixing schemes that could maintain the exact $Z_2$ symmetry; however, these attempts have not yet been successful. In this study, we developed modified TM1 and TM2 mixing schemes that involve a certain degree of $Z_2$ symmetry breaking to realize the desired simultaneous reproducibility. 

We observed that unitarity of the modified mixing matrix $\tilde{U}$ could be maintained by modifying the original mixing matrix $U$ using $\tilde{U} = AU$, where $A$ has only one real parameter. The two successful modification are based on the following two matrices
\begin{eqnarray}
A_{12,21}^{(2)} 
= \left(
\begin{matrix}
\alpha & -\sqrt{1-\alpha^2} & 0  \\
\sqrt{1-\alpha^2} & \alpha  &0\\
0 & 0  & 1  \\
\end{matrix}
\right),
\end{eqnarray}
and
\begin{eqnarray}
A_{13,31}^{(2)} = \left(
\begin{matrix}
\alpha & 0&-\sqrt{1-\alpha^2}  \\
 0 & 1  &0\\
\sqrt{1-\alpha^2}& 0  & \alpha  \\
\end{matrix}
\right),
\end{eqnarray}
where $\alpha$ denotes the real parameter. We have shown that the modified TM1 and TM2 mixing matrices can simultaneously predict the best-fit values of $\theta_{12}$ and $\theta_{13}$ using $A_{12,21}^{(2)}$, with $\alpha=0.99973$, and $A_{13,31}^{(2)}$, with $\alpha=0.99847$.

\vspace{3mm}







\end{document}